\documentclass[11pt]{JHEP3}
\input epsf.tex
\epsfclipon

\UseRawInputEncoding
\usepackage{color}
\let\normalcolor\relax
\usepackage{epsfig}
\usepackage{epstopdf}
\usepackage{epsf}
\input{epsf.sty}
\usepackage{graphicx,amsmath,amssymb}
\usepackage{epsfig,multicol}
\usepackage{multirow}
\allowdisplaybreaks

\usepackage{bbm,bm,amsmath,amssymb}
 
\newcommand{\red}{\color{red}}
\usepackage[normalem]{ulem}

\usepackage{comment}
\def\be{\begin{equation}}
\def\ee{\end{equation}}
\def\figs/B{B}
\def\bea{\begin{eqnarray}}
\def\eea{\end{eqnarray}}
\def\bg{\begin{eqnarray}}
\def\nd{\end{eqnarray}}
\def\sin{{\rm sin}}
\def\cos{{\rm cos}}
\def\tan{{\rm tan}}
\def\log{{\rm log}}
\def\ln{{\rm log}}
\renewcommand{\d}{{\mathrm{d}}}

\renewcommand{\(}{\left(}
\renewcommand{\)}{\right)}

\def\H{\mathrm{H}}

\def\e{\mathrm{e}}
\def\be{\begin{equation}}
\def\ee{\end{equation}}

\def\doi{http://doi.org}

\def\d{\mathrm{d}}

\title{de Sitter Space as a Glauber-Sudarshan State}
\author{Suddhasattwa Brahma$^{1}$, Keshav Dasgupta$^{1}$,  Radu Tatar$^{2}$\\
	\vskip.03in
	${}^1$ Department of Physics, McGill University, Montr\'{e}al, Qu\'{e}bec, H3A 2T8, Canada \\
	${}^2$ Department of Mathematical Sciences,
University of Liverpool,  Liverpool, L69 7ZL, ~~United Kingdom \\
	{\tt suddhasattwa.brahma@gmail.com}
	
{\tt keshav@hep.physics.mcgill.ca, Radu.Tatar@Liverpool.ac.uk}}

\date{\today}

\maketitle

\abstract{Glauber-Sudarshan states, sometimes simply referred to as Glauber states, or alternatively as coherent and squeezed-coherent states, are interesting states in the configuration spaces of any quantum field theories, that closely resemble classical trajectories in space-time. In this paper, we identify four-dimensional de Sitter space as a coherent state over a supersymmetric Minkowski vacuum.  Although such an identification is not new, what is new however is the claim that this is realizable in full string theory, but only in conjunction with temporally varying degrees of freedom and quantum corrections resulting from them.  Furthermore, fluctuations over the de Sitter space is governed by a generalized graviton (and flux)-added coherent state, also known as the Agarwal-Tara state.
The realization of de Sitter space as a state, and \textit{not} as a vacuum, resolves many issues associated with its entropy, zero-point energy and trans-Planckian censorship, amongst other things.}

\begin{document}

\section{Introduction and summary}	
\label{sec:intro}

The path towards an actual realization of a four-dimensional de Sitter {\it vacuum} in string theory is long and arduous with obstacles coming, for example, from the no-go conditions with increasing sophistications.  The original no-go condition, given by Gibbons \cite{gibbons}, ruled out the possibility of realizing four-dimensional de Sitter vacuum with supergravity fluxes. This was followed by a more refined version from Maldacena and Nunez \cite{malnun}, that excluded branes and anti-branes. The conclusion was that neither branes nor anti-branes can provide positive cosmological constant solutions, although anti-branes could break supersymmetry spontaneously. More recently, however, it was proposed in \cite{nogo} that even the O-planes {\it cannot} help, so eventually the sole burden of realizing four-dimensional de Sitter vacua in string theory rested on the shoulder of the quantum terms. Unfortunately, as also shown in \cite{nogo, togo}, this is easier said than done: the quantum terms are more delicate as they are constrained by a condition that seemed to be satisfied only if there existed an inherent {\it hierarchy}. In the analysis of \cite{nogo}, which was done from M-theory point of view, the hierarchy could be between $g_s$, the type IIA string coupling, and ${\rm M}_p$, the Planck scale. Generically, as in any quantum system, one would expect that at least the ${\rm M}_p$ hierarchy could be easily attained as supergravity operators typically have different ${\rm M}_p$ scalings. All in all, the belief till early 2018 was that, quantum corrections could in principle do the job. Subtleties like fine-tunings etc should go hand in hand, but at the deepest level, there would be no further obstructions that could prohibit four-dimensional de Sitter vacua to exist in string theory. With a little stretch, solutions like \cite{kklt}, should then be part of the ensemble of models that demonstrate the existence of stringy de Sitter vacua.

Such a rosy picture did not last long as objections were raised from many directions. Early objections, specifically regarding \cite{kklt}, were either aimed mostly at the usage of anti-D3 branes  \cite{bena, westphal}, or at the choice of the supersymmetric AdS vacua \cite{sethi}, in the constructions of stringy IIB de Sitter vacua. The former objections were shown to not exist if anti-D3 branes (in the presence of O3-planes) broke supersymmetry spontaneously \cite{kalloshevan}, and, in fact, in \cite{HamadaS, massprod} numerous constructions of de Sitter vacua in IIA were shown using anti-branes (plus O-planes and quantum corrections \cite{HamadaS}) as the primary ingredients.

The latter objections, primarily from \cite{sethi}, were more subtle. It questioned the fundamental edifice on which the whole construction of \cite{kklt} rested, namely the four-dimensional supersymmetric AdS vacua. The claim in \cite{sethi} was that an AdS vacuum simply could not be a good starting point for uplifting to a non-supersymmetric de Sitter vacuum. Subtleties come from the quantum terms, mostly involving the quartic order curvature terms, that  typically lead to rolling solutions. Thus backgrounds involving no AdS vacua, somewhat along the lines of \cite{linde1}, might be more productive.

The situation by mid-2018 was then an atmosphere of hopeful optimism. While the objections against the existence of  four-dimensional de Sitter vacua from string theory were acknowledged \cite{String_No_dS}, the way out of these did not seem hard either. Again with a little extra fine tunings, and some minor adjustments, the construction of \cite{kklt} still seemed to hold water.

A more severe blow to this construction, and in particular to {\it any} constructions that aimed to reproduce a positive cosmological solution from string theory, appeared from a series of papers starting roughly from mid-June of 2018 \cite{vafa1, vafa2, vafa22}. These set of papers challenged the very existence of a positive cosmological constant solution in a consistent theory of quantum gravity citing contradictions coming from the weak gravity conjecture \cite{motl} as well as from bounds on derivatives of the potential on the landscape \cite{vafa1} and other related issues. This was followed by a slew of papers, a subset of which is cited in \cite{swampland} (see \cite{Review_Swampland} for reviews), that either provided support or challenged these ideas. The latter mainly questioned the ad hoc nature of the conjectures in \cite{vafa1, vafa2, vafa22} and pointed out issues where the derivative bounds were too stringent to match with the experimental values. Questions were also raised, for example in \cite{trivkachru}, against the non-existence of effective field theory (EFT) descriptions propagated by \cite{vafa1, vafa2, vafa22}, although in retrospect it is now clear from \cite{togo, desitter2, desitter3, desitter4} that in the {\it absence} of time-dependent degrees of freedom, a four-dimensional EFT description fails to arise. This does lend some credence to the swampland conjectures of \cite{vafa1, vafa2, vafa22}, but the works of \cite{togo, desitter2, desitter3, desitter4} approached the issue of the existence of an EFT following a very different path by actually analyzing the infinite class of interactions, both local and non-local, in a systematic way. In this paper, and especially in sections \ref{sec3.1} and \ref{sec3.2}, we will analyze the non-perturbative effects from instantons and world-volume fermions that were not discussed in \cite{togo, desitter2, desitter3, desitter4}. In the absence of time-dependent degrees of freedom, all our computations lead to one conclusion: the loss of both $g_s$ and ${\rm M}_p$ hierarchies from the quantum loops, implying a severe breakdown of an EFT description in four-dimensions. 

The ad hoc nature of the conjectures in \cite{vafa1, vafa2, vafa22} were soon given a slightly better theoretical motivation in \cite{tcc}. The aim of \cite{tcc} was basically to blend the trans-Planckian issue raised in \cite{martin} with the swampland conjectures of \cite{vafa1, vafa2, vafa22}, and rechristen them in a packaged form as the Trans-Planckian Censorship Conjecture (TCC). The trans-Planckian issue of cosmology, especially in the presence of accelerating backgrounds (de Sitter space being the prime example), challenges the very notion of Wilsonian effective action because any fluctuations over these backgrounds create modes that have time-{\it dependent} frequencies. Since frequencies are related to the energies, and Wilsonian method involve {\it integrating out} the high energy modes, we face a severe conundrum: how to integrate out high energy modes when the energies themselves are changing with respect to time? One way would be to integrate out the modes at every instant of time. This is not a very efficient way, because the integrated out UV modes will become IR at a later time, leading to a {\it waterfall} like structure where the depleted IR modes get continually replaced by the red-shifted UV modes. Additionally, in a theory of gravity the far UV modes, {\it i.e.} the modes beyond ${\rm M}_p$, are not well defined because there is no UV completion like we have in string theory. These modes will show up at a later time, thus bringing in the UV issue now at low energies, unless we have a way to {\it censor} these modes. Such censoring will rule out inflation, or at least sufficiently long-lasting inflation to avoid fine-tuning issues, along with all the advantages that we have assimilated through inflationary dynamics.  

The trans-Planckian problem \cite{martin} or it's new guise, the TCC \cite{tcc}, relies on two things: one, the time-dependent frequencies in an accelerating background, and two, the far UV modes that do not have well-defined dynamics (\textit{i.e.} they could even be non-unitary). However if we view string theory as the UV completion of gravity (and other interactions), there appears no basis to the second point as the UV modes clearly have well-defined unitary dynamics! The first point however is still a concern: the modes do change with respect to time, so Wilsonian method will still be hard to perform in any accelerating background. Is there a way out of this? 

It turns out, if we view de Sitter space to be a {\it state} instead of a {\it vacuum}, then there is a way out. Such a state should, at best, replace the classical configuration to something that closely resembles it. This means what we are looking for is a Glauber-Sudarshan state \cite{sudarshan, glauber}, alternatively, known as a coherent state \cite{schrodinger}. This identification is not entirely new as the {\it first} construction of de Sitter space as a coherent state in four-dimensional quantum gravity has appeared in \cite{dvaliG} (see also \cite{guistia}), but what is {\it new} here is the claim that such a state may be realized in full string theory and is quantum mechanically {\it stable}. The stability is crucial as there are literally an infinite number of possible local and non-local, including their perturbative, non-perturbative and topological, quantum corrections. One of our aims here is to justify the stability of the Glauber-Sudarshan state amidst all these corrections.    

Viewing de Sitter space as a {\it state} instead of a {\it vacuum} resolves other issues related to zero point energy, supersymmetry breaking, entropy etc that we will concentrate on here. The first two are easy to understand. The zero point energies from the bosonic and the fermionic degrees of freedom {\it cancel} once we take a supersymmetric warped-Minkowski background. Once supersymmetry is broken spontaneously by the Glauber-Sudarshan state, the cosmological constant $\Lambda$ is determined from the fluxes and the quantum corrections, with no contributions from the zero-point energies \cite{desitter2}. The spontaneous breaking of supersymmetry arises from the expectation values of the fluxes $-$ in this case G-fluxes in the M-theory uplift of the IIB model $-$ over the eight-manifold when they are no longer self-dual. These fluxes are self-dual over the vacuum eight-manifold, so it's the expectation values that break supersymmetry. 

The issue of entropy is bit more non-trivial, and we shall elaborate on this in section \ref{sec4.3}. The important question here is the reason for a {\it finite} entropy of the de Sitter space, when a Minkowski space generically has an infinite entropy.  Since we define our de Sitter space as a Glauber-Sudarshan state over a warped-Minkowski background, should this not be a concern now? The answer has to do with the interacting Hamiltonian, as well as the finiteness of the number of gravitons in the Glauber-Sudarshan state defined over some Hubble patch; including other criteria that we shall elaborate in section \ref{sec4.3}. 

There are also other related issues that enter the very definition of the interacting Hamiltonian that is useful in resolving the entropy puzzle, and they have to do with the infinite collections of perturbative and non-perturbative corrections. These corrections appear as series, and so convergence of the series is important. We will discuss them in section \ref{sec3.2}.

\subsection{Why getting de Sitter space is a hard problem? \label{hard}}

Getting de Sitter is a hard problem not just because of the no-go conditions \cite{gibbons, malnun, nogo}, or because of the swampland constraints \cite{vafa1, vafa2, vafa22} $-$ the former can be easily overcome by taking quantum corrections, and the latter by switching on time-dependent degrees of freedom or by viewing de Sitter as a Glauber-Sudarshan state $-$ but because of the fact that the analysis to {\it show} the existence of a de Sitter space, either as a state or by solving the Schwinger-Dyson's equations (see details in section \ref{sec3.3}), is technically challenging. Why is that so? 

The reason is not hard to see. Imagine we want to express \eqref{betta3}, or more appropriately the M-theory uplifted background \eqref{vegamey3}, as a consequence of the Glauber-Sudarshan state in say M-theory (what this means will be elaborated later. Here we simply sketch the picture). A background like \eqref{vegamey3} can exist if it solves some equations of motion (EOMs). These EOMs turn out be the Schwinger-Dyson's equations, which are like the Ehrenfest's equations in quantum mechanics, meaning that they appear as EOMs for {\it expectation values}. Thus M-theory equations appear rather surprisingly as Schwinger-Dyson's equations (details are in section \ref{sec3.3}).  Such simple-minded statement entails some important consequences that have been largely ignored in the literature. In the following, we list some of these consequences.

The first and the foremost of them is to solve the time-dependent EOMs, either as supergravity EOMs in the presence of all the quantum corrections mentioned above, or as a consequence of the Schwinger-Dyson's equations. This, by itself, is challenging because, unless we have a way to control the infinite set of quantum terms, there is no simple way to express them. The EOMs however can only provide  a local picture, but the existence of a solution, or even the Glauber-Sudarshan state, relies heavily on global constraints too. The global constraints come from flux quantizations, anomaly cancellations etc, and therefore we have to (a) not only solve the time-dependent EOMs, but also (b) explain how fluxes remain quantized with time-dependences, (c) how anomaly cancellations work, (d) how moduli stabilization may be understood when the moduli themselves are varying with time, (e) how the no-go conditions are satisfied, (f) how the null, weak and the strong energy conditions are overcome, (g) how the generic perturbative corrections may be analyzed, (h) how the non-perturbative corrections may be analyzed, (i) how the non-local quantum terms may be understood, (j) how the four-dimensional Newton's constant may be kept 
constant\footnote{An interesting related question is whether the four-dimensional Newton's constant gets {\it renormalized}. What is easy to infer is that the four-dimensional Newton's constant remains time-independent, but it's renormalization (or it's {\it running}) solely depends on the effective action whose perturbative and the non-perturbative parts at a given {\it scale}, as discussed in sections \ref{sec3.1} and \ref{sec3.2}, will be elaborated in section \ref{sec3.3}.
Somewhat intriguingly, as we shall discuss in section \ref{sec3.3}, the internal metric components {\it do not} receive corrections to any orders in ${g_s^a\over {\rm M}_p^b}$ for appropriate choice of the Glauber-Sudarshan state. \label{newton}}, 
(k) how the {\it positive} cosmological constant may be generated by quantum corrections, (l) how the zero point energy gets renormalized in a  non-supersymmetric background, (m) how the geometry and the topology of the internal {\it compact} space, which is now a highly non-K\"ahler manifold, may be expressed (n) how the Bianchi identities are satisfied in a time-varying background, (o) how the swampland criteria are averted, (p) how the early-time physics should be understood, (q) how the inflationary paradigm may be recovered from our analysis, (r) how other related solutions like Kasner de Sitter or dipole-deformed de Sitter could be studied, etc.; all in a top-down (not bottom up!) string theory set-up. 

The situation is further complicated by the fact that one needs to solve almost {\it all} of the above problems to justify consistency of our background either as a supergravity solution or as a Glauber-Sudarshan state. There doesn't exist a simple solution that only answers parts of the above set of questions, because then it will not lead to a well-defined solution to the system. This {\it all-or-none} criterion makes the finding of de Sitter space in string theory a really {\it hard} problem. In our earlier works \cite{desitter2, desitter3, desitter4}, we have managed to answer most of the essential questions, so here we re-interpret all of them as a consequence of being a Glauber-Sudarshan state. This re-interpretation turns out to be a completely different beast in the sense that, as the reader herself or himself will find out, a complete re-evaluation of the scenario is called for because a new set of rules needs to be laid out and a new set of computations needs to be performed. In our opinion, these have hitherto never been attempted, so the analysis will naturally get involved. These computations are essential to understand the full picture, or at least to appreciate the consistency of the full framework, but the take-home message is surprisingly simple: viewing de Sitter as a Glauber-Sudarshan state and {\it not} as a vacuum,  alleviates most of the problems associated to entropy, TCC, stability etc. 
 
Let us concentrate on one such point from above, namely (d), {\it i.e.} how moduli stabilization may be understood when we expect the moduli themselves to vary with respect to time. Other points related to the non-perturbative corrections and the EOMs emanating from the Schwinger-Dyson's equations will be explained later as we progress in the text. The remaining points, related to anomaly cancellations, flux quantizations etc., have already been discussed in details in \cite{desitter2} so we will not dwell on them here.  

The reason for singling out (d) as opposed to the other points is because of its subtlety. Moduli stabilization requires a more careful analysis here because of the underlying Dine-Seiberg runaway problem \cite{dines}. Dine and Seiberg said that, once string vacua are left with unfixed moduli, they {\it decompactify} and quickly go to strong coupling. However, once moduli are fixed, vacua could be easily studied using perturbative string theory.  However, note that all of these discussions, as presented in \cite{dines}, are for time-independent compactifications. Question is, how does this translate into the case when there are time dependences?

In our case, as we mentioned in (d) above, once we fix the moduli at {\it every instant} of time, the Dine-Seiberg runaway can be stopped. The fixing of the moduli works in the following way. Starting with the solitonic vacuum configuration, as in \eqref{betbab3} with the corresponding G-flux components to support it, the non-trivial quantum corrections captured by ${\bf H}_{\rm int}$, as described in section \ref{sec2.4} onwards,  the moduli get fixed without breaking supersymmetry. This time-independent configuration with no running moduli forms our supersymmetric vacuum configuration on which we study the fluctuations. The vacuum however is not a {\it free} vacuum: it's an {\it interacting} vacuum from which we can construct our Glauber-Sudarshan state (exactly how this is done will form the basis of sections \ref{sec2.4} and \ref{sec2.5}). Expectation values of the metric and the G-flux components will govern the behavior of the moduli for the de Sitter case. We will call this the {\it dynamical moduli stabilization}. 

Interestingly,  what we find in the time-independent case is that to order ${g_s^0 \over {\rm M}_p^0}$, {\it i.e.} to the zeroth orders in $g_s$ (string coupling) and ${\rm M}_p$ (Planck mass), although the Dine-Seiberg runaway is apparently stopped, there are still an {\it infinite} number of operators with no hierarchy when the internal degrees of freedom are time-independent. This clearly show us that there are no solutions to the EOMs and the vacua do not exist. Note that this happens to {\it any} orders in ${g_s^a\over {\rm M}_p^b}$, and in particular for small $g_s$, so is not a strong-coupling question {\it at all}!

Thus the difficulty in generating a de Sitter solution lies not just on overcoming the no-go and the swampland constraints, but also on the various technical challenges  that we encounter along the path towards an actual realization of a background data either as expectation values or as supergravity solutions. However as our earlier works \cite{togo, desitter2, desitter3, desitter4} and the present paper will justify, this is not an insurmountable problem. Solutions do exist and, with some efforts, may be determined precisely.

\subsection{An \`etude on solitons, fluctuations and quantizations \label{etude}} 

Let us start with a simple example from quantum mechanics of a potential $V(x)$ in $1+1$ dimensions that has at least one local minimum at $x = a$. The potential, near the vicinity of $x = a$, may be represented in the following standard way:
\bg\label{pot1}
V(x) = V(a) + {1\over 2} \omega^2 (x - a)^2  + \sum_{n = 3}^\infty{1\over n!} \lambda_n (x - a)^n, \nd
where $\lambda_n$ characterize the anharmonic terms. For $\lambda_n$ sufficiently small   the low lying 
eigenstates near $x = a$ will satisfy the consistency conditions:
\bg\label{manara}
\lambda_n \langle (x - a)^n \rangle << \omega^2 \langle (x - a)^2\rangle, \nd
with $n = 3, 4, ..$, such that the wave-functions are that of a simple harmonic oscillator. The details are rather well-known so we will avoid repeating them here, except to point out that this simple analysis will form the basis of our construction in full IIB supergravity in section \ref{sec2}. The $x = a$ point herein will form the {\it solitonic} vacuum for us. 

The simple analysis however hides a subtlety that is typically not visible from quantum mechanics. Assuming the condition \eqref{manara} to hold, {\it i.e.} we can ignore higher order terms, one would naively think that \eqref{pot1} is simple harmonic potential for a {\it free} theory. This is not correct: the simple harmonic oscillator term can in fact  hide an infinite number of interactions with the soliton itself, as it becomes clear if we go to the field theory case in the presence of an interacting Hamiltonian ${\bf H}_{\rm int}$.  

We made a brief mention of the interacting Hamiltonian ${\bf H}_{\rm int}$ in section \ref{hard},  but it is an absolutely essential quantity to even construct the solitonic vacuum \eqref{betbab3} and fluctuations over it. All of these will be elaborated as we go along, but here we would like to construct a toy example in quantum field theory that captures some of the salient features of 
the actual construction in a simpler fashion.  

The basics of the toy example resides in the soliton physics, and there are many excellent textbook 
treatments on the subject $-$ for example \cite{rajaraman} $-$ but here we will generalize this a bit to 
capture the influence of the interacting Hamiltonian. To avoid conflicting with the Derrick's theorem 
\cite{derrick}, we will concentrate only in $1+1$ dimensions with a single scalar field $\varphi(x)$ where 
$x = ({\bf x}, t)$. Let us assume that the potential for the scalar field is:

{\footnotesize
\bg\label{lib1}
V(\varphi) & = & \sum_{n, m} {C}_{nm} \{\partial^n\} \varphi^m \equiv \sum_{n, k, m} {C}^{(k)}_{nm} \varphi^n 
\left(\partial^k \varphi\right)^m \\
& = & \sum_m {C}_{0m} \varphi^m + \sum_{n, m} {C}^{(1)}_{nm} \varphi^n \left(\partial \varphi\right)^m
+ \sum_{n, m} {C}^{(2)}_{nm} \varphi^n \left(\partial^2 \varphi\right)^m 
+ \sum_{n, m} {C}^{(3)}_{nm} \varphi^n \left(\partial^3 \varphi\right)^m + ..., \nonumber \nd}
which is basically a very simplified version of \eqref{selahran} that we shall encounter in section 
\ref{sec3.1}. In constructing \eqref{lib1} we have ignored both time derivatives and supersymmetry 
as neither are very essential to understand the dynamics here. The derivative terms are all suppressed by 
the coupling constants $C^{(k)}_{nm}$ which, in the language of \eqref{selahran}, are proportional to 
${g_s^{|a|}\over {\rm M}_p^b}$. Here of course we will simply assume that $C^{(k)}_{nm} << 1$ and not worry about either $g_s$ or ${\rm M}_p$. We will also assume that there is a solitonic vacuum given by $\varphi = \varphi_0({\bf x})$. 

Existence of the solitonic vacuum\footnote{Note that the M-theory background \eqref{betbab3}, or it's IIB counterpart, is a soliton in only a loose sense. It is of course clear that there does exists a vacuum configuration of the form ${\bf R}^{2, 1} \times \mathbb{T}^8$ with no G-flux, so \eqref{betbab3}, with it's corresponding G-flux, forms the nearby vacuum configuration with non-zero energy. \label{soldiff}} 
implies that we are looking at the minima of the total potential, which is the combination of the potential \eqref{lib1} and the contribution from the kinetic term. The fluctuation over the solitonic vacuum can be represented as $\eta({\bf x}, t)$ which is a function of both space and time. In the presence of the fluctuation, we can express the field as:
\bg\label{lib2}
\varphi({\bf x}, t) &=& \varphi_0({\bf x}) + \eta({\bf x}, t) \nonumber\\
 &\equiv& \varphi_0({\bf x}) 
+ \int d{\bf k} ~f_{\bf k}(t) \psi_{\bf k}({\bf x}) \approx \varphi_0({\bf x}) 
+ \sum_{\bf k} f_{\bf k}(t) \psi_{\bf k}({\bf x}), \nd  
where in the last equality we have assumed discrete momenta, and $f_{\bf k}(t)$ is generic amplitude not necessarily on-shell. Note that the integral is over $d{\bf k}$ and not over $d^2k = d{\bf k} dk_0$, which is of course what we expect. Question is whether we can determine the function $\psi_{\bf k}({\bf x})$. It turns out that the function $\psi_{\bf k}({\bf x})$ satisfies the following 
Schr\"odinger equation: 

{\footnotesize
\bg\label{lib3}
\left[-\nabla^2 + \sum_m C_{0m} {}^m{\rm C}_2\varphi_0^{m-2} + \sum_{n,m,p} C^{(p)}_{nm}  \varphi^n_0
\left(\partial^p  \varphi_0\right)^m \left({{}^n{\rm C}_2\over  \varphi^2_0} + {nm \partial^p \over \varphi_0
\partial^p \varphi_0} 
- {{}^m{\rm C}_2(\partial^p)^2 \over (\partial^n \varphi_0)^2}\right)\right]\psi_{\bf k}({\bf x}) =
\omega^2_{\bf k}~ \psi_{\bf k}({\bf x}), \nonumber\\ \nd}
with eigenvalue $\omega_{\bf k}^2$ and $p \ge 1$. We have used a flat metric, and the powers of the derivatives, i.e $\partial^p$, are only along ${\bf x}$, so there should be no confusion. A missing factor of 2 
in the first term of \eqref{lib3}, to allow $-{1\over 2} \nabla^2$ so that \eqref{lib3} may indeed look like a 
Schr\"odinger equation, can be easily inserted in by a simple redefinition of the ${\bf x}$ coordinate. This is a standard manipulation (see \cite{rajaraman}), so we will not worry about it and call \eqref{lib3} simply as the Schr\"odinger equation with a highly non-trivial potential. 

The {\it potential} that enters the Schr\"odinger equation is not the potential in \eqref{lib1}, although it is related to it, but the crucial question is how do the eigenstates $\psi_{\bf k}$ behave in this potential. Before we discuss this, we should point out that, plugging \eqref{lib3} in the $1+1$ dimensional field theory action immediately reproduces the standard simple harmonic oscillator action in the following way:
\bg\label{lib4}
{\bf S} = {1\over 2} \int dt d{\bf k} \Bigg(\Big\vert \dot{f}_{\bf k}(t)\Big\vert^2 - \omega^2_{\bf k} \Big\vert f_{\bf k}(t)\Big\vert^2 -  V(\varphi_0) + ....\Bigg), \nd
where the dotted terms are ${\cal O}\left(\vert f_{\bf k}\vert^3\right)$ interactions. In this form the potential part of the action matches precisely with the quantum mechanical result from \eqref{pot1}, but the difference should be clear: \eqref{lib4} is derived from an highly interacting theory whereas \eqref{pot1} is a simple quantum mechanical result. Despite that we will call the vacuum from \eqref{lib4} as the {\it free} vacuum (or sometime the {\it harmonic} vacuum) to distinguish it from the {\it interacting} vacuum to be discussed later.

The eigenstates $\psi_{\bf k}({\bf x})$ and the eigenvalues $\omega^2_{\bf k}$ are important because they will decide the subsequent behavior of the $1+1$ dimensional field theory. Typically eigenstates of a Schr\"odinger equation are divided into three categories: (a) zero modes with $\omega^2_{\bf k} = 0$, (b) discrete levels with $\omega^2_{\bf k}$ given by a set of discrete integers, and (c) continuum levels where
$\omega^2_{\bf k}$ is related to ${\bf k}^2$ by an {\it on shell} condition. The latter is what we want, but we need to worry about the zero modes and the discrete states. What do they mean here?

In standard solitonic theory, the zero modes are related to the motion of the soliton itself, and they are typically used to {\it quantize} the soliton. For our case, the soliton will be related to the vacuum metric configuration \eqref{betbab3} and the zero modes should appear as the translation or the rotational modes of the internal metric that {\it do not change the energy of the system}. Do they exist? For the simple scalar field theory case one can easily show that if the solitonic solution $\varphi_0({\bf x})$ belongs to an equipotential curve 
consisting of the family of mutually translated solitons $\varphi_0({\bf x} - a)$, then there does appear one zero mode $\psi_0({\bf x})$ that takes the following form:
\bg\label{lib5}
\psi_0({\bf x}) = {\partial \varphi_0({\bf x}) \over \partial {\bf x}}, \nd
which should solve the equation \eqref{lib3} with $\omega^2_{\bf k} = 0$. For the case we want to concentrate on, i.e the metric \eqref{betbab3} with the corresponding G-fluxes, one needs to check whether such zero modes can appear. There are also the K\"ahler and the complex structure moduli of the internal metric which tell us that we can {\it change} the metric without costing any energy to the system. 
These moduli are governed by a Lichnerowicz type of equation (now in the presence of metric, fluxes and quantum corrections). Fluctuations over these metric (and flux) configurations would now have their corresponding Schr\"odinger equations like \eqref{lib3}. All these class of equations should be related to each other and therefore one could associate a combination of the zero modes of these Schr\"odinger equations to these moduli themselves.  Clearly this will make the system much more complicated, so to avoid this as well as the Dine-Seiberg runaway \cite{dines}, we want the moduli to be fixed.  The moduli are fixed in the presence of quantum terms, and therefore we expect those zero modes that correspond to the K\"ahler and the complex structure moduli to 
{\it not} arise in the presence of sufficient number of quantum terms in \eqref{lib3}. For a generic potential like the one that appears in \eqref{lib3} this may be hard to show, but there are alternative ways (see for example \cite{DRS, GKP})  to justify this (more on this a bit later). 

The {\it discrete} modes correspond to the bound state of the mesons, which are basically the fluctuations of the scalar field,  with the soliton. These mesons get trapped inside the potential of the soliton and the discrete states show that the probability amplitudes peak near the centre of the soliton. For the background \eqref{betbab3}, again there is no such simple interpretation because we will treat the fluctuations separately and not consider them as getting bound by the soliton. Of course these differences arise because the classical background \eqref{betbab3} and the corresponding G-flux shares many similarities with a soliton, but is a soliton only in a loose sense (see footnote \ref{soldiff}).

All our above discussion points out that it is the continuum level of the Schr\"odinger equation \eqref{lib3} that concerns us here, although we will interpret these wave-functions as the fluctuation wave-functions and not scattering states with the soliton. Again the slight difference in interpretation stems from our loose identification of \eqref{betbab3} to an actual soliton. The set of the continuum wave-functions 
$\psi_{\bf k}({\bf x})$ takes the following form:
\bg\label{lib6}
&&~~\psi_{\bf k}({\bf x}) \equiv {\rm exp}\left(i{\bf kx}\right) {\rm G}_{\bf k}({\bf x})\nonumber\\
&&\lim_{{\bf x} \to \pm\infty} \psi_{\bf k}({\bf x}) = {\rm exp}\left[i{\bf kx} \pm i \sigma({\bf k})\right], \nd
where ${\rm G}_{\bf k}({\bf x})$ is a non-trivial complex function of $({\bf k}, {\bf x})$ which for large ${\bf x}$ 
approximates to a phase $\sigma({\bf k})$. In this sense the continuum level resembles somewhat the scattering states. The analysis for the actual case with \eqref{betbab3} will be much more involved, but fortunately, as we shall see, we will not have to determine the functional forms for $\psi_{\bf k}({\bf x})$ explicitly.

\subsection{Organization and a brief summary of the paper \label{sumary}}

The paper is organized in the following way. Broadly, section \ref{sec2} studies the background from solitonic point of view, {\it i.e.} from the M-theory uplift of the supersymmetric warped Minkowski background 
in \eqref{betbab3}; whereas sections \ref{sec3.1} and \ref{sec3.2} studies the same directly using 
\eqref{vegamey3}, which is the M-theory uplift of the IIB de Sitter background of \eqref{betta3}. In section 
\ref{sec3.3} we derive the de Sitter results in two ways, one, by taking expectation values over the Glauber-Sudarshan state and two, by solving the Schwinger-Dyson's equations. Thus section \ref{sec3.3} serves as a culmination and synthesis of the results accumulated from sections \ref{sec2}, \ref{sec3.1} and \ref{sec3.2}.  
Section \ref{sec4} serves as a vantage point to analyze many of the important properties of de Sitter space from the point of view of the Glauber-Sudarshan state, namely trans-Planckian censorship, quantum swampland and de Sitter entropy. 

Let us now go to a more detailed survey of various sections of the paper. In section \ref{sec2.1} we lay out the formalism of the Glauber-Sudarshan state and discuss how one should construct it using the 
momentum modes of the theory. In particular we discuss the precise wave-function of the Glauber-Sudarshan state using configuration space variables associated with the metric components of the solitonic 
vacuum.  Other properties, like the Schr\"odinger wave-functions, oscillatory behavior, and the study of the zero modes are all discussed here.

Section \ref{sec2.2} studies the fluctuations over a de Sitter background directly from our solitonic configuration. Since the de Sitter space itself is a state over the solitonic background, the fluctuations should also appear from a corresponding state. In this section we discuss how such state should be constructed 
by combining the various oscillatory states over the solitonic vacuum. We also discuss how the fluctuations over a de Sitter {\it vacuum} could be reinterpreted as an artifact of Fourier transform over the solitonic vacuum. This means the time-dependent frequencies that we see from the fluctuations over a de Sitter {\it vacuum} is a consequence of the linear combinations of the modes over the solitonic vacuum. This provides
not only an answer to the trans-Planckian issue because the frequencies over  the solitonic vacuum are time-independent, but also provides a way to tackle the Wilsonian integration. 

In section \ref{sec2.3} we identify the state that defines the fluctuations over de Sitter space $-$ viewed as a Glauber-Sudarshan state $-$ as the Agarwal-Tara \cite{agarwal} state. The Agarwal-Tara state, sometime also called as the Agarwal state, is constructed by adding gravitons to the Glauber-Sudarshan state and is therefore popularly known as the graviton added coherent state (GACS). We construct a generic operator, controlled by a coupling parameter, that when acting on the Glauber-Sudarshan state creates the necessary Agarwal-Tara state.   

So far we have used the {\it free} (or the {\it harmonic}) vacuum to construct the Glauber-Sudarshan state. This cannot quite be the full picture because there is no free vacuum in an interacting theory like M-theory. Thus we have to construct our Glauber-Sudarshan from an {\it interacting} vacuum $\vert \Omega\rangle$.  
Such a construction is laid out in section \ref{sec2.4}, we we start by first shifting the interacting vacuum using a displacement operator. The construction of the displacement operator is in itself a non-trivial exercise because of the interactions that mix all the momentum modes of the metric and the flux components. Expectation values of the metric operators over such shifted interacting vacuum, which we call as the {\it generalized Glauber-Sudarshan} state, are carried out using the path integral formalism. This is elaborated in section \ref{sec2.5}. The path integral formalism is particularly useful in analyzing the expectation values and we show that we can reproduce the expected metric configuration in \eqref{vegamey3} from there, up to 
${\cal O}\left({g_s^{|a|}\over {\rm M}_p^b}\right)$ corrections. These corrections are sub-leading, and their presence is because of our choice of the generic form of the displacement operator. There does exist a specific choice of the displacement operator that reproduces \eqref{vegamey3} as expectation value precisely, without any extra corrections. To develop that requires more preparation, and we postpone it till section \ref{sec3.3}.  

The path integral formalism is also powerful to study the expectation values of the metric components over the Agarwal-Tara state, viewed as an operator acting on the generalized Glauber-Sudarshan state. The answer that we get clearly shows not only that we can reproduce the fluctuation spectrum over a de Sitter space, but also the fact that the fluctuation spectrum is indeed an artifact of the Fourier transform over our solitonic vacuum. 

What we haven't tackled so far is the contributions from the G-flux components. This is discussed in 
section \ref{sec2.6}. Although these contributions make the system a bit more complicated, their presence is necessary for the self-consistency of the Glauber-Sudarshan state. For example they help us to understand how supersymmetry is broken spontaneously by the state while the solitonic vacuum remains perfectly supersymmetric.  The new modes that are created by fluctuations over the G-flux background in the solitonic vacuum now mix non-trivially with the modes from the metric components. This leads to complicated interactions and therefore the stability of the Glauber-Sudarshan state becomes an issue. 

These interactions are the main focus of sections \ref{sec3.1} and \ref{sec3.2}, and the question of the stability of the Glauber-Sudarshan state is finally resolved in section \ref{sec3.3}. To study the interactions we change the gear a bit by using the background \eqref{vegamey3} directly instead of going via the configuration \eqref{betbab3}. There is a definite advantage of using such a procedure as will become clear from the computations in sections \ref{sec3.1} and \ref{sec3.2}.

The quantum corrections are a bit non-trivial to deal with because there are an {\it infinite} number of possible local and non-local $-$ that include the perturbative, non-perturbative and topological $-$ interactions. In section \ref{sec3.1} we classify the perturbative contributions using the formalism developed earlier in \cite{desitter2, desitter3}. The non-perturbative corrections have not been discussed before, and we detail them in section \ref{sec3.2}. These corrections generically come from the branes and the instantons, and we show that certain aspects of these corrections may be extracted from the non-local interactions discussed in \cite{desitter2, desitter3}.   

The non-perturbative contributions from the instantons come from both M2 and M5-instantons, and we show that it's only the M5-instantons contribute here. The M5-instantons' contributions are further classified by the 
BBS \cite{bbs} and KKLT \cite{kklt} type instantons. Their contributions are discussed in sections 
\ref{sec3.2.1} and \ref{sec3.2.3} respectively. The contributions from the seven-branes and in particular the {\it fermionic} terms on the seven-branes are discussed in details in section \ref{sec3.2.2}. 

All these perturbative and non-perturbative quantum corrections contribute to the equations of motion (EOMs). In section \ref{sec3.3} we show that these EOMs appear from a class of Schwinger-Dyson's equations (SDEs) \cite{dyson}. Interestingly, the SDEs for our case split into two sets of equations. One set of equations are completely expressed in terms of expectation values over the solitonic vacuum. Since, as we discussed in sections \ref{sec2.4} and \ref{sec2.5}, the expectation values of the metric and the G-flux components over the solitonic configuration \eqref{betbab3} reproduces the background \eqref{vegamey3}, along with it's G-flux components from section \ref{sec2.6}, these SDEs give rise to the M-theory EOMs for the background \eqref{vegamey3}! This immediately implies that all the computations that we did in 
sections \ref{sec3.1} and \ref{sec3.2} appear now as quantum contributions to the SDEs. 

There is also a second set of SDEs that relate the Faddeev-Popov ghosts, the displacement operator 
and the expectation values of the variations of the {\it total action} with respect to the field variables. These SDEs are in general hard to solve and we leave them for future works. 
In section \ref{sec3.4} we give a brief discussion on supersymmetry breaking and other related effects.  One of the important question that we tackle in this section is the connection between the supersymmetry breaking scale and the cosmological constant $\Lambda$. We show that in general there is a large hierarchy between them, implying that the supersymmetry breaking scale could be large yet $\Lambda$ could remain relatively small. Of course the exact value of $\Lambda$ would depend on the values of the fluxes and the quantum terms, which would only be determined if we solve all the SDEs exactly, but what we argue here is that the precise form, which may be extracted from our earlier work \cite{desitter2}, appears to have no contributions from the ground state energies of the bosonic and the fermionic degrees of freedom.
We also elaborate on the moduli stabilization and discuss what happens when we go to the strong coupling limit of type IIB.

Once we construct our de Sitter solution using the Glauber-Sudarshan state, we examine some of its properties, especially with respect to the swampland. The aim of the swampland has been to argue against the existence, and especially the stability, of de Sitter solutions from general quantum gravity arguments that having nothing to do with string theory constructions in particular. In our work, we first establish how our solution manages to escape the so-called (refined) de Sitter swampland condition and, more importantly, the TCC. We show that the time for which our solutions remain well-defined is compatible with the time-limit set by the TCC. We do not argue for or against the criterion set by the TCC but rather find that, quite remarkably, our Glauber-Sudarshan state naturally satisfies it. However, more pertinently for us, we shall show why the trans-Planckian problem does not even apply to our coherent state description. This is because our setup is already incorporates a UV-complete theory -- string theory -- and have an underlying Minkowski space-time. In section \ref{sec4.1}, we shall elaborate on how our solution escapes the TCC based on these two conditions. More general field-theoretic obstructions against the stability of de Sitter space-times essentially arise from the arbitrariness of the choice of the vacuum for fluctuations on top of de Sitter. It has long been extensively debated what is the {\it right} choice for this vacuum. While some have argued in favour of the Bunch-Davies vacuum, other have advocated for, say, de Sitter-invariant vacuum like the $\alpha$-vacuum \cite{kaloper, Quantum_swampland}. We shall demonstrate in section \ref{sec4.2} is that the main reason for the problems emerges for these vacua is due to the complicated  time-dependence of the mode functions corresponding to them. And thankfully, this is exactly what is \textit{not} the case for our solution since even fluctuations over de Sitter is constructed over the interacting vacuum in Minkowski spacetime and, therefore, the time-dependences arise as an artifact of Fourier transforms over the Glauber-Sudarshan state.

Finally, in section \ref{sec4.3}, we interpret the usual Gibbons-Hawking entropy for de Sitter as an entanglement entropy between the modes, on top of the warped-Minkowski background, which construct the Glauber-Sudarshan state itself. We then show how this entanglement entropy remains for our de Sitter solution. This is mainly due to two factors -- (a) having the de Sitter symmetries being emergent in our scenario which remain valid for a finite time-period, and (b) there is necessarily an interaction Hamiltonian ${\bf {\bf H}_{\rm int}}$ for us. If we take the limit ${\bf {\bf H}_{\rm int}} \rightarrow 0$, we find that the entanglement entropy goes to infinity and, on the other hand, for the same limit, our Glauber-Sudarshan state cannot be constructed and we get back flat space-time. In this sense, we find that there is a nice consistency for why we manage to find a finite entropy for de Sitter, while laying down the path for getting corrections beyond the semiclassical result.

\subsection{Notations and conventions}

Throughout the paper we have used the mostly-plus convention although in a couple of field theory computations we have used the mostly-minus convention so as to comply with known results. They will be indicated as we go along. Similarly, the Hubble parameter will be denoted by ${\rm H}$ whereas the warp-factor will be denoted by $H = H(y)$. The Hubble parameter features prominently in section \ref{sec4}, so this should not be a cause of confusion (in any case it'll be properly indicated which is which). The eleven-dimensional Planck constant will be denoted by ${\rm M}_p$ whereas we will use ${\rm M}_{\rm Pl}$ to denote four-dimensional Planck constant.

More importantly, in section \ref{sec2} we will mostly use the solitonic background \eqref{betbab3}, and therefore the {\it fields} will be denoted by $(g_{{\rm MN}}, {\rm C}_{\rm MNP})$, whereas the {\it operators} will be denoted by {\bf bold} faced letters, {\it i.e.} $({\bf g}_{{\rm MN}}, {\bf C}_{\rm MNP})$, unless mentioned otherwise.

In sections \ref{sec3.1} and \ref{sec3.2} we will use the background \eqref{vegamey3} and therefore they will involve $g_s$ dependent quantities. Both these sections only involve fields and to distinguish them from the field variables of the solitonic background, we use {\bf bold} faced letters to write them, i.e  
$({\bf g}_{{\rm MN}}, {\bf G}_{\rm MNPQ})$ will denote fields associated with the uplifted de Sitter background  \eqref{vegamey3} in M-theory. 

In sections \ref{sec3.3} and \ref{sec3.4} we have to use both fields and operators of the solitonic background as well as the fields of the uplifted de sitter background \eqref{vegamey3}. Our convention for this section then is the following: {\it all} fields and operators over the solitonic vacuum are expressed using un-bolded letters, for example $g_{\rm MN}$ denotes a {\it field} and $\langle g_{\rm MN}\rangle_\sigma$ denotes the expectation value of an {\it operator} over some state $\vert\sigma\rangle$ constructed over the solitonic vacuum. Again which is which should be clear from the context. The {\bf bold} faced letters are reserved for the {\it fields} associated with the uplifted de Sitter background  \eqref{vegamey3} in M-theory as we had in sections \ref{sec3.1} and \ref{sec3.2}. This can be made clear by an example: 
$\langle g_{\rm MN}\rangle_\alpha =  {\bf g}_{\rm MN}$, which implies that the expectation value of the metric {\it operator} over the Glauber-Sudarshan state $\vert\alpha\rangle$ gives the warped metric component of the de Sitter space, namely \eqref{vegamey3}.

\section{de Sitter space as a Glauber-Sudarshan state \label{sec2}}

Our analysis of the toy example in section \ref{etude} for a $1+1$ dimensional field theory provided the necessary groundwork  on which we can build our theory. Our aim would to allow for a solitonic configuration in IIB, which would be stable supersymmetric solution. Due to certain technical efficiency, as discussed below,
it is better to uplift the configuration to M-theory.
This is as discussed in \cite{desitter2}, wherein we will realize the solitonic vacuum from M-theory instead of type IIB, as:    
\bg\label{betbab3}
ds^2= {1\over \sqrt[3]{h_2^2(y, x_i)}}\left(-dt^2+ dx_1^2+ dx_2^2\right) + \sqrt[3]{h_1(y)}{g}^{(0)}_{MN}dy^M dy^N,
\nd 
where $h_1(y)$ and $h_2(y, x_i)$, $i = 1, 2$, are the warp-factors and $g^{(0)}_{MN}(y)$ is the metric of the internal eight-manifold. Although the choice of M-theory over type IIB is mainly because of the compactness of the representations of the degrees of freedom (the total degrees of freedom remains the same on either sides),
the fact that M-theory allows a well-defined low energy effective action whereas there is no simple action formalism for the IIB side, provides a better motivation to dwell on the M-theory side. Additionally, since our analysis will rely heavily on the path integral formalism, that in turn relies on the presence of an effective action, the M-theory uplift is more useful here. Minor compromise, like compactifying the $x_3$ direction, will not have any effect on the late time physics that we want to study here.  

Coming back to our simple harmonic oscillator from section \ref{etude}, we see that, in addition to the solitonic solution $x = a$, there are other solutions of the form $x = a + A~ {\rm cos}~\omega t$, for arbitrary choices of the constant $A$. These are time-dependent solutions having an oscillatory part, but they solve the EOMs.  On a solitonic vacuum $x = a$, the oscillatory part may be realized as a {\it coherent} state \cite{schrodinger}. In QFT, such configurations 
are the Glauber-Sudarshan states \cite{sudarshan, glauber} and their presence are augmented by the fact that they solve the EOMs.
In fact it is the combination 
$x = a + A~{\rm cos}~\omega t$ that solves the EOMs, so we can try to realize the type IIB de Sitter solution of the form:

{\footnotesize
\bg\label{betta3}
ds^2=\frac{1}{\Lambda(t)\sqrt{h}}(-dt^2+dx_1^2
+dx_2^2+dx_3^2)+ \sqrt{h} \Big(F_1(t){g}_{\alpha\beta}(y)dy^\alpha dy^\beta
+ F_2(t){g}_{mn}(y)dy^m dy^n\Big), \nd}
as a Glauber-Sudarshan state over a Minkowski background. Note that the four-dimensional part of 
\eqref{betta3} is a de Sitter space with a flat-slicing when $\Lambda(t) = \Lambda \vert t\vert^2$
and therefore the temporal coordinate covers $-\infty \le t \le 0$, implying $t \to 0$ to be the late time. The internal space is typically a compact non-K\"ahler six manifold whose details, and especially the decomposition in \eqref{betta3},  will be discussed below.

There are many issues that need clarifications before such a claim could be presented. In the  following we will go through them carefully. First, the solitonic solution \eqref{betbab3} is expectedly an uplift of a warped 
four-dimensional Minkowski background in the type IIB side. As such we expect the warp-factors 
$h_i$ in \eqref{betbab3} to be related. The relation is rather simple: $h_2(y) = h_1(y)$, but here we generalize this somewhat by keeping $h_2 \equiv h_2(y, x_i)$ and $h_1 \equiv h_1(y)$ unequal. The equality is a special case dealt in great details in \cite{becker} and \cite{DRS}. 

Secondly, the solitonic background \eqref{betbab3}, with a compact internal 
eight-dimensional space ${\cal M}_8$, can only be supported in the presence of G-fluxes. If we denote 
($y^m, y^\alpha$) respectively to be the coordinates of the six-dimensional base, with  $(\alpha, \beta) = (4, 5)$
and $(m, n) = (6, 7, 8, 9)$ as
${\cal M}_4 \times {\cal M}_2$ of ${\cal M}_8$ such that:
\bg\label{anonymous} 
{\cal M}_8 \equiv {\cal M}_4 \times \left({\cal M}_2 \times {\mathbb{T}^2\over {\cal G}}\right), \nd
with ($y^a, y^b$) denoting the coordinates of the fiber torus (${\cal G}$ is the isometry group of the torus), then we need G-fluxes with components 
on both the base and the fiber, as well as components like ${\bf G}_{0ijm}$ and ${\bf G}_{0ij\alpha}$, all functions of the six-dimensional base coordinates \cite{becker, DRS, desitter2}. 

All these imply that the fluctuations over the background \eqref{betbab3} as well as over the G-flux components that satisfy the EOMs allow more non-trivial time-independent Schr\"odinger type equations
whose solutions provide the {\it fluctuation} modes of the spectra, at least if we want to bring them in the 
simple harmonic form satisfying the condition \eqref{manara}\footnote{As it happens in any quantum field theory, the fluctuations {\it never} satisfy the full equations of motion. They only satisfy linear equations appearing from the quadratic parts of the action when expressed in terms of the fluctuating fields. As an example for a scalar field fluctuation $\delta\varphi$, written as $\delta\varphi(x, y, z) = \int d^{10}{\bf k}~ f_{\bf k}(t) \psi_{\bf k}({\bf x}, y, z)$, generically it's only the $\psi_{\bf k}({\bf x}, y, z)$ piece that becomes non-trivial over the solitonic background \eqref{betbab3}. See section \ref{etude} for more details.}. 
The identification to \eqref{pot1}, or even to 
\eqref{manara}, is more subtle as fluctuations along both space-time as well as the internal directions need to be accounted for, implying that the Glauber-Sudarshan states are not as simple as they were for the 
case with electromagnetic fields \cite{sudarshan, glauber}. Nevertheless, once we know the corresponding 
Schr\"odinger wave-functions we can at least try to use them to determine the Fourier modes of the 
metric and the G-flux components. The additional leverage that we get here is from the existence of the coherent states themselves (which we will justify a bit later). Assuming that the coherent states may be constructed 
for all modes, at least in the case where we can bring the individual mode-dynamics to the simple-harmonic case, this will justify that in the configuration spaces of each modes there are simple oscillatory motion (at least for both real and the complex parts). This is then obviously consistent with the time-dependent parts of the corresponding Schr\"odinger wave-functions which become $\psi_k(x), \eta_k(y, t), \xi_k(y, t)$ and 
$\zeta_k(z, t)$ respectively for the $2+1$ dimensional space-time, the internal six-manifold   
${\cal M}_4 \times {\cal M}_2$ and the fiber torus for every mode $k$ and at any given time $t$. 

\subsection{Metric and Glauber-Sudarshan wave-function \label{sec2.1}}

There is one issue that we kept under the rug so far, and has to do with the $F_i(t)$ factors in \eqref{betta3}. These factors are essential if we want to realize the IIB configuration \eqref{betta3} as a Glauber-Sudarshan state, simply because coherent states generically cannot produce time-independent configurations! Thus what we want in M-theory is a configuration of the following form\footnote{A small puzzle appears now that is worth mentioning at this stage. In M-theory, the internal eight-manifold is always time-dependent if the four-dimensional space in the type IIB side has de Sitter isometries.  As such a coherent state construction in M-theory should work whether or not the internal six-dimensional space in IIB is time-dependent. Why do we then need the internal space metric in IIB to take the form \eqref{betta3}? The resolution of the puzzle will require a more detailed understanding of the {\it existence} of a coherent state in M-theory, that we shall indulge in later (see section \ref{sec3.3}). For the time-being we will assume both the internal six and the eight manifolds in IIB and M-theory respectively to be time-dependent. \label{dorritt}}:

{\footnotesize
\bg\label{vegamey3}
ds^2= {1 \over \left(\Lambda\vert t\vert^2\sqrt{h}\right)^{4/3}} (-dt^2+dx_1 ^2+dx_2^2)
+e^{2{B_1}(y,t)}
{g}_{\alpha \beta}dy^\alpha dy^\beta+
e^{2{B_2}(y,t)}{g}_{mn}dy^m dy^n + e^{2{C}(y,t)} g_{ab} dx^a dx^b, \nonumber\\  \nd}
that encapsulates all the essential features of the IIB background \eqref{betta3} with appropriate choices of the coefficients ($B_i(y, t), C(y, t)$). The way we have expressed \eqref{vegamey3} suggests generalities beyond \eqref{betta3}, although one expects:
\bg\label{novha}
C(y, t) \equiv {1\over 2} {\rm log}\Big(\left[\Lambda(t)\right]^{2/3} \left[h(y)\right]^{1/3}\Big), \nd
if one wants to preserve the full de-Sitter isometries in $3+1$ dimensional space-time in the IIB side.
With this in mind, our first guess for the four set of 
Fourier components are:
\bg\label{cannon}
&&\widetilde{g}_{\mu\nu}(k) = \int {d^3 x} \left[{1\over \left(\Lambda\vert t\vert^2\sqrt{h}\right)^{4/3}}-
{1\over h_2^{2/3}}\right] h_2^{-1}
\psi^\ast_k(x)\eta_{\mu\nu}\nonumber\\
&& \widetilde{g}_{\alpha\beta}(k) = \int d^2 y dt \sqrt{g^{(0, 2)}_{\rm base}}\left(e^{2B_1(y, t)}g_{\alpha\beta}
 - h_1^{1/3}g^{(0)}_{\alpha\beta}\Big\vert_{\rm base}\right)h_2^{-1/3} \eta^\ast_k(y, t) \nonumber\\
&&\widetilde{g}_{mn}(k) = \int d^4 y dt \sqrt{g^{(0, 4)}_{\rm base}}\left(e^{2B_2(y, t)}g_{mn}
 - h_1^{1/3}g^{(0)}_{mn}\Big\vert_{\rm base}\right) h_2^{-1/3}\xi^\ast_k(y, t) \nonumber\\
&&\widetilde{g}_{ab}(k) = \int d^2z dt \sqrt{g^{(0)}_{\rm fibre}}\left(h^{1/3} \Lambda^{2/3}\vert t\vert^{4/3}
\delta_{ab}- h_1^{1/3}g^{(0)}_{ab}\bigg\vert_{\rm fibre}\right)h_2^{-1/3} \zeta^\ast_k(z, t), \nd
with $g_{\rm base}^{(0, q)}$ denoting the classical base metric of the six-manifold expressed as 
${\cal M}_4 \times {\cal M}_2$.
In writing \eqref{cannon} we have ignored many subtleties that we should clarify. First, the reason for taking  Fourier transforms is because the construction of Glauber-Sudarshan state is most easily expressed in terms of the Fourier components, as we shall see soon. Second, because of the solitonic pieces, \eqref{cannon} is partly off-shell. Third,
the modes 
$(\psi_k(x), \eta_k(y, t), \xi_k(y, t), \zeta_k(z, t))$ aren't necessarily as simple as we presented here. If we write 
$x \equiv ({\bf x}, t)$, {\it i.e.} separate the spatial and temporal coordinates, then the four modes that actually appear from the underlying Schr\"odinger type equation are typically of the form\footnote{We have used a simplifying normalization condition for the modes in \eqref{cannon}. The correct on-shell normalization condition for all the modes in \eqref{hudson} should be: $$\int d^{11}x ~\Psi_{\bf k}({\bf x}, y, z, t) \Psi^\ast_{{\bf k}'}({\bf x}, y, z, t)
h_2^{-1}({\bf x}, y) h_1^{4/3}(y) \equiv  \delta^{10}({\bf k}- {\bf k}') \delta(\omega^{(a)}_{\bf k} - \omega^{(a)}_{{\bf k}'})$$ 
\noindent However the compactness of the internal eight-manifold will simplify this and one can bring it in the 
form used for \eqref{cannon} if one further restricts to a slice in the internal space. As mentioned later, such restriction is not essential. \label{sarabig}}:
\bg\label{hudson} 
\Psi_{\bf k}({\bf x}, y, z, t) \equiv
\psi_{\bf k}({\bf x}, y, z, t), ~\eta_{\bf k}({\bf x}, y, z, t), ~ \xi_{\bf k}({\bf x}, y, z, t), ~ \zeta_{\bf k}({\bf x}, y, z, t), \nd
where $\psi_{\bf k}({\bf x}, y, z, t)$ denotes the {\it set} of modes governing the dynamics in $2+1$ dimensional space-time. Similarly $\xi_{\bf k}({\bf x}, y, z, t)$ denotes the set of modes governing the dynamics on 
${\cal M}_4$ internal space, and so on. These subtleties will only become relevant in section \ref{sec3.3}, so for the time being we will avoid over-complicating the system by assuming only single set of modes representing the directions respectively. 
Note that we have isolated the $t$ dependence on each modes and $y \equiv (y^\alpha, y^m)$. 
The reason is that the $t$ dependences of each modes should be of the form 
${\rm exp}\left(i\omega_{\bf k}^{(a)} t\right)$ with $a \equiv (\psi, \eta, \xi, \zeta)$ signifying the different modes. Such a temporal dependence is absolutely essential if the system has to allow for a coherent state description.  Alternatively, this boils down to the familiar decomposition of the metric fluctuations as:
\bg\label{kberry}
g_{\mu\nu}(x, y, z) = {\eta_{\mu\nu}\over h_2^{2/3}(y, {\bf x})} + \int d^{10}{\bf k}~ \widetilde{g}_{\mu\nu}({\bf k}, t) 
\psi_{\bf k}({\bf x}, y, z), \nd
with similar decompositions for the other internal components. Note two things: one, the integral is over 
$d^{10}{\bf k}$ and not 
over $d^{11}k$, and two, the appearance of  a generic $\widetilde{g}_{\mu\nu}({\bf k}, t)$ and not just 
$\widetilde{g}_{\mu\nu}({\bf k})$ from \eqref{cannon}.
They are of course expected consequences of any standard field theory so we will refrain from elaborating further on them. Additionally, fixing the values of ($\mu, \nu$) would lead to 
{\it three} set of fields, with each field having an infinite set of modes with an infinite number of harmonic oscillator states for each modes. All these follow standard results and if we, with some abuse of notations, define $\psi_k(x, y, z) = \psi_{\bf k}({\bf x}, y, z) {\rm exp}\left(i\omega_{\bf k}^{(\psi)} t\right)$, then the ``Fourier" modes\footnote{Thus naturally identifying $k$ in \eqref{cannon} (and also in \eqref{cannon2}) as 
$k \equiv ({\bf k}, \omega_{\bf k})$.}
in the first line of \eqref{cannon} follow naturally once we fix $y = y_0$ and $z = z_0$ to some {\it slice} in the internal space. Such a choice of slice simplifies the underlying analysis but has no physical consequence. Thus we could easily replace $\psi_k(x)$ by $\psi_k(x, y, z)$ etc in 
\eqref{cannon} giving us the following {\it on-shell} pieces (we will deal with the off-shell part soon):
\bg\label{cannon2}
&&\widetilde{g}_{\mu\nu}(k) = \int {d^{11} x} {\sqrt{g^{(0)}_{\rm base}}\over \left(\Lambda\vert t\vert^2\sqrt{h}\right)^{4/3}}~
 h_2^{-1}({\bf x}, y)h_1^{4/3}(y) 
\psi^\ast_k(x, y, z)\eta_{\mu\nu}\\
&& \widetilde{g}_{\alpha\beta}(k) = \int d^{11} x \sqrt{g^{(0)}_{\rm base}}\left(e^{2B_1(y, t)}g_{\alpha\beta}\right)h_2^{-1}({\bf x}, y)h_1^{4/3}(y) \eta^\ast_k(x, y, z) \nonumber\\
&&\widetilde{g}_{mn}(k) = \int d^{11} x \sqrt{g^{(0)}_{\rm base}}\left(e^{2B_2(y, t)}g_{mn}\right) h_2^{-1}({\bf x}, y) h_1^{4/3}(y) \xi^\ast_k(x, y, z) \nonumber\\
&&\widetilde{g}_{ab}(k) = \int d^{11} x \sqrt{g^{(0)}_{\rm fibre}}\left(h^{1/3} \Lambda^{2/3}\vert t\vert^{4/3}
\delta_{ab}\right)h_2^{-1}({\bf x}, y) h_1^{4/3}(y) \zeta^\ast_k(x, y, z), \nonumber \nd
where the determinant $g^{(0)}_{\rm base} \equiv  g^{(0, 4)}_{\rm base} g^{(0, 2)}_{\rm base}g^{(0)}_{\rm fibre}$ as defined for \eqref{cannon}. The slight differences from \eqref{cannon} are significant because they would eventually determine the Fourier components. 

The above discussion tells us that the modes in the theory are typically the time {\it independent} modes for any ${\bf k}$, and the time-dependences appear from the harmonic oscillator states that have energies 
in odd and even multiples of ${\omega^{(a)}_{\bf k}\over 2}$.
However there are other subtleties that appear here that need some elaborations. First, let us concentrate on the zero modes, {\it i.e.} the possibility of modes satisfying:
\bg\label{sarbb}
\omega^{(\psi)}_{\bf k}  = \omega^{(\eta)}_{\bf k}  = \omega^{(\xi)}_{\bf k}   = \omega^{(\zeta)}_{\bf k} = 0. \nd  
These are the troublesome modes in the theory that would lead to the Dine-Seiberg \cite{dines} runaway problem, even at the level of the solitonic vacuum \eqref{betbab3}. These zero modes appear from the complex and the K\"ahler structure moduli and therefore they have to be fixed to make sense of the underlying theory. In the standard solitonic theory, these zero modes help us to {\it quantize} the soliton itself, but here we will have to deal with them differently. In fact the time-independent G-fluxes that we added in the theory would help us achieve our goal by creating the necessary superpotential \cite{DRS, GKP, GVW}\footnote{As elaborated in section \ref{etude}, this identification of the K\"ahler and the complex structure moduli to the zero modes \eqref{sarbb} is more subtle. To illustrate the point, let us consider the wave-function $\xi_{\bf k}({\bf x}, y, z)$ whose zero-modes may be denoted by the set 
$\xi^{(0)}_{{\bf k}(l)}$ with $l$ denoting the parameter associated with the moduli (for example in the Calabi-Yau case $l$ will parametrize $(h_{11}, h_{21})$ moduli). The set of zero modes satisfy Schr\"odinger equations of the form:
\bg\label{dadima}
\left[-\nabla^2 + \mathbb{V}_{(l)}({\bf x}, y, z)\right]\xi^{(0)}_{{\bf k}(l)} = 0, \nonumber \nd
which is similar to the Schr\"odinger equation \eqref{lib3} with two main differences: one, the potential 
$\mathbb{V}_{(l)}({\bf x}, y, z)$ is more involved than what appears in \eqref{lib3}, and two, there are not one set of Schr\"odinger equations, but $l$ set of them. This representation is not unique, and in fact there are an infinite possible such choices, depending on what values of the moduli we choose (i.e what {\it values} of the K\"ahler and the complex structure moduli we choose). Clearly a linear combination of 
$\xi^{(0)}_{{\bf k}(l)}$ will contain the information of these moduli, although extracting them might be harder. 
However if the moduli are {\it stabilized} then the only zero modes that we need to worry about are the translation and the rotations of the background \eqref{betbab3} with it's G-flux components.}.

Such an approach now ties two things together. One, the fact that the ($\psi_{\bf k}, \eta_{\bf k}, \xi_{\bf k}, \zeta_{\bf k}$)
 modes in the solitonic background are {\it not} of the kind ${\rm exp}\left(\pm i{\bf k}\cdot {\bf x}\right), {\rm exp}\left(\pm i{\bf k}\cdot {y}\right)$ and ${\rm exp}\left(\pm i{\bf k}\cdot {z}\right)$; and two, the presence of the G-fluxes. The latter leads to $-$ and the former is a consequence of $-$ {\it interactions}. Thus the underlying theory that we discuss here cannot be a free theory! This also means that the coherent states that we study here are {\it not} the coherent states of a free theory. 

This is where we differ from the standard coherent state constructions in Quantum Field Theories, but now the pertinent question is the source of the interactions themselves.
Where are the interactions coming from? This is already answered in some sense in \cite{becker, DRS} and more recently in \cite{sethi}. The answer lies in the compactness of the internal manifold, both in the IIB and in the M-theory sides. From M-theory, it is easy to argue that a generic compact internal space cannot be supported in the absence of the G-fluxes \cite{becker, SVW, DM, DRS}. Once G-fluxes are switched on, they automatically allow higher curvature topological terms like ${\bf C} \wedge {\bf X}_8$, where ${\bf X}_8$ involve fourth order curvature terms \cite{becker}. To the same order, non-topological terms, like:
\bg\label{nontop}
S_{\rm ntop} =  {\rm M}_p^3\int d^{11}x \sqrt{g} \left(t_8 t_8 - {1\over 24} \epsilon_{11} \epsilon_{11}\right) {\bf R}^4 + {\cal O}\left[\left(\partial {\bf G}\right)^4\right] + ...., \nd
are switched on simultaneously, where the dotted terms are the mixed terms. Similar story unfolds in the type IIB side also as emphasized recently in \cite{sethi}. The outcome of our discussion then is the following. There is no simple {\it free} field theory description in the presence of G-fluxes and/or compact internal manifolds. The latter is absolutely necessary to allow for a finite Newton's constant in the non-compact directions. As a further consequence of our statement above, the modes  ($\psi_{\bf k}, \eta_{\bf k}, 
\xi_{\bf k}, \zeta_{\bf k}$) then become highly non-trivial but thankfully remain 
time-independent\footnote{As pointed out in section 
\ref{etude}, the harmonic oscillator regime will henceforth be termed as the {\it free} vacuum $\vert 0 \rangle$ here, and $\vert \Omega\rangle$, which we will deal in section \ref{sec2.4}, will be the {\it interacting} vacuum with the full anharmonicity brought in, unless mentioned otherwise. \label{freevac}}. 

The underlying Schr\"odinger equation that appears from the interacting Lagrangian in the M-theory side, could also allow {\it bound} state solutions, in addition to the zero mode spectra \eqref{sarbb}. These, if  present, should be interpreted as the discrete excited states\footnote{Much like the excited states of an electron in a hydrogen atom. The bound state spectra of the electron precisely spell out these states. The continuum levels of the electron, on the other hand, have an interpretation of scattering states. See also 
section \ref{etude}.}  
of the full solitonic backgrounds 
\eqref{betbab3}, with various choices of the internal manifolds, themselves.  We will not worry about them for the time being, and concentrate only on the continuum of levels parametrized by the eigenfunctions 
($\psi_{\bf k}, \eta_{\bf k}, \xi_{\bf k}, \zeta_{\bf k}$). These are essentially the modes that appear, once we juxtapose them with the temporal behavior ${\rm exp}\left(i\omega^{(a)}_{\bf k} t\right)$, in the Fourier transforms 
\eqref{cannon2}. Once we go to the Euclidean picture, where 
${k} \to {k}_E$, we can integrate out the modes lying between $M \le {k}_E \le \Lambda_{\rm UV}$ to write the Wilsonian action at a given scale $M$ and with a UV cut-off $\Lambda_{\rm UV}$. This scale $M$ could in principle be identified with ${\rm M}_p$, and  in that case the effective action at that scale would match-up with the effective action that we elaborated in great details in \cite{desitter2, desitter3} (see \cite{desitter4} for a review on this).  

More subtlety ensues, mostly associated with the oscillatory behavior of the coherent states that allowed us to interpret the Fourier modes in \eqref{cannon} as ($\psi_{\bf k}, \eta_{\bf k}, \xi_{\bf k}, \zeta_{\bf k}$) in the first place. For example, the way we have expressed \eqref{cannon}, once added to the solitonic background \eqref{betbab3}, it appears to simply {\it replace} the solitonic background \eqref{betbab3} by the time-dependent background \eqref{vegamey3}, which is the uplift of the IIB de Sitter background \eqref{betta3}. How is this then any different from simply taking \eqref{betta3} as the non-supersymmetric {\it vacuum} in type IIB? The answer lies in the distinction between classical solution and quantum states. The M-theory background \eqref{vegamey3}, or it's IIB counter-part \eqref{betta3}, does not appear just as a classical solution here. Rather the probablity amplitude in a given coherent state for any {\it given} mode derived from \eqref{cannon} 
{\it peaks}  at a specific value which once added to the classical solitonic solution \eqref{betbab3} reproduces the background \eqref{vegamey3}, or equivalently \eqref{betta3} in IIB. This reinforces the main point of our paper, namely, \eqref{vegamey3} in M-theory, or equivalently 
\eqref{betta3} in type IIB, cannot appear as a vacuum configuration but can only appear as the most probable value in the coherent state. 

The above discussion then raises the following question. How about choosing delta function states for any 
${\bf k}$ in $\Psi_{\bf k}$ given in \eqref{hudson}? Clearly the delta function states in the configuration 
space\footnote{One needs to be careful to {\it not} interpret these delta function states as localized states in space-time. In space-time we will continue to have the standard Schr\"odinger wave-functions 
$\Psi_{\bf k}({\bf x}, y, z, t)$ \eqref{hudson}. Here, in a similar vein as with the realization of  the coherent states in the configuration space, the delta function states will also be realized in the configuration space.} 
would appear to serve as better candidates because they would exactly reproduce the background 
\eqref{vegamey3} with probability 1. Unfortunately however, this configuration doesn't survive long and the delta function states expand very fast in the presence of the interacting M-theory Hamiltonian. This ruins our hope of realizing \eqref{vegamey3} (or equivalently \eqref{betta3} in IIB) purely as a classical solution with zero quantum width, reinforcing, yet again, the coherent state nature of the background.

The coherent state description for any given mode ${\bf k}$ for a given Schr\"odinger wave-function in 
\eqref{hudson} is clearly the best possible description we can have for our case. Let us now see how we can quantify this further. For simplicity we will look at one specific mode, $\widetilde{g}_{\mu\nu}({\bf k})$, with the spatial part of the 
Schr\"odinger wave-function $\psi_{\bf k} \equiv \psi_{\bf k}({\bf x}, y, z)$.  The Fourier mode decomposition \eqref{cannon2} will tell us precisely the most probable amplitude for 
$\widetilde{g}_{\mu\nu}({\bf k})$, and let us denote it by 
$\alpha^{(\psi)}_{\mu\nu}({\bf k}, t)$. The temporal behavior of $\alpha^{(\psi)}_{\mu\nu}({\bf k}, t)$ is clearly oscillatory with frequency 
$\omega^{(\psi)}_{\bf k}$, which is the same $\omega_{\bf k}^{(a)}$ that appeared earlier in our discussion. 
The subtlety with the off-shell parts of \eqref{cannon} will be dealt soon. Meanwhile we define: 
\bg\label{cnelson2}
\alpha^{(\psi)}_{\mu\nu}({\bf k}, t) \equiv \alpha^{(\psi)}_{\mu\nu}({\bf k}, \omega_{\bf k}) {\rm exp}\left(-i\omega_{\bf k}^{(\psi)}t\right), \nd
where $\alpha^{(\psi)}_{\mu\nu}({\bf k}, \omega_{\bf k})$ can be defined up to a possible phase of 
${\rm exp}\left(i\sigma_{\bf k}^{(\psi)}\right)$. Collecting everything together then provides the following wave-function in the configuration phase for the mode $\widetilde{g}_{\mu\nu}({\bf k})$:
\bg\label{spolley}
{\bf \Psi}^{\left[\alpha^{(\psi)}_{\mu\nu}({\bf k}, t)\right]} \left(\widetilde{g}_{\mu\nu}({\bf k}), t\right) &=& 
\left({\omega_{\bf k}^{(\psi)}\over \pi}\right)^{1/4} {\rm exp}\left[-{\omega_{\bf k}^{(\psi)}\over 2}
\left(\widetilde{g}_{\mu\nu}({\bf k})- {1 \over 2 \omega_{\bf k}^{(\psi)}} {\bf Re}\left[\alpha^{(\psi)}_{\mu\nu}({\bf k}, t)\right]\right)^2 \right]\nonumber\\
& \times & 
{\rm exp}\Bigg({i\over 2} ~{\bf Im}\left[ \alpha^{(\psi)}_{\mu\nu}({\bf k}, t) \right] \widetilde{g}^{\mu\nu}({\bf k}) + i \theta_{\bf k}^{(\psi)}(t) \Bigg), \nd
where no sum over repeated indices are implied above; and $\theta_{\bf k}^{(\psi)}(t)$ is yet another phase that appears in the definition of the wave-function for the mode $\widetilde{g}_{\mu\nu}({\bf k})$. This phase can be determined by demanding that 
${\bf \Psi}^{\left[\alpha^{(\psi)}_{\mu\nu}({\bf k}, t)\right]} \left(\widetilde{g}_{\mu\nu}({\bf k}), t\right)$ solves the harmonic oscillator wave-equation that appears from the interacting Lagrangian in the M-theory picture. As discussed earlier, the interactions are essential and they make the Schr\"odinger wave-function\footnote{To remind the readers again, there are at least two set of Schr\"odinger equations that can appear here. The Schr\"odinger equations whose 
solutions are the four set of wave-functions in \eqref{hudson} are {\it never} of the simple harmonic oscillator form. In fact these Schr\"odinger equations allow highly non-trivial potentials (for example \eqref{lib3}). On the other hand, the Schr\"odinger equations allowing wave-functions like \eqref{spolley} have simple harmonic oscillator potentials.} 
in \eqref{spolley} simple but render the wave-functions in \eqref{hudson} rather complicated. Nevertheless 
the form for ${\bf \Psi}^{\left[\alpha^{(\psi)}_{\mu\nu}({\bf k}, t)\right]} \left(\widetilde{g}_{\mu\nu}({\bf k}), t\right)$ may be determined precisely even if 
$\psi_{\bf k}$ may be involved as we saw above. The phase $\theta_{\bf k}^{(\psi)}(t)$ then becomes:
\bg\label{naskinski}
 \theta_{\bf k}^{(\psi)}(t) =  -{1\over 2} \left[ \omega_{\bf k}^{(\psi)} t - \vert \alpha^{(\psi)}_{\mu\nu}({\bf k}, 0)\vert^2 
 {\rm sin}\left(2\omega_{\bf k}^{(\psi)} t - 2 \sigma_{\bf k}^{(\psi)}\right)\right], \nd
 where $\sigma_{\bf k}^{(\psi)}$ is the initial phase  of the eigenvalue $\alpha^{(\psi)}_{\mu\nu}({\bf k}, 0)$ discussed  above. From \eqref{spolley}, it is easy to infer that 
 $\left\vert {\bf \Psi}^{\left[\alpha^{(\psi)}_{\mu\nu}({\bf k}, t)\right]} \left(\widetilde{g}_{\mu\nu}({\bf k}), t\right)
 \right\vert^2$ peaks exactly at ${\bf Re}\left[\alpha^{(\psi)}_{\mu\nu}({\bf k}, t)\right]$, which in turn oscillates as 
 ${\rm sin}\left(\omega_{\bf k}^{(\psi)} t\right)$, as one would expect.  On the other hand, a wave-function of the form:
 \bg\label{kmeagain}
 {\bf \Psi}^{\left[\alpha^{(\psi)}_{\mu\nu}({\bf k}, 0)\right]} \left(\widetilde{g}_{\mu\nu}({\bf k}), 0\right) 
 = \prod_{\bf k} \delta\left(\widetilde{g}_{\mu\nu}({\bf k}) -  \alpha^{(\psi)}_{\mu\nu}({\bf k}, 0)\right), \nd
 defined specifically at $t = 0$, would reproduce the background \eqref{vegamey3} in the exact classical form 
 with zero quantum width at $t = 0$. However as discussed above, this immediately expands to become highly quantum with no trace of the classical picture left resembling anything close to \eqref{vegamey3}. 

Interestingly, the wave-function \eqref{spolley}, itself has a gaussian form with the center of the gaussian behaving in a simple harmonic oscillator fashion, up-to an overall phase factor. This phase factor can be simplified a bit if $\alpha^{(\psi)}_{\mu\nu}({\bf k}, t)$ becomes real, but not too much. The overall phase of 
${\rm exp}\left(i\theta_{\bf k}^{(\psi)}(t)\right)$ given in \eqref{naskinski} cannot be made to vanish, and will survive. The probablity however remains peaked at the desired value. In terms of Schr\"odinger {\it operator} formalism this implies:
\bg\label{chukkam}
\langle \delta{\bf g}_{\mu\nu}(x, y, z)\rangle_{\alpha^{(\psi)}} = {\bf Re}\left(\int {d^{10} {\bf k}\over 
{2\omega^{(\psi)}_{\bf k}}} ~
\alpha^{(\psi)}_{\mu\nu}({\bf k}, t)\psi_{\bf k}({\bf x}, y, z)\right), \nd
where the expectation value of the {\it fluctuation} of metric operator, $\delta{\bf g}_{\mu\nu}(x, y, z)$ is taken over the coherent state 
\eqref{spolley}. Expectedly the integral is done over $d^{10} {\bf k}$ because $\omega^{(\psi)}_{\bf k}$ 
appearing from \eqref{cnelson2} is related to ${\bf k}$ {\it on-shell}. The puzzle however is the {\it off-shell} part from \eqref{cannon}. What would be the simplest way of reproducing that? This is in general tricky, so  let us propose the following expression for $\alpha^{(\psi)}_{\mu\nu}({\bf k}, t)$:
\bg\label{cnelson}
\alpha^{(\psi)}_{\mu\nu}({\bf k}, t) \equiv \left[\alpha^{(1; \psi)}_{\mu\nu}({\bf k}, \omega_{\bf k}^{(\psi)}) + 
\lim_{\omega_o \to 0}
{2\omega_{\bf k}^{(\psi)}\alpha^{(2; \psi)}_{\mu\nu}({\bf k})\over \vert \omega_o\vert \sqrt{\pi}}~ 
e^{-\left({\omega_{\bf k}^{(\psi)}/\omega_o}\right)^2}\right] e^{-i\omega_{\bf k}^{(\psi)}t}, \nd
where the first term, $\alpha^{(1; \psi)}_{\mu\nu}({\bf k}, \omega_{\bf k}^{(\psi)})$ is the on-shell piece and the second term, 
$\alpha^{(2; \psi)}_{\mu\nu}({\bf k})$, is related to the off-shell piece. The extra factor of  
$2\omega_{\bf k}^{(\psi)}$ is essential to eliminate the measure of the integral \eqref{chukkam}, with the limiting value of $\omega_o \to 0$ localizing the second function in \eqref{cnelson} to 
$\delta(\omega_{\bf k}^{(\psi)})$. One advantage of such an approach is that we can continue using the on-shell integral form \eqref{chukkam} to express the metric components even if there are off-shell pieces in the Fourier transforms. The disadvantage however is that the {\it inverse} Fourier transform is a bit tricky: one will have to resort back to $d^{11}k$ to get all the factors correctly. 

Let us make a few more observations. First, 
the way we have expressed \eqref{cnelson}, tells us that the off-shell piece is in general arbitrarily small, but does appear to provide non-zero value once integrated over an on-shell integral of the form 
\eqref{chukkam}. This is good because it'd mean that we don't have to worry too much about the off-shell 
parts from \eqref{cannon} and continue using our on-shell analysis. Such a point of view will become very useful when we have to extract values from path-integral computations in section \ref{sec2.4}. Of course there does exist other ways to deal with the off-shell parts, but here we will pursue the simplest one. Second, 
we haven't added the solitonic background ${\eta_{\mu\nu}h_2^{-2/3}(y, {\bf x})}$ to it. We could in principle add this to \eqref{chukkam} using the completeness condition of the coherent states. This way the full on-shell metric configuration \eqref{vegamey3} appears from our coherent state description. On the other hand, in terms of Feynman {\it field} formalism, this on-shell relation is not required (in fact fields are maximally {\it off-shell}),  and therefore we expect:
\bg\label{poladom}
g_{\mu\nu}(x, y, z) = {\eta_{\mu\nu}\over h_2^{2/3}(y, {\bf x})} + \int d^{11}k ~\widetilde{g}_{\mu\nu}({\bf k}, k_0) ~\psi_{\bf k}({\bf x}, y, z) e^{-ik_0 t}, \nd
where with some abuse of notation we used $\widetilde{g}_{\mu\nu}({\bf k}, k_0)$ to denote the 
field amplitudes\footnote{Needless to say, the notations ${\bf g}_{\mu\nu}$ will henceforth denote metric {\it operator} and $g_{\mu\nu}$ the field.}. Thus here, 
$\widetilde{g}_{\mu\nu}({\bf k}, k_0)$ is generic and should not be confused with the Fourier decomposition \eqref{cannon2}. As expected, there is also no relation between ${\bf k}$ and $k_0$, and the integral spans over the full eleven-dimensional momentum space. 

There is an immediate advantage of expressing the fields in terms of the off-shell form \eqref{poladom}, that is not there in the on-shell form \eqref{chukkam}. The form \eqref{poladom} allows us to study the dynamics of the system more consistently than the evolution of the most probable state in \eqref{chukkam}. The expression \eqref{poladom} captures {\it any} quantum width of a state, no matter how sharply peaked the
configuration space wave-functions are. In fact for a state like \eqref{kmeagain}, where any field configuration would go wildly off-shell, \eqref{poladom} is well-suited to tackle the dynamics. Additionally, the time-independencies of the modes $\psi_{\bf k}({\bf x}, y, z)$, are essential to study the Wilsonian effective action 
that formed the basis of our analysis in \cite{desitter2} and \cite{desitter3}.   

\subsection{Metric fluctuations and quantum states \label{sec2.2}}

In the {\it operator} formalism,  where $g_{\mu\nu}(x, y, z)$ becomes an operator, then can be expanded in terms of the corresponding creation and annihilation operators for the modes $\psi_{\bf k}({\bf x}, y, z)$. It is instructive to present this formalism and compare the result in terms of the modes expanded around de Sitter 
{\it vacuum}. First let us express the modes over the solitonic background \eqref{betbab3}. This takes the form:

{\footnotesize
\bg\label{virgmads}
{\bf g}_{\mu\nu}({\bf x}, y, z; t) - {\eta_{\mu\nu}\over h_2^{2/3}(y, {\bf x})}= \int {d^{10}{\bf k}\over (2\pi)^{10}} {1\over \sqrt{2\omega^{(\psi)}_{\bf k}}}&&
\sum_{s = \pm}\Big[a_s({\bf k}) e_{\mu\nu}({\bf k}, s) \psi_{\bf k}({\bf x}, y, z) {\rm exp}\left(-i \omega^{(\psi)}_{\bf k}t\right) \nonumber\\
 &&+ ~a_s^\dagger({\bf k}) e^\ast_{\mu\nu}({\bf k}, s) \psi^\ast_{\bf k}({\bf x}, y, z) {\rm exp}\left(i \omega^{(\psi)}_{\bf k}t\right)\Big], \nd} 
where $e_{\mu\nu}({\bf k}, s)$ is the polarization tensor for a given momentum ${\bf k}$ and $s = \pm$; with ($a_s({\bf k}), a_s^\dagger({\bf k})$) forming the standard annihilation and the creation operators. Note however the difference from usual QFT mode expansion: the spatial modes are not simple and they are given in terms of the Schr\"odinger wave-functions 
$\psi_{\bf k}({\bf x}, y, z)$, whereas the temporal modes take the usual form of 
${\rm exp}\left(\pm i \omega^{(\psi)}_{\bf k} t\right)$ with $t$ being the time coordinate used here.  The integral
in \eqref{virgmads} is over $d^{10}{\bf k}$ and not over $d^{11}k$ as one would expect and we have 
expressed the temporal behavior using the on-shell value of $\omega^{(\psi)}_{\bf k}$. We could have used also the off-shell form ${\rm exp}(\pm ik_0 t)$, but then we will have to specify the pole at $\omega_{\bf k}^{(\psi)}$. Of course, this is all very standard, so we won't elaborate it further.

We will however compare the above mode expansion \eqref{virgmads}, which is over the solitonic background \eqref{betbab3}, to the one over the uplifted background \eqref{vegamey3}. The mode expansion now takes the following form\footnote{A word of caution here. The metric has 66 degrees of freedom, and in the presence of other fields, namely the G-fluxes (we will discuss them soon), some of these degrees of freedom have to be gauged. This gauging sometimes lead to propagating ghosts, but one may choose a gauge choice where the propagating ghosts are absent (this is easy to achieve either in the effective four-dimensional case in type IIB or in the effective three-dimensional case in M-theory with {\it only} metric degrees of freedom). Subtlety appears when both metric and fluxes are present, but we will not worry about them right now, and consider them only in section \ref{sec3.3}.}:

{\footnotesize
\bg\label{racadams}
{\bf g}_{\mu\nu}({\bf x}, y, z; t) - {\eta_{\mu\nu}\over \left(\Lambda |t|^2\sqrt{h(y)}\right)^{4/3}}= \int {d^{10}{\bf k}\over (2\pi)^{10}} {1\over \sqrt{2\omega^{(\psi)}_{\bf k}}}&&
\sum_{s = \pm}\Big[{\hat a}_s({\bf k}) e_{\mu\nu}({\bf k}, s) {\hat\psi}_{\bf k}({\bf x}, y, z) 
{\rm exp}\left(-i {\hat\omega}^{(\psi)}_{\bf k}(t)t\right) \nonumber\\
 &+&{\hat a}_s^\dagger({\bf k}) e^\ast_{\mu\nu}({\bf k}, s) {\hat\psi}^\ast_{\bf k}({\bf x}, y, z) 
 {\rm exp}\left(i {\hat\omega}^{(\psi)}_{\bf k}(t)t\right)\Big], \nd} 
where as before $e_{\mu\nu}({\bf k}, s)$ denotes the polarization tensor, but now there are quite a few noticeable differences. The spatial modes ${\hat\psi}_{\bf k}({\bf x}, y, z)$ are more involved than the spatial modes 
${\psi}_{\bf k}({\bf x}, y, z)$ encountered earlier. This is expected, because the background \eqref{vegamey3} is different from the solitonic background \eqref{betbab3}. A more crucial thing is the appearance of both 
${\hat\omega}^{(\psi)}_{\bf k}(t)$ and ${\omega}^{(\psi)}_{\bf k}$ in \eqref{racadams}. In fact 
${\hat\omega}^{(\psi)}_{\bf k}(t)$ is a much more complicated function of ${\bf k}$ {and} $t$, and we expect:
\bg\label{robbiem}
{\hat\omega}^{(\psi)}_{\bf k}(t) = {\omega}^{(\psi)}_{\bf k} + {i \over t}~{\rm log}\left[g^{(\psi)}({\bf k}, t)\right], \nd
where the second term implies that the frequencies themselves are time-dependent. In fact we are not restricted to real values now and $g^{(\psi)}({\bf k}, t)$ can be imaginary, leading to 
${\hat\omega}^{(\psi)}_{\bf k}(t)$ becoming imaginary, although ${\omega}^{(\psi)}_{\bf k}$ remains real throughout. This difference is important and spells out the fact that the frequency of a given mode can change with time, both in terms of the modulus and argument of a complex number, leading to all kinds of Trans-Planckian issues encountered in the literature (see for example \cite{martin}).

One may quantify the above mode expansion directly from an effective $2+1$ dimensional point of view. In such an effective picture\footnote{For example the case with $C^{(p)}_{nm} << 1$ in \eqref{lib3}.}, the internal degrees of freedom at least to first approximation do not effect the mode expansions, implying that ${\hat\psi}_{\bf k}({\bf x}, y, z) = {\rm exp}(i{\bf k}\cdot{\bf x})$. This could also be interpreted as though we have taken a slice of the internal manifold with fixed ($y, z$). The mode expansion over the $2+1$ dimensional space-time \eqref{vegamey3} can be expressed as in \eqref{racadams} with fixed ($y, z$), such that:
\bg\label{horlat}
{\hat\omega}_{\bf k}^{(\psi)} = {i\over t} ~{\rm log}\left[\vert{\bf k}\vert t^{7/6}
\Big(c_1 {\bf J}_{7/6}(\vert{\bf k}\vert t) + c_2 {\bf Y}_{7/6}(\vert{\bf k}\vert  t)\Big)\right], \nd
where $\vert{\bf k} \vert = \sqrt{{\bf k}\cdot{\bf k}}$; and 
${\bf J}_n(\vert{\bf k}\vert t)$ and ${\bf Y}_n(\vert{\bf k}\vert t)$ are the Bessel functions of the first and the second kinds respectively. The $c_i$ are constants, but they cannot both be real, because we want to extract a factor of ${\rm exp}(-i\vert{\bf k} \vert t)$ from \eqref{horlat}. 
Their precise value can be easily determined by demanding that 
\eqref{horlat} takes the form \eqref{robbiem}.  Note that the dimensions are taken care of inside the logarithm by introducing appropriate powers of the Hubble constant ${\rm H}$. The exact form for \eqref{horlat} is not important, but what is important however is to note that \eqref{horlat} implies time-dependent frequency for the modes expanded over the up-lifted de sitter background \eqref{vegamey3} in M-theory. Dimensionally reducing to IIB, which is the same as expanding over the four-dimensional part of \eqref{betta3}, the fluctuating modes have a frequency given by:
\bg\label{haipulat} 
{\hat\omega}_{\bf k}^{(\psi)} = {i\over t} ~{\rm log}\left\{{1\over {\vert {\bf k} \vert}}
\Big[c_1\Big(\sin~{\vert {\bf k} \vert} t  - {\vert {\bf k} \vert}t ~\cos~{\vert {\bf k} \vert} t\Big) + 
c_2\Big({\vert {\bf k} \vert} t~\sin~{\vert {\bf k} \vert} t  + \cos~{\vert {\bf k} \vert} t\Big)\Big]\right\} \nd
which although expectedly differs from \eqref{horlat}, carries the same information as above. Again, the $c_i$ constants cannot be all real, and comparing to \eqref{robbiem}, it is easy to infer that $c_1 = i$ and 
$c_2 = -1$.  This then leads to the familiar result in the literature with $g^{(\psi)}({\bf k}, t)$ in 
\eqref{robbiem} taking the 
form:
\bg\label{hlhplat}
g^{(\psi)}({\bf k}, t) = \sqrt{t^2 + {1\over {\vert {\bf k} \vert}^2}}~
{\rm exp}\Big(i \tan^{-1}\left({\vert {\bf k} \vert} t\right)\Big), \nd 
where one would have to again insert the Hubble constant ${\rm H}$ to make the quantity under the square-root to have the right dimension (to avoid clutter, we will henceforth take ${\rm H} = 1$ unless mentioned otherwise). Both the results, \eqref{horlat} and \eqref{hlhplat}, show that the frequencies in M-theory and IIB respectively are time varying frequencies, thus would in principle create problems if we try to integrate out the high energy modes in the Wilsonian way. Such an issue {\it do not} exist when we express 
the de Sitter space as a coherent state because the modes are described using  \eqref{virgmads}. In fact {\it any} fluctuations, even the ones that could be interpreted as over the coherent state themselves, should be expressed using the modes over the solitonic vacuum \eqref{betbab3}.

Let us quantify the above statement more carefully. The {\it effective} fluctuation spectra over a de Sitter {\it vacuum} 
in IIB or its uplift in M-theory typically go like 
${\rm exp}\Big(i{\bf k}\cdot {\bf x} - i{\hat\omega}_{\bf k}^{(\psi)}(t) t\Big)$, as we saw above. Our concern is
with the frequencies of the modes as they depend on the conformal time\footnote{This is an abuse of notation. The dimensionless time parameter is always $\sqrt{\Lambda} t$, where $\Lambda$ is the cosmological constant.}
 $t$. Our goal would be to express this as a {\it linear} combination of the modes over the solitonic vacuum, either in M-theory or in IIB.  In other words, we expect:
\bg\label{jlewis}
{\rm exp}\Big(- i{\hat\omega}_{\bf k}^{(\psi)}(t) t\Big) = \int dk_0~ f({\bf k}, k_0) ~{\rm exp}\left(-ik_0 t\right), \nd
where the RHS is expressed with modes over the solitonic vacuum. At this stage we don't care whether the modes $k_0$ take the on-shell values $\omega_{\bf k}$, and typically this may not always be possible. For the modes taking the form given in \eqref{robbiem}, the Fourier coefficients $f({\bf k}, k_0)$ can be expressed in the following way:
\bg\label{lizshue}
f({\bf k}, k_0) = \int_{-T}^{+T} dt~{\rm exp}\bigg[-i\Big(\omega_{\bf k}^{(\psi)} 
+ i t^{-1}{\rm log}\big(g^{(\psi)}({\bf k}, t)\big) - k_0\Big) t\bigg], \nd
where $g^{(\psi)}({\bf k}, t)$ appears in \eqref{robbiem}, and $T$ (or more appropriately $\sqrt{\Lambda}T$) denotes temporal boundary. The Fourier coefficient $f({\bf k}, k_0)$ is an effective way to {\it interpret} the fluctuations over a coherent state, but the problem with an expression like \eqref{lizshue} is that it may not always be {\it convergent}. The issue of convergence stems from the fluctuating modes themselves. For example the temporally varying frequencies in IIB as given in \eqref{haipulat} tend to blow-up as $\vert {\bf k} \vert t \to \pm \infty$, and this is reflected directly in the Fourier coefficient $f({\bf k}, k_0)$ once we plug-in \eqref{hlhplat} in \eqref{lizshue}
to get:

{\footnotesize
\bg\label{kirkoo}
f({\bf k}, k_0) = {1\over {\vert{\bf k}\vert}}~\delta(\vert{\bf k}\vert - k_0) 
- {2 \over \vert{\bf k}\vert  - k_0}~ T ~\cos\Big[(\vert{\bf k}\vert - k_0)T\Big]  + 
{2 \over (\vert{\bf k}\vert - k_0)^2} ~\sin\Big[(\vert{\bf k}\vert - k_0)T\Big], \nd} 
where the expected non-convergence appears from the $T$ dependence of the second term in 
\eqref{kirkoo}.  This may not quite be an issue because we could keep $T$ large but not necessarily 
infinite (a careful study with path integrals in section \ref{sec2.5} will show that this problem does not arise). Note also that the way we have represented $f({\bf k}, k_0)$, it is a real function so the standard relation between $f^\ast({\bf k}, k_0)$ and $f(-{\bf k}, \pm k_0)$ do not hold here because the transformation \eqref{jlewis} is arranged to reproduce a complex function, not a real one. In a similar vein, in M-theory, one will have to reproduce \eqref{horlat} using the Fourier transform
\eqref{jlewis}. The issue of convergence appears here too, as both $x^{7/6} {\bf J}_{7/6}(x)$ and 
$x^{7/6} {\bf Y}_{7/6}(x)$ in \eqref{horlat} blow-up for $x \equiv \vert{\bf k}\vert t \to \infty$. As before we can bound $T$ to a large but finite value so that an expression like \eqref{kirkoo}, but now for M-theory, makes 
sense. 

Thus it appears from an effective $d+1$ dimensional point of view ($d = 2$ for M-theory and $d = 3$ for IIB), the fluctuations over de Sitter {\it vacuum}, can be represented here as a linear combination of the modes 
over the solitonic vacuum. In other words:

{\footnotesize
\bg\label{dholi}
\delta g_{\mu\nu}(x) &= & \sum_{s = \pm} \int d^{d+1}k ~e_{\mu\nu}({\bf k}, s) f({\bf k}, k_0) \psi_{\bf k}({\bf x})~ e^{ik_0 t} \\
& = & \sum_{s = \pm} \int d^d{\bf k} ~e_{\mu\nu}({\bf k}, s)~\psi_{\bf k}({\bf x}) \int dk_0 f({\bf k}, k_0)~e^{ik_0 t} 
\equiv \sum_{s = \pm} \int d^d {\bf k} ~ e_{\mu\nu}({\bf k}, s)  f_{\bf k}(t) \psi_{\bf k}({\bf x}), \nonumber \nd} 
where $\psi_{\bf k}({\bf x}) = \int d^{D-d-1} {\bf k}~ \psi_{\bf k}({\bf x}, y_0, z_0)$ for $D$ space-time dimensions; and with some abuse of notations we have used ${\bf k}$ to also signify the momenta along the internal directions. Which is which, should be clear from the context.   
 
The point of the above exercise \eqref{dholi} is simple. It is to show that the fluctuations may be controlled by 
a time varying amplitude for any $d$-dimensional momentum ${\bf k}$, and thus takes the expected standard form. In fact demanding reality of $\delta g_{\mu\nu}({\bf x})$, reproduces the familiar constraint: 
$f^\ast_{\bf k}(t) = (-1)^{d+1} f_{-{\bf k}}(t)$ with real $e_{\mu\nu}({\bf k}, s)$ and $\psi^\ast_{\bf k}({\bf x}) = 
\psi_{-{\bf k}}({\bf x})$. Combining all these we can then perform the following series of manipulations from \eqref{dholi}:
\bg\label{pimuse}
\delta g_{\mu\nu}(x) &=& {1\over 2} \sum_{s = \pm}\int_{-\infty}^{+\infty} d^d{\bf k}~ e_{\mu\nu}({\bf k}, s)\left[\left(f_{\bf k} 
+ b^2 {\partial \over  \partial f^\ast_{\bf k}} + f_{\bf k} 
- b^2 {\partial \over  \partial f^\ast_{\bf k}}\right)\psi_{\bf k}({\bf x})\right] \\
& = & {1\over 2} \sum_{s = \pm}\int_{-\infty}^{+\infty} d^d{\bf k}~ e_{\mu\nu}({\bf k}, s)\left[\left(f^\ast_{\bf k} 
+ b^2 {\partial \over  \partial f_{\bf k}}\right)\psi_{-\bf k}({\bf x}) + 
\left(f_{\bf k} 
- b^2 {\partial \over  \partial f^\ast_{\bf k}}\right)\psi_{\bf k}({\bf x})\right], \nonumber \nd
where in the first line we simply added and subtracted a derivative piece, but in the second line a redefinition of the variable ${\bf k}$ puts it in a much more suggestive format. The quantity $b^2$ appearing in 
\eqref{pimuse} is related to the ground state wave-function of the configuration space (not to be confused with the wave-functions in spacetime!), in the following way:
\bg\label{hkpgkm}
\Psi_0(f_{\bf k}) \equiv \langle f_{\bf k} \vert 0 \rangle = {\cal N}_{\bf k}~ 
{\rm exp} \left(- {\vert f_{\bf k}\vert^2 \over b^2}\right), \nd
with ${\cal N}_{\bf k}$ forming the normalization constant, and $b^2$ provides the Gaussian width of the ground state. Note that this wave-function is only for the mode ${\bf k}$, and the generic wave-function
for the ground state $\vert 0 \rangle$ is a matrix product of the wave-functions of the form \eqref{hkpgkm}. 
Clearly {\it shifting} \eqref{hkpgkm} in the configuration space should give us the required 
coherent state for the mode ${\bf k}$, so the question is whether we can quantify the shift in a precise 
way\footnote{A more precise identification of the coherent states to the shifted {\it interacting} vacuum will be discussed in section \ref{sec2.4}. Meanwhile what we have here should suffice.}. This is where the decomposition \eqref{pimuse} pays off, because:
\bg\label{kkkbkb}
&&\left(f^\ast_{\bf k} + b^2 {\partial \over  \partial f_{\bf k}}\right) \Psi_0(f_{\bf k}) = 0 \nonumber\\
&& \left(f_{\bf k} - b^2 {\partial \over  \partial f^\ast_{\bf k}}\right) \Psi_0(f_{\bf k}) = 2{\cal N}_{\bf k} f_{\bf k} 
~{\rm exp} \left(- {\vert f_{\bf k}\vert^2 \over b^2}\right), \nd
 implying that the first operator {\it annihilates} the vacuum wave-function whereas the second operator 
 creates a new wave-function. The new wave-function is exactly proportional to the first excited state of a 
 harmonic oscillator, implying that the second operator acts as a {\it creation} operator! These are then the 
 Schr\"odinger representations of the familiar creation and the annihilation operators. Thus we can identify:
 \bg\label{janem}
 a_{\bf k} \equiv  \sqrt{\omega^{(\psi)}_{\bf k}\over 2}\left(f^\ast_{\bf k} + b^2 {\partial \over  \partial f_{\bf k}}\right),~~~~ a^\dagger_{\bf k} &\equiv & \sqrt{\omega^{(\psi)}_{\bf k}\over 2}
 \left(f_{\bf k} - b^2 {\partial \over  \partial f^\ast_{\bf k}}\right), \nd
 connecting us with the mode expansion \eqref{virgmads} proposed earlier provided we identify $a_{\bf k}$ 
 and $a^\dagger_{\bf k}$ from \eqref{janem} with $a_s({\bf k})$  and $a^\dagger_s({\bf k})$ respectively from 
 \eqref{racadams}, and go to the Heisenberg picture. The spin $s$ informations in the definitions of the creation and the annihilation operators 
 in \eqref{racadams} are redundant because for either choices of $s$ in \eqref{racadams} the creation or the annihilation operators remain unchanged (so we will ignore them in the subsequent discussion).  Note that imposing 
 $[a_{\bf k}, a^\dagger_{{\bf k}'}] = \delta^{10}({\bf k} - {\bf k}')$ makes 
 $b = \left(\omega_{\bf k}^{(\psi)}\right)^{-1/2}$, which is consistent from \eqref{spolley}.
 Finally the effective {\it space-time} wave-function for a given mode appears naturally from the one-point function (treating $\delta {\bf g}_{\mu\nu}(x, s)$ as an operator):
 \bg\label{sondre}
 \langle 0 \vert \delta {\bf g}_{\mu\nu}(x, s) \vert {\bf k}\rangle = e_{\mu\nu}({\bf k}, s) \psi_{-{\bf k}}({\bf x}) 
 ~{\rm exp}\left(i\omega_{\bf k}^{(\psi)}t\right), \nd
 which is as one would expect for any fluctuation over a solitonic background provided we identify 
$\psi_{-{\bf k}}({\bf x}) = \psi^\ast_{{\bf k}}({\bf x})$. This again confirms the fact that modes over a solitonic background may have non-trivial spatial {\it wave-functions}, if $h_2 = h_2(y, {\bf x})$ in \eqref{betbab3}, but the temporal dependences remain simple with time-independent frequencies.  
  
On the other hand, in the same space-time, {\it i.e.} over the solitonic background \eqref{betbab3},  we can construct fluctuations \eqref{dholi} that could have different interpretations. For example a kind of fluctuation in eleven
dimensional space-time that we want to reproduce would be:
\bg\label{thequiet}
 \langle \delta {\bf g}_{\mu\nu}(x, y, z)\rangle_{\Psi^{(\psi)}}  &=&  \sum_{s = \pm} \int {d^{10}{\bf k}\over 
 (2\pi)^{10}}~ 
 b^{(\psi)}_{\bf k}e_{\mu\nu}({\bf k}, s) ~{\hat\psi}_{\bf k}({\bf x}, y, z) ~{\rm exp}\Big(-i {\hat\omega}^{(\psi)}_{\bf k}(t) t\Big)\\
&+& \sum_{s = \pm} \int {d^{10}{\bf k}\over (2\pi)^{10}}~ c^{(\psi)}_{\bf k} e^\ast_{\mu\nu}({\bf k}, s)~{\hat\psi}_{-{\bf k}}({\bf x}, y, z) ~{\rm exp}\Big(+ i {\hat\omega}^{(\psi)}_{\bf k}(t) t\Big), \nonumber 
 \nd
where the subscript $\Psi^{(\psi)}$ denotes some state over the solitonic vacuum  \eqref{betbab3}; and 
($b^{(\psi)}_{\bf k}, c^{(\psi)}_{\bf k}$) are coefficients that only depend on ${\bf k}$. As we saw before for the effective case in \eqref{dholi}, fluctuations on the RHS of \eqref{thequiet} could be achieved with a choice of $f({\bf k}, k_0)$ taking the form \eqref{lizshue}. However in quantum theory, it is more important to find a {\it state} $\Psi^{(\psi)}$ that reproduces the fluctuations as an expectation value over the state itself. Question then is: what would be the form of $\Psi^{(\psi)}$? From the RHS of \eqref{thequiet}, one might presume
$\Psi^{(\psi)}$, for a given mode ${\bf k}$, to be a state of the following form: 
\bg\label{bsimple}
\left\vert \Psi^{(\psi)}_{\bf k}(t)\right\rangle = \sum_n c_{n}^{(\psi)}({\bf k}, t) 
~{\rm exp}\left[-i\left(n + {1\over 2}\right) \omega_{\bf k}^{(\psi)} t\right] 
\vert n; {\bf k}, \psi_{\bf k}\rangle, \nd
in the configuration space, expressed in terms of a linear combinations of the eigenstates with time-dependent coefficients. This extra time-dependence is necessary 
because the expectation value \eqref{thequiet} involve mode expansions from 
\eqref{virgmads} that could only relate one up or one down states in the configuration space.
Thus only a linear combination of eigenstates for a given momentum ${\bf k}$ with time-dependent coefficients could provide the temporal behavior with frequencies 
${\hat\omega}^{(\psi)}_{\bf k}(t)$ as in \eqref{robbiem}. However we will also need to worry about an overlap 
integral of the form:
\bg\label{tapsi}
\int d^{10}x ~{\psi}_{\mp{\bf k}}({\bf x}, y, z)  {\hat\psi}_{\pm{\bf k}'}({\bf x}, y, z)~h^{-1}_2({\bf x}, y) ~h_1^{4/3}(y)
 \equiv r({\bf k}, {\bf k}'), \nd 
 between the spatial wave-function ${\psi}_{{\bf k}}({\bf x}, y, z)$ over the solitonic background 
 \eqref{betbab3} and the spatial wave-function ${\hat\psi}_{{\bf k}'}({\bf x}, y, z)$ over the M-theory uplifted background 
 \eqref{vegamey3}. This overlap condition should in turn be compared to the orthogonality condition that appeared in footnote \ref{sarabig}. The question now is whether we can quantify the function
 $r({\bf k}, {\bf k}')$. For this, note that at any given time $\sqrt{\Lambda} \vert t \vert \equiv T_0$, the uplifted 
 metric \eqref{vegamey3} resembles the solitonic background \eqref{betbab3} by some redefinitions of the coordinates by constant factors, implying that the wave-function ${\hat\psi}_{{\bf k}}({\bf x}, y, z)$ cannot be very different from ${\psi}_{\bf k}({\bf x}, y, z)$. In other words, we can expect:
 \bg\label{sabibut}
{\hat\psi}_{\bf k}({\bf x}, y, z) = \int d^{10}{\bf k}' ~r({\bf k}, {\bf k}') {\psi}_{{\bf k}'}({\bf x}, y, z) \approx
\sum_{{\bf k}'} r_{{\bf k}{\bf k}'}~ {\psi}_{{\bf k}'}({\bf x}, y, z), \nd
where in the second equality we have assumed the wave-functions are discrete ({\it i.e.} in a box). Our discussions above will tell us that the function $r({\bf k}, {\bf k}')$ is sharply peaked near ${\bf k}' = {\bf k}$, 
implying that the overlap is sub-leading when ${\bf k}' \ne {\bf k}$, in other words 
$r({\bf k}, {\bf k}') \approx r_{\bf k} \delta^{10}({\bf k} - {\bf k}')$.  
This further means that $c_{n}^{(\psi)}({\bf k}, t)$ coefficients in \eqref{bsimple} are related by the following set of equations, the first one being\footnote{Note that for the generic relation, without imposing any specific condition on the overlap function $r({\bf k}, {\bf k}')$, one has to start with the following state: 
$$\left\vert \Psi^{(\psi)}(t) \right\rangle \equiv \left\vert \Psi^{(\psi)}_{{\bf k}_1}(t) \right\rangle \otimes
\left\vert \Psi^{(\psi)}_{{\bf k}_2}(t) \right\rangle .... \equiv \otimes_{\bf k} \left\vert \Psi^{(\psi)}_{{\bf k}}(t) \right\rangle$$
which is constructed out of the product of all allowed momenta ${\bf k}$. Here we take them as discrete to give some meaning to the otherwise infinite product of states labelled by the continuous variable ${\bf k}$. 
Using this as the input on the LHS of \eqref{thequiet}, and comparing the terms proportional to 
$e_{\mu\nu}({\bf k}, s)$, gives us the following relation:
$$b_{\bf k}^{(\psi)} {\rm exp}\left(i\omega_{\bf k}^{(\psi)}t\right)~g^{(\psi)}({\bf k}, t) = \mathbb{Z}(t)
\int {d^{10}{\bf k}' \over \sqrt{2 \omega^{(\psi)}_{{\bf k}'}}} \left({{}^{\sum}_n~{}^{\sqrt{n}~r({\bf k}, {\bf k}')~ c_{n}^{(\psi)}({\bf k}', t) 
c_{n - 1}^{\ast(\psi)}({\bf k}', t)}_{} \over {}^{\sum}_n~ {}^{\left\vert c_n^{(\psi)}({\bf k}', t)\right\vert^2}_{}}\right)$$
which boils down to the first equality in \eqref{cambella} when $r({\bf k}, {\bf k}') \approx r_{\bf k} \delta^{10}({\bf k} - {\bf k}')$, upto the normalization factor of $\mathbb{Z}$. This factor is defined as:
$$\mathbb{Z}(t) \equiv {\rm exp}\left[\int d^{10} {\bf k} ~{\rm log} 
\left(\sum_n \left\vert c_n^{(\psi)}({\bf k}, t)\right\vert^2\right)\right]$$
which would not appear if we normalize the {\it overall} state  $\left\vert \Psi^{(\psi)}(t) \right\rangle$ from the start itself. Interestingly this normalization factor is time-dependent, and so is 
$\sum_n \vert c_n^{(\psi)}({\bf k}, t)\vert^2$.
We will assume such normalization is defined even for the state \eqref{bsimple}, so that 
with the delta function choice for $r({\bf k}, {\bf k}')$ we can reproduce \eqref{cambella} without any extra factors. 
Note however that the second equality in \eqref{cambella} is not much effected by choosing a generic form of $r({\bf k}, {\bf k}')$.}: 
\bg\label{cambella}
\sqrt{2\omega^{(\psi)}_{\bf k}} ~b_{\bf k}^{(\psi)} {\rm exp}\left(i\omega_{\bf k}^{(\psi)}t\right)~g^{(\psi)}({\bf k}, t) &=& \sum_n \sqrt{n}~{r}({\bf k}, t)~ c_{n}^{(\psi)}({\bf k}, t) 
c_{n - 1}^{\ast(\psi)}({\bf k}, t)\\
& \equiv & \sum_l  d_{l}^{(\psi)}({\bf k})
~{\rm exp}\left[-i\left(l + {1\over 2}\right)\omega_{\bf k}^{(\psi)} t\right], \nonumber \nd
where $g^{(\psi)}({\bf k}, t)$, as defined in \eqref{robbiem}, could in principle be a complex number; and 
$r({\bf k}, t) = {r_{\bf k}\over {}^{\sum}_n ~{}^{\vert c_n^{(\psi)}({\bf k}, t)\vert^2}_{}}$. Note that 
in the second line of \eqref{cambella}, we have expressed the sum of the products of the $c_{n}^{(\psi)}({\bf k}, t)$ coefficients as another linear combination expressed in terms of time-independent coefficients. 
In a similar vein, the other set becomes:
\bg\label{luss}
\sqrt{2\omega^{(\psi)}_{\bf k}} ~c_{\bf k}^{(\psi)} {\rm exp}\left(-i\omega_{\bf k}^{(\psi)}t\right) 
\left[g^{(\psi)}({\bf k}, t)\right]^{-1} &=& \sum_n \sqrt{n + 1}~ r({\bf k}, t)~ c_{n}^{(\psi)}({\bf k}, t) 
c_{n + 1}^{\ast(\psi)}({\bf k}, t)\nonumber\\
& \equiv & \sum_l e_{l}^{(\psi)}({\bf k}) 
~{\rm exp}\left[-i\left(l + {1\over 2}\right)\omega_{\bf k}^{(\psi)} t\right], \nonumber\\ \nd
where again we have expressed the sum of the products of $c_{n}^{(\psi)}({\bf k}, t)$ coefficients in terms of a series defined with coefficients $e_l^{(\psi)}$.
These 
coefficients are not hard to find, and one can show that:
\bg\label{ecuthbert}
&&d_l^{(\psi)}({\bf k}) \equiv \sqrt{2} \vert \omega^{(\psi)}_{\bf k}\vert^{3/2} b_{\bf k}^{(\psi)} \int_{-\infty}^{+\infty} dt ~g^{(\psi)}({\bf k}, t)~{\rm exp}\left[i\left(l + {3 \over 2}\right)\omega^{(\psi)}_{\bf k}t\right] \nonumber\\
&& e_l^{(\psi)}({\bf k}) \equiv \sqrt{2} \vert \omega^{(\psi)}_{\bf k}\vert^{3/2} c_{\bf k}^{(\psi)} \int_{-\infty}^{+\infty} dt \left[g^{(\psi)}({\bf k}, t)\right]^{-1} {\rm exp}\left[i\left(l - {1 \over 2}\right)\omega^{(\psi)}_{\bf k}t\right] , \nd
where the modulus keeps only the positive frequencies $\omega_{\bf k}^{(\psi)}$; with $g^{(\psi)}({\bf k}, t)$ 
 as in \eqref{robbiem} and $l \in \mathbb{Z}$. Note, because of the $3/2$ and $1/2$ modings, the integrands 
 above are not inverse of each other, and therefore the two coefficients $d_l^{(\psi)}({\bf k})$ and 
 $e_l^{(\psi)}({\bf k})$ are different functions of the momenta ${\bf k}$.  In IIB the only changes would be 
 $d^9{\bf k}$ instead of $d^{10}{\bf k}$ in \eqref{thequiet}, and $g^{(\psi)}({\bf k}, t)$ taking the form 
 \eqref{hlhplat} instead of the one derived from \eqref{horlat}. For example in IIB, $d_l^{(\psi)}({\bf k})$ becomes:
 \bg\label{annaM}
 d_l^{(\psi)}({\bf k}) = {\sqrt{2} b_{\bf k}^{(\psi)}\over \left\vert l + {3\over 2}\right\vert} \left\{ \delta({\bf k}) - 
 2T~\cos\left[\left(l+{3\over 2}\right){\bf k} T\right] + {2~\sin\left[\left(l+{3\over 2}\right){\bf k} T\right] \over 
 \left\vert l + {3\over 2}\right\vert {\bf k}}\right\}, \nd
where we have defined $\omega_{\bf k}^{(\psi)} = {\bf k}$, and $T$ as before is to be taken to be very large, but not infinite.  Expectedly the $T$ behavior is similar to what we saw earlier in \eqref{kirkoo}. Note however that the first term fixes ${\bf k}$ to ${\bf k} = 0$. Thus they are zero momentum states with arbitrary energy, so will appear as off-shell states.

There is yet another condition in addition to \eqref{cambella} and \eqref{luss}, which has to do with certain orthogonality relations between the states \eqref{spolley} and \eqref{bsimple}. To quantify this, let us express 
the wave-function \eqref{spolley} as  
$\left\langle \widetilde{g}_{\mu\nu}({\bf k})\Big\vert \Psi^{(\alpha)}_{\bf k}(t)\right\rangle$, where the ket would correspond to the coherent state for momentum ${\bf k}$. They are {\it not} orthogonal states, in addition to being over-complete, but we want to demand at least the following orthogonality conditions:
\bg\label{lucliu}
\left\langle\Psi^{(\psi)}_{{\bf k}'}(t) \Big\vert \delta{\bf g}_{\mu\nu}(x, y, z)\Big\vert \Psi^{(\alpha)}_{\bf k}(t)\right\rangle =  \left\langle\Psi^{(\alpha)}_{{\bf k}'}(t) \Big\vert \delta{\bf g}_{\mu\nu}(x, y, z)\Big\vert \Psi^{(\psi)}_{\bf k}(t)\right\rangle =   0, \nd
for all momenta ${\bf k}$ and for all time.  The above orthogonality is a bit harder to achieve in the light of the two additional constraints \eqref{cambella} and \eqref{luss}, but is not impossible given that the number of 
$c_n^{(\psi)}({\bf k}, t)$ coefficients are infinite in the definition of the state \eqref{bsimple}. In fact since 
the ket $\left\vert \Psi^{(\alpha)}_{\bf k}(t)\right\rangle$ can be expressed as a linear combinations of eigen-states, \eqref{lucliu} imposes two linear relations between the coefficients $c_n^{(\psi)}({\bf k}, t)$, which become one when the operator $\delta{\bf g}_{\mu\nu}(x, y, z)$ becomes real. In either case then, a generic state of the form:
\bg\label{elementary}
\left\vert \Psi_{\bf k}^{(c_1c_2)}(t)\right\rangle \equiv 
c_1 \left\vert \Psi_{\bf k}^{(\alpha)}(t)\right\rangle + c_2 \left\vert \Psi_{\bf k}^{(\psi)}(t)\right\rangle, \nd
where ($c_1, c_2$), which are constants\footnote{Note that the states  $\left\vert \Psi_{\bf k}^{(\alpha)}(t)\right\rangle$ and $\left\vert \Psi_{\bf k}^{(\psi)}(t)\right\rangle$ are not necessarily {\it orthogonal} to each other. 
Additionally two coherent states $\left\vert \Psi_{\bf k}^{(\alpha)}(t)\right\rangle$ and 
$\left\vert \Psi_{\bf k}^{(\beta)}(t)\right\rangle$ are also {\it not} orthogonal to each other unless 
$\left\vert \alpha_{\bf k}^{(\psi)} - \beta_{\bf k}^{(\psi)}\right\vert \gg 0$.}
independent of (${\bf k}, t$), 
would succinctly capture all the information that we  want once we take an expectation value of the operator 
 $\delta{\bf g}_{\mu\nu}(x, y, z)$ over \eqref{elementary}  and add the background solitonic value in the following way:
 \bg\label{tapseep}
   {\eta_{\mu\nu} \over h_2({\bf x}, y)} + {1\over \mathbb{N}} \int d^{10} {\bf k} \left\langle \Psi_{\bf k}^{(c_1c_2)}(t)\Big\vert    
   \delta{\bf g}_{\mu\nu}(x, y, z) \Big\vert  \Psi_{\bf k}^{(c_1 c_2)}(t) \right\rangle, \nd
   where $c_1 = 1, c_2 = 0$ reproduces the {\it classical} de Sitter background and $c_1 = 1, c_2 \ne 0$ reproduces the 
   {\it fluctuation} spectra over the classical de Sitter background with 
   $\mathbb{N} = \langle \Psi_{\bf k}^{(c_1c_2)}(t)\big\vert    
    \Psi_{\bf k}^{(c_1 c_2)}(t) \rangle$.   The miraculous thing is that all these are over the solitonic background \eqref{betbab3} and we are able to reproduce the de Sitter results as expectation values.   
   
The other metric modes of the theory, namely $g_{mn}, g_{\alpha\beta}$ and $g_{ab}$ would also be described using coherent state wave-functions of the form \eqref{spolley}, although specific details of the constructions might differ. The {\it interactions} that are required to construct the fluctuation wave-functions ($\eta_{\bf k}, \xi_{\bf k}, \zeta_{\bf k}$), {\it i.e.} the solutions of the corresponding Schr\"odinger equations, are necessarily with the soliton themselves and all interactions between the modes (including self-interactions) go in the definition of the interacting Hamiltonian. This interacting Hamiltonian  becomes more complicated once the UV degrees of freedom are further integrated out. One could also spell out equivalent operator and 
Feynman prescription as in \eqref{chukkam} and \eqref{poladom} respectively. These description do become simpler from $2+1$ dimensional perspective because the internal metric appear as scalar fields there. We will come back to this a bit later.

\subsection{Graviton numbers and excited coherent states \label{sec2.3}}

Let us take this opportunity to discuss two related topics, one, dealing with the number of gravitons in a coherent state and two, dealing with the excitation of a coherent state. The second case, {\it i.e.} the one related to excited coherent states, is a rich subject in itself and basically deals with the dynamics of a coherent state once we add $m$ number of gravitons. This was originally developed for the photon case by Agarwal and Tara \cite{agarwal}, and unfortunately here we will only be able to elaborate the bare minimum required for our purpose. Interested readers may want to go to the original papers in the subject starting with 
\cite{agarwal}. 

The question that we want to ask here is what happens when we fluctuate the coherent state, with wave-function ${\bf \Psi}^{\left[\alpha^{(\psi)}_{\mu\nu}({\bf k}, t)\right]} \left(\widetilde{g}_{\mu\nu}({\bf k}), t\right)$
as in \eqref{spolley}, by adding $m$ number of gravitons of momenta ${\bf k}$. The goal of the exercise is to see whether there is any tangible connection between graviton-added coherent states (GACS) and the state 
$\left\vert \Psi_{\bf k}^{(c_1c_2)}(t)\right\rangle$ constructed in \eqref{elementary}. 

To proceed, certain redefinitions of the coordinates in the configuration space might ease our computations. For example in the configuration space wave-function \eqref{spolley}, we can redefine the Fourier modes 
$\widetilde{g}_{\mu\nu}({\bf k})$ and $\widetilde{\alpha}_{\mu\nu}({\bf k}, t)$ as $e_{\mu\nu} f_{\bf k}$ and 
$e_{\mu\nu} \alpha^{(\psi)}_{\bf k}(t)$ respectively with the condition $e_{\mu\nu}e^{\mu\nu} \equiv 1$ with no sum over repeated indices implied. The $f_{\bf k}$ appearing here is the same $f_{\bf k}$ that appears in the definition of the creation and the annihilation operators in \eqref{janem}. Thus our way of expressing the Fourier modes would then be generic.
However the condition on the polarization tensor is not generic and there exists 
other ways to fix the product, but for our purpose we will stick to the simplest case here.  With these definitions, the form of our wave-function \eqref{spolley} simplifies from:
\bg\label{puchipu}
{\bf \Psi}^{\left[\alpha^{(\psi)}_{\mu\nu}({\bf k}, t)\right]} \left(\widetilde{g}_{\mu\nu}({\bf k}), t\right) \equiv 
\left\langle \widetilde{g}_{\mu\nu}({\bf k})\Big\vert \Psi^{(\alpha)}_{\bf k}(t)\right\rangle ~ \longrightarrow~
\Psi^{(\alpha)}_{\bf k}\left(f_{\bf k}, t\right) \equiv \left\langle f_{\bf k} \Big\vert \Psi^{(\alpha)}_{\bf k}(t)\right\rangle, \nd
implying that the coherent state that control the dynamics for any momentum ${\bf k}$ and any instant of time $t$ is $\left\vert \Psi^{(\alpha)}_{\bf k}(t)\right\rangle$. Question we want to ask is what happens when the coherent state is acted on by an operator of the form:
\bg\label{horlil}
{\cal G}^{(\psi)}\left(a_{\bf k} + a^\dagger_{\bf k}; t\right) \equiv \sum_{n = 0}^\infty C_{n{\bf k}}^{(\psi)}(t) 
\left(z_1a_{\bf k} + z_2 a^\dagger_{\bf k}\right)^n, \nd
where $C_{n{\bf k}}^{(\psi)}(t)$ are generic time-dependent coefficients and $z_i$ are time-independent constants. Note that this operation will lead to a state more generic than the usual Agarwal-Tara \cite{agarwal} type state\footnote{Recall that the Agarwal-Tara state \cite{agarwal} is for the limit where 
$(z_1, z_2) = (0, 1)$ and $C_{n{\bf k}}^{(\psi)}(t) = \delta_{nm}$ with $m \in \mathbb{Z}$. Here we will explore a more generalized version of this by keeping both $z_i$ non-zero and switching on time-dependent coefficients.}, but more importantly the time-dependent coefficients in \eqref{horlil} might tie up with what we discussed in the previous section. The state then becomes:

{\footnotesize
\bg\label{ckijwala}
\left\vert \Psi^{(\alpha g)}_{\bf k}\right\rangle \equiv {\cal G}^{(\psi)}\left(a_{\bf k} + a^\dagger_{\bf k}; t\right)  
\left\vert \Psi^{(\alpha)}_{\bf k}(t)\right\rangle = \sum_{n=0}^\infty C_{n{\bf k}}^{(\psi)}(t) 
\left(-{i\over \sqrt{2}}\right)^n \mathbb{H}_n\left[{i\left(z_2a^\dagger_{\bf k} + z_1\alpha^{(\psi)}_{\bf k}(t)\right) \over 
\sqrt{2z_1 z_2}}\right] \left\vert \Psi^{(\alpha)}_{\bf k}(t)\right\rangle, \nonumber\\ \nd}
where $\mathbb{H}_n(x)$ are the Hermite polynomials,  now expressed in terms of the creation operator 
$a^\dagger_{\bf k}$ for any given mode ${\bf k}$. The coherent states $\left\vert \Psi^{(\alpha)}_{\bf k}(t)\right\rangle$, on the other hand, may be expressed as  linear combinations of the eigenstates
${\rm exp}\left[-i\left(n + {1\over 2}\right) \omega_{\bf k}^{(\psi)} t\right] 
\vert n; {\bf k}, \psi_{\bf k}\rangle$, which immediately ties up \eqref{ckijwala} to \eqref{bsimple}. This means the coefficients $C_{n{\bf k}}^{(\psi)}(t)$ of \eqref{ckijwala} should be {\it related} to the coefficients 
$c_n^{(\psi)}({\bf k}, t)$ of \eqref{bsimple}. The relation is not too hard to find, and may be expressed as:

{\footnotesize
\bg\label{vanpelt}
c_m^{(\psi)}({\bf k}, t) = \sum_{n = 0}^\infty (-i)^n C_{n{\bf k}}^{(\psi)}({t})~
\mathbb{H}_n\left[{i\sqrt{\omega_{\bf k}^{(\psi)}\over z_1 z_2}}\left(z_2 f_{\bf k} - z_2 b^2 {\partial\over \partial f^\ast_{\bf k}} 
+ {z_1 \alpha_{\bf k}^{(\psi)}(t)\over \sqrt{2\omega_{\bf k}^{(\psi)}}}\right)\right] ~
{\rm exp}\left(-{\left\vert \alpha_{\bf k}^{(\psi)}(t)\right\vert^2 \over 2}\right)
{\left(\alpha_{\bf k}^{(\psi)}(0)\right)^m\over \sqrt{2^nm!}}, \nonumber\\ \nd}
where $b^2$ is the same parameter that appears in the vacuum wave-function \eqref{hkpgkm} and we have expressed the Hermite polynomial of operators in terms of Schr\"odinger representation so that it can directly act on the wave-functions in the configuration space. In this sense \eqref{vanpelt} is more like an 
{\it operator} relation where the LHS should be thought of as an identity operator modulated by the constant 
factor of $c_m^{(\psi)}({\bf k}, t)$ whereas the RHS is a sum of operators in Schr\"odinger formalism. One could in principle work out an operator free relation between the coefficients $c_m^{(\psi)}({\bf k}, t)$
and $C_{n{\bf k}}^{(\psi)}({t})$ by acting the creation operator $a_{\bf k}^\dagger$ inside the Hermite polynomial iteratively from \eqref{ckijwala}, but since this will lead to no new physics beyond the fact that the coefficients are connected, we will refrain from indulging in a more convoluted exercise here. Instead, from the fact that the two states $\left\vert \Psi^{(\alpha g)}_{\bf k}\right\rangle$ from \eqref{ckijwala} and 
$\left\vert \Psi^{(\psi)}_{\bf k}\right\rangle$ from \eqref{bsimple} are related, we can then propose that the state 
$\left\vert \Psi^{(c_1c_2)}_{\bf k}\right\rangle$ from \eqref{elementary} may be directly related to the coherent state $\left\vert \Psi^{(\alpha)}_{\bf k}\right\rangle$ via the following relation:
\bg\label{ferelisa}
\left\vert \Psi^{(c_1c_2)}_{\bf k}(t)\right\rangle = \left[c_1 + c_2 ~{\cal G}^{(\psi)}(a_{\bf k} + a^\dagger_{\bf k}; t)\right] \left\vert \Psi^{(\alpha)}_{\bf k}(t) \right\rangle, \nd
where the operator ${\cal G}^{(\psi)}(a_{\bf k} + a^\dagger_{\bf k}; t)$ is defined in \eqref{horlil}. The relation 
\eqref{ferelisa} is valid for all time $t$, and we get pure coherent state for vanishing\footnote{Interestingly, for non-vanishing $c_2$ but vanishing $z_2$ in \eqref{horlil}, we get back the coherent state.}
$c_2$. The relation 
\eqref{ferelisa} also leads us to conclude the following:
\newline
\newline
\noindent\fbox{%
    \parbox{\textwidth}{%
    Four-dimensional de Sitter space is a {\bf Glauber-Sudarshan} state \cite{sudarshan, glauber} in string theory, or alternatively,  a coherent state in string theory. Similarly {\it fluctuations} over a de Sitter space 
    appear from a generalized {\bf Agarwal-Tara} state \cite{agarwal}, or alternatively, from a generalized graviton-added coherent state. Both these descriptions are over  supersymmetric 
    Minkowski backgrounds, or more generically, over  supersymmetric solitonic backgrounds.    
   }%
}

\vskip.15in

\noindent In the rest of the paper we will make the above statement more precise and concrete by answering many 
issues that may arise in an actual realization of de Sitter space as a coherent state. 

The realization 
\eqref{ferelisa} tells us that the fluctuations over a de Sitter space in our analysis can be inferred by adding extra gravitons to our coherent state. Recall that the coherent state is already a condensation of gravitons, so the natural question is to ask about the number of gravitons in a state like \eqref{ferelisa}. Such an analysis should shed some light not only on the entropy of the de Sitter space itself but also on how the entropy changes by adding fluctuations over the de Sitter space. First however we should determine the total number of gravitons packed in the coherent states for all the allowed {\it accessible} modes ${\bf k}$.  The wave-function for such a state may be represented by the following integral representation:
\bg\label{tutusubal}
\Psi^{(\alpha)}(f, t) = \left\langle f \Big\vert \Psi^{(\alpha)}(t)\right\rangle \equiv
{\rm exp}\left(\int_{-\infty}^{+\infty} d^{10}{\bf k} 
~{\rm log}\left\langle f_{\bf k} \Big\vert \Psi^{(\alpha)}_{\bf k}(t)\right\rangle\right), \nd
where the ket $\vert f\rangle$ denotes the coordinates in the configuration space for all the modes ${\bf k}$. We have also used the simplifying notation as in \eqref{puchipu} which, although useful to avoid clutter, loses the information about the fact that it is only the $\psi_{\bf k}({\bf x}, y, z)$ part of the whole system. This means   
the above wave-function is still {\it not} the full wave-function in the configuration space that reproduces the M-theory space-time metric configuration 
\eqref{vegamey3} as the most probable outcome. Nevertheless it is a useful guide for what is about to follow. For example, 
the number $N^{(\psi)}$ of gravitons in such a state is then the standard expectation value of the {\it number operator} over the state \eqref{tutusubal}. Since we know the precise wave-functions for every mode ${\bf k}$ from \eqref{spolley}, the number of gravitons becomes:
\bg\label{triplets}
N^{(\psi)} \equiv \int_{-\infty}^{+\infty} d^{10}{\bf k} ~\left\vert \alpha_{\bf k}^{(\psi)}(0) \right\vert^2, \nd
which, as we warned before, is not the full answer yet. It only tells us about the number of gravitons with space-time wave-functions as in \eqref{sondre}. What about the number of gravitons in a state like 
\eqref{ckijwala}? Can we pack arbitrary number of gravitons in such a state? This is where the issue of back-reaction comes in.

A necessary requirement for the GACS to be identified as fluctuations over a de Sitter space-time would be that the back-reaction corresponding to this state is under control. In usual perturbation theory over de Sitter, the standard back-reaction constraint for such a system would be to imply that the energy in the fluctuation fields are much smaller than the energy density of the background. In our case, the analogous relation would imply that the energy density of the fluctuations in the GACS state has to be significantly smaller compared to the energy density of our Glauber-Sudarshan state. The subtlety for our construction lies in the fact that both these states are constructed over a solitonic vacuum and one cannot use a simple expression, such as ${\rm M}_{p}^2 {\rm H}^2$, to characterize the Hubble scale of the background\footnote{Although, note that, we expect such an effective description to emerge from our Glauber-Sudarshan wave-function just as a cosmological constant is emergent in this case.}. 

Having said this, it is easy to see that the energy in the  wave-function \eqref{spolley} can be calculated analogously to the way the number of gravitons were calculated in \eqref{triplets}, and is given by:
\begin{eqnarray}\label{BG_energy}
	E^{(\psi)} \equiv \int_{-\infty}^{\infty} d^{10} {\bf k}~ \omega^{(\psi)}_{\bf k}\,\left\vert \alpha_{\bf k}^{(\psi)}(0) \right\vert^2, 
\end{eqnarray}
keeping in mind, as usual, that this only corresponds to gravitons with space-time wave-functions as in 
\eqref{sondre}.
In a similar vein, one can calculate the energy of the gravitons packed in the GACS state \eqref{tapsi} as:
\begin{eqnarray}\label{Fluc_energy}
E^{(\alpha\, g)} \equiv \sum_{n}\,\int_{-\infty}^{\infty} d^{10} {\bf k}~ \omega^{(\psi)}_{\bf k}\,n\,\left\vert c_{n}^{(\psi)}({\bf k}, 0) \right\vert^2 .
\end{eqnarray}
At first sight, it might seem odd that there is no $\alpha^{(\psi)}_{{\bf k}}$ dependence of this expression since we have stressed that fluctuations over de Sitter space as a GACS. However, keep in mind that while expressing the state \eqref{tapsi} as a generalized Agarwal-Tara state in \eqref{ferelisa},  we have expressed the coefficients $c_n^{(\psi)}$ in terms of the coefficients  $C_{n\,{\bf k}}^{(\psi)}$ (see \eqref{vanpelt}), we had to include several terms which depend on $\alpha^{(\psi)}_{{\bf k}}$. Now that we are assured of the consistency of our treatment, let us finally recall that our generic state from \eqref{elementary}, or in \eqref{ferelisa}, comes with (constant) pre-factors $c_1$ and $c_2$. 

This is all good, and points towards the consistency of our treatment, both for the de Sitter space viewed as a coherent state and fluctuations over the de Sitter space viewed as a GACS. Therefore,
given these results, it is easy to express our backreaction constraint as:
\begin{eqnarray}\label{Backreaction}
	c_1^2\; E^{(\psi)} \gg ~ c_2^2\; E^{(\alpha\, g)}\,.
\end{eqnarray}
It is important for our generic state defined in \eqref{ferelisa} to always satisfy the above condition for it to be able to describe small fluctuations over de Sitter space-time. The limiting condition of $c_1 \rightarrow 1,\; c_2 \rightarrow 0$ is trivially satisfied, implying that there exists no extra condition for the existence of our Glauber-Sudarshan de Sitter state, as it should. Note that we are not suggesting that the above condition must always be valid for the state \eqref{ferelisa}, but rather that if such a violation occurs, one cannot interpret the system as a de Sitter state, with small fluctuations on it, over a solitonic vacuum. 

Another thing to note from \eqref{triplets}, \eqref{BG_energy} and \eqref{Fluc_energy} is that we have integrated over {\it all} allowed momenta, implying an access to arbitrarily short distances. Such an analysis has to be reconsidered in the light of the Wilsonian effective action, which allows us to access momenta $\vert {\bf k} \vert \le {\rm M}_p$. The modes lying between ${\rm M}_p \le \vert{\bf k}\vert \le \Lambda_{\rm UV}$, where $\Lambda_{\rm UV}$ is the short distance cut-off, are integrated out resulting in a non-trivial effective action. Does our de Sitter space, resulting from the coherent state construction, survive the tower of quantum corrections coming from integrating out momentum shell from the cut-off $\Lambda_{\rm UV}$ to ${\rm M}_p$? This question clearly {\it cannot} be answered from what we did so far: while certain interactions were entertained in the construction of the solitonic vacuum and from there the coherent states, our analysis no way mixes the ${\bf k}$ and ${\bf k}'$ modes in any way. In fact not only the modes don't mix, the various sectors represented by the four set of wave-functions in \eqref{hudson} do not mix either. However this is not the only short-coming: the worse is yet to come. There are also modes coming from the G-fluxes, that will have their own sectors represented by similar spatial wave-functions. These modes should mix amongst each other, and they should also mix with the modes of the gravitational sector that we made meticulous efforts to construct in the previous sections. In addition to that there are also higher order perturbative and non-perturbative, as well as local and non-local, quantum corrections (including topological ones!).  How do we know, when we let everything mix amongst each other, the de Sitter state would survive in the final theory?    

The above question looks almost like an impossible question to answer, but if we carefully analyze the situation, this may not be that difficult. In the following section we will start by discussing one possible approach to address this question.

\subsection{The interacting quantum vacuum and coherent states \label{sec2.4}}

A way to address question like this is to start by laying out all the contents of our interacting configuration space. Needless to say, from eleven-dimensional point of view, these interactions are going to be highly non-trivial.   
First, however there is some light at the end of the tunnel: the Wilsonian analysis {\it can} be performed because the modes, at least what we argued from \eqref{hudson}, do not have time-dependent frequencies although their spatial wave-functions could be non-trivial implying, in turn, that there are {\it no} trans-Planckian issues plaguing our analysis. Secondly, there is a possibility that mixing of the modes from each sector ({\it i.e.} from the gravitational as well as the G-fluxes)  will allow us to create new sectors with their own mixed spatial wave-functions, and with new creation and annihilation operators, on which we can have our coherent states. The vacuum of the mixed sector will be non-trivial, which is nothing but the {\it interacting vacuum} generated from:
\bg\label{novocaine}
\vert \Omega(t) \rangle ~\propto ~ \lim_{T \to \infty(1-i\epsilon)}{\rm exp} \left(-i\int_{-T}^t d^{11}x ~{\bf H}_{\rm int}\right) \vert 0 \rangle, \nd 
where ${\bf H}_{\rm int}$ is the interacting Hamiltonian in M-theory that we will specify soon. In fact 
${\bf H}_{\rm int}$ contains all information about the local and non-local, that include the perturbative, non-perturbative and topological, quantum corrections.  The state that we are looking for, in light of what we discussed earlier, and in the fully interacting theory, may be expressed as:
\bg\label{laudern}
\Psi_\Omega^{(\alpha)}(f, t) \equiv \big\langle f \big\vert \mathbb{D}(\alpha(t)) \big\vert \Omega(t) \big\rangle, \nd
where the ket $\vert f \rangle$, as before, is the coordinate of the interacting configuration space, 
$\vert\Omega(t)\rangle$ is the same interacting vacuum as in \eqref{novocaine}, and 
$\mathbb{D}(\alpha(t))$ is the displacement operator that {\it shifts} the interacting vacuum in the configuration space. The question then is: does this create a coherent state in the interacting theory?

Even before we start answering this question, the very meaning of a {\it displacement operator} in an interacting theory is not clear. From the mode-by-mode analysis that we did above, an interacting theory will not only be highly anharmonic, to say the least, but will also have interactions between the modes themselves.  Such interaction would typically take us away from the simple-harmonic-oscillator regime, but if the interactions have perturbative expansions then we can at least formally write:

{\footnotesize
\bg\label{15movs}
\mathbb{D}(\alpha(t)) \big\vert \Omega(t) \big\rangle \propto \mathbb{D}(\alpha(t))\vert 0 \rangle 
+ \sum_{n= 1}^\infty {(-i)^n\over n!} \mathbb{D}(\alpha(t)) \int_{-T}^t dt_1....dt_n 
\mathbb{T} \left\{\prod_{i = 1}^n
\int d^{10} x_i {\bf H}_{\rm int}\left(t_i, {\bf x}_i, y_i, z_i\right)\right\}\vert 0 \rangle, \nonumber\\ \nd}
where in the second term we still have to allow $T \to \infty(1 - i\epsilon)$ to avoid other interacting states to emerge in the sum; and $\mathbb{T}$ denotes time-ordering. Here $\vert 0 \rangle$ is the solitonic vacuum whose wave-function may be described as:
\bg\label{carriep}
\Psi_0(f) \equiv \langle f\vert 0\rangle = {\rm exp}\left(\int_{-\infty}^{+\infty} d^{10}{\bf k} 
~{\rm log} ~\Psi_0(f_{\bf k})\right), \nd
with $\vert f\rangle$ denoting the coordinates in the configuration space with a wave-function 
$\Psi_0(f_{\bf k})$ for every mode ${\bf k}$. This wave-function will have further finer sub-divisions\footnote{Even further  if we want to incorporate all the spatial wave-functions describing the components of G-fluxes. We will avoid these complications for the time being and deal with them a bit later.}  if we want to describe all the spatial wave-functions in \eqref{hudson} for any given mode ${\bf k}$. Once we know the ground state wave-function, the constants of proportionalities 
in \eqref{novocaine} and \eqref{15movs} are both related to the overlap function 
$\langle\Omega(t)\vert 0\rangle$. 

The above identification \eqref{15movs} would still make no sense unless we identify the {\it displacement} operator $\mathbb{D}(\alpha(t))$ for the interacting theory. In our earlier analysis with coherent states, the displacement operator is defined by exponentiating the creation and the annihilation operators. For a highly interacting theory, there is no simple description of the creation and the annihilation operators, but we can define two operators that take the following form:
\bg\label{adler}
&&a_{\rm eff}({\bf k}, t) = a_{\bf k} + \sum_{l, n, m} \int d^{10}{\bf k}_1.....d^{10}{\bf k}_nd^{10}{\bf k}'_1.....
d^{10}{\bf k}'_m~c_{lnm} ~f_{l{\bf k}}\left(a_{{\bf k}_1}.....a_{{\bf k}_n}a^\dagger_{{\bf k}'_1}.....
a^\dagger_{{\bf k}'_m}; t\right)\nonumber\\
&&a^\dagger_{\rm eff}({\bf k}, t) = a^\dagger_{\bf k} + \sum_{l, n, m} \int d^{10}{\bf k}_1.....d^{10}{\bf k}_n
d^{10}{\bf k}'_1.....
d^{10}{\bf k}'_m~c^\ast_{lnm} ~f^\dagger_{l{\bf k}}\left(a_{{\bf k}_1}.....a_{{\bf k}_n}a^\dagger_{{\bf k}'_1}.....
a^\dagger_{{\bf k}'_m}; t\right), \nonumber\\ \nd
such that $a_{\rm eff}({\bf k}, t)$ annihilates\footnote{This will still {\it not} fix the form of $a_{\rm eff}$ unambiguously unless more conditions are specified. For the time being it will suffice to assume that at least a particular $a_{\rm eff}$ exists in the theory that mixes all the harmonic creation and annihilation operators for each modes ${\bf k}$ in the simplest possible way. More elaborations on this will be dealt in section \ref{sec3.3}. \label{bribeac}.} the interacting vacuum $\vert\Omega(t)\rangle$. Here
($a_{\bf k}, a^\dagger_{\bf k}$) are as defined in \eqref{janem}; and 
the conjugate transpose action on $f_{l{\bf k}}(...)$ acts in a standard way by complex conjugating the coefficients and converting $a_{\bf k} \to a^\dagger_{\bf k}$ with due considerations to the ordering of the operators. The operator definition in \eqref{adler} makes sense if all the dimensionless coefficients multiplying the operator products in the definition of $f_{l{\bf k}}$ are smaller than $c_{lnm}$ and $c^\ast_{lnm}$; and additionally 
${\bf Re}\left(c_{lnm}\right) << 1$  and ${\bf Im}\left(c_{lnm}\right) << 1$.  
 If this be the case\footnote{The coefficients $c_{lnm}$ are related to the coupling constants of the theory and therefore one would expect non-perturbative corrections like 
${1\over c_{lnm}}$ to appear too ($n, l$ and $m$ are not Lorentz indices!). In string theory, $c_{lnm}$, or any other coefficients, can only be proportional to $g_s$, the string coupling (which we shall specify later). As such one expects non-perturbative series in ${1\over g_s}$ to appear. 
However these inverse $g_s$ factors can be resummed as a resurgent trans series to  
${\rm exp}\left(-{1\over g^{1/3}_s}\right)$, or to
${\rm exp}\left(-{1\over c_{lnm}}\right)$ here. They become arbitrarily {\it smaller} than any polynomial powers of $g_s$ or $c_{lnm}$ in the limit $g_s << 1$ (or ${\bf Re}\left(c_{lnm}\right) << 1$  and ${\bf Im}\left(c_{lnm}\right) << 1$), and therefore can be ignored in \eqref{adler}. We will discuss more about this when we study the interacting Hamiltonian in M-theory in section \ref{sec3}.}, then the commutator brackets:
\bg\label{irenead}
&&\left[a_{\rm eff}({\bf k}, t), a^\dagger_{\rm eff}({\bf k}', t)\right] = \delta^{10}({\bf k} - {\bf k}') + 
{\cal O}\left(c_{lnm}\right)\nonumber\\
&& \left[a_{\rm eff}({\bf k}, t), a_{\rm eff}({\bf k}', t)\right] = 
{\cal O}\left(c^2_{lnm}\right) = \left[a^\dagger_{\rm eff}({\bf k}, t), a^\dagger_{\rm eff}({\bf k}', t)\right],
\nd 
would tell us how far are we from a simple-harmonic-oscillator description in the configuration space.  Note three things: one, the first commutator in \eqref{irenead} has ${\cal O}(c_{lnm})$ correction term whereas the second commutator has ${\cal O}(c^2_{lnm})$ correction term; two, due to the ${\cal O}(c_{lnm})$ correction term, $a^\dagger_{\rm eff}({\bf k}, t)$ cannot be defined as a standard creation operator like 
$a_{\bf k}^\dagger$;  
and three,  the appearance of $t$ in the definitions of the operators in \eqref{adler}. The former implies stronger suppression of the second commutator whereas the latter means the commutation relations \eqref{irenead} continue for the range of time that keeps all the dimensionless coefficients entering in the definition \eqref{adler} under perturbative control implying, in turn, that the two operators in the second set of commutators of \eqref{adler} remain orthogonal when ${\bf k} \ne {\bf k}'$ up-to  ${\cal O}\left(c^2_{lnm}\right)$. Therefore using \eqref{irenead}
we can give the following operator definition of 
$\mathbb{D}(\alpha(t))$:

{\footnotesize
\bg\label{maristone}
\mathbb{D}(\alpha(t)) & \equiv & {\rm exp}\left(-{1\over 2}\vert \alpha\vert^2\right) {\rm exp}\left(\alpha a^\dagger_{\rm eff}\right)
{\rm exp}\left(-\alpha^\ast a_{\rm eff}\right)\\
&= & {\rm exp}\left[\alpha a^\dagger_{\rm eff} - \alpha^\ast a_{\rm eff}
-{1\over 2}\vert \alpha\vert^2 + \sum_{n = 1}^\infty {1 \over n(n+1)} \int_0^1 dh \left(\mathbb{I} - 
{\rm exp}(\mathbb{L}_{\alpha a^\dagger_{\rm eff}}) 
{\rm exp}\left(h\mathbb{L}_{-\alpha^\ast a_{\rm eff}}\right)\right)^n \alpha^\ast a_{\rm eff}\right] \nonumber
\nd}
which would be similar in spirit of the definition of a displacement operator that was used to generate the coherent states earlier\footnote{Another definition of the displacement operator may be given by taking only the first two terms from \eqref{maristone}, namely 
$\mathbb{D}(\alpha(t)) = {\rm exp}\left(\alpha a^\dagger_{\rm eff} - \alpha^\ast a_{\rm eff}\right)$. This definition, although closer in spirit to the definition of displacement operator in free theory, does not reproduce the simple product relation in the first equality of \eqref{maristone}. In other words, if we want to express this definition of the displacement operator also as $e^{\bf A} e^{\bf B}$ where (${\bf A}, {\bf B}$) are 
two operators, they cannot be some simple combinations of $a_{\rm eff}$ and $a^\dagger_{\rm eff}$. Clearly either definitions would work in the limit $c_{lnm} \to 0$, but since we wish to explore dynamics governed by 
$c_{lnm} < 1$, the choice \eqref{maristone} is better suited for us.}.  We will call the shifted interacting vacuum $\mathbb{D}(\sigma)\vert \Omega\rangle$ as the {\it generalized Glauber-Sudarshan} state to distinguish it from the original Glauber-Sudarshan state created out of the shifted harmonic vacuum 
$\mathbb{D}_0(\sigma) \vert 0\rangle$.
 The second equality in \eqref{maristone} comes from using the generic form of the Baker-Campbell-Hausdorff relation.
The other terms may be defined as follows.
$\mathbb{I}$ is the identity operator (in the relevant basis), and we have used the following definitions of the products of operators and parameters in \eqref{maristone}:
\bg\label{moriarty}
&& \alpha a^\dagger_{\rm eff} = \int d^{10} {\bf k} ~e^{\mu\nu}({\bf k}, s)\alpha^{(\psi)}_{\mu\nu}({\bf k}, t) 
a^\dagger_{\rm eff}({\bf k}, t), ~
\vert\alpha\vert^2 = \int d^{10}{\bf k} ~\alpha^{(\psi)}_{\mu\nu}({\bf k}, t) \alpha^{\ast(\psi)\mu\nu}({\bf k}, t)
\\
&& \mathbb{L}_{\alpha a^\dagger_{\rm eff}} \left(\alpha^\ast a_{\rm eff}\right) = \int d^{10}{\bf k} d^{10}{\bf k}' 
e^{\mu\nu}({\bf k}, s) e^{\ast\sigma\rho}({\bf k}', s) \alpha_{\mu\nu}^{(\psi)}({\bf k}, t) 
\alpha_{\rho\sigma}^{\ast(\psi)}({\bf k}', t) \left[a^\dagger_{\rm eff}({\bf k}, t), a_{\rm eff}({\bf k}', t)\right],
 \nonumber \nd
where, as before, the repeated indices are {\it not} summed over as we are only dealing with the wave-functions (both spatial and in the configuration space) associated with the metric mode 
$g_{\mu\nu}(x, y, z)$. Note that the operator action of $\mathbb{L}_{\alpha a^\dagger_{\rm eff}}$ 
on $\alpha^\ast a_{\rm eff}$ from \eqref{moriarty}, when combined with \eqref{irenead}, generates one term that exactly cancels the $-{1\over 2} \vert \alpha\vert^2$ piece in \eqref{maristone}. This is as it should be, but now we see that more involved operator products also emerge in addition to the expected answers. However one would also like to compare the {\it difference} operator:
\bg\label{evikant}
\mathbb{Q}_1^{(\psi)} \equiv \left\vert {\rm exp}\left(\alpha a^\dagger_{\rm eff}\right) - \prod_{\bf k} 
~{\rm exp}\left(e^{\mu\nu}({\bf k}, s)\alpha^{(\psi)}_{\mu\nu}({\bf k}, t) 
a^\dagger_{\rm eff}({\bf k}, t)\right)\right\vert, \nd
where for simplicity we have taken discrete momenta ${\bf k}$ to give meaning to the second term above. 
We can similarly define $\mathbb{Q}_2^{(\psi)}$ by the following replacement: 
$a^\dagger_{\rm eff} \to a_{\rm eff}, \alpha \to \alpha^\ast$ and 
$\alpha^{(\psi)}_{\mu\nu} \to \alpha^{\ast(\psi)}_{\mu\nu}$ appropriately in \eqref{evikant}. The operator
$\mathbb{Q}_1^{(\psi)}$ is identically {\it zero} for free theory and can be made arbitrarily small when 
$c_{lnm} \to 0$, and therefore serves as a signature of how interacting the theory is. One may similarly define the conjugate difference operator 
$\vert \mathbb{Q}_2^{(\psi)} - \mathbb{Q}_1^{\dagger(\psi)}\vert$ which do not vanish either. Either of these serves as a good focal point to study interactions, but more importantly they signify how efficiently one may study the coherent states in an interacting theory using the operator definition \eqref{maristone}.

With these, we are almost ready to write down the wave-function in the configuration space for the interacting vacuum $\vert \Omega(t)\rangle$ when it is displaced by an 
amount\footnote{Here as a trial example we are displacing by an amount 
$\alpha^{(\psi)}_{\mu\nu}({\bf k}, t)$ from \eqref{cnelson}. As we shall see soon, such a choice reproduces 
correct answers {\it up to}  ${\cal O}\left({g_s^{|a|}\over {\rm M}_p^b}\right)$ corrections. The actual choice of shift that does not allow extra corrections terms is more subtle and we will have to wait till section \ref{sec3.3} to get more exact results. For the time being this will suffice. \label{joelle}}  
$\alpha^{(\psi)}_{\mu\nu}({\bf k}, t)$
for any mode ${\bf k}$. However since the modes mix non-trivially, the wave-function should be expressed, not in terms of individual modes ${\bf k}$, but in terms of the field itself. The field in question is the space-time metric component $g_{\mu\nu}({\bf x}, y, z)$ at any given instant of time. In the limit where 
$a_{\rm eff}({\bf k}, 0) \to a_{\bf k}$ and $a^\dagger_{\rm eff}({\bf k}, 0) \to a^\dagger$, {\it i.e.} when $c_{lnm} = 0$ (see footnote \ref{freevac}) the displacement operator is the free-field one and will be denoted by $\mathbb{D}_0(\alpha(t))$. This means we can define an operator\footnote{Expressing the displacement operator \eqref{maristone} as a linear combination of the free-field displacement operator $\mathbb{D}_0 (\alpha)$ and perturbatively controlled corrections $\delta\mathbb{D}(\alpha)$,
has an additional advantage of elucidating some of the properties expected of this operator. They are listed as follows. 
\bg\label{angeld}
&&\mathbb{D}^{-1}(\alpha) = \mathbb{D}_0^{-1}(\alpha) - \mathbb{D}_0^{-1}(\alpha) 
 \delta \mathbb{D}(\alpha) \mathbb{D}_0^{-1}(\alpha) + ....\nonumber\\
&&\mathbb{D}^\dagger(\alpha) = \mathbb{D}_0^\dagger(\alpha) + (\delta \mathbb{D})^\dagger(\alpha), ~~
 \mathbb{D}(-\alpha) = \mathbb{D}_0(-\alpha) + \delta \mathbb{D}(-\alpha),\nonumber\nd
which would tell us that while the equalities $\mathbb{D}_0^\dagger(\alpha) = \mathbb{D}_0^{-1}(\alpha) = 
\mathbb{D}_0(-\alpha)$ are exact, similar equalities for $\mathbb{D}(\alpha)$ are only true to 
${\cal O}\left(\delta \mathbb{D}(\alpha)\right)$, implying that $\mathbb{D}(\alpha)$ is only approximately unitary. In a similar vein one may easily see that:
\bg\label{alux}
\mathbb{D}^\dagger(\alpha) a_{\rm eff} \mathbb{D}(\alpha) & = & \left(\mathbb{D}_0^\dagger(\alpha) + (\delta \mathbb{D})^\dagger(\alpha)\right)\left(a + {\cal O}(c_{lnm})\right)\left(\mathbb{D}_0(\alpha) + \delta \mathbb{D}(\alpha)\right)\nonumber\\
& = & a + \alpha + {\cal O}\left(\delta \mathbb{D}(\alpha)\right)  + {\cal O}\left(\delta \mathbb{D}^\dagger(\alpha)\right)  + 
{\cal O}\left(c_{lnm}, \delta \mathbb{D}(\alpha)\right) + {\cal O}\left(\vert\delta \mathbb{D}(\alpha)\vert^2\right), \nonumber \nd
where the first two terms in the second equality would relate to the expected identity if one replaces 
$\mathbb{D}(\alpha)$ by $\mathbb{D}_0(\alpha)$. Taking the conjugate of the operator relation above would provide yet another identity. Clearly since the RHS of the above equation do not amount to 
$a_{\rm eff} + \alpha$, or,  from its conjugate identity, $a^\dagger_{\rm eff} + \alpha^\ast$, the state created by the action of $\mathbb{D}(\alpha)$ on the {\it interacting} vacuum $\vert\Omega\rangle$
will {\it not} be a coherent state, at least not exactly. This will also become clear from the wave-function \eqref{siortegas} discussed below.}: 
\bg\label{shelort}
\delta \mathbb{D}(\alpha(t)) = \mathbb{D}(\alpha(t)) - 
\mathbb{D}_0(\alpha(t)),\nd
which is by construction perturbatively controlled by the coefficients $c_{lnm}$, with $\mathbb{D}(\alpha(t))$ as in \eqref{maristone}.
It is also clear that there is a free field displacement operator for any mode ${\bf k}$, and they generically 
{\it commute} amongst each other. Therefore putting everything together, the wave-function for the shifted interacting vacuum can be expressed as:

{\footnotesize
\bg\label{siortegas}
&&\Psi^{(\alpha)}_{\Omega}\left(g_{\mu\nu}, t\right)  =  {\rm exp}\left[\int_{-\infty}^{+\infty} d^{10}{\bf k} ~{\rm log}
\left(\Psi^{(\alpha)}_{\bf k}\left(\widetilde{g}_{\mu\nu}({\bf k}), t\right)\right)\right]  + 
\int {\cal D}g'_{\mu\nu} \langle g_{\mu\nu} \big\vert \delta\mathbb{D}(\alpha(t))\big\vert g'_{\mu\nu}\rangle 
\Psi_0(g'_{\mu\nu})\\
& &~~ +  \sum_n {(-i)^n\over n!}\int {\cal D}g'_{\mu\nu} {\cal D}\hat{g}_{\mu\nu} 
\langle\mathbb{D}(\alpha(t))\rangle
\int_{-T}^t dt_1.....dt_n 
\langle \hat{g}_{\mu\nu}\big\vert \mathbb{T} \left\{\prod_{i = 1}^n
\int d^{10} x_i {\bf H}_{\rm int}\left(t_i, {\bf x}_i, y_i, z_i\right)\right\}\big\vert g'_{\mu\nu}\rangle
\Psi_0\left(g'_{\mu\nu}\right), \nonumber \nd}
where $\langle\mathbb{D}(\alpha(t))\rangle \equiv \langle g_{\mu\nu} \big\vert \mathbb{D}(\alpha(t))\big\vert \hat{g}_{\mu\nu}\rangle$, which is similar to how we defined the expectation value of $\delta\mathbb{D}(\alpha)$ above. The other parameters appearing above are the wave-functions that are defined as follows: 
$\Psi_0\left(g'_{\mu\nu}\right)$ is the vacuum wave-function in the configuration space defined over the free vacuum exactly as in \eqref{carriep}; and $\Psi^{(\alpha)}_{\bf k}\left(\widetilde{g}_{\mu\nu}({\bf k}), t\right)$
is the coherent state wave-function that we derived earlier in \eqref{spolley}. Here, as promised above, we have defined the wave-functions for {\it all} modes, but restricted ourselves to only the space-time component $g_{\mu\nu}$. There exists a more complete wave-function for all modes and all components, including the ones from the G-fluxes, but we will not discuss this right now. In fact the full wave-function is not necessary as \eqref{siortegas} is enough to elucidate the main point of our analysis: the {\it dominant} part of the shifted vacuum wave-function for the interacting theory is indeed given by the Glauber-Sudarshan wave-function for the harmonic theory. The remaining two terms of \eqref{siortegas} are the perturbative corrections 
controlled by $c_{lnm}$ from \eqref{maristone}.

The third term that appears in \eqref{siortegas} is interesting by itself. It involves the interacting Hamiltonian 
${\bf H}_{\rm int}$ from M-theory and is thus a much more complicated object. The action of this interacting Hamiltonian, acting on the field states $\vert g_{\mu\nu}\rangle$, is an integrated action, namely, that we integrate from $-T = -\infty$ (in a slightly imaginary direction) to the present time $t$ (or more appropriately 
$\sqrt{\Lambda} t$ to make this dimensionless). There are clearly three possible outcomes of such an integrated action: 

\vskip.1in

\noindent (1) The integrated action of the Hamiltonian ${\bf H}_{\rm int}$ over the range of time 
from $-T = -\infty$ till the present epoch $\sqrt{\Lambda} t$, along-with the effects from \eqref{shelort}, 
exactly {\it cancels} out so that the system continues to evolve classically over an indefinite period of time. 

\vskip.1in

\noindent (2) The combined action of the interacting Hamiltonian and the operator \eqref{shelort}, over the integrated period of time, allows the classical feature to persist exactly for a certain interval of time beyond which the system becomes truly quantum. The point of time at which such classical to quantum transition happens 
should be related to the quantum {\it break time} of the system\footnote{It is interesting to compare the 
quantum {\it break time} as proposed in \cite{dvaliG} and the one that we want to consider in \eqref{spamca}. Our choice of the temporal domain is motivated by the behavior of the string coupling, or more appropriately ${g_s^2\over \sqrt{h(y)}}$, till it hits strong coupling because beyond that we have no control on the dynamics. The behavior of the Glauber-Sudarshan state also becomes out of control once strong coupling sets in. This noticeable different approach of dealing with the quantum break time from \cite{dvaliG} has it's roots in the actual string dynamics underlying the system.}.

\vskip.1in

\noindent (3) The combined action of the interacting Hamiltonian and the operator \eqref{shelort}, do contribute non-zero values over the integrated period of time, but never {\it enough} to change substantially the behavior of the mode-by-mode wave-function \eqref{spolley}. As such the system continues to allow classical configurations, like \eqref{vegamey3} {\it with} perturbative corrections, either as most probable outcomes or as  expectation values.    

\vskip.1in

\noindent It is clear that we can safely disregard option 1, as it is highly unlikely that quantum corrections could cancel {\it exactly} to allow for the quantum system to evolve classically and over an indefinite period of time. The choice then is between options 2 and 3. In the following we will discuss which of the two options would favor our system.



\subsection{A path integral approach for the displaced vacuum \label{sec2.5}}

We will start by revisiting the computation of the expectation value in \eqref{chukkam}, but now using the path-integral formalism so that we can generalize this to extract relevant information from the displaced interacting vacuum \eqref{15movs}.

We will start with a simple example from free massive scalar field theory in $3+1$ dimensions with a mass term given by $m$. Our aim would be to reproduce the expectation value of the scalar field $\varphi$ on a coherent state 
$\vert\alpha\rangle$, using path integral formalism. The vacuum of this theory is simple, its given by 
$\varphi = 0$, which makes the analysis even simpler. In fact it also makes 
$\psi_{\bf k}({\bf x}) = {\rm exp}\left(-i{\bf k}\cdot {\bf x}\right)$. We could of course generalize this to allow more non-trivial solitonic solution, and go to the harmonic oscillator regime where the spatial wave-functions are given by a non-trivial function $\psi_{\bf k}({\bf x})$, but will avoid these complications for the time being. However since we want to use path integrals, we will resort to a more off-shell description of the fields as the following Fourier expansion:
\bg\label{aletex}
\varphi(x) = \int d^4 k ~\varphi({\bf k}, k_0) ~{\rm exp}\left(ik\cdot x\right), \nd
where we have used the mostly minus signature for this example, and $k \equiv ({\bf k}, k_0)$. There is no relation between $k_0$ and ${\bf k}$ right now, so the fields are off-shell.  We will also need to express the creation and the annihilation operators \eqref{janem} in terms of the Fourier modes \eqref{aletex} without involving any derivatives because the path integral formalism prefers fields instead of operators. In fact this is not hard and the answer is to replace these operators by:
\bg\label{janem2}
a_{\bf k} & = & \sqrt{\omega_{\bf k}\over 2}\left(\hat{\varphi}_{k} + {i\hat{\pi}_{k} \over \omega_{\bf k}} \right) ~\to ~
\left({\omega_{\bf k} + k_0}\right) \varphi_k
 \nonumber\\
 a^\dagger_{\bf k} & = & \sqrt{\omega_{\bf k}\over 2}\left(\hat{\varphi}^\ast_{k} - {i\hat{\pi}^\ast_{k} \over 
 \omega_{\bf k}} \right) ~\to ~
\left({\omega_{\bf k} + k_0}\right) \varphi^\ast_k, \nd
where, the hatted quantities are the operators\footnote{Note the absence of $\sqrt{2\omega_{\bf k}}$ in the second equalities of \eqref{janem2}, although the first equalities demand it. This is because the standard mode expansion in field theory appears with a measure ${d^3{\bf k} \over \sqrt{2\omega_{\bf k}}}$, and thus the creation and the annihilation operators are required to have this as prefactors (see \eqref{janem}). Once we go to the {\it field} description we allow off-shell quantities; however this is a bit puzzling from 
\eqref{pimuse} which is expressed in terms of on-shell quantities in the Heisenberg formalism (the integral is also $d^d{\bf k}$ and not $d^{d+1} k$). A simple way out is to define the measure as ${d^3 {\bf k} \over 2\omega_{\bf k}} \equiv \int_{k_0} d^4k ~\delta(k^2 - m^2)$ so that it may be expressed as a four-dimensional integral. One can then claim that the operator form, {\it i.e.} using \eqref{janem}, is a special case where we take a four-dimensional integral with fields and put it on-shell when we replace the fields by the operators \eqref{janem}. This introduces a pre-factor of 
$\omega_{\bf k}$ in the second equalities of \eqref{janem2} which cancel against the inverse $\omega_{\bf k}$ factors in the definition of the conjugate momentum. The extra factor of ${1\over 2}$, which appears from \eqref{pimuse}, cancels out also leaving us with the two rightmost quantities in \eqref{janem2} without any other numerical factors. It also matches up with \eqref{spolley} as well as with \eqref{ryacone}. As we shall see below, this helps us to get the appropriate residue at the pole $k_0 = \omega_{\bf k}$.} ; 
and we have used a slightly different definition from \eqref{janem} because of the change of signature, and $\varphi_k \equiv \varphi({\bf k}, k_0)$. The conjugate momentum\footnote{Here regarded as a field and not  as an operator.} ${\pi}_k \equiv -ik_0 {\varphi}_k$, and note the appearance of $\omega_{\bf k} + k_0$, $\omega_{\bf k}$ being the usual $\sqrt{{\bf k}^2 + m^2}$,
which is the sign that this is an off-shell definition. In fact these definitions help us to 
express\footnote{The expression inside the exponential factor in \eqref{ryacone} appears to break Lorentz invariance, because we chose to insert the factor $\omega_{\bf k} + k_0$ from the definition \eqref{janem2}.  
In fact Lorentz invariance is actually not broken because the quantity that appears on the exponential is not $\alpha_k$, rather 
$\widetilde{\alpha}_k$. They are related by: $$\widetilde{\alpha}_k \equiv (k_0 - \omega_{\bf k}) \alpha_1(k) + {\alpha_k \over k_0 + \omega_{\bf k}}$$
\noindent where $\alpha_1(k)$ is an arbitrary function of $k^2$. For the cases that we would be interested in, $\alpha_1(k) = 0$, so the term in the exponential will be simply $\alpha_k \varphi^\ast_k$, which in turn is a Lorentz invariant quantity. \label{vlake}}:
\bg\label{ryacone}
{\rm exp}\left(\alpha a^\dagger\right) \to  {\rm exp}\left[\int_{-\infty}^{+\infty}d^4 k 
\left({\omega_{\bf k} + k_0}\right) \widetilde{\alpha}_k \varphi^\ast_k\right], \nd
and similar description for the conjugate operator ${\rm exp}\left(\alpha^\ast a\right)$. Such a description is useful because we can define the coherent states as an action of $\hat{\mathbb{D}}_0(\alpha)$ over the free vacuum in the following way: 
\bg\label{phoen}
\hat{\mathbb{D}}_0(\alpha)\vert 0 \rangle \equiv
{\rm exp}\left(\alpha a^\dagger -{\vert\alpha\vert^2\over 2}\right) \vert 0 \rangle,\nd
with $\vert 0 \rangle$ being the free vacuum, with $\hat{\mathbb{D}}_0(\alpha)$ differing from 
${\mathbb{D}}_0(\alpha)$ by not being unitary. This will necessitate a division by $\langle\alpha\vert\alpha\rangle$ in the path-integrals to keep the normalization straight.
 Combining everything together, the expectation of the scalar field on the coherent state may be expressed as:
 \bg\label{urensis}
\langle{\bf\varphi}(x)\rangle_\alpha = {\langle\alpha\vert {\bf\varphi}(x) \vert \alpha\rangle \over 
\langle \alpha \vert \alpha \rangle} =  {\int {\cal D}\varphi ~e^{iS_0} \hat{\mathbb{D}}^\dagger_0(\alpha)~ \varphi(x) ~\hat{\mathbb{D}}_0(\alpha) \over 
\int {\cal D}\varphi ~e^{iS_0} \hat{\mathbb{D}}^\dagger_0(\alpha) \hat{\mathbb{D}}_0(\alpha)}, \nd
where $S_0$ is the action for the free scalar field and here it may be written as an integral of the form 
$S_0 \equiv \int d^4k (k^2 - m^2) \vert \varphi_k\vert^2$. The path integral \eqref{urensis} involve various nested integrals and therefore to perform the integrals efficiently it will be advisable to use discrete sum. 
This is all very standard, so we simply show one step of how to express the numerator of the path integral (the second equality in \eqref{urensis}) in the following way\footnote{We will analyze the path integral using Lorentzian signature although, from Wilsonian sense Euclidean signature serves better. We will resort to this a bit later when we actually discuss the Wilsonian effective action.}:

{\footnotesize
\bg\label{meyepolce}
\int {\cal D}\varphi ~e^{iS_0} \hat{\mathbb{D}}^\dagger_0(\alpha)~ \varphi(x) ~\hat{\mathbb{D}}_0(\alpha)
 &=&  \left(\prod_{k} \int d\left({\bf Re}~\varphi_k\right) d\left({\bf Im}~\varphi_k\right)\right)
  {\rm exp}\left[{i\over V}\sum_{k} \left(k^2 - m^2\right) \vert \varphi_k\vert^2\right] \\
&\times&
 {\rm exp} \left[{2\over V} \sum_{k'} \left({\omega_{{\bf k}'} + k'_0}\right)
\left({\bf Re}~\widetilde{\alpha}_{k'}~{\bf Re}~\varphi_{k'} + {\bf Im}~\widetilde{\alpha}_{k'}~{\bf Im}~\varphi_{k'}\right)\right]
\nonumber\\
&\times & {1\over V} \sum_{k''} ~{\rm exp}\left(ik''\cdot x\right)
\left({\bf Re}~\varphi_{k''} + i{\bf Im}~\varphi_{k''}\right)~{\rm exp}\left(-{1\over V}\sum_{k'} \alpha_{k'}
\alpha^\ast_{k'}\right), \nonumber \nd}
where $V$ is the volume of the four-dimensional space (which is finite and that makes all momenta discrete so that the sum could be performed). Even without introducing any simplifications, we see that if the momenta ($k, k', k''$) are all different the integral (or more appropriately the sum) {\it vanishes}. Thus the above integral can only give non-zero values if the momenta at every stage of the sum are related to each other. This is a crucial point and deserves some explanation. In the standard path integral, integrating a term {\it linear} in $\varphi_k$ would vanish because the gaussian functions from the action are all even functions. Here however there is an additional exponential piece which is expressed by linear powers of 
$\varphi_k$. As such this would shift the center of the gaussian giving a non-zero value for the integral.  
In a similar vein the denominator of the path integral in \eqref{urensis} will be exactly similar to 
\eqref{meyepolce} {\it without} the last term.  Now to determine the precise value of the integral \eqref{meyepolce}  we can make a simplifying assumption that $\widetilde{\alpha}_k$'s are real.  The quadratic pieces in $\varphi_k$'s then produce the necessary poles at $k_0 = \omega_{\bf k}$, rendering:
\bg\label{melmonr}
\langle{\bf\varphi}(x)\rangle_\alpha = \int {d^{3}{\bf k}\over 2\omega_{\bf k}}~{\bf Re}~\alpha_k ~
{\rm exp}\left(i\omega_{\bf k} t - i {\bf k}\cdot {\bf x}\right), \nd 
where the $dk_0$ integral takes care of the poles by inserting appropriate residues, with the off-shell part 
coming from an expression similar to \eqref{cnelson}. The result is close to what we had in \eqref{chukkam} once we replace the 
graviton field $g_{\mu\nu}$ by the scalar field and go to four-dimensions. The graviton description is over a solitonic background and thus we see that ${\rm exp}\left(i{\bf k}\cdot {\bf x}\right)$ in \eqref{melmonr} is to be replaced by $\psi_{\bf k}({\bf x})$ (the two results \eqref{melmonr} and \eqref{chukkam} appear to be complex conjugates of each other, once we replace $\psi_{\bf k}({\bf x})$ by $\psi_{-{\bf k}}({\bf x})$, because of the change of signature used here, as defined earlier). 

This roundabout way of getting the simple result \eqref{melmonr} (or \eqref{chukkam}) is not without its merit.
It prepares us to tackle a much more involved problem associated to the displaced interacting vacuum 
$\vert\Omega(t) \rangle$. However before moving ahead it will be instructive to resolve one puzzle associated to the path integral computation that we presented above. The puzzle involve the usage of 
$\hat{\mathbb{D}}_0(\alpha)$ instead of $\mathbb{D}_0(\alpha)$ in \eqref{urensis}. If we had used 
$\mathbb{D}_0(\alpha)$, the integral \eqref{urensis} or \eqref{meyepolce} would have vanished because 
of the unitary nature of $\mathbb{D}_0(\alpha)$ (it is no longer an operator in the Feynman formalism as seen from \eqref{ryacone}). The reason for this apparent discrepancy lies on the following {\it redundancy} of the free vacuum $\vert 0 \rangle$:
\bg\label{oshomsara}
f(a) \vert 0 \rangle \equiv \left(1 + \sum_{n = 1}^\infty c_n a^n\right)\vert 0 \rangle = \vert 0 \rangle, \nd
for $n \in \mathbb{Z}^+$ but $c_n$ any positive or negative number, with $a$ being the annihilation operator. This means that there is always an ambiguity when we choose $\mathbb{D}_0(\alpha)$: 
both $\mathbb{D}_0(\alpha)$  and $\mathbb{D}_0(\alpha) \mathbb{G}\left(f(a)\right)$, with $\mathbb{G}(f)$ being any functional of $f(a)$, would create the same coherent state implying, we can always use 
$\mathbb{G}(f)$ to eliminate the $a$ dependence of $\mathbb{D}_0(\alpha)$ and instead use the 
definition \eqref{phoen}, which is non-unitary. If we make the same change in \eqref{shelort}, then 
$\mathbb{D}(\alpha(t))$ in \eqref{maristone} may be re-expressed as $\mathbb{D}(\alpha) = 
\hat{\mathbb{D}}_0(\alpha) + \delta\mathbb{D}(\alpha)$, which is again non-unitary. 

Our aim now is to figure out which of the two options, options (2) or (3), in the previous sub-section, would qualify our system. For that we want to compare the expectation value of the metric operator on the coherent state $\vert\alpha\rangle$. In other words we want to determine $\langle {\bf g}_{\mu\nu} \rangle_\alpha$ with the assumption that:
\bg\label{bwstefen}
\mathbb{D}(\alpha(t)) \vert \Omega(t)\rangle \equiv \vert\alpha\rangle, \nd
and compare it with \eqref{urensis}. Our aim would be to see how much the expectation value differs from 
the expected answer \eqref{chukkam}. Any deviation from \eqref{chukkam} would imply the deviation of the configuration space wave-function from the Glauber-Sudarshan wave-function due to the 
interaction Hamiltonian ${\bf H}_{\rm int}$.  As such this might help us to quantify the contributions from the interaction Hamiltonian.

We will start by assigning an off-shell description of the metric component $g_{\mu\nu}$ as in 
\eqref{poladom}. This is similar to what we did for the scalar field case in \eqref{aletex}. One immediate use of such a description is that it helps us to express the operators by fields, in the same vein as in 
\eqref{ryacone}. In our case we expect\footnote{See footnote \ref{vlake} and \ref{joelle}.}:
\bg\label{ryacone2}
{\rm exp}\left(\alpha a_{\rm eff}^\dagger\right) \to  {\rm exp}\left[\int_{-\infty}^{+\infty}d^{11} k ~
\alpha^{(\psi)}_{\mu\nu}({\bf k}, k_0) \widetilde{g}^{\ast\mu\nu}({\bf k}, k_0) + ...... \right], \nd
where no sum over the repeated indices are implied here and the dotted terms are higher orders in 
$\alpha^{(\psi)}_{\mu\nu}({\bf k}, k_0)$ and  $\widetilde{g}^{\mu\nu}({\bf k}, k_0)$ that could mix the momenta as well as the spatial indices as expected from \eqref{adler}. One should also compare \eqref{ryacone2} to \eqref{moriarty} where $e^{\mu\nu}({\bf k}, s) a^\dagger({\bf k}, t)$ in \eqref{moriarty} goes in the definition of the {\it operator} 
$\widetilde{\bf g}^{\mu\nu}({\bf k}, t, s)$ which is then converted to the off-shell {\it field} 
$\widetilde{g}^{\mu\nu}({\bf k}, k_0)$ that appears in \eqref{ryacone2}. 

Our next couple of steps will be similar to what we had for the simple scalar field theory, namely, we define the operator $\mathbb{D}(\alpha)$ as in \eqref{phoen} with appropriate replacements from \eqref{ryacone2} and \eqref{moriarty}. This is of course motivated from the fact that $a_{\rm eff}({\bf k}, t) \vert \Omega(t)\rangle
= 0$ and therefore the interacting vacuum has similar redundancy as in \eqref{oshomsara}. This reproduces\footnote{Its interesting to note that the form of the displacement operator $\mathbb{D}(\alpha)$, as given by the first term in \eqref{juanivan}, is somewhat similar to exponentiating the {\it vertex} operator $\mathbb{V}$ to create gravitons in string perturbation theory, namely:
$${\rm exp}\left(\mathbb{V}\right) = {\rm exp}\left({1\over 4\pi\alpha'}\int d^2\sigma \sqrt{h} ~h^{ab} \partial_a {\bf X}^\mu 
\partial_b {\bf X}^\nu~{\bf g}_{\mu\nu}\right) \equiv {\rm exp}\left(\int d^2k ~\alpha^{\mu\nu}(k) 
\widetilde{g}_{\mu\nu}(k)\right)$$
\noindent where $\alpha'$ is related to the string length, ($a, b$) denote the two-dimensional world-sheet coordinates, and ${\bf X}^\mu$ is the standard space-time coordinate. This similarity is of course not accidental because exponentiating the vertex operator in string theory does create a coherent state of gravitons! Our construction here is from eleven-dimensional point of view, and therefore more generic because of the absence of world-sheet in M-theory. \label{vertex}}:

{\footnotesize
\bg\label{juanivan}
\mathbb{D}(\alpha) = {\rm exp}\left[\int_{-\infty}^{+\infty}d^{11} k ~ 
\alpha^{(\psi)}_{\mu\nu}({\bf k}, k_0) \widetilde{g}^{\ast\mu\nu}({\bf k}, k_0) -{1\over 2} \int_{-\infty}^{+\infty} 
d^{11}k~\alpha^{(\psi)}_{\mu\nu}({\bf k}, k_0) \alpha^{\ast(\psi)\mu\nu}({\bf k}, k_0)
+ .... \right], \nd}
where the repeated indices are not summed over, and the dotted terms carry over from \eqref{ryacone2}.
The eleven-dimensional integral tells us that the 
quantities appearing in \eqref{juanivan} is maximally off-shell. Putting everything together, the expectation value for metric component over the coherent state becomes:
\bg\label{camber}
\langle {\bf g}_{\mu\nu}(x, y, z)\rangle_{\alpha} = {\int \left[{\cal D}g_{\mu\nu}\right] ~e^{iS} ~
\mathbb{D}^\dagger(\alpha)~g_{\mu\nu}(x, y, z) ~\mathbb{D}(\alpha) \over 
\int \left[{\cal D}g_{\mu\nu}\right] ~e^{iS} ~
\mathbb{D}^\dagger(\alpha)~\mathbb{D}(\alpha)}, \nd
where $S$ is now the full interacting lagrangian of M-theory (which we will specify a bit later),  and the first term from evaluating the path integral is the solitonic background ${\eta_{\mu\nu} \over h_2^{2/3}(y, {\bf x})}$
as one would have expected. The other terms may be derived by computing the path integral carefully. As in 
\eqref{meyepolce}, the numerator of \eqref{camber} takes the following form:

{\footnotesize
\bg\label{meyepolce2}
&&\int \left[{\cal D}g_{\mu\nu}\right] ~e^{iS} {\mathbb{D}}^\dagger(\alpha)~ \delta g_{\mu\nu}  ~{\mathbb{D}}(\alpha)
 =  \left(\prod_{k} \int d\left({\bf Re}~\widetilde{g}_{\mu\nu}(k)\right) d\left({\bf Im}~\widetilde{g}_{\mu\nu}(k)\right)\right)
  {\rm exp}\left[{i\over V}\sum_{k} k^2 \vert \widetilde{g}_{\mu\nu}(k)\vert^2 + iS_{\rm sol} + ...\right] \nonumber\\
&&~~~~~~~~~~~~~~~~~~~~~~~~~~~~~~~~~~~~\times
 {\rm exp} \left[{2\over V} \sum_{k'} ~
\left({\bf Re}~\alpha^{(\psi)}_{\mu\nu}(k')~{\bf Re}~\widetilde{g}^{\mu\nu}(k') + {\bf Im}~\alpha^{(\psi)}_{\mu\nu}(k')~{\bf Im}~\widetilde{g}^{\mu\nu}(k')\right) + ...\right]
\nonumber\\
&&~~~~~~~~~~~~~~~~~~~~~~~~~ \times  {1\over V} \sum_{k''} ~\psi_{{\bf k}''}({\bf x}, y, z) e^{-ik''_0t}
\left({\bf Re}~\widetilde{g}_{\mu\nu}(k'') + i{\bf Im}~\widetilde{g}_{\mu\nu}(k'')\right)~{\rm exp}\left(-{1\over V}\sum_{k'} \vert\alpha^{(\psi)}_{\mu\nu}(k')\vert^2\right), \nonumber\\ \nd}
where $S_{\rm sol}$ is the action for the solitonic background \eqref{vegamey3} (plus the contributions from the G-fluxes that we will specify in the next sub-section), and $\delta g_{\mu\nu}$ is the part of $g_{\mu\nu}$ without the solitonic piece. Comparing \eqref{meyepolce} with \eqref{meyepolce2}, we see that there are few differences. The action integral in \eqref{meyepolce2} does not terminate to the quadratic term because of the presence of interactions. However the simplified form of the 
kinetic term hides an {\it infinite} set of interactions with the soliton itself. These interactions arrange themselves to form a time-independent Schr\"odinger equation with a highly non-trivial potential. In fact this is exactly the Schr\"odinger equation whose wave-functions are $\psi_{\bf k}({\bf x}, y, z)$ discussed earlier and which appears in the Fourier decomposition \eqref{cannon}!
 Thus this wave-function appears in the third line above replacing the $e^{-i{\bf k}''\cdot {\bf x}}$ in \eqref{meyepolce}.

The interactions, denoted by the dotted terms in the second and the third lines of \eqref{meyepolce2} are 
important. They are classified by ${g_s^{\vert c\vert}\over {\rm M}_p^d}$ where 
$(c, d) \in \left({\mathbb{Z}\over 3}, 
\mathbb{Z}\right)$, the $1/3$ moding has been explained in \cite{desitter2, desitter3, desitter4}, and we will dwell on this later. In the absence of the interactions, or in the limit where $g_s << 1$, the path integral \eqref{meyepolce2} can be exactly evaluated and the result is:
\bg\label{mcapopo}
\langle {\bf g}_{\mu\nu}(x, y, z)\rangle_{\alpha} &=& {\eta_{\mu\nu} \over h_2^{2/3}(y, {\bf x})} + 
{\bf Re}\left(\int {d^{10} {\bf k}\over 2\omega^{(\psi)}_{\bf k}} ~
\alpha^{(\psi)}_{\mu\nu}({\bf k}, t)\psi_{\bf k}({\bf x}, y, z)\right) 
+ {\cal O}\left({g_s^{c}\over {\rm M}_p^d}\right) \nonumber\\
&=& 
{\eta_{\mu\nu}\over \left(\Lambda\vert t\vert^2 \sqrt{h}\right)^{4/3}} + {\cal O}\left({g_s^{\vert c\vert}\over {\rm M}_p^d}\right) + {\cal O}\left[{\rm exp}\left(-{1\over g_s^{1/3}}\right)\right], \nd
which is exactly the M-theory uplift \eqref{vegamey3} of the de Sitter metric (at least the space-time part of it), and matches well with 
what we had earlier in \eqref{chukkam} using the Glauber-Sudarshan wave-function 
\eqref{spolley}, and \eqref{cnelson}. One could also work out the expectation values of all the internal components of the metric, namely $g_{mn}, g_{\alpha\beta}$ and $g_{ab}$, from the path integral as \eqref{meyepolce2} and show that they match, up to corrections of ${\cal O}\left({g_s^{\vert c\vert}/{\rm M}_p^d}\right)$, with \eqref{vegamey3}. We will not pursue this here: it's a straight-forward exercise and may be easily performed. Instead we will discuss other immediate questions related to the path integral \eqref{meyepolce2}. Before going into this,  note two things: one, the integral in \eqref{mcapopo} is over $d^{10}{\bf k}$ and not $d^{11}k$. This is because the kinetic term in \eqref{meyepolce2} creates a pole at $k_0 = \omega_{\bf k}^{(\psi)}$, and once we take the residue at that pole the $dk_0$ integral goes away. Two, any extra terms, other than what appears in \eqref{mcapopo}, exactly cancel out from the denominator of \eqref{camber}, as one might have expected.  

Another important point has to do with the ${\cal O}\left({g_s^{c}\over {\rm M}_p^d}\right)$ corrections in \eqref{mcapopo}. For 
$c > 0$ the corrections are perturbative in $g_s$, but become non-perturbative when $c < 0$. Such non-perturbative series in inverse powers of $g_s$ may be resummed as a resurgent trans-series to  take the expected non-perturbative form ${\rm exp}\left(-{1\over g_s^{1/3}}\right)$.  This means when $g_s << 1$, both the perturbative and the non-perturbative corrections are {\it small}.  The string coupling (which is the IIA string coupling) is given by ${g^2_s\over \sqrt{h(y)}} = \Lambda \vert t \vert^2$, and therefore as long as:
\bg\label{spamca}
-{1\over \sqrt{\Lambda}} < t \le 0, \nd
we are at weak coupling. This rather awkward choice of the temporal domain results from our type IIB  metric \eqref{betta3} that allows a flat-slicing of de Sitter where $-\infty < t \le 0$. We can easily see that once we venture beyond the domain \eqref{spamca}, both the perturbative and the non-perturbative terms go out of control and we lose the simple {\it classical} feature \eqref{mcapopo}. The shifted interacting vacuum 
$\mathbb{D}(\alpha(t))\vert\Omega(t)\rangle$ no longer evolves as a coherent state, not even approximately, and the system truly becomes quantum (or at least we lose quantitative control over it). It might be interesting to see if there are other expansion parameters, for example a S-dual description in the IIB side, that could shed light for $t < -{1\over \sqrt{\Lambda}}$. However a more pertinent question is: what happens in the domain \eqref{spamca}? Do the 
${\cal O}\left({g_s^c\over {\rm M}_p^d}\right)$ corrections provide small but finite contributions, or do they just cancel out\footnote{The result \eqref{mcapopo} {\it deviates} from an exact de Sitter background of 
\eqref{vegamey3}, albeit in a very small way, implying option (3). While this also justifies the point raised in footnote \ref{joelle}, there does exist a particular choice of $\mathbb{D}(\alpha)$ for which these extra correction terms {\it do not} appear thus implying option (2) instead. To see this we need to develop the story a bit more, and will be elaborated in section \ref{sec3.3}. \label{ladkill}}?  

Answering these questions will require us to introduce the other players in the field, namely the G-fluxes and quantum corrections. We will introduce them shortly, although not before we analyze few other details pertaining to the path integrals discussed above. The first has to do with the {\it fluctuations} over the coherent state background. How do we study them using path integrals? 

The analysis will involve a careful manipulation of the operator \eqref{horlil}, because we have interpreted the fluctuations over the de Sitter coherent state as the Agarwal-Tara \cite{agarwal} state \eqref{ckijwala}. The operator \eqref{horlil}, on the other hand, cannot have free Lorentz indices, so in the field formalism one way to express this would be to use even powers of the graviton field {\it i.e.} even powers of 
$g^{\mu\nu}g_{\mu\nu}$ with no sum over the repeated indices. Thus what we are looking for is an expectation value of the form:

{\footnotesize
\bg\label{hitgtko}
\langle {\bf g}_{\mu\nu}\rangle_{\Psi^{(c_1c_2)}} &\equiv&  
{\langle \Omega\vert \mathbb{D}^\dagger(\alpha(t))\left(c_1^\ast + c_2^\ast {\cal G}^{\dagger(\psi)}(a, a^\dagger; t)\right) {\bf g}_{\mu\nu} \left(c_1 + c_2 {\cal G}^{(\psi)}(a, a^\dagger; t)\right)~\mathbb{D}(\alpha(t)) \vert \Omega \rangle
\over 
\langle \Omega\vert \mathbb{D}^\dagger(\alpha(t))\left(c_1^\ast + c_2^\ast {\cal G}^{\dagger(\psi)}(a, a^\dagger; t)\right) \left(c_1 + c_2 {\cal G}^{(\psi)}(a, a^\dagger; t)\right)~\mathbb{D}(\alpha(t))\vert \Omega \rangle} \\
& = & {\int \left[{\cal D}g_{\mu\nu}\right] ~e^{iS} ~
\mathbb{D}^\dagger(\alpha)~\left\vert c_1 + c_2 \int d^{11}k~\sum_m C_{m}^{(\psi)}(k)(\widetilde{g}^{\mu\nu}(k) 
\widetilde{g}_{\mu\nu}(k))^{m}\right\vert^2~   
g_{\mu\nu}(x, y, z) ~\mathbb{D}(\alpha) \over 
\int \left[{\cal D}g_{\mu\nu}\right] ~e^{iS} ~
\mathbb{D}^\dagger(\alpha)~\left\vert c_1 + c_2 \int  d^{11}k~\sum_m C_{m}^{(\psi)}(k)(\widetilde{g}^{\mu\nu}(k) 
\widetilde{g}_{\mu\nu}(k))^{m}\right\vert^2 ~\mathbb{D}(\alpha)} \nonumber \nd}
where the first line is expressed using operators and the second line is expressed using fields. With some abuse of notation we have used $\mathbb{D}(\alpha(t))$ to denote the operator and $\mathbb{D}(\alpha)$ 
to denote the field. The latter is defined as in \eqref{juanivan}. We have also chosen a specific form of the 
${\cal G}^{(\psi)}$ function, as an integral over all momenta modes; and $c_i$ are the coefficients that appear in \eqref{ferelisa} which, in turn, are bounded by \eqref{Backreaction}. Note that the factor of
$\omega^{(\psi)}_{\bf k} + k_0$ does not appear explicitly. This is in fact redundant and is therefore absorbed in the definition of $C_{m}^{(\psi)}(k)$ (see footnote \ref{vlake}). The coefficient $C^{(\psi)}_m(k)$
is related to the coefficient $C_{m{\bf k}}^{(\psi)}(t)$ in \eqref{horlil}.

The path integral expression in \eqref{hitgtko} is rather complicated because it involves an infinite sum of powers of the metric factor, so naturally a simplifying scheme is warranted for. We will start by making $c_1 = 1$ and $C_m^{(\psi)}(k) = \delta_{m2} C^{(\psi)}(k)$, as a toy example to see what kind of answer we expect from the path integral. With this in mind, the numerator takes the following form:

{\footnotesize
\bg\label{meyepolce3}
&&{\rm Num}\left[\langle {\bf g}_{\mu\nu}\rangle_{\Psi^{(c_1c_2)}}\right] 
=  \left(\prod_{k} \int d\left({\bf Re}~\widetilde{g}_{\mu\nu}(k)\right) d\left({\bf Im}~\widetilde{g}_{\mu\nu}(k)\right)\right)
  {\rm exp}\left[{i\over V}\sum_{k} k^2 \vert \widetilde{g}_{\mu\nu}(k)\vert^2 + iS_{\rm sol} + ...\right] \nonumber\\
&&~~~~~~~~~~~~~~~~~~~~~~~~~~~~~~~~~~~~\times
 {\rm exp} \left[{2\over V} \sum_{k'} ~
\left({\bf Re}~\alpha^{(\psi)}_{\mu\nu}(k')~{\bf Re}~\widetilde{g}^{\mu\nu}(k') + {\bf Im}~\alpha^{(\psi)}_{\mu\nu}(k')~{\bf Im}~\widetilde{g}^{\mu\nu}(k')\right) + ...\right]
\nonumber\\
&&~~~~~~~~~~~~~~~~~~~~~~~ \times  {1\over V} \sum_{k''} ~\psi_{{\bf k}''}({\bf x}, y, z) e^{-ik''_0t}
\left({\bf Re}~\widetilde{g}_{\mu\nu}(k'') + i{\bf Im}~\widetilde{g}_{\mu\nu}(k'')\right)~{\rm exp}\left(-{1\over V}\sum_{k'} \vert\alpha^{(\psi)}_{\mu\nu}(k')\vert^2\right), \nonumber\\ 
&&~~~~~~~~~~~~~~~~~~~~~~~ \times \Bigg\vert 1 + {c_2 \over V} \sum_{k'''}~C^{(\psi)}(k''')~
\left[\left({\bf Re}~\widetilde{g}^{\mu\nu}(k''') + i{\bf Im}~\widetilde{g}^{\mu\nu}(k''')\right)
\left({\bf Re}~\widetilde{g}_{\mu\nu}(k''') + i{\bf Im}~\widetilde{g}_{\mu\nu}(k''')\right)\right]^2\Bigg\vert^2, 
\nonumber\\ \nd}
that still involves a complicated set of nested integrals (which we express as sum over ($k. k', k'', k'''$) assuming discrete momenta). The way we have expressed \eqref{meyepolce3}, all the alternate lines involve real quantities. Even without doing anything we see that some structure is evolving from the integrals. For example, when $c_2 = 0$, we reproduce \eqref{mcapopo}. This is no surprise. Now taking 
only the quartic term from the last line of \eqref{meyepolce3}, we see that the numerator takes the following form:

{\footnotesize
\bg\label{numerator}
&& {\rm Num}\left[\langle {\bf g}_{\mu\nu}\rangle_{\Psi^{(c_1c_2)}}\right] = \prod_k
 \left({\pi e^{-\alpha^2(k)/4k^2}\over k^2}\right)
\left[{\eta_{\mu\nu} \over h_2^{2/3}(y, {\bf x})} + 
{\bf Re}\left(\int {d^{10} {\bf k}\over 2\omega^{(\psi)}_{\bf k}} ~
\alpha^{(\psi)}_{\mu\nu}({\bf k}, t)\psi_{\bf k}({\bf x}, y, z)\right)\right]\\
&+& \prod_k \left({\pi e^{-\alpha^2(k)/4k^2}\over k^2}\right) {c_2\over 8}\left[ \int {d^{11} k' \over k^{'6}} ~C^{(\psi)}(k')~\alpha_{\mu\nu}^{(\psi)}(k') ~\psi_{{\bf k}'}({\bf x}, y, z)~e^{ik'_0 t}\left(15 + {\alpha^4 (k') \over 4k^{'4}} + 
{5 \alpha^2 (k')\over k^{'2}}\right) + ...... \right]\nonumber \nd}
where the infinite factor in front comes from the gaussian integrals\footnote{This can in-fact be brought in a more manageable form $\mathbb{Q} \equiv {\rm exp}\left[\int_{-\infty}^{+\infty} d^{11}k
~{\rm log}\left({\pi e^{-\alpha^2(k)/4k^2}\over k^2}\right)\right]$, but since this will cancel out eventually, we won't have to evaluate it.}, 
and in the second line we see that the 
term is suppressed by $c_2$ because of the condition \eqref{Backreaction} imposed earlier. The dotted terms in the bracket involve other powers of the metric integrals from the last line in \eqref{meyepolce3}, including the ${\cal O}\left({g_s^c\over {\rm M}_p^d}\right)$ corrections coming from the interaction Hamiltonian. We have also defined:
\bg\label{mcaconsti}
 \alpha^2 (k) \equiv \alpha^{(\psi)}_{\mu\nu}(k)\alpha^{(\psi)\mu\nu}(k), \nd
and similarly $\alpha^4 (k)$. The pole structure of the second line is more complicated, as one would have expected, and in addition there appears the factor $C^{(\psi)}(k')$ which is the remnant of the similar coefficient in \eqref{horlil}. The denominator, to the same order in the graviton expansion, takes the following form:

{\footnotesize
\bg\label{denominator}
{\rm Den}\left[\langle {\bf g}_{\mu\nu}\rangle_{\Psi^{(c_1c_2)}}\right]  =  \prod_k
 \left({\pi e^{-\alpha^2(k)/4k^2}\over k^2}\right)\left[1 + {c_2\over 4} \int {d^{11} k' \over k^{'4}}
 \left(3 + {\alpha^4 (k') \over 4k^{'4}} + 
{3 \alpha^2 (k')\over k^{'2}}\right)C^{(\psi)}(k') + .......\right], \nonumber\\ \nd}
where as before, the dotted terms appear from two sources, one, from the higher order terms in the graviton expansion, and two, from the ${\cal O}\left({g_s^c\over {\rm M}_p^d}\right)$ corrections. Expectedly, the pole structures of the $k^2$ integrals are different here, but the $c_2$ suppression is consistent to what we had before. There is no appearance of the space-time wave-function $\psi_{\bf k}(x, y, z)$ and the Lorentz invariance is perfectly maintained. Dividing \eqref{numerator} by \eqref{denominator}, we get:

{\footnotesize
\bg\label{kimbas}
&&\langle {\bf g}_{\mu\nu}\rangle_{\Psi^{(c_1c_2)}} = {\eta_{\mu\nu} \over h_2^{2/3}(y, {\bf x})} + 
{\bf Re}\left(\int {d^{10} {\bf k}\over 2\omega^{(\psi)}_{\bf k}} ~
\alpha^{(\psi)}_{\mu\nu}({\bf k}, t)\psi_{\bf k}({\bf x}, y, z)\right) + c_2 \int d^{11} k~f_{\mu\nu}({\bf k}, k_0) 
~\psi_{{\bf k}}({\bf x}, y, z)~e^{ik_0 t} + ...\nonumber\\
&&~~~~~~~~ = {\eta_{\mu\nu}\over \left(\Lambda\vert t\vert^2 \sqrt{h}\right)^{4/3}} + 
c_2 \int d^{11} k~f({\bf k}, k_0)~\alpha^{(\psi)}_{\mu\nu}({\bf k}, k_0) 
~\psi_{{\bf k}}({\bf x}, y, z)~e^{ik_0 t} + {\cal O}\left({g_s^{\vert c\vert}\over {\rm M}_p^d}\right) + {\cal O}\left[{\rm exp}\left(-{1\over g_s^{1/3}}\right)\right],  \nonumber\\ \nd}
where we see that the infinite factors in front of \eqref{numerator} and \eqref{denominator} have cancelled out. The second line of \eqref{kimbas} tells us the Agarwal-Tara state \cite{agarwal}, or the GACS \eqref{ferelisa} indeed reproduces the {\it fluctuations} over our de Sitter space, realized as a Glauber-Sudarshan state. The Fourier coefficient $f({\bf k}, k_0)$ appearing above is now defined as:
\bg\label{sabinde}
 f({\bf k}, k_0) = {C^{(\psi)}(k)\over 8k^6}\left(15 + {\alpha^4 (k) \over 4k^{4}} + 
{5 \alpha^2 (k)\over k^{2}} + ...\right) + {\cal O}(c_2) + {\cal O}\left({g_s^c\over {\rm M}_p^d}\right), \nd
which should be compared to \eqref{jlewis}, \eqref{lizshue} and \eqref{kirkoo}. The pole structures 
in \eqref{sabinde} are different from \eqref{kirkoo}, which is expected because \eqref{kirkoo} is for IIB and 
\eqref{sabinde} is  for M-theory. Nevertheless, the awkward factors of $T$ that renders \eqref{kirkoo} somewhat non-convergent, do not appear in \eqref{sabinde}. This means, the way we have expressed our fluctuations, \eqref{kimbas} do provide an answer to the Trans-Planckian issue, namely:
\newline
\newline
\noindent\fbox{%
    \parbox{\textwidth}{%
The time-dependent frequencies that we encounter from fluctuations over a de Sitter {\it vacuum} are actually artifacts of Fourier transforms over a de Sitter {\it state}, viewed as a Glauber-Sudarshan state. 
}%
}

\vskip.2in

\noindent There are however two questions that may arise in interpreting \eqref{kimbas} as fluctuations over de Sitter space. The first is the form of $f({\bf k}, k_0)$ given in \eqref{sabinde} and in \eqref{kirkoo}. There appears to be $\alpha^{(\psi)}_{\mu\nu}(k)$ dependence in \eqref{sabinde} whereas none appears in 
\eqref{kirkoo}. The reason is simple: the form \eqref{sabinde} is an integrand, and once we insert the functional form for $\alpha^{(\psi)}_{\mu\nu}(k)$ from \eqref{cannon2}, and sum over higher values of $m$, we should be able to reproduce \eqref{kirkoo}. However since \eqref{sabinde} is always a function of  $k$, awkward factors like $T$ may not appear in the final expression. This suggests that the path-integral is much more powerful way to analyze the expectation values here. 

The second question is the form of $C^{(\psi)}_m(k)$ in \eqref{horlil}. Will it create new poles in \eqref{sabinde}? The answer depends on the way we construct the Agarwal-Tara state \eqref{ferelisa} and 
\eqref{horlil}. The coefficient $C^{(\psi)}_m(k)$ is the Fourier transform of $C^{(\psi)}_{m{\bf k}}(t)$ from \eqref{horlil}, so depending on our choice of \eqref{horlil} there could be higher order zeroes or poles. However this could be fixed from the very beginning, so that the integral \eqref{sabinde} may determine the eventual behavior of $f({\bf k}, k_0)$.

Now that we have understood how to interpret fluctuations over the de Sitter state, it is time to discuss the second issue, namely the existence of the Wilsonian effective action. In other words we would like to ask the following question: how do the background \eqref{vegamey3} get affected once we integrate out the momentum modes from a cut-off $\Lambda_{\rm UV}$ to say ${\rm M}_p$?  

To proceed, and as it goes for the Wilsonian integration procedure, we will have to go to the Euclidean formalism. We will also define the field configuration, as given in \eqref{poladom}, to the following:

{\footnotesize
\bg\label{poladom2}
g_{\mu\nu}(x, y, z) &=& {\eta_{\mu\nu}\over h_2^{2/3}(y, {\bf x})} + \int_{-{\rm M}_p}^{+{\rm M}_p} d^{11}k ~\widetilde{g}_{\mu\nu}({\bf k}, k_0) ~\psi_{\bf k}({\bf x}, y, z) e^{-ik_0 t}\\
&+& \int_{-\Lambda_{\rm UV}}^{-{\rm M}_p} d^{11}k ~\widetilde{g}_{\mu\nu}({\bf k}, k_0) ~\psi_{\bf k}({\bf x}, y, z) e^{-ik_0 t} + 
\int_{+ {\rm M}_p}^{+\Lambda_{\rm UV}} d^{11}k ~\widetilde{g}_{\mu\nu}({\bf k}, k_0) ~\psi_{\bf k}({\bf x}, y, z) e^{-ik_0 t}, 
\nonumber \nd} 
where all dynamics bounded by $\vert k \vert < {\rm M}_p$ are used to define the effective field theory. The modes lying between ${\rm M}_p < \vert k\vert \le \Lambda_{\rm UV}$ are integrated out, but now we see that this procedure can indeed be explicitly performed because the modes themselves have no trans-Planckian issues. 
We will start by asking how this effects the action $S$, the action that appears in say \eqref{meyepolce2}. In fact all we need is $\langle\alpha\vert\alpha\rangle$ which appears in the denominator of 
\eqref{meyepolce2}  and is defined as:
\bg\label{josjame}
\langle\alpha\vert\alpha\rangle \equiv {\int \left[{\cal D}g_{\mu\nu}\right] ~e^{iS} ~
\mathbb{D}^\dagger(\alpha)~\mathbb{D}(\alpha) \over \int \left[{\cal D}g_{\mu\nu}\right] ~e^{iS}}, \nd
which is not identity, as we saw earlier, because $\mathbb{D}(\alpha)$ is not unitary. The reason why  
\eqref{josjame} suffices is because, the expectation value of the space-time metric in \eqref{meyepolce2} may be re-written as the following integral:

{\footnotesize
\bg\label{sisphool} 
\langle {\bf g}_{\mu\nu}(x, y, z)\rangle_\alpha = 2\lim_{\Lambda_{\rm UV} \to \infty} \int_{-\Lambda_{\rm UV}}^{+\Lambda_{\rm UV}} d^{11} k~\psi_{\bf k}({\bf x}, y, z) e^{ik_0 t}~ 
{\partial \over \partial \alpha^{\ast(\psi)\mu\nu}(k)} {\rm log} \left[{\rm exp}
\left(\int_{-\Lambda_{\rm UV}}^{+\Lambda_{\rm UV}} d^{11} k' \vert \alpha^{(\psi)}(k')\vert^2\right) \langle \alpha\vert \alpha\rangle\right], \nonumber\\ \nd}
where the $\vert\alpha\vert^2$ part is defined in the same way as in \eqref{mcaconsti} but now with
$\alpha^{(\psi)}_{\mu\nu}$ and its complex conjugate. The integrals are bounded by the cut-off 
$\Lambda_{\rm UV}$ which, here, we assume would approach infinity although it could in principle be any scale. The LHS is a physical quantity so it's scale dependence could be subtle. However on the RHS, once we change the scale, other quantities would change accordingly, for example the action $S$ will become $S_{\rm eff}$ and all the fields would get re-scaled,  to keep the zeroth order (in $g_s$) result, {\it i.e.} \eqref{mcapopo} for $g_s \to 0$, unchanged. We will see later, once we do explicit computations, that this is indeed the case\footnote{This however does not immediately imply that there is no scale dependence. In fact this is also tied up to the question of the renormalization of the four-dimensional Newton's constant. We will discuss this towards the end of section \ref{sec3.3}.}.

\subsection{G-fluxes, susy breaking and consistency conditions \label{sec2.6}}

There are also G-fluxes whose dynamics we haven't discussed yet. The solitonic solution \eqref{betbab3} 
require non-trivial G-fluxes to allow for the compact internal eight-manifold ${\cal M}_8$ in 
\eqref{anonymous}. These G-fluxes are time-independent quantities that that have non-trivial functional behavior over the eight-manifold \eqref{anonymous}, including components along the $2+1$ dimensional space-time itself. If (${\rm M, N})$ denotes components in the internal eight-manifold, and ($\mu, \nu$) the components along the space-time directions, then we require flux components of the form 
${\bf G}_{\rm MNPQ}(y)$  
and ${\bf G}_{\mu\nu\rho {\rm M}}(y)$ to allow for the  solitonic background like \eqref{betbab3} to exist (see 
\cite{becker, DRS, SVW, DM} for the special case when $h_2(y, {\bf x}) = h_1(y)$). Supersymmetry is preserved in the special case when the internal G-fluxes are {\it self-dual} over the eight-manifold 
\eqref{anonymous} \cite{becker}. Our present aim is to search for fluxes that can go along with a background like 
\eqref{vegamey3}. Clearly now we are looking for coherent states, or Glauber-Sudarshan wave-functions, that may capture the specific fluxes or flux components as most probable values (or alternatively, as expectation values). Two questions follow here: one, how do we construct the corresponding Glauber-Sudarshan wave-functions? And two, what are the consistency conditions that allow such states to exist in the first place? 

From the start we know that all flux components have to be time-{\it dependent} so that coherent state representations may at least be constructed. This is however easier said than done: one needs to overcome numerous subtleties to even start attempting a construction of this kind. For example, fluxes in a compact manifold have to be quantized, otherwise we will face problems with the Dirac quantization procedure. If now the fluxes become time-dependent, how do we even justify that there is a quantization procedure? Even more seriously, what would such quantization procedure mean: quantization at every instant of time? Or more like quantization over some interval of time? 

Even if we find some meaning to the quantization procedure, the fluxes have to satisfy 
Gauss' law\footnote{Sometime also called the anomaly cancellation condition.}, otherwise consistent construction cannot even be attempted. Here we want the fluxes to be time-dependent. How do we satisfy Gauss' law? Do we want Gauss' law to be satisfied at every instant of time or, as again before, more like over an interval? 

The subtleties don't end here. Time-dependent fluxes have to break supersymmetries so that we could consistently support a non-supersymmetric Glauber-Sudarshan state that provide the metric configuration \eqref{vegamey3} as expectation value (see \eqref{mcapopo}). Breaking supersymmetry in either the metric or the flux sectors will switch on an uncontrolled plethora of {\it time-dependent} quantum corrections. How do we control them?    

Surprisingly the answer to all these questions come from an unexpected corner:  from the large plethora of time-dependent quantum corrections! These have been answered in great details in 
\cite{desitter2} so we will be brief here. Both the flux quantization and the Gauss' law are related to the Bianchi identities and the G-flux EOMs. As shown in sections 4.2.1 and 4.2.2 of \cite{desitter2}, these set of equations get corrected, order by order in $g^c_s/{\rm M}_p^d$, from the infinite towers of quantum corrections (both local and non-local). The question then is: how do these corrections help us to balance the flux-quantization and Gauss' law when the fluxes, as well as the underlying manifold, are varying with time?

The answer, as shown in \cite{desitter2} is as simple as it is instructive. The plethora of quantum corrections  are known to scale in very specific ways with respect to time\footnote{Recall we have identified $g_s$ with time $t$ via the following relation ${g_s\over H} = \sqrt{\Lambda} \vert t\vert$. Note also that $H = H(y)$ denotes the warp-factor whereas ${\rm H}$ denotes the Hubble parameter. The Hubble parameter ${\rm H}$ will only feature prominently in section \ref{sec4} whereas the warp-factor $H = H(y)$ appears in section 
\ref{sec2} and \ref{sec3}.}, or here, with respect to $g_s/H$ where $H(y) = h_1^4(y) = h^4_2 (y)$. On the other hand, we have claimed that the G-fluxes themselves are time-dependent, {\it i.e.} they too scale with respect to $g_s/H$ in very specific ways (that we will elaborate below). This means, in either Bianchi identities or EOMs, augmented by the quantum corrections, all we need is to identify equivalent powers of $g_s/H$ from flux components and from the quantum corrections!  For flux quantizations, we need some integrated form of the identities (see details in \cite{desitter2}), but the moral of the story should be clear: order-by-order in $g_s/H$ equations could be balanced and one could give meaning to both flux-quantizations and anomaly-cancellations in a time-dependent background.     

However, how do we know that such balancing of the $g_s/H$ terms is consistent with the EOMs? The answer to this, as shown in \cite{desitter2}, is again rather simple. The flux-quantization procedures, or the {\it integrated} forms of the Bianchi identities, are in-fact the EOMs of the {\it dual-forms}, {\it i.e.} the EOMs of the seven-form\footnote{The bold faced flux components either denote the $g_s$ dependent G-fluxes from 
\cite{desitter2, desitter3}, or fields/operators here. To avoid confusion, only one notation will be used throughout.}  
${\bf G}_{\rm MNPQRST}$ that are Hodge-dual of the four-form ${\bf G}_{\rm MNPQ}$. Thus the very act of balancing the equations order-by-order in $g_s/H$, we are in-fact solving the dual seven-form EOMs! Such construction is then self-consistent to solving the four-form EOMs, implying an overall self-consistency of the system in the presence of time-dependent degree of freedom.

Our aim then is to re-interpret a part of the story as appearing from the Glauber-Sudarshan wave-functions.
For that we will require similar Fourier modes as in \eqref{cannon} which we got from the metric configurations \eqref{betbab3} and \eqref{vegamey3}. However new subtleties appear regarding what to choose: the three-form fields ${\bf C}_{\rm MNP}$ or the four-form G-fluxes ${\bf G}_{\rm MNPQ}$ to study the dynamics? The reason is, if we choose the three-form to be $g_s$ dependent (or time-dependent), then this will inadvertently lead to a four-form G-flux components of the form ${\bf G}_{0{\rm MNP}}$. Such components will break de Sitter isometries in the type IIB side so cannot be allowed. How can we find a way to not allow such components to arise here? 

The answer, as shown in section 4.2.3 of \cite{desitter2}, again lies in the quantum corrections.  In the absence of M5-branes, the EOM for a seven-form flux component becomes the Bianchi identity of the corresponding four-form G-flux component. Solving this will generically give ${\bf G}_4 = d{\bf C}_3 ~+$ quantum corrections, implying that ${\bf G}_4$ is not just $d{\bf C}_3$ but comes with an additional baggage of quantum terms. This freedom can be used to set:
\bg\label{coutumey}
{\bf G}_{0{\rm MNP}} = 0, \nd
thus saving us from breaking the four-dimensional de Sitter isometries in the type IIB side. Therefore it appears either ${\bf C}_{\rm MNP}$ or ${\bf G}_{\rm MNPQ}$ may be used to study the background from M-theory. 

It turns out, unfortunately that this is not {\it quite} true. There is yet another level of subtlety that we have kept 
under the rug and it has to do with G-flux components like ${\bf G}_{{\rm MN}ab}$ where 
$(M, N) \in {\cal M}_4 \times {\cal M}_2$ and $(a, b) \in {\mathbb{T}^2\over {\cal G}}$ as in 
\eqref{anonymous}. Such components clearly violate the type IIB de Sitter isometries, unless:
\bg\label{pikilia}
{\bf G}_{{\rm MN}ab}(y^m, y^\alpha, y^a, g_s) \equiv {\bf F}_{\rm MN}(y^m, g_s) 
\otimes \Omega_{ab}(y^\alpha, y^a, g_s), \nd
where $\Omega_{ab}$ is a localized two-form on ${\cal M}_2 \times {\mathbb{T}^2\over {\cal G}}$ and ${\bf F}_{\rm MN}$ appears as type IIB gauge field on seven-branes. How and why such seven-branes  appear have been explained in \cite{desitter2, desitter3}, so we will avoid going into it here. Instead we will point out two 
aspects of \eqref{pikilia}. The first one is the obvious one:  the decomposition {\it does not} produce a three-form field, rather a one-form gauge field. Such a gauge field can become non-abelian, but that's another story that we will avoid getting into\footnote{Here it will suffice to say that wrapped M2-branes on vanishing two-cycles of the internal eight-manifold \eqref{anonymous} play an important role in making the system non-abelian. For more details see \cite{nonabel}.}. The second one is not very obvious: the $g_s$ dependences of
${\bf F}_{MN}$ and $\Omega_{ab}$.  It turns out there is a natural way to impart both $g_s$ and ${\rm M}_p$ dependence to $\Omega_{ab}$ that come from the metric \eqref{vegamey3} along the toroidal directions. As a first approximation then we can keep ${\bf F}_{\rm MN}$ to be $g_s$ independent. 

Our discussion above might have convinced that a uniform description with G-fluxes appear once we take field strengths and not fields. The field strengths are also gauge invariant and in many cases they capture 
quantum corrections that neither the three-form ${\bf C}_{\rm MNP}$ nor the one-form ${\bf A}_{\rm M}$ captures. It would then appear that the coherent states may be constructed for the G-flux components directly. In fact this would fit well with the EOMs themselves, as EOMs involve only field strengths and not fields themselves \cite{desitter2}. Typically then we expect:

{\footnotesize
\bg\label{prcard}
\widetilde{\bf G}_{\rm MNPQ}(k) = \int d^{11}x \left[\sum_{p, n \ge 0} (-1)^{\delta_{n0}}~
{\cal G}^{(p, n)}_{\rm MNPQ}(y) 
\left({g_s\over H}\right)^{2p/3} {\rm exp}\left(-{nH^{1/3}\over g_s^{1/3}}\right)\right]
{h^{4/3}_1\over h_2}~ \gamma_{\bf k}(x, y, z), \nonumber\\ \nd} 
where ${\cal G}^{(0, 0)}_{\rm MNPQ}(y)$ is the self-dual G-flux that we switch on to support the solitonic background \eqref{betbab3}. The sum is over ($p, n$) and typically $p \in {\mathbb{Z}\over 2}$ with 
$p \ge {3\over 2}$ as shown in \cite{desitter2}. Near $g_s \to 0$, the dominant contributions from the non-perturbative piece come from $n = 0$, and therefore the higher order G-flux contributions, alluded to earlier, appear from ${\cal G}^{(p, 0)}_{\rm MNPQ}(y)$ for $p \ge {3\over 2}$. The other important ingredient of 
\eqref{prcard} is the Schr\"odinger wave-function $\gamma_{\bf k}(x, y, z)$ that should be compared to the
wave-functions appearing in \eqref{hudson}, and notably $\psi_{\bf k}(x, y, z)$. A similar wave-function also appears in the Fourier decomposition of the following G-flux components:
\bg\label{kwinter}
\widetilde{\bf G}_{\mu\nu\rho {\rm M}} = \int d^{11} x~\epsilon_{\mu\nu\rho}~\partial_{\rm M}\left[{1 \over h_1(y)}
\left({1\over \Lambda^2 \vert t \vert^4} - 1\right)\right] {h^{4/3}_1\over h_2}~ \beta_{\bf k}(x, y, z), \nd
with the corresponding Schr\"odinger equation for $\beta_{\bf k}(x, y, z)$. As mentioned earlier, the temporal behavior of these wave-functions with spatial form 
$\gamma_{\bf k}({\bf x}, y, z)$ and $\beta_{\bf k}({\bf x}, y, z)$, would be captured for the corresponding 
coherent state wave-functions with frequencies $\omega^{(\gamma)}_{\bf k}$ and 
$\omega^{(\beta)}_{\bf k}$ respectively. 

There are a few subtleties that we should point out regarding the constructions \eqref{prcard} and 
\eqref{kwinter}. The first has to do with the corresponding fields ${\bf C}_{\rm MNP}$ and 
${\bf C}_{\mu\nu\rho} \equiv {\bf C}_{0ij}$. Modulo the issues mentioned earlier, we could also expand them in powers of $g_s/H$, and take their Fourier transforms in terms of two other set of wave-functions. However since the three-form fields are not gauge invariants, there would be issues when we want to study {\it fluctuations} over the solitonic background. In the path-integral analysis this will introduce Faddeev-Popov ghosts, thus complicating the ensuing analysis. To avoid this we will stick with the wave-functions 
$\gamma_{\bf k}({\bf x}, y, z)$ and $\beta_{\bf k}({\bf x}, y, z)$, keeping in mind that they represent the field-strengths and therefore differ somewhat with the other set of wave-functions \eqref{hudson}. Note that such an argument also extends to the G-flux components of the form \eqref{pikilia}, although there appears a possibility that $k < {3\over 2}$ for such cases (see details in \cite{desitter3}). 

The second subtlety has to do with the Glauber-Sudarshan wave-function. What is the form of the wave-function now? The answer remains the same as before: it is related to the shifted interacting vacuum 
\eqref{novocaine}. In other words, we expect the configuration space wave-function \eqref{laudern}, much along the lines of \eqref{siortegas}, to still capture the essential dynamics of the system, except now the interaction Hamiltonian ${\bf H}_{\rm int}$ include interactions with the G-fluxes too. The displacement operator
$\mathbb{D}(\alpha)$ in \eqref{maristone} will naturally become more involved because the annihilation and the creation operators $a_{\rm eff}$ and $a^\dagger_{\rm eff}$ respectively will involve the flux sector too. The interacting vacuum is still annihilated by $a_{\rm eff}$, but the rest of the construction simply gets more involved retaining, however, the essential features that helped us to cement the coherent state construction. Expectation values over the coherent state, for example\footnote{Or more generically
$\langle {\bf G}_{\rm MNPQ}\rangle_{(\alpha, \beta)}$ and $\langle {\bf G}_{0ij{\rm M}}\rangle_{(\alpha, \beta)}$ as we shall explain in section \ref{sec3.3}.}  
$\langle {\bf G}_{\rm MNPQ}\rangle_\alpha$ and $\langle {\bf G}_{0ij{\rm M}}\rangle_\alpha$, would provide the required time-{\it dependent} fluxes to support a background like \eqref{vegamey3}.  

Finally, how is the supersymmetry broken? The answer, in the language of the expectation values, is simple. We demand:
\bg\label{sosie}
\left\langle {\bf G}_{\rm MNPQ}\left({\bf G}^{\rm MNPQ} - \left(\ast_8 {\bf G}\right)^{\rm MNPQ}\right)\right\rangle_\sigma
\ne 0, \nd
where $\ast_8$ is the Hodge dual of the four-form over the solitonic internal metric \eqref{betbab3} and 
$\sigma$ is the generalized coherent states that we will elaborate in section \ref{sec3.3}. The un-warped internal metric is a non-K\"ahler one (including a non-K\"ahler metric on the base ${\cal M}_4 \times 
{\cal M}_2$ of \eqref{anonymous}), and the non-zero value on the RHS of \eqref{sosie} comes from the quantum terms alluded to earlier. A generic derivation\footnote{An alternative, and probably more useful, way to express supersymmetry breaking is to demand $\vert \langle {\bf G}_{{\rm MN}ab}\rangle_\sigma 
- \langle\left(\ast_8 {\bf G}\right)_{{\rm MN}ab}\rangle_\sigma\vert > 0$ where we have restricted 
$({\rm M, N}) \in {\cal M}_4 \times {\cal M}_2$ and $(a, b) \in {\mathbb{T}^2\over {\cal G}}$ (as taken from
\eqref{anonymous}). The reason for choosing this specific component of G-flux is because of it's appearance in the Schwinger-Dyson's equations (see section \ref{sec3.3} for details). \label{ADM}} of \eqref{sosie} is a bit technical, so we will only discuss it later in section \ref{sec3.3}. The readers could also refer to  \cite{nogo, pisin} and \cite{desitter2} for details.


\section{Quantum effects and the Schwinger-Dyson equations \label{sec3}}  

We have by now more or less specified all the essential ingredients that go in the construction of the 
Glauber-Sudarshan state, namely the metric and the G-flux components. All the time-dependent quantities, for example the metric \eqref{vegamey3}, appear as expectation values over the Glauber-Sudarshan states. What we haven't specified yet are the actual form of the quantum terms that go in the interaction Hamiltonian ${\bf H}_{\rm int}$ in 
\eqref{novocaine}. In the following two sections, \ref{sec3.1} and \ref{sec3.2}, we will revisit and hence elaborate further how these quantum terms shaped our construction of the time-dependent background 
\eqref{vegamey3}, and the corresponding G-flux components from \eqref{prcard} and \eqref{kwinter}. This means the metric and the G-flux components entering the discussion in sections \ref{sec3.1} and \ref{sec3.2} are those of the metric \eqref{vegamey3} and the corresponding $g_s$ dependent flux components, and not of the solitonic background \eqref{betbab3} (the bold-faced symbols signify those, so should not be confused with operators in the earlier sections).  In section \ref{sec3.3} we will see how the results of the sections \ref{sec3.1} and \ref{sec3.2} appear from the
Glauber-Sudarshan states. This is where the Schwinger-Dyson equations would become useful.

\subsection{A plethora of perturbative local and non-local quantum  terms \label{sec3.1}}

Our starting point would be the perturbative series of quantum effects that include both local and non-local 
terms. 
These quantum terms will in turn determine the ${\cal O}\left({g_s^c\over {\rm M}_p^d}\right)$ 
corrections to the Glauber-Sudarshan wave-function from \eqref{siortegas}. In the following we keep 
$c > 0$ but $d$ could take any signs. 
All of these are of course embedded inside the eleven-dimensional action that we write in the following way:
\bg\label{sheela}
{\bf S}_1 &=& {\rm M}_p^9 \int d^{11} x \sqrt{-{\bf g}_{11}}\Big({\bf R}_{11} + {\bf G}_4 \wedge \ast {\bf G}_4 + 
{\bf C}_3 \wedge {\bf G}_4 \wedge {\bf G}_4 + {\rm M}_p^2 ~{\bf C}_3 \wedge \mathbb{Y}_8\Big)\\
&+& \sum_{\{l_i\}, n_i}  \int d^{11} x \sqrt{-{\bf g}_{11}}\left(
\mathbb{Q}_{\rm T}\left(\{l_i\}, n_0, n_1, n_2, n_3\right) \over {\rm M}_p^{\sigma(\{l_i\}, n_i) - 11}\right)
+ {\rm M}_p^3 \sum_{r = 1}^\infty \int d^3x \sqrt{-{\bf g}_3} ~ c_{(r)} \mathbb{W}^{(r)}\nonumber\\
&-& {n_4 {\rm T}_2\over 2}\int d^3\sigma \left\{\sqrt{-\gamma_{(2)}}\Big(\gamma^{\mu\nu}_{(2)} 
\partial_\mu X^{\rm M}\partial_\nu X^{\rm N} {\bf g}_{\rm MN} - 1\Big) + {1\over 3} \epsilon^{\mu\nu\rho} 
\partial_\mu X^{\rm M}\partial_\nu X^{\rm N}\partial_\rho X^P {\bf C}_{\rm MNP}\right\},
\nonumber \nd
where in the first line we denote the standard kinetic terms and the interactions in M-theory, in the second line we denote other possible interactions, and in the third line we introduce $n_4$ number of integer and fractional M2-branes\footnote{Here we are simply {\it adding} the action of the individual M2-brane assuming well separation between them. When they are on top of each other such a simple addition is not possible and one should look for something more along the lines of the Bagger-Lambert form \cite{bagger}. Such construction will make the system even more complicated than it already is, so we will avoid it here.}. 
The ${\rm M}_p$ scalings of each of the terms are denoted 
carefully, including the ones that involve complicated interactions (see \eqref{chatchat} below). These interactions will involve polynomial powers of the curvature tensors, G-flux components and possible derivative actions. To quantify them we will have to get down to more finer space-time notations, advocated early in the sections. For example if ($m, n$) denote the coordinates of ${\cal M}_4$, ($\alpha, \beta$) the coordinates of 
${\cal M}_2$ and ($a, b$) the coordinates of ${\mathbb{T}^2\over {\cal G}}$ in \eqref{anonymous}, the 
quantum term $\mathbb{Q}_{\rm T}\left(\{l_i\}, n_0, n_1, n_2\right)$ may be expressed as:
\bg\label{selahran}
\mathbb{Q}_{\rm T}^{(\{l_i\}, n_i)} & = & {\bf g}^{m_i m'_i}.... {\bf g}^{j_k j'_k} 
\{\partial_m^{n_1}\} \{\partial_\alpha^{n_2}\} \{\partial_a^{n_3}\}\{\partial_0^{n_0}\}
\left({\bf R}_{mnpq}\right)^{l_1} \left({\bf R}_{abab}\right)^{l_2}\left({\bf R}_{pqab}\right)^{l_3}\left({\bf R}_{\alpha a b \beta}\right)^{l_4} \nonumber\\
&\times& \left({\bf R}_{\alpha\beta mn}\right)^{l_5}\left({\bf R}_{\alpha\beta\alpha\beta}\right)^{l_6}
\left({\bf R}_{ijij}\right)^{l_7}\left({\bf R}_{ijmn}\right)^{l_8}\left({\bf R}_{iajb}\right)^{l_9}
\left({\bf R}_{i\alpha j \beta}\right)^{l_{10}}\left({\bf R}_{0mnp}\right)^{l_{11}}
\nonumber\\
& \times & \left({\bf R}_{0m0n}\right)^{l_{12}}\left({\bf R}_{0i0j}\right)^{l_{13}}\left({\bf R}_{0a0b}\right)^{l_{14}}\left({\bf R}_{0\alpha 0\beta}\right)^{l_{15}}
\left({\bf R}_{0\alpha\beta m}\right)^{l_{16}}\left({\bf R}_{0abm}\right)^{l_{17}}\left({\bf R}_{0ijm}\right)^{l_{18}}
\nonumber\\
& \times & \left({\bf R}_{mnp\alpha}\right)^{l_{19}}\left({\bf R}_{m\alpha ab}\right)^{l_{20}}
\left({\bf R}_{m\alpha\alpha\beta}\right)^{l_{21}}\left({\bf R}_{m\alpha ij}\right)^{l_{22}}
\left({\bf R}_{0mn \alpha}\right)^{l_{23}}\left({\bf R}_{0m0\alpha}\right)^{l_{24}}
\left({\bf R}_{0\alpha\beta\alpha}\right)^{l_{25}}
\nonumber\\
&\times& \left({\bf R}_{0ab \alpha}\right)^{l_{26}}\left({\bf R}_{0ij\alpha}\right)^{l_{27}}
\left({\bf G}_{mnpq}\right)^{l_{28}}\left({\bf G}_{mnp\alpha}\right)^{l_{29}}
\left({\bf G}_{mnpa}\right)^{l_{30}}\left({\bf G}_{mn\alpha\beta}\right)^{l_{31}}
\left({\bf G}_{mn\alpha a}\right)^{l_{32}}\nonumber\\
&\times&\left({\bf G}_{m\alpha\beta a}\right)^{l_{33}}\left({\bf G}_{0ijm}\right)^{l_{34}} 
\left({\bf G}_{0ij\alpha}\right)^{l_{35}}
\left({\bf G}_{mnab}\right)^{l_{36}}\left({\bf G}_{ab\alpha\beta}\right)^{l_{37}}
\left({\bf G}_{m\alpha ab}\right)^{l_{38}},
\nd
which is written completely in terms of all the possible non-zero components of curvature tensors and 
G-flux components that would appear from the background \eqref{vegamey3}. The sum is over all 
$l_i$ and $n_i$, so this would be an exhaustive collections of all possible interactions in the system. However one might question the {\it absence} of other possible interactions. Aren't they important? The answer is: not here, because 
as we shall see in section \ref{sec3.3}, these are exactly the interactions that would enter the Schwinger-Dyson equations which would help us to express
the EOMs as expectation values over the Glauber-Sudarshan states. In the Wilsonian analysis that we did earlier, the other components could be thought of as being integrated out. Interestingly, in the limit $(n_1, n_2, n_3) \to (\infty, \infty, \infty)$ the interactions start to become 
{\it non-local}\footnote{The limit $n_0 \to \infty$ is non-locality in time and is discussed as case 4 in section 3.2.6 of \cite{desitter2}. This is surprisingly harmless. We could also absorb $n_0$ by shifting $n_1, n_2$ and $n_3$.}. In the next set of interactions in \eqref{selahran}, we consider a more advanced form of non-local interactions using $\mathbb{W}^{(r)}(y)$ which are nested integrals of the form:

{\footnotesize
\bg\label{romiran} 
\mathbb{W}^{(r)}_{(\{l_i\}, n_i)}(y) &=& {\rm M}_p^8 \int d^8y' \sqrt{{\bf g}_8(y')} ~\mathbb{F}^{(r)}(y - y') \mathbb{W}^{(r-1)}_{(\{l_i\}, n_i)}(y')\\
& = & {\rm M}_p^{16} \int d^8 y' \sqrt{{\bf g}_8(y')}~\mathbb{F}^{(r)}(y - y') 
\int d^8y'' \sqrt{{\bf g}_8(y'')} ~\mathbb{F}^{(r-1)}(y' - y'') \mathbb{W}^{(r-2)}_{(\{l_i\}, n_i)}(y''). \nonumber
 \nd}  
with $r$ and $\mathbb{F}^{(r)}(y - y')$, the latter depending implicitly on $(g_s, {\rm M}_p)$, denoting the level of non-localities and the non-locality functions respectively with $\mathbb{F}^{(0)}(y - y') \equiv 1$ (see details in section 3.2.6 of \cite{desitter2} and in section 3 of \cite{desitter3}). 
$c_{(r)}$ are numerical constants, and the lowest order non-local interaction $\mathbb{W}^{(1)}(y)$ may be expressed in terms of \eqref{selahran} as:
\bg\label{gingersh}
\mathbb{W}^{(1)}_{(\{l_i\}, n_i)}(y) \equiv
 \int d^8 y' \sqrt{{\bf g}_8(y')} \left({ \mathbb{F}^{(1)}(y - y') 
\mathbb{Q}_{\rm T}^{(\{l_i\}, n_i)} (y') \over {\rm M}_p^{\sigma(\{l_i\}, n_i) - 8}}\right).  
 \nd
The list of interactions mentioned above may not still be the full set of {\it perturbative} interactions that theory could allow for a generic choice of $g_s$, but for $g_s < 1$ we have pretty much an exhaustive list. 
In the following sub-section we will discuss some of the {\it non-perturbative} effects although
clearly higher order {\it perturbative} interactions\footnote{Typically D-branes and instantons in type II theories contribute terms of order ${1\over g_s}$ and ${1\over g_s^{2}}$ respectively. The perturbative contributions then come from the higher-order world-volume terms that scale as $g_s^{2+\theta}$ with $\theta > 0$. In a charge neutral configuration, the factor of 2 in the exponent off-sets the inverse $g_s$ factors, and therefore they can contribute to \eqref{sheela}.} 
from M2 and M5 instantons are captured by 
\eqref{selahran}, and so are the effects from the wrapped branes in \eqref{romiran}.  The topological corrections are assimilated in: 
\bg\label{mcawalk}
\mathbb{Y}_8 \equiv a_1 {\rm tr}~\mathbb{R}^4  + a_2 \left({\rm tr}~\mathbb{R}^2\right)^2 + 
a_3 \left({\rm tr}~\mathbb{R}^2\right) \left({\rm tr}~\mathbb{G}^2\right) + a_4 {\rm tr}~\mathbb{G}^4
+ a_5 \left({\rm tr}~\mathbb{G}^2\right)^2 + ...
, \nd
with $(a_1, a_2)$ taking numerical values while $(a_3, a_4, a_5, ..)$ proportional to powers of ${\rm M}_p$ so that \eqref{mcawalk} remains 
dimensionless. The dotted terms are additional possible traces.
All the parameters appearing above are {\it two-forms} that should not be confused with the corresponding tensors\footnote{Here we define the two-form in the standard way from the corresponding curvature tensors and G-flux components using the vielbeins $e^{a_o P}$ and the holonomy matrices ${\bf M}_{a_ob_o}$, as:
\bg\label{kolabaaz}
&&\mathbb{R} \equiv {\bf R}_{\rm MN}^{a_ob_o} {\bf M}_{a_ob_o}~ dy^{\rm M} \wedge dy^{\rm N}, ~~~~
\mathbb{G} \equiv {\bf G}_{\rm MN}^{a_ob_o} {\bf M}_{a_ob_o} ~dy^{\rm M} \wedge dy^{\rm N} \nonumber\\
&&{\bf R}_{\rm MN}^{a_ob_o} \equiv {\bf R}_{\rm MNPQ}~ e^{a_o P} e^{b_o Q}, ~~~~~~~ 
{\bf G}_{\rm MN}^{a_ob_o} \equiv {\bf G}_{\rm MNPQ}~ e^{a_o P} e^{b_o Q}. \nonumber \nd
By construction $\mathbb{R}$ is dimensionless, but $\mathbb{G}$ has a dimension of length so to make it dimensionless we need to insert ${\rm M}_p$. This in fact will determine the ${\rm M}_p$ scalings of $(a_3, a_4)$ in 
\eqref{mcawalk}. Note that the Wilson surface will become ${\rm exp}\left(i{\rm M}_p^3 \int \mathbb{C}\right)$, and so would be the charge of the M2-branes. The latter is captured by ${\rm T}_2$ in \eqref{sheela}.}. 

Another interesting thing to note is the ubiquity of four-form G-fluxes in the action \eqref{sheela}, although there are a few places where the three-form ${\bf C}_{\rm MNP}$ appears. For most of these cases we could alternatively use the four-form fluxes to rewrite in the following way:

{\footnotesize
\bg\label{cibelmell}
\int_{{\cal M}_{12}} {\bf G}_4 \wedge {\bf G}_4 \wedge {\bf G}_4 = 
\int_{\partial{\cal M}_{12}} {\bf C}_3 \wedge {\bf G}_4 \wedge {\bf G}_4, ~~~~ 
\int_{{\cal M}_{11}} {\bf C}_3 \wedge \mathbb{Y}_8  = 
-\int_{{\cal M}_{11}} {\bf G}_4 \wedge \mathbb{Y}_7, \nd}  
where the eleven-dimensional space-time ${\cal M}_{11} \equiv \mathbb{R}^{(1, 2)} \times {\cal M}_4 
\times {\cal M}_2 \times {\mathbb{T}^2\over {\cal G}}$ is considered as a boundary of a {\it twelve}-dimensional space-time  ${\cal M}_{12}$. This clearly suggests a F-theory uplift of our construction, as, such a coupling do appear there (see \cite{minasian}),  connecting in turn directly to the type IIB dual description. 

For the second case in \eqref{cibelmell}, we are assuming that $\mathbb{Y}_8$ is a locally exact form 
$d\mathbb{Y}_7$. In the limit $a_3 = a_4 = a_5 = 0$ in \eqref{mcawalk}, and with appropriate choice of 
$(a_1, a_2)$ (see \cite{desitter2}), the locally exact form is easy to demonstrate. Once we switch on G-flux components, we expect similar feature to show up. However the three-form ${\bf C}_{\rm MNP}$ do appear as the charge term for the M2-branes, so we will still need to deal with this. Additionally the EOMs for the G-fluxes are variations of the action \eqref{sheela} with respect to the three-form, so it would be a worth-while exercise to express the three-form in terms of Fourier components using the Schr\"odinger wave-functions
$(\gamma'_{\bf k}, \beta'_{\bf k})$, much like $(\gamma_{\bf k}, \beta_{\bf k})$ used in \eqref{prcard} and \eqref{kwinter} respectively. The connections between $(\gamma'_{\bf k}, \beta'_{\bf k})$ and 
$(\gamma_{\bf k}, \beta_{\bf k})$ are not so straightforward and may be formally presented as:
\bg\label{mellchat}
\gamma'_{\bf k} \equiv \gamma'_{\bf k}\left(\gamma_{\bf k}, \beta_{\bf k}, \Psi_{\bf k}\right), ~~~~
\beta'_{\bf k} \equiv \beta'_{\bf k}\left(\gamma_{\bf k}, \beta_{\bf k}, \Psi_{\bf k}\right), \nd
thus mixing with both $(\gamma_{\bf k}, \beta_{\bf k})$ and even involving the other set of 
Schr\"odinger wave-functions $\Psi_{\bf k}$ as in \eqref{hudson} from the gravitational sector. Fortunately however, as we shall see a bit later, we will not have to deal with this here.

The energy-momentum tensors from the interacting part of the action \eqref{sheela} are now important. It is also important to keep track of all the space-time directions carefully so we will follow the finer subdivision, namely $(m, n), (\alpha, \beta)$ and $(a, b)$ for the internal space \eqref{anonymous}, and $(\mu, \nu)$ for the remaining $2+1$ dimensional space-time.  Let us now define the following variables:
\bg\label{cibelmoja}
&&\mathbb{N}_1 \equiv 2\sum_{i = 1}^{27} l_i, ~~~~\mathbb{N}_2 \equiv n_0 + n_1 + n_2 - 2n_3 + 
l_{34} + l_{35} \nonumber\\
&& \mathbb{N}_3 \equiv 2(l_{28} + l_{29} + l_{31}), ~~~ \mathbb{N}_4 \equiv l_{30} + l_{32} + l_{33},
~~~ \mathbb{N}_5 \equiv 2(l_{36} + l_{37} + l_{38}), \nd
where $l_i$ are the powers of the curvature tensors and G-flux components appearing in \eqref{selahran}; and $n_i$, with $i = (0, 1, 2, 3)$ are the derivatives along the temporal, ${\cal M}_4$, ${\cal M}_2$ and the toroidal directions ${\mathbb{T}^2\over {\cal G}}$ respectively. We can typically take $n_3 = 0$. This way the fields remain independent of the toroidal directions and therefore do not jeopardize the duality to the IIB side.  With this we can define the energy-momentum tensor from the quantum terms along the 
$({\rm M}, {\rm N}) \in \Big((m, n), (\alpha, \beta)\Big)$ directions in the following way:

{\footnotesize
\bg\label{cibelchut} 
 \mathbb{C}^{(q, 2p)}_{\rm MN} \equiv - {2\over \sqrt{-{\bf g}_{11}}} {\partial {\bf S}_{\rm int} \over \partial 
 {\bf g}^{\rm MN}} & = &
 \sum_{\{l_i\}, n_i}{1\over {\rm M}_p^{\sigma(\{l_i\}, n_i)}}
\left(-2 {\bf g}_{\rm MN} \mathbb{Q}_{\rm T}^{(\{l_i\}, n_i)} + {\partial \mathbb{Q}_{\rm T}^{(\{l_i\}, n_i)} \over \partial {\bf g}^{\rm MN}}\right)\\
&\times & \delta\Big(\mathbb{N}_1 + \mathbb{N}_2 + (p+2)\mathbb{N}_3 + (2p+1)\mathbb{N}_4
+ (p-1)\mathbb{N}_5 - q - 2\Big), \nonumber \nd}
where for a given value of $q \ge 0$ and $p \ge 3/2$, the delta function above gives an equation in terms of 
$\mathbb{N}_i$, or alternatively, in terms of ($l_i, n_i$). The integer solutions of these are then summed over to provide the full contributions from the quantum corrections. The ${\rm M}_p$ scalings of each of these quantum terms, given here and in \eqref{sheela} by $\sigma(\{l_i\}, n_i)$, also get fixed because:
\bg\label{chatchat}
\sigma(\{l_i\}, n_i) \equiv \mathbb{N}_1 + \mathbb{N}_2 + \mathbb{N}_4 + {1\over 2}\left(\mathbb{N}_3 +
\mathbb{N}_5\right), \nd
showing that all the quantum terms entering in \eqref{cibelchut} have in general {\it different} ${\rm M}_p$ scalings. So at the face value there is a ${\rm M}_p$ hierarchy in the quantum terms. When $p = 0$, we lose the $g_s$ hierarchy completely because there are an infinite number of terms for any values of $q \ge 0$ in 
$\mathbb{C}_{mn}^{(q, 0)}$. What about ${\rm M}_p$ hierarchy? From \eqref{chatchat} we do however seem to retain the ${\rm M}_p$ hierarchy, although a careful consideration with the {\it localized} fluxes 
\eqref{pikilia} as discussed in section 2.2 of \cite{desitter3}, show that this is not true. Since the $p = 0$ case is the time-{\it independent} case from \eqref{prcard}, it suggests that there are an infinite number of quantum corrections with neither $g_s$ nor ${\rm M}_p$ hierarchies contributing to the system. This is a clear sign of a breakdown of an effective field theory (EFT) description in the time-independent case (see also \cite{desitter2, desitter3} for more details).

In analyzing the energy-momentum tensor for the quantum term \eqref{cibelchut} we have ignored the non-local terms. They can be easily accommodated in because the non-local counter-terms \eqref{romiran} and \eqref{gingersh} are defined using nested integrals from \eqref{selahran} and using non-locality functions. These functions tend to become very small at low energies (see section 3.2.6 of \cite{desitter2}), so we can ignore their contributions. Nevertheless, even if we do take their contributions into account, as done in section 3 of \cite{desitter3}, the problems associated with $p = 0$ case {\it do not} get alleviated, implying that an EFT description cannot be restored in the time-independent case. Once time-dependences are switched on, {\it i.e.} when $p \ge 3/2$, an EFT description with a finite set of local and non-local quantum terms and well-defined  ($g_s, {\rm M}_p$) hierarchies, magically appear. 

The energy-momentum tensor along the toroidal direction, {\it i.e.} along ${\mathbb{T}^2\over {\cal G}}$, has a similar structure to \eqref{cibelchut} but with some minor, and crucial, changes. It may be expressed as:

{\footnotesize
\bg\label{melischut} 
 \mathbb{C}^{(q, 2p)}_{ab} \equiv - {2\over \sqrt{-{\bf g}_{11}}} {\partial {\bf S}_{\rm int} \over \partial 
 {\bf g}^{ab}} & = &
 \sum_{\{l_i\}, n}{1\over {\rm M}_p^{\sigma(\{l_i\}, n_i)}}
\left(-2 {\bf g}_{ab} \mathbb{Q}_{\rm T}^{(\{l_i\}, n_i)} + {\partial \mathbb{Q}_{\rm T}^{(\{l_i\}, n_i)} \over \partial {\bf g}^{ab}}\right)\\
&\times & \delta\Big(\mathbb{N}_1 + \mathbb{N}_2 + (p+2)\mathbb{N}_3 + (2p+1)\mathbb{N}_4
+ (p-1)\mathbb{N}_5 - q + 4\Big), \nonumber \nd}  
with $p$ still bounded below by $p \ge 3/2$, but now, more importantly, $q \ge 6$. This keeps the last two terms negative definite, which is what we want for the system to make sense. This awkward factor of 6 may be easily explained from the metric configuration \eqref{vegamey3} that scales as $g_s^{4/3}$ along the 
toroidal direction. Finally, the energy-momentum tensor for the space-time directions differ from \eqref{cibelchut} and \eqref{melischut} in the following way:

{\footnotesize
\bg\label{novharr} 
 \mathbb{C}^{(q, 2p)}_{\mu\nu} \equiv - {2\over \sqrt{-{\bf g}_{11}}} {\partial {\bf S}_{\rm int} \over \partial 
 {\bf g}^{\mu\nu}} & = &
 \sum_{\{l_i\}, n}{1\over {\rm M}_p^{\sigma(\{l_i\}, n_i)}}
\left(-2 {\bf g}_{ab} \mathbb{Q}_{\rm T}^{(\{l_i\}, n_i)} + {\partial \mathbb{Q}_{\rm T}^{(\{l_i\}, n_i)} \over \partial {\bf g}^{\mu\nu}}\right)\\
&\times & \delta\Big(\mathbb{N}_1 + \mathbb{N}_2 + (p+2)\mathbb{N}_3 + (2p+1)\mathbb{N}_4
+ (p-1)\mathbb{N}_5 - q - 8\Big), \nonumber \nd}  
with the same lower bound on $p$ as before, and we expect $q \ge 0$. There are however a few subtleties  that need to be elaborated on before we can fix the value of $q$. This is what we turn to next.

\begin{table}[tb]
 \begin{center}
\renewcommand{\arraystretch}{1.5}
\begin{tabular}{|c|c|c|}\hline orientation & ${g_s}$ scaling & 
IIB dual \\ \hline\hline
${\cal M}_4 \times {\cal M}_2$ & ${1\over g_s^2}$ & Taub-NUT instanton along ${\bf R}^{3, 1}$ \\ \hline
${\cal M}_4 \times {\mathbb{T}^2 \over {\cal G}}$ & $g_s^0$ & D3 instanton along ${\cal M}_4$ \\ \hline
${\cal M}_4 \times ({\bf S}^1)_\alpha \times ({\bf S}^1)_{a = 3}$ & ${1\over g_s}$ & NS5 instanton along 
${\cal M}_4 \times ({\bf S}^1)_\alpha \times ({\bf S}^1)_{a = 3}$ \\ \hline
${\cal M}_4 \times ({\bf S}^1)_\alpha \times ({\bf S}^1)_{b = 11}$ & ${1\over g_s}$ & D5 instanton along 
${\cal M}_4 \times ({\bf S}^1)_\alpha \times ({\bf S}^1)_{a = 3}$\\ \hline
  \end{tabular}
\renewcommand{\arraystretch}{1}
\end{center}
 \caption[]{The ${g_s}$ scalings of the wrapped M5-instantons along various six-cycles inside the 
 non-K\"ahler eight-manifold 
 \eqref{anonymous} in M-theory; with ($\alpha, \beta$) parametrizing the coordinates of ${\cal M}_2$ and 
 ($a, b$) parametrizing the coordinates of ${\mathbb{T}^2 \over {\cal G}}$. The last two configurations, wrapped on local one-cycles, break the four-dimensional de Sitter isometries in the IIB side so 
 cannot contribute to the energy-momentum tensors.}
  \label{instabeta}
 \end{table}

\subsection{New non-perturbative effects from non-local counter-terms \label{sec3.2}} 

Before we derive the subtleties associated with the energy-momentum tensor along the $2+1$ dimensional space-time, $\mathbb{C}^{(q, 2p)}_{\mu\nu}$, we need to re-visit the local and the non-local quantum terms to search for non-perturbative effects that {\it can} contribute. Recall that the non-perturbative effects were 
Borel summed to ${\rm exp}\left(-{1\over g_s^{1/3}}\right)$, so in the limit $g_s \to 0$, they simply decouple. Is this always true? In the following we want to argue that this may not always be true and there could be terms that do contribute. 


\subsubsection{Case 1: The BBS type instanton gas \label{sec3.2.1}}


Our starting point is the perturbative series of quantum terms in \eqref{selahran}. We can ask whether we can 
generalize this further. For example, how about {\it powers} of the individual terms in \eqref{selahran}? The answer is that it not necessary to do this because:
\bg\label{brazimey}
\mathbb{Q}_{\rm T}^{(\{l_i\}, n_i)}\left(y \right) \otimes  \mathbb{Q}_{\rm T}^{(\{l_j\}, m_j)}\left( y \right) 
  \equiv \mathbb{Q}_{\rm T}^{(\{l_i + l_j\}, n_i+ m_j)}\left(y \right), \nd
so in principle \eqref{selahran} does capture the most generic perturbative quantum effects, and any arbitrary modifications to it is already embedded in the series itself. This is of course an expected property of the underlying renormalization group itself, so its appearance here should not be of any surprise. 

However such a generalization unfortunately {\it do not} extend to the non-local counter-terms. The non-local counter-terms are expressed in terms of nested integrals \eqref{romiran}, and we do not expect products of two nested integrals could produce another equivalent integral. For example:

{\footnotesize
\bg\label{melisromi}
\mathbb{W}^{(r_i)}_{(i)}(y) \otimes \mathbb{W}^{(r_j)}_{(j)}(y)
&=& {\rm M}_p^{16} \int d^8y' d^8 y''\sqrt{{\bf g}_8(y'){\bf g}_8(y'')} ~\mathbb{F}^{(r_i)}(y - y')
\mathbb{F}^{(r_j)}(y - y'') \mathbb{W}^{(r_i-1)}_{(i)}(y')  \mathbb{W}^{(r_j-1)}_{(j)}(y'') \nonumber\\
& \ne & \mathbb{W}^{(r_i + r_j)}_{(i+j)}(y) \equiv 
{\rm M}_p^{8} \int d^8z \sqrt{{\bf g}_8(z)} ~\mathbb{F}^{(r_i + r_j)}(y - z)
~ \mathbb{W}^{(r_i + r_j-1)}_{(i+ j)}(z), \nd}
with $\mathbb{W}^{(r_i)}_{(i)}(y) = \mathbb{W}^{(r_i)}_{(\{l_i\}, n_i)}(y)$;
which remains true unless the non-locality function, given by $\mathbb{F}^{(r)}(y - z)$, becomes a localized function of the form ${\delta^8(y - z) \over {\rm M}^8_p\sqrt{{\bf g}_8(z)}}$. In the latter case this reduces to 
\eqref{brazimey}. Taking \eqref{melisromi} and \eqref{brazimey} into account suggests that there could be 
additional quantum terms of the form:
\bg\label{romirpa}
\mathbb{U}^{(r)}_{(\{l_i\}, n_i)}(y) &= & \sum_{n = 1}^\infty d_n ~{\rm M}_p^{8n} \left(\int d^8y' \sqrt{{\bf g}_8(y')} ~\mathbb{F}^{(r)}(y - y') \mathbb{W}^{(r-1)}_{(\{l_i\}, n_i)}(y')\right)^n \nonumber\\
& = &  \sum_{n = 1}^\infty (-1)^n d_n ~{\rm M}_p^{8n} \left(\int d^8 z \sqrt{{\bf g}_8(y - z)} ~\mathbb{F}^{(r)}(z) \mathbb{W}^{(r-1)}_{(\{l_i\}, n_i)}(y - z)\right)^n, \nd
for any choice of the level of non-locality $r$. In the second equality we simply redefined the coordinated to shift the non-localities to the metric and the quantum terms. At the lowest level of non-locality, {\it i.e.} for $r = 1$, we can simplify \eqref{romirpa} in the following way:
\bg\label{ginagers}
\mathbb{U}^{(1)}_{(\{l_i\}, n_i)}(y) \equiv \sum_{n = 1}^\infty  d_n 
\left[ \int d^6 y' \sqrt{{\bf g}_6(y')} \left({ \mathbb{F}^{(1)}(y - y') 
\mathbb{Q}_{\rm T}^{(\{l_i\}, n_i)} (y') \over {\rm M}_p^{\sigma(\{l_i\}, n_i) - 6}}\right)\right]^n  
 \nd
where we have restricted the coordinate dependences only on the base ${\cal M}_4 
\times {\cal M}_2$ of \eqref{anonymous}, and therefore both the determinant of the metric and the 
${\rm M}_p$ scaling change accordingly. The {\it absence} of the warped torus volume $\mathbb{V}_{{\bf T}^2}$, that appeared prominently in \cite{desitter2}, is important to get the scaling right. We will also take 
$\Omega_{ab} = \epsilon_{ab}$ in \eqref{pikilia}.
Thus one should interpret \eqref{ginagers} as a separate class of non-local interactions. Later however we will consider the case where the torus volume does show up. The other term appearing above is 
$\mathbb{Q}_{\rm T}^{(\{l_i\}, n_i)} (y')$ which is given in \eqref{selahran}. The $g_s$ scaling of the 
$n$-th term in the series expansion may be written as $g_s^{\theta_p}$ where $\theta_p$ is given by:
\bg\label{relishlu}
\theta_p = {n\over 3} \Big(\mathbb{N}_1 + \mathbb{N}_2 + (2+p) \mathbb{N}_3 + (2p+1)\mathbb{N}_4 
+ (p-1)\mathbb{N}_5 - 6\Big), \nd
with $\mathbb{N}_i$ defined as in \eqref{cibelmoja} and $\mathbb{F}^{(1)}(y - y')$ do not explicitly depend on $g_s$. Since $p \ge 3/2$ most of the terms are positive definite 
in \eqref{relishlu}, except for the last term. If we only keep $\mathbb{N}_5 \ne 0$ by switching on quantum terms associated with the G-flux components ${\bf G}_{{\rm MN}ab}$ in \eqref{selahran}, the quantum term 
\eqref{ginagers} blows up in the limit $g_s \to 0$. This is because of the ${1\over g_s^2}$ factor from the 
determinant of the metric in \eqref{ginagers}. Studying the EOMs order by order in powers of $g_s$, as shown in \cite{desitter2, desitter3}, we switch on smaller values of $\mathbb{N}_1, \mathbb{N}_2$ and $\mathbb{N}_5$ and for these values \eqref{ginagers} tends to blow-up. However such a series, for appropriate choices of $d_n$, could be summed as a trans series to take the following form\footnote{One concern is whether this can always be done. If $d_n = {(-1)^n\over n!}$ then there is no doubt that \eqref{lushreli} is always true with $c_k = 0$ for $k > 1$. Question is what happens in the generic case. Generically however we can expect: $$ d_n \equiv {1\over n!} \sum_{l = 1}^\infty c_l (-l)^n = {(-1)^n c_1 \over n!}  + 
{(-2)^n c_2 \over n!} + {(-3)^n c_3 \over n!} + ....$$
\noindent where $d_n$ is defined in \eqref{romirpa} and $c_l$ are positive or negative integers. We can now go to the limit where the terms inside the bracket in \eqref{ginagers} are {\it smaller} than 1. Plugging this in say \eqref{romirpa}, every term of $d_n$, when summed over $n$, will produce an exponentially decaying contribution like \eqref{lushreli} when the integrals therein become much {\it bigger} than 1. Solutions would exist if the following matrix:
\bg\label{hkpilla}
\mathbb{M} = \left(\begin{matrix} -1 & -2 & -3 & -4 & .... \cr ~~{1\over 2} & ~~2 & ~~{9\over 2} & ~~ 8 & ....\cr
-{1\over 6} & -{4\over 3} & -4 & -{32\over 3} & .... \cr
~~{1\over 24} & ~~{2\over 3} & ~~ {27\over 8} & ~~{32\over 3} & ... \cr
.... & ....& {} & {} & ...\cr  \end{matrix}\right) \nonumber \nd
has an inverse. Unfortunately $\mathbb{M}$ is {\it infinite} dimensional so in practice it will be impossible to ascertain the full inverse of such a matrix. However since higher values of $c_l$ and $n$ also correspond to smaller and smaller contributions, it would make sense to terminate $\mathbb{M}$ to  large but finite dimensions. For such cases, one may verify that the inverses continue to exist thus giving more practicality to the series of $d_n$ above. Additionally, 
assuming wide separation between 
two consecutive $c_i$ and $c_j$, the dominant term will always be the first few terms of 
\eqref{lushreli}. Relatively, therefore it makes sense to keep only the first few terms of \eqref{lushreli} as others will die-off faster than this when $g_s \to 0$ and $\theta < 2$, for every choice of $(\{l_i\}, n_i)$. When  $\theta > 2$, one may perturbatively expand the exponential to arbitrary orders and study the corresponding $g_s$ scalings. \label{luisis}}:

{\footnotesize
\bg\label{lushreli}
\mathbb{U}^{(1)}(y) &\equiv & \sum_{\{l_i\}, n_i, k} c_k~ {\rm exp}\left[- k \int d^6 y' \sqrt{{\bf g}_6(y')} 
\left({ \mathbb{F}^{(1)}(y - y') 
\mathbb{Q}_{\rm T}^{(\{l_i\}, n_i)} (y') \over {\rm M}_p^{\sigma(\{l_i\}, n_i) - 6}}\right)\right] \\
&\approx & \sum_{\{l_i\}, n_i, k} c_k~{\rm exp}\left[- k \int d^6 y' \sqrt{{\bf g}_6(y')} \left({ \mathbb{F}^{(1)}(- y') 
\mathbb{Q}_{\rm T}^{(\{l_i\}, n_i)} (y') \over {\rm M}_p^{\sigma(\{l_i\}, n_i) - 6}}\right) 
+ k {\cal O}\left(y {\partial \mathbb{F}^{(1)}(- y') \over \partial y'}\right)\right], \nonumber \nd}
where the validity of the second line rests on the smallness of the derivative of the function 
$\mathbb{F}^{(1)}(- y')$. Interestingly, in this limit, the first term looks suspiciously close\footnote{Close, but not exactly the same: the integral structure here is a bit different from an actual BBS instanton gas contribution.}
 to the action of a gas of 
{\it neutral} BBS M5-instantons \cite{bbs} wrapped on the base ${\cal M}_4 \times {\cal M}_2$ of our eight-manifold 
\eqref{anonymous} once we absorb the function $\mathbb{F}^{(1)}(- y')$ in the quantum series 
\eqref{selahran} (see {\bf Table \ref{instabeta}}). The tell-tale sign of ${1\over g_s^2}$ from the determinant of the base metric, which signals the presence of the wrapped instantons, in fact {\it deters} it to contribute to the energy-momentum tensor unless the $g_s$ scaling from $\mathbb{Q}_{\rm T}^{(\{l_i\}, n_i)} (y')$ for {\it any} choices of 
($\{l_i\}, n_i$) is always bigger than 2. For the two energy-momentum tensors that we derived in  
\eqref{cibelchut} and \eqref{melischut}, the lowest order quantum corrections contributing to the EOMs scale as $g_s^{2/3}$ \cite{desitter2, desitter3}. For these cases, even though the non-local counter-terms 
like \eqref{romiran} do contribute, there appears to be no contribution from \eqref{lushreli}. Can higher order in $g_s$ contribute? Let us infer it from the following quantitative analysis, without using any approximations:

{\footnotesize
\bg\label{notiyell}
\mathbb{T}^{({\rm np}; 1)}_{\rm MN}(z) & = & -{2{\rm M}_p^{11} \over \sqrt{-{\bf g}_{11}(z)}} 
~{\delta\over \delta {\bf g}^{\rm MN}(z)} \left(\int d^{11}x \sqrt{-{\bf g}_{11}(x)} \mathbb{U}^{(1)}(x)\right)
\\
& = &  \sum_{\{l_i\}, n_i, k} c_k~{\bf g}_{\rm MN}(z)~{\rm exp}\left[- k \int d^6 y' \sqrt{{\bf g}_6(y')} 
\left({ \mathbb{F}^{(1)}(z- y') 
\mathbb{Q}_{\rm T}^{(\{l_i\}, n_i)} (y') \over {\rm M}_p^{\sigma(\{l_i\}, n_i) - 6}}\right)\right]\nonumber\\
&+& \sum_{\{l_i\}, n_i, k} \left({2k c_k \over {\rm M}_p^{\sigma(\{l_i\}, n_i) - 11}}
\int d^{11} x ~\mathbb{F}^{(1)}(x - z)~{\delta \left(\sqrt{{\bf g}_6(z)}\mathbb{Q}_{\rm T}^{(\{l_i\}, n_i)}(z) \right)\over 
\delta {\bf g}^{\rm MN}(z)}\sqrt{{\bf g}_{11}(x) \over {\bf g}_{11}(z)}\right) \nonumber\\
&&~~~~~~~~~~ \times {\rm exp}\left[- k \int d^6 y' \sqrt{{\bf g}_6(y')} 
\left({ \mathbb{F}^{(1)}(x - y') 
\mathbb{Q}_{\rm T}^{(\{l_i\}, n_i)} (y') \over {\rm M}_p^{\sigma(\{l_i\}, n_i) - 6}}\right)\right], \nonumber \nd}
where we have used the fact that ${\delta {\bf g}^{\rm MN}(w) \over \delta {\bf g}^{\rm PQ}(z)} = 
{1\over {\rm M}^{d}_p}~\delta^{\rm M}_{\rm P} \delta^{\rm N}_{\rm Q} \delta^d(w - z)$, with no extra 
$\sqrt{{\bf g}_d}$ factor with $d = 11, 6$, the latter because we have assumed the internal coordinates 
$y' = (y^m, y^\alpha)$ to only span the coordinates of the base ${\cal M}_4 \times {\cal M}_2$ of eight-manifold \eqref{anonymous}. The first term of \eqref{notiyell} is interesting: its the metric component suppressed by an exponentially decaying factor (for small $g_s$ scaling of \eqref{selahran}). We will not worry about this term as we shall show later that such a term is cancelled by a counter-term. The second term, on the other hand, contains the quantum pieces 
\eqref{selahran}, again suppressed by the exponential factor. It is instructive to note the $g_s$ scalings of
every components of the energy-momentum tensor:
\bg\label{yellstop}
&& \mathbb{T}^{({\rm np}; 1)}_{ab}(z) = \sum_k c_k \left[g_s^{4/3} + {k \over g_s^2} \left(g_s^{\theta + 4/3}\right)\right]
~{\rm exp} \left(-{k \over g_s^2} \cdot g_s^\theta\right) \nonumber\\
&& \mathbb{T}^{({\rm np}; 1)}_{\mu\nu}(z) = \sum_k c_k \left[g_s^{-8/3} + {k \over g_s^2} \left(g_s^{\theta - 8/3}\right)\right]
~{\rm exp} \left(-{k \over g_s^2} \cdot g_s^\theta\right) \nonumber\\
&& \mathbb{T}^{({\rm np}; 1)}_{mn}(z) = \mathbb{T}^{({\rm np}; 1)}_{\alpha\beta}(z)  = \sum_k c_k \left[g_s^{-2/3} + {k \over g_s^2} \left(g_s^{\theta - 2/3}\right)\right]
~{\rm exp} \left(-{k \over g_s^2} \cdot g_s^\theta\right), \nd
where $\theta \equiv {1\over n}(\theta_p + 2n)$ and $\theta_p$ as in \eqref{relishlu}. Notice that the only way 
\eqref{yellstop} can contribute is when $\theta \ge 2$. From our earlier studies in \cite{desitter2, desitter3}, 
$\theta \ge 2/3$ for the components $(a, b), (m, n)$ and $(\alpha, \beta)$,  and therefore naively there seems to be  no contributions from their corresponding energy-momentum tensors \eqref{yellstop}. The only energy-momentum tensor that appears to contribute seems to be $\mathbb{T}^{({\rm np}; 1)}_{\mu\nu}(z)$.

The above conclusion is not quite correct once we look at the 
case corresponding to $\theta = 8/3$ for the components $(a, b), (m, n)$ and $(\alpha, \beta)$. The four components now scale as $(g_s^2, g_s^{-2}, g_s^0, g_s^0)$ with the adjoining exponential pieces going as 
${\rm exp}\left(-k g_s^{2/3}\right)$ for all of them. Recall from \cite{desitter2} and \cite{desitter3}, these $g_s$ scalings are exactly how the classical EOMs scale, and therefore it appears that there are non-perturbative corrections that scale as $\theta = 8/3$, with the corresponding energy-momentum tensor going as:

{\footnotesize
\bg\label{notiyello}
\hskip-.11in\mathbb{T}^{({\rm np}; 1)}_{\rm MN}(z)
& = &  \sum_{\{l_i\}, n_i, k} c_k~{\rm exp}\left[- k \int d^6 y' \sqrt{{\bf g}_6(y')} 
\left({ \mathbb{F}^{(1)}(- y') 
\mathbb{Q}_{\rm T}^{(\{l_i\}, n_i)} (y') \over {\rm M}_p^{\sigma(\{l_i\}, n_i) - 6}}\right)
+ k {\cal O}\left({\partial \mathbb{F}^{(1)}(- y') \over \partial y'}\right)\right]\\
&\times &\left({\bf g}_{\rm MN}(z) + {2k \over {\rm M}_p^{\sigma(\{l_i\}, n_i) - 11}}
\int d^{11} x ~\mathbb{F}^{(1)}(- z)~{\delta \left(\sqrt{{\bf g}_6(z)}\mathbb{Q}_{\rm T}^{(\{l_i\}, n_i)}(z) \right)\over 
\delta {\bf g}^{\rm MN}(z)}\sqrt{{\bf g}_{11}(x) \over {\bf g}_{11}(z)} + ...\right), \nonumber \nd}
once we impose the approximation of slowly varying $\mathbb{F}^{(1)}(-y)$. Our exact expression, 
\eqref{notiyell}, tells us that this approximation is {\it not} necessary but is useful nevertheless because it appears to isolate the instanton effects from all the non-local factors. The problem however with this approximation is the appearance of the eleven-dimensional volume factor for the second term in 
\eqref{notiyello}, unless we use a box normalization condition. Clearly this issue does not arise in 
\eqref{notiyell} where the eleven-dimensional volume factor is regulated by the function 
$\mathbb{F}^{(1)}(x - z)$, for example as:
\bg\label{Ebanks}
\int d^{11}x \sqrt{-{\bf g}_{11}(x)}~\mathbb{F}^{(1)}(x - z) = g_s^{-14/3} ~\mathbb{F}_e(z), \nd
where $\mathbb{F}_e(z)$ is a well-defined {\it finite} function even if we did not include the contributions from the exponential factor. The latter would further improve the value of the function, so its exclusion from 
\eqref{Ebanks} does not change anything (see \eqref{Ebanks2}). 
The interesting take-home point from such an integral is that the effects of the non-localities from $\mathbb{F}^{(1)}(z - y')$, 
$\mathbb{F}^{(1)}(x - z)$ and $\mathbb{F}^{(1)}(x - y')$ in \eqref{notiyell} are effectively removed by the 
nested integrals so that the final result for the energy-momentum tensor is a consistent quantity.  Expanding perturbatively the exponential factors in \eqref{notiyell}, we get:

{\footnotesize
\bg\label{mcamahal}
\mathbb{T}^{({\rm np}; 1)}_{{\rm M}_a{\rm N}_a}(z) = g_s^{l_a}\left[1 - g_s^{2/3}\sum_k k c_k  + 
{\cal O}(g_s^{4/3})\right] + {g_s^{8/3 + l_a} \over g_s^2} \sum_k \left[ k c_k(1 - k g_s^{2/3}) + 
{\cal O}(g_s^{4/3})\right],\nd}
where the subscript $a$ specifies the components, with $l_a$ taking the corresponding values,  in \eqref{yellstop}. We have also expanded the exponential piece ${\rm exp}\left(-k g_s^{2/3}\right)$ perturbatively in powers of $g^{2/3}_s$. The integer $k$ can be arbitrarily large and let us assume that it approaches infinity as $k \to \epsilon^{-1}$ with $\epsilon \to 0$. In that case as long as $g_s$ goes to zero as $g_s \to \epsilon^b$, 
with $b$ bounded by:
\bg\label{mcabuddhi}
{3\over 2} < b < {3\over 2}\left\vert 3 - \left({{\rm log}~\vert c_{\rm max}\vert \over {\rm log}~\epsilon}\right)\right\vert, \nd
and $c_{\rm max}$ being the largest value of $c_k$,
most of the series appearing above should be convergent, except the one without $g_s$ suppression.
Such a coefficient can in principle be absorbed in the definition of $\mathbb{F}^{(1)}(x - z)$ which we have left unspecified so far\footnote{We have used \eqref{mcabuddhi} to allow for a convergent series by restricting ourselves to ${\cal O}(g_s^{2/3})$. One can do slightly better than this by noting that the series in $k$ may be bounded in the following way: 
\bg\label{lemurg}
&& \sum_k k c_k < {1\over 2} k(k+1) \vert c_{\rm max} \vert < k^2 \vert c_{\rm max}\vert \nonumber\\
&& \sum_k k^2 c_k < {1\over 6} k(k+1)(2k +1) \vert c_{\rm max} \vert < k^3 \vert c_{\rm max}\vert, ~~~~
\sum_k k^n c_k < k^{n+1} \vert c_{\rm max} \vert, \nd
which determines the generic bound that one could place on the perturbative expansion of the exponential term. To see what value of $n$ should suffice, we note that the perturbative expansion of the exponential factor from \eqref{notiyell}, to order $n$ typically involve a term of the form:
\bg\label{mcakhoj}
\mathbb{S}(z) = \sum_k {c_k k^{n+1}\over n!}\left({\mathbb{F}^{(1)}(w_1 - w_2) g_s^{2/3}\over {\rm M}_p^{\sigma_{(i)} -6}}\right)^n
\equiv \sum_k {c_k k^{n+1}\over n!} {\bf \Gamma}^n(w_1, w_2), \nd
which is heavily suppressed by various factors like $g_s, {\rm M}_p$ and 
$\mathbb{F}^{(1)}(w_1 - w_2)$ including $n!$ so long as $ \sigma_{(i)} > 6$ where $\sigma_{(i)} 
\equiv \sigma(\{l_i\}, n_i)$ is defined in \eqref{chatchat}. Such suppression factors tell us that we need not go beyond some order of expansion for the exponential part in \eqref{notiyell}. Thus if we go up to order $p$, then it is easy to see that the bounds in \eqref{mcabuddhi} become:
\bg\label{chukmukh}
{3\over 2} < b <  {3\over 2p}\left\vert p + 2 - \left({{\rm log}~\vert c_{\rm max}\vert \over {\rm log}~\epsilon}\right)\right\vert,  \nd
which reproduces \eqref{mcabuddhi} as a special case when $p = 1$. To see what happens for arbitrary values of ($k, n$), let us  assume that $k$ goes to a large value $k_{\rm max} \equiv \epsilon^{-1}$. We can now use the lower bound on $b$ from \eqref{chukmukh} to first sum over $k$ in \eqref{mcakhoj} and then sum over $n$.  In this case, it is clear that the series in \eqref{mcakhoj} may be bounded from above by:
\bg\label{mcaposter}
\mathbb{S}(z) < ~ \vert c_{\rm max} \vert k^2_{\rm max} ~{\rm exp}\Big(-k_{\rm max} 
{\bf \Gamma}(w_1, w_2)\Big), \nd
which is how we can control the series right to the point where $g_s$ goes to zero as $g_s = \epsilon^b$, with $b$ as in \eqref{chukmukh}. Thus for $\epsilon^b < g_s < 1$, both $\vert c_{\rm max}\vert$ and 
$k_{\rm max}$ can be arbitrarily large, yet $\mathbb{S}(z)$ in \eqref{mcakhoj} can still be {\it finite}. \label{knauf}}.

It is important to note that we have imposed one condition on $\theta$ to \eqref{cibelchut} and \eqref{melischut}, and a different condition on $\theta$ to the corresponding components in \eqref{notiyell}. The reason is simple: although the quantum series appearing in both these set of energy-momentum tensors are the same, {\it i.e.} the series 
\eqref{selahran}, the latter has an extra $1/g_s^2$ suppression going with it, creating the necessary difference in the outcome. 
If we go beyond $r = 1$ and sum the trans-series in \eqref{romirpa}, much like how we did before\footnote{Implying that the same conditions used earlier apply here too.}, the contributions to the EOMs from any of the corresponding energy-momentum tensors are very small. The generic action may be expressed as:

{\footnotesize
\bg\label{sheela2}
{\bf S}_2 = \sum_{\{l_i\}, n_i, k} \int d^{11} x \sqrt{-{\bf g}_{11}} \sum_{r = 1}^\infty  c_k~
{\rm exp}\Big[- k {\rm M}_p^6 \int d^6 y \sqrt{{\bf g}_6(y)}
~\mathbb{F}^{(r)}(x - y) \mathbb{W}^{(r - 1)}\left(y; \{l_i\}, n_i\right)\Big], \nd}
where we again assume dependence on the coordinates of ${\cal M}_4 \times {\cal M}_2$ of  the internal eight-manifold \eqref{anonymous} although, as we shall discuss in section \ref{sec3.2.3}, the warped toroidal volume could be inserted in the integrand.  The action that appeared in \eqref{notiyell} is the restrictive piece with $r = 1$, with no additional dependence on the toroidal volume. The other function appearing above is defined in \eqref{cibelmoja}. 


\subsubsection{Case 2: F-theory seven-branes \label{sec3.2.2}}


So far we discussed various ways of generating non-perturbative terms that contribute as $1/g_s^2$ to the energy-momentum tensors. From {\bf Tables \ref{instabeta}} and {\bf \ref{instabeta2}} it seems other wrapped instantons either do not contribute\footnote{Except for a class of instantons, that we will discuss a bit later.} $-$ by breaking the de Sitter isometries in the IIB side $-$ or they just contribute perturbatively, as positive powers of $g_s$. What about type IIB seven-branes wrapped on ${\cal M}_4$? These seven-branes would map to Taub-NUT spaces oriented along ${\cal M}_2 \times {\mathbb{T}^2\over {\cal G}}$ {\it i.e.} along
$(\alpha, \beta)$ and $(a, b)$ directions. We could even include more generic seven-branes that are not necessarily D7-branes. Their dual, in M-theory, would be warped Taub-NUT spaces whose properties are not too hard to ascertain (see \cite{rajukaju, nonabel}). Unfortunately two obstacles forbid a naive realization of this scenario: one, a Taub-NUT space cannot have a simple product geometry as ${\cal M}_2 \times 
{\mathbb{T}^2\over {\cal G}}$; and two, we cannot allow non-trivial axio-dilaton charge in the type IIB side, as this will change the type IIA coupling completely, ruining our basic $g_s$ scaling behavior. 

The only way out is to allow for a charge {\it neutral} configuration of the seven-branes in the IIB side. In fact this is exactly the F-theory scenario with 24 seven-branes wrapping ${\cal M}_4$ and stretched along the $3+1$ dimensional space-time. The orthogonal space to the seven-branes is a ${\bf P}^1$ with 24 points where the F-theory fibre torus degenerates \cite{stringy}. This  space could be identified with ${\cal M}_2$, allowing us to realize such a configuration from M-theory. 
In fact one could even go a step further: write an equivalent seven-dimensional action using the normalizable forms on the base manifold, that involves the higher order quantum terms in the following 
way\footnote{As it happens in the usual case, the coordinates of the eight-manifold \eqref{anonymous}
are represented in \eqref{sheela3} by: $y \equiv y(\sigma), y^\alpha \equiv y^\alpha(\sigma)$ and 
$y^a \equiv y^a(\sigma)$. We can find a gauge where $\sigma^{\rm M}$ is identified with the coordinates of ${\bf R}^{2, 1} \times {\cal M}_4$.}:
\bg\label{sheela3}
{\bf S}_3 = \sum_{\{l_i\}, n_i} {\rm T}_7 \int d^7 \sigma \sqrt{-{\bf g}_7} ~{\bf g}^{ab} ~\partial_a \partial_b 
\left({\hat{\mathbb{Q}}_{\rm T}^{(\{l_i\}, n_i)} (y, y^\alpha, y^a) \over {\rm M}_p^{\sigma(\{l_i\}, n_i) - 7}}\right), \nd
where the quantum terms are similar to the ones that we encountered in \eqref{selahran}, the difference now being their dependence on the coordinates of the fibre torus. Concerns about the charges of the seven-branes are no longer there because of the charge neutrality in the IIB side. Additionally, in the weak-coupling limit, 
\eqref{sheela3} reproduces the higher order action on the IIA six-branes. Expectedly a T-duality along $x_3$ direction then provides us a configuration of space-filling, but charge neutral, seven-branes in the IIB side that do not break any of the de Sitter isometries. Clearly the gauge fields on the seven-brane would be generated from the M-theory G-flux component via a decomposition similar to \eqref{pikilia}, where $\Omega_{ab}(y^a, y^\alpha)$ is now a function of the ${\cal M}_2 \times {\mathbb{T}^2\over {\cal G}}$ coordinates. 

The inclusion of ${\bf g}_{ab}$ and the derivatives along the toroidal directions tell us that we are now taking 
$n_3 = 2$ in \eqref{cibelmoja} (which we had put to zero earlier). The fact that this comes with a negative sign in \eqref{cibelmoja} can be turned to our advantage here. To see this we shall first assume that the time dependences of the gauge field  ${\bf F}_{\rm MN}$ and the two-form $\Omega_{ab}$ can be collected  together as
$\left({g_s\over H}\right)^{2p/3}$ so that it remains consistent with the generic temporal-dependences advocated in \eqref{prcard} or in \cite{desitter2}. This way:
\bg\label{duitagra}
\Omega_{ab}(x_{11}) = \sum_{n = 1}^\infty {\rm B}_n~{\rm exp}\left(-{\rm M}_p^{2n} x_{11}^{2n}\right)
\epsilon_{ab}, \nd
where ${\rm B}_n$ are ${\rm M}_p$ and $g_s$ {\it independent} constants. Note also the absence of 
any $g_s$ dependent factors as they have already been accounted for.  
One could in principle deviate from this to allow for a slightly different option where the $g_s$ dependence appear in the exponent, as considered in \cite{desitter3}, but then this forms a different class of solution that we will discuss a bit later. Putting everything together, the energy-momentum tensor from \eqref{sheela3} becomes:

{\footnotesize
\bg\label{mutachat}
\mathbb{T}^{({\rm np}; 2a)}_{\rm MN}(z) = - \sum_i {2{\bf T}_7  \over \sqrt{-{\bf g}_{11}(z)}} \int {d^7 y' \over  
{\rm M}_p^{\sigma^{(i)} - 7}}\left[ 
\sqrt{{\bf g}_7(y')} ~{\bf g}^{bb} \partial^2_{b} \cdot
{\delta \over \delta {\bf g}^{\rm MN}(z)} +
{\delta \left(\sqrt{-{\bf g}_7(y')}~{\bf g}^{bb}\right) \over \delta {\bf g}^{\rm MN}(z)} 
\partial^2_{b}\right]{\hat{\mathbb{Q}}}_{\rm T}^{(i)} (y, y^b), \nonumber\\ \nd}
where the sum is over all $(\{l_i\}, n_i)$ and $y^b \equiv x_{11}(y')$, thus ${\bf g}_{bb} = {\bf g}_{11, 11}$. Note that we have restricted the dependence on the coordinates of ${\cal M}_4$ and $x_{11}$, but a more generic dependence with the coordinates of ${\cal M}_2$ should not be too hard. The simplified 
dependence helps us to express the world-volume metric by ${\bf g}_7(y')$; and $\partial_b^2$ depends on the embedding $y^b = y^b(y')$ via:
\bg\label{naamki}
\partial_b^2 = \left({\partial y^{'{\rm P}} \over \partial y^b} 
{\partial y^{'{\rm Q}} \over \partial y^b}\right) {\partial^2 \over \partial y^{'{\rm P}}\partial y^{'{\rm Q}}}. \nd
If we take ${\rm B}_n = 0$
for $n > 1$ in \eqref{duitagra}, then the two-derivative action brings down a factor of ${\rm M}^4_p$ so that this terms scales with respect to ${\rm M}_p$ in the same way as  \eqref{notiyell}. Regarding the $g_s$ scalings, the various components of the energy-momentum tensor scale in the following way:
\bg\label{brazinaam}
&& \mathbb{T}^{({\rm np}; 2a)}_{mn}(z) =  \mathbb{T}^{({\rm np}; 2a)}_{\alpha\beta}(z) 
= {1\over g_s^2}\Big(g_s^{\theta - 2/3}\Big) \nonumber\\
&&\mathbb{T}^{({\rm np}; 2a)}_{\mu\nu}(z) = {1\over g_s^2}\Big(g_s^{\theta - 8/3}\Big), ~~
\mathbb{T}^{({\rm np}; 2a)}_{ab; 2}(z) = {1\over g_s^2}\Big(g_s^{\theta + 4/3}\Big), \nd
similar to the second terms in \eqref{notiyell}, with $\theta$ being the same as the one appeared there. Interestingly, when $\theta = 8/3$, the four set of energy-momentum tensors scale as 
$(g_s^0, g_s^0, g_s^{-2}, g_s^2)$, with no relative suppressions between them  as we had for
 \eqref{notiyell}. However their contributions rely crucially on the
embedding  $y^b = y^b(y')$, and therefore also on the derivative constraint 
$\partial_b^2 {\hat{\mathbb{Q}}}_{\rm T}^{(i)} (y, y^b) \ne 0$. Choosing a standard embedding of the
seven-branes would make these contributions vanish.

\begin{table}[tb]
 \begin{center}
\renewcommand{\arraystretch}{1.5}
\begin{tabular}{|c|c|c|}\hline orientation & ${g_s}$ scaling & 
IIB dual \\ \hline\hline
${\bf S}^3 \in {\cal M}_4$ & ${1\over g_s}$ & D3 instanton along $\left({\bf S}^3 \in {\cal M}_4\right) \times 
({\bf S}^1)_{a = 3}$ \\ \hline
$\left({\bf S}^2 \in {\cal M}_4\right) \times ({\bf S}^1)_\alpha$ & ${1\over g_s}$ & D3 instanton along 
$\left({\bf S}^2 \in {\cal M}_4\right) \times ({\bf S}^1)_\alpha \times ({\bf S}^1)_{a = 3}$ \\ \hline
$({\bf S}^1)_m \times {\cal M}_2$ & ${1\over g_s}$ & D3 instanton along 
$({\bf S}^1)_m \times {\cal M}_2 \times  ({\bf S}^1)_{a = 3}$ \\ \hline
${\cal M}_2 \times ({\bf S}^1)_{a = 3}$ & $g_s^0$ & D1 instanton along ${\cal M}_2$ \\ \hline
$\left({\bf S}^2 \in {\cal M}_4\right) \times ({\bf S}^1)_{a = 3}$ & $g_s^0$ & D1 instanton along ${\bf S}^2 \in {\cal M}_4$ \\ \hline
${\cal M}_2 \times ({\bf S}^1)_{b = 11}$ & $g_s^0$ & F1 instanton along ${\cal M}_2$ \\ \hline
$\left({\bf S}^2 \in {\cal M}_4\right) \times ({\bf S}^1)_{b = 11}$ & $g_s^0$ & F1 instanton along ${\bf S}^2 \in {\cal M}_4$ \\ \hline
$({\bf S}^1)_\alpha \times {\mathbb{T}^2\over {\cal G}}$ & $g_s^{1/3}$ & Kaluza-Klein instanton along $({\bf S}^1)_\alpha$ \\ \hline
$({\bf S}^1)_m \times {\mathbb{T}^2\over {\cal G}}$ & $g_s^{1/3}$ & Kaluza-Klein instanton along $({\bf S}^1)_m$ \\ \hline  \end{tabular}
\renewcommand{\arraystretch}{1}
\end{center}
 \caption[]{The ${g_s}$ scalings of the wrapped M2-instantons along various three-cycles inside the non-K\"ahler eight-manifold 
 \eqref{anonymous} in M-theory; with ($m, n$) parametrizing the coordinates of ${\cal M}_4$, ($\alpha, \beta$) the coordinates of ${\cal M}_2$ and 
 ($a, b$) the coordinates of ${\mathbb{T}^2\over {\cal G}}$. Again most cases break the four-dimensional de Sitter isometries in the IIB side so they 
 cannot contribute to the energy-momentum tensors. Some of the above configurations rely on the existence of local one-cycles inside the non-K\"ahler sub-manifolds. Absence of these will remove their contributions further.}
  \label{instabeta2}
 \end{table}

Although the seven-branes themselves don't seem to contribute non-perturbatively to the energy-momentum tensor, the world-volume fields on the seven-branes in principle could. One specific set of contributions could come from the {\it fermionic} terms on the seven-branes. In M-theory we should then look for possible fermionic completions of the quantum terms like \eqref{selahran}.  Fermionic terms imply introducing the Gamma matrices, and since two Gamma matrices anti-commute to the corresponding metric components, the Gamma matrices themselves should become time-{\it dependent} (because the metric components are). What does that mean? 

It means that the eleven-dimensional Gamma matrices should be expressed in terms of the eleven-dimensional vielbeins as ${\bf \Gamma}_{\rm M} \equiv \Gamma_{\bar a} {\bf e}^{\bar a}_{\rm M}$ and 
${\bf e}^{\bar a}_{\rm M} {\bf e}_{{\bar a}{\rm N}} = {\bf g}_{\rm MN}$, where $\Gamma_{\bar a}$ are the standard constant Gamma matrices that now anti-commute to the flat metric $\eta_{{\bar a}{\bar b}}$. 
The fermionic completion of the four-form G-flux in M-theory, may be expressed using anti-symmetric products of Gamma functions and fermions, somewhat along the lines of \cite{fermions}, in the following way:  
\bg\label{mcatrason}
\hat{{\bf G}}_{{\rm MN}ab}(y^m, y^\alpha, y^a, g_s) \equiv e_1\bar{\bf\Upsilon}^{\rm P} {\bf \Gamma}_{{\rm MNPQ}ab} {\bf\Upsilon}^{\rm Q}(y, g_s) 
+ e_2\bar{\bf\Upsilon}_{[{\rm M}}{\bf \Gamma}_{ab} {\bf\Upsilon}_{{\rm N}]}(y, g_s), \nd
where ${\bf\Upsilon}_{{\rm M}}$ is the eleven-dimensional gravitino and $e_i$ are just constants.
Despite its complicated appearance compared to 
what we encountered in \cite{fermions},  due solely to the presence of 
eleven-dimensional gravitino, this is not new (see for example \cite{bergroo}). 
Once we decompose 
${\bf\Upsilon}_{\rm M}(y, g_s) = {\bf\Psi}(y^m, g_s) \otimes \Theta_{\rm M}(y^\alpha, y^a, g_s) 
+ {\bf\Psi}_{\rm M}(y^m, g_s) \otimes \Theta'(y^\alpha, y^a, g_s)$,  
then, compared to \eqref{pikilia}, we have the following forms: 

{\footnotesize
\bg\label{tarachand} 
&&\hat{\Omega}_{ab} \propto \bar{\Theta}' {\bf \Gamma}_{ab} \Theta', ~~~
\hat{\Omega}_{{\rm MN}ab} \propto \bar{\Theta}_{[{\rm M}} {\bf \Gamma}_{ab} \Theta_{{\rm N}]}, ~~~ 
\hat{\Omega}_{{\rm M}ab} \propto \bar{\Theta}_{[{\rm M}} {\bf \Gamma}_{{ab}]} \Theta' \\
&&\hat{\Omega}'_{{\rm MN}ab} \propto \bar{\Theta}^{\rm P} {\bf \Gamma}_{{\rm MNPQ}ab} \Theta^{\rm Q}, ~~~
\hat{\Omega}'_{{\rm MNQ}ab} \propto \bar{\Theta}^{\rm P} {\bf \Gamma}_{{\rm MNPQ} ab} \Theta', ~~~ 
\hat{\Omega}'_{{\rm MNPQ}ab} \propto \bar{\Theta}'{\bf \Gamma}_{{\rm MNPQ}ab} \Theta', \nonumber \nd}
alongwith $\bar{\hat\Omega}_{{\rm M}ab}$ and $\bar{\hat\Omega}_{{\rm MNQ}ab}$;
which would make them functions of ($y^m, y^\alpha, g_s$) from the vielbeins. We expect the internal fermions and gravitinos, {\it i.e.} 
$\Theta$ and $\Theta_{\rm N}$, to reproduce a behavior like 
\eqref{duitagra}, from the corresponding Dirac and Rarita-Schwinger equations on ${\cal M}_2 \times {\mathbb{T}^2\over {\cal G}}$. This is bit more complicated scenario so some simplification is warranted for.
In the generic scenario we place no constraints on $\Theta$ and $\Theta_{\rm N}$ right now. 
With these in mind, let us consider the following variation of \eqref{pikilia}:
\bg\label{mink}
{\hat{\bf G}}_{{\rm MN}ab}(y^m, y^\alpha, y^a, g_s) &= & \bar{{\bf\Psi}} 
\Big(e_{11} \hat{\Omega}_{{\rm MN}ab} + e_{12} \hat{\Omega}'_{{\rm MN}ab}\Big){\bf\Psi} \nonumber\\
& + & e_{21} \bar{\bf\Psi}_{[{\rm M}}  ~\hat{\Omega}_{ab} {\bf\Psi}_{{\rm N}]}+ 
e_{22} \bar{\bf\Psi}^{[{\rm P}}  ~\hat{\Omega}'_{{\rm MNPQ}ab}{\bf\Psi}^{{\rm Q}]}\nonumber\\
& + & e_{31} \bar{\bf\Psi} ~\hat{\Omega}_{[{\rm M}ab}  {\bf\Psi}_{{\rm N}]}+ 
e_{32} \bar{\bf\Psi}  ~\hat{\Omega}'_{{\rm MNQ}ab} {\bf\Psi}^{{\rm Q}}+ {\rm h. c}, \nd
where (${\bf\Psi}, {\bf\Psi}_{{\rm M}}$) are the two kinds of fermions in M-theory, now localized along 
${\bf R}^{2, 1} \times {\cal M}_4$, and $e_{ij}$ are constants. 
If we assume that the fermions do not have any $g_s$ dependences, then 
$\bar{\bf\Psi} = {\bf\Psi}^\dagger {\bf \Gamma}^0 = 
{\bf\Psi}^\dagger \Gamma^{\bar a} {\bf e}^{0}_{\bar a}$, as well as $\bar{\bf\Psi}_{\rm M}$, would naturally scale as 
$g_s^{4/3}$. If we restrict $({\rm M}, {\rm N}) \in {\cal M}_4$, then ${\bf \Gamma}_{\rm MN}$ scales as 
$g_s^{-2/3}$, implying that the coefficients of $e_{ij}$ terms in \eqref{mink} all
scale as $g_s^{8/3}$.  The G-flux component ${\bf G}_{{\rm MN}ab}$ in \eqref{pikilia}, on the other hand, scales as $g_s^{2k/3}$. 

This mismatch has a natural explanation. In a time-dependent background, if we switch on fermionic bilinears, they cannot be time-independent. The simplest bilinear function will involve two fermions without any Gamma functions, implying that the fermions themselves should have some $g_s$ dependences. With
this in mind, let us arrange for the following $g_s$ dependences for the two kinds of fermions in M-theory:

{\footnotesize
\bg\label{chand}
{\bf\Psi}({\bf x}, y^m, g_s) = \sum_{k'} {\bf\Psi}^{(2k' + 4)}({\bf x}, y^m) \left({g_s\over H}\right)^{2k'/3}, ~~
{\bf\Psi}_{\rm M}({\bf x}, y^m, g_s) = \sum_{k'} {\bf\Psi}_{\rm M}^{(2k' + 4)}({\bf x}, y^m) 
\left({g_s\over H}\right)^{2k'/3}, \nd}
which is arranged so that \eqref{mink} or \eqref{mcatrason} transform exactly as the corresponding G-flux components ${\bf G}_{mnab}, {\bf G}_{\alpha\beta ab}$ and ${\bf G}_{m\alpha ab}$ when 
$k' = {1\over 2}(k - 4)$. Note that this implies that $\Theta(y^\alpha, y^a)$ and $\Theta_{\rm M}(y^\alpha, y^a)$ do not have any $g_s$ dependences. 
We have also inserted a spatial dependence on ${\bf x}$, to the already expected spatial dependence on  
$y^m$,  in anticipation of the following decomposition: 
\bg\label{tara}
{\bf\Psi}({\bf x}, y^m) = \psi_1({\bf x}) \otimes \chi^{(4)}(y^m), ~~
{\bf\Psi}_{\rm M}({\bf x}, y) = 
\psi_2({\bf x}) \otimes \chi^{(4)}_m(y^m), \nd
with $m \in {\cal M}_4$ and $\psi_i({\bf x})$ being fermionic degrees of freedom in ${\bf R}^2$. Clearly this decomposition leads to variety of fermionic degrees of freedom but the problem arises when we try to restore the de Sitter isometries in the IIB side.  One way out is to remove the ${\bf x}$ dependence altogether and view the fermions to be completely on the sub-manifold ${\cal M}_4$ and consider simply the bilinear \eqref{mink}. On one hand, if we consider the bilinear $\bar{\bf\Psi}{\bf\Psi}(y^m, g_s)$, this would scale as $\left({g_s\over H}\right)^{2(k' + k'' + 2)/3}$, which seems to contribute for $(k', k'')  >  (0, 0)$. On the other hand, the $g_s$ expansions for ${\bf\Psi}$ and ${\bf\Psi}_{\rm M}$ in \eqref{chand} would make sense if 
$k > 4$. More importantly however in the decomposition \eqref{mink}, the gravitino degrees of freedom 
${\bf\Psi}_{\rm M}$ {\it cannot} appear because type IIB seven-branes do not have gravitino fields on their world-volumes. This can be made possible if from the start we impose $\Theta'(y^\alpha, y^a) = 0$. In that case there exists the following fermionic completion of the four-form flux:
\bg\label{lebantagra}
{\bf G}^{\rm tot}_{{\rm MN}ab} \equiv  {\bf G}_{{\rm MN}ab} + {\rm M}_p~ {\hat{\bf G}}_{{\rm MN}ab}, \nd
where the first term, from \eqref{pikilia}, provides the world-volume gauge fields in the IIB side; and the second term, from the first line of \eqref{mink} with $e_{2i} = e_{3i} = 0$, provides the fermionic terms. Both these contributions 
provide $U(1)$ degrees of freedom, but  can be made {\it non-abelian} if we incorporate wrapped M2-branes. These non-abelian degrees of freedom will reside on multiple seven-branes at a point on 
${\cal M}_2$, and could therefore form a structure similar to the one discussed for the heterotic theories 
in \cite{dinewitten}. We have also assumed that the fermion $\Psi$ is dimensionless so that \eqref{lebantagra} remains dimensionless also. 
All these informations may be inserted in the following quantum action:
\bg\label{sheela4}
{\bf S}_4 = \sum_{\{l_i\}, n_i, q} {\rm T}_7 \int d^7 \sigma \sqrt{-{\bf g}_7}~\Big(\bar{\bf\Psi}{\bf\Psi}\Big)^q \times  
\left({\widetilde{\mathbb{Q}}_{\rm T}^{(\{l_i\}, n_i)} (y^m, y^\alpha, g_s) \over {\rm M}_p^{\sigma(\{l_i\}, n_i) - 7}}\right)\Bigg\vert_{l_{28+r} = 0}^{0 \le r \le 7}, \nd
where $\widetilde{\mathbb{Q}}_{\rm T}^{(\{l_i\}, n_i)} (y^m, y^\alpha, g_s)$ is the same one as in \eqref{selahran} except ${\bf G}_{{\rm MN}ab}$ therein is replaced by the fermionic bilinear 
\eqref{mink}. We could have instead replaced ${\bf G}_{{\rm MN}ab}$ by ${\bf G}^{\rm tot}_{{\rm MN}ab}$, but then we have to keep track of the ${\rm M}_p$ scalings a bit more carefully.  Additionally we have augmented the quantum piece with a bilinear without extra Gamma-functions. The quantum terms in \eqref{sheela4} provide us an expression with higher powers of curvature and the fermionic bilinears, but there are no interactions with G-flux components (which in principle could be constructed by relaxing the $l_{28 + r} = 0$ constraints, or by inserting ${\bf G}^{\rm tot}_{{\rm MN}ab}$ in \eqref{selahran}). If the $g_s$ scaling of the full quantum term in \eqref{sheela4} is given by $g_s^{\theta_k}$, and the ${\rm M}_p$ scaling by 
${\rm M}_p^{\sigma(\{l_i\}, n_i)}$, then $\theta_k$ and $\sigma(\{l_i\}, n_i)$, take the following form:
\bg\label{desmey}
\sigma(\{l_i\}, n_i) = \sum_{i = 0}^2 n_i  + \mathbb{N}_1, ~~
\theta_k = {1\over 3}\Big(\sum_{i = 0}^2 n_i + \mathbb{N}_1 + (k - 1) \mathbb{N}_5 + 2q(k - 2)\Big), \nd
 with no dependence on $\mathbb{N}_5$ for $\sigma(\{l_i\}, n_i)$, and we have used $2k' = k - 4$ in 
 \eqref{chand} to match with the standard G-flux scalings.  There is a slight difference though: the ${\rm M}_p$ scalings of the actual G-flux components do depend on $\mathbb{N}_5$. This can be rectified by giving a dimension of ${\rm M}_p^{-1/2}$ to the fermion $\Psi$ itself and zero dimensions to the internal fermions $\Theta$ and $\Theta_{\rm M}$. Note that $\theta_k$, computed without incorporating contribution from the ${\bf g}_7$ in \eqref{sheela4},
would be positive definite for $k > 2$. The other quantities, namely 
 $\mathbb{N}_i, n_i$, are defined in \eqref{cibelmoja}. The energy-momentum tensor from \eqref{sheela4} becomes:
 
{\footnotesize
\bg\label{mutachat2}
\mathbb{T}^{({\rm np}; 2b)}_{\rm MN}(z) = - \sum_{i, q} {2{\bf T}_7  \over \sqrt{-{\bf g}_{11}(z)}} 
\int {d^7 y' \left(\bar{\bf\Psi}{\bf\Psi}\right)^q  \over  
{\rm M}_p^{\sigma^{(i)} - 7}}\left[ 
\sqrt{-{\bf g}_7(y')} ~
{\delta \over \delta {\bf g}^{\rm MN}(z)} +
{\delta \left(\sqrt{-{\bf g}_7(y')}\right) \over \delta {\bf g}^{\rm MN}(z)} 
\right]{\widetilde{\mathbb{Q}}}_{\rm T}^{(i, r)} (y', y^b), \nonumber\\ \nd}
which should be compared to what we had in \eqref{mutachat}. One thing to note is that while 
\eqref{mutachat} depends crucially on the embedding of the seven-branes, \eqref{mutachat2} is independent of the embedding. Once we insert the G-flux components, {\it i.e.} insert the anti-symmetric tensors along with their fermionic bilinears in \eqref{mink}, we can have gauge fields and fermions interacting with each other and contribution to the potential. Generalizations aside,  the $g_s$ scalings of the  various components of the energy momentum tensors may be written as:
\bg\label{brazinaam2}
&& \mathbb{T}^{({\rm np}; 2b)}_{mn}(z) =  \mathbb{T}^{({\rm np}; 2b)}_{\alpha\beta}(z) 
= {1\over g_s^{2/3}}\Big(g_s^{\theta_k - 2/3}\Big) \nonumber\\
&&\mathbb{T}^{({\rm np}; 2b)}_{\mu\nu}(z) = {1\over g_s^{2/3}}\Big(g_s^{\theta_k - 8/3}\Big), ~~
\mathbb{T}^{({\rm np}; 2b)}_{ab}(z) = {1\over g_s^{2/3}}\Big(g_s^{\theta_k + 4/3}\Big), \nd
with $\theta_k$ as in \eqref{desmey}. This {\it contributes} to the classical EOMs in \cite{desitter2} 
at $\theta_k \ge 4/3$ which, in turn, leads to contributions from terms like 
$\left(\bar{\bf\Psi} \Omega_{{\rm MN} ab}{\bf\Psi}\right)^4$ and other higher-order terms for $k \ge {3/ 2}$ and $q = 0$. However if we had taken $q > 0$, then $k \ge {5/2}$ otherwise would be issues with the existence of a EFT description \cite{desitter2, desitter3}. For this case the non-perturbative contributions come from:
\bg\label{firealarm}
2\left(n_1 + n_2 + \mathbb{N}_1 + q\right) + 3\mathbb{N}_5 = 8, \nd
which can allow terms like $\left(\bar{\bf\Psi}{\bf\Psi}\right)^4$, amongst other possible contributions coming from
the curvature, gauge fields and their derivatives. They are finite in number because $k > 2$. Interestingly for $k = 0$, and $\theta_k = 4/3$ in \eqref{desmey} there would have been  an infinite number of terms, a subset of which are given by various derivatives on powers of fermion bilinears. However $k = 0$ is {\it not} the time-independent case for the fermions which may be seen from our scaling \eqref{chand} above.

Before ending this section, let us clarify one worrisome feature related to the possibility of {\it vanishing} contribution from a term like $\left(\bar{\bf\Psi} \Omega_{{\rm MN} ab}{\bf\Psi}\right)^2$ and higher powers. This is much like what happens in the ten-dimensional case with gauge singlet Majorana-Weyl fermions \cite{bergroo}. However such concerns are alleviated once we go to the {\it non-abelian} case by including wrapped ${\rm M2}$-branes on vanishing cycles. In the dual IIB side these wrapped M2-branes become 
tensionless strings between coincident F-theory seven-branes. The non-abelian enhancement occurs geometrically via \cite{zwiebacH} or algebraically via the Tate's algorithm \cite{tate}. In the non-abelian case this would convert:
\bg\label{wonderwoman}
\left(\bar{\bf\Psi} \Omega_{{\rm MN} ab}{\bf\Psi}\right)^2 ~ \to ~ 
 \left({\rm tr}_{\rm adj}~\bar{\bf\Psi} \Omega_{{\rm MN} ab}{\bf\Psi}\right)^2, \nd
 which is in general non-zero. Here we have taken the trace in the adjoint representation, although one could in principle construct it for arbitrary representation of the gauge group.
 In \eqref{lebantagra}, the total G-flux ${\bf G}^{\rm tot}_{{\rm MN}ab}$ then is comprised of the usual tensorial flux components ${\bf G}_{{\rm MN}ab}$, and the traces over the fermionic bilinears instead of the abelian ones discussed in \eqref{mink}, much like what we know in the heterotic side from \cite{dinewitten}.


\subsubsection{Case 3: The KKLT type instanton gas \label{sec3.2.3}}


So far our concentrations have mostly been towards scenario where $g_s \to 0$. In this limit many of the 
non-perturbative contributions vanish because of their dependence on either 
${\rm exp}\left(-{1\over g_s^{1/3}}\right)$ or on ${\rm exp}\left(-{1\over g_s^{2}}\right)$, the latter being from the wrapped instantons. What happens for $g_s < 1$, where some of the exponential factors do not go to zero as fast as it were when $g_s \to 0$? To study the implications of these, let us express \eqref{duitagra} alternatively as (see also \cite{desitter3}):

{\footnotesize
\bg\label{duitagra2}
\Omega_{ab}(x_{11}, y^\alpha, g_s) = \sum_{n, k = 1}^\infty {\rm B}_{nk}~
{\rm exp}\left(-{\rm M}_p^{2n} g_s^{4n/3} x_{11}^{2n}\right) 
{\rm exp}\left[-\left(y_\alpha y^\alpha\right)^k g_s^{-2k/3} {\rm M}_p^{2k}\right] \epsilon_{ab}, \nd}
where $y_\alpha y^\alpha = g_{\alpha\beta} y^\alpha y^\beta$ with un-warped metric $g_{\alpha\beta}$. 
In \cite{desitter3} we studied the case where all ${\rm B}_{nk} = 0$ except ${\rm B}_{10}$. Here we would like to concentrate on ${\rm B}_{11} \ne 0$ in addition to  ${\rm B}_{10}$. Note the appearance of $g_s$ in the exponents themselves. This may seem to take us away from the generic time-dependences of the G-flux components in \eqref{prcard}, giving us:

{\footnotesize
\bg\label{duitagra3}
\Omega_{ab}(x_{11}, y^\alpha, g_s) = {\rm B}_{10}~
{\rm exp}\left(-{\rm M}_p^{2} g_s^{4/3} x_{11}^{2}\right)\left[1 + {{\rm B}_{11}\over {\rm B}_{10}}~{\rm exp} 
\left(-{{\rm M}_p^2~ y^2_\alpha \over g_s^{2/3}}\right)\right]\epsilon_{ab}
 \equiv  \epsilon_{ab} + {\rm B}_{11}~{\rm exp} \left(-{{\rm M}_p^2~ y^2_\alpha \over g_s^{2/3}}\right)
 \epsilon_{ab}, \nonumber\\ \nd}
in the decomposition \eqref{pikilia}. The second equality appears from restricting the coordinate dependence to the base ${\cal M}_4 \times {\cal M}_2$.  However such a choice of the two-form do not change the $g_s$ scaling 
of \eqref{selahran} with $\theta = {1\over n}(\theta_p + 2n)$, and $\theta_p$ as in \eqref{relishlu}, unless the ($n_2, n_3$) derivative actions in \eqref{selahran} act on \eqref{duitagra3}. It is this action, in particular the one associated with $n_2$ in \eqref{selahran}, is what we are interested in here.

To see how this develops, let us start with $\mathbb{U}^{(1)}(y)$ as in \eqref{lushreli}, with two differences: one, the integral is over the six-manifold ${\cal M}_4 \times {\mathbb{T}^2\over {\cal G}}$, and two, we choose \eqref{duitagra3}, with $({\rm B}_{10}, {\rm B}_{11}) \ne (0, 0)$, instead of ${\rm B}_{11} = 0$. In the IIB side this will be instantons wrapping ${\cal M}_4$ (see {\bf Table \ref{instabeta}}), and we will call these the {\it delocalized} KKLT type instanton gas \cite{kklt}. Thus the limit we are looking at here is:
\bg\label{notiyell2}
g_s ~\to ~ \epsilon, ~~~~{\rm M}_p ~ \to ~ \epsilon^{-2/3}, ~~~~ \mathbb{T}^{({\rm np}; 1)}_{\rm MN}(z) = 
\mathbb{T}^{({\rm np}; 1)}_{\rm MN}(z; {\rm B}_{10} = 1), \nd 
with $\mathbb{T}^{({\rm np}; 1)}_{\rm MN}(z)$ is as given in \eqref{notiyell} and $\epsilon < 1$. Interestingly, in this limit, even if we had entertained a non-zero ${\rm B}_{11}$, the exponential factor would have gone to zero as 
${\rm exp}\left(-{1\over \epsilon^2}\right)$ for $\epsilon \to 0$. When $\epsilon < 1$, the coefficient of ${\rm B}_{11}$ doesn't go to zero as fast. The energy-momentum tensor on the other hand, deviates from \eqref{notiyell} in the following instructive way:

{\footnotesize
\bg\label{notiyell3}
\hskip-.11in\mathbb{T}^{({\rm np}; 3)}_{\rm MN}(z) 
& = &  \sum_{\{l_i\}, n_i, k} c_k~{\bf g}_{\rm MN}(z)~ {\rm exp}\left[- k \int d^6 z' {\sqrt{{\bf {\red g}}_{\red 6}(z')}\over 
\red{g_s^{2(1 + n_2)/3}}} 
\left({ \mathbb{F}^{(1)}(z - z') 
\mathbb{Q}_{\rm T}^{(\{l_i\}, n_i)} (z') \over {\rm M}_p^{\sigma(\{l_i\}, n_i) - 6}}\right)
{\red \hat{\mathbb{V}}_{2}}\right]\nonumber\\
&+ &\sum_{\{l_i\}, n_i, k}\left({2k c_k ~{\red{{\rm M}_p^{2n_2}~ \mathbb{V}_{2}}} \over {\rm M}_p^{\sigma(\{l_i\}, n_i) - {11}}}
\int d^{11} x ~{{\red (-2z_\alpha)^{n_2}}~\mathbb{F}^{(1)}(x - z)\over {\red g_s^{2(1+n_2)/3}}}\cdot {\delta \left(\sqrt{{\bf {\red g}}_{\red 6}(z)}\mathbb{Q}_{\rm T}^{(\{l_i\}, n_i)}(z)\right)\over 
\delta {\bf g}^{\rm MN}(z)}\sqrt{{\bf g}_{11}(x) \over {\bf {g}}_{11}(z)}\right) \nonumber\\ 
&&~~~~~~ \times 
{\rm exp}\left[- k \int d^6 z' {\sqrt{{\bf {\red g}}_{\red 6}(z')}\over 
\red{g_s^{2(1 + n_2)/3}}} 
\left({ \mathbb{F}^{(1)}(x - z') 
\mathbb{Q}_{\rm T}^{(\{l_i\}, n_i)} (z') \over {\rm M}_p^{\sigma(\{l_i\}, n_i) - 6}}\right)
{\red \hat{\mathbb{V}}_{2}}\right], \nd}
where the dotted terms are now the less dominant ones with powers of $n_2$ that may be derived from 
\eqref{selahran}. There are a few differences from \eqref{notiyell}, as shown in ${\rm \red{red}}$ above: one, the determinant of the metric of the six-dimensional base $\sqrt{{\bf g}_6(y')}$ in \eqref{notiyell} scales as ${1\over g_s^2}$, whereas here it scales as $g_s^0$; two, the ${\rm M}_p$ scaling in \eqref{notiyell} is different from what we have here;  three, there is an extra factor of $z^{n_2}_\alpha$ in \eqref{notiyell3} that is absent in \eqref{notiyell}; and four, the appearance of the volume factors $\mathbb{V}_2$ and 
$\hat{\mathbb{V}}_2$
compared to their absence in \eqref{notiyell}. The volume factor $\hat{\mathbb{V}}_2$
is important, and is related to the {\it effective} volume of the base sub-manifold ${\cal M}_2$. Its presence here signifies the fact that we have integrated over the full internal space \eqref{anonymous}, including the contributions from the derivatives along ${\cal M}_2$ in \eqref{selahran}. On the other hand, 
$\mathbb{V}_2$ is extracted from an integral 
$\int d^2 y^\alpha g(y^\alpha) \sqrt{{g}_2} \approx g(y^\alpha) \mathbb{V}_2$ and is the {\it unwarped} volume of ${\cal M}_2$. 

\begin{table}[tb]
 \begin{center}
\renewcommand{\arraystretch}{1.5}
\begin{tabular}{|c|c|c|c|c|}\hline instanton type & $\mathbb{V}^a_2 \mathbb{V}^b_{{\bf T}^2}$  & ${g_s}$ scaling & $\mathbb{T}^{({\rm np})}_{\rm MN}$ & $\theta_{\rm min}$ \\ \hline\hline
BBS & $a = b = 0$ & ${g_s^{\theta + l_a}\over g_s^2}$ & \eqref{notiyell} & ${8\over 3}$\\ \hline
Delocalized BBS & $a = 0, b = 1 $ & ${g_s^{\theta + l_a}\over g_s^{2/3}}$ & \eqref{notiyell5} with $n_2 = 0$ 
& ${4\over 3}$ \\ \hline
KKLT & $a = b = 0$ & ${g_s^{\theta + l_a}}$ & \eqref{notiyell7} with $n_2 = 0$ & ${2\over 3}$ \\ \hline
Delocalized KKLT & $a =1,  b = 0$ & ${g_s^{\theta + l_a}\over g_s^{2/3}}$ & \eqref{notiyell4} & ${4\over 3}$\\ \hline
\end{tabular}
\renewcommand{\arraystretch}{1}
\end{center}
 \caption[]{The ${g_s}$ scalings of the four kind of wrapped M5-instantons that contribute to the non-perturbative energy-momentum tensor $\mathbb{T}^{({\rm np})}_{\rm MN}$.  The two-form $\Omega_{ab}$
 that defines the G-flux components in \eqref{pikilia}, is chosen to be 
 $\Omega_{ab} = \Big(1 + {\rm B}_{11}{\rm exp}\left(-{\rm M}_p^2 y_\alpha^2\right)\Big)\epsilon_{ab}$, with 
 ${\rm B}_{11}$ a constant independent of ($g_s, {\rm M}_p$). The $g_s$ scalings remain unchanged for the two cases with ${\rm B}_{11} = 0$ and ${\rm B}_{11} \ne 0$, although their corresponding ${\rm M}_p$ scalings change. The other parameter $\theta$ is defined in the text as 
 $\theta \equiv {1\over n}(\theta_p + 2n)$ and $\theta_p$ as in \eqref{relishlu} with $\theta_{\rm min}$ being the minimum value of $\theta$ that contributes to the non-perturbative energy-momentum tensors; $l_a = {4/3}, 
 -{2/3}, - {8/3}$ depending on the space-time directions; $\mathbb{V}_2$ is the unwarped volume of the sub-manifold ${\cal M}_2$; and $\mathbb{V}_{{\bf T}^2}$ is the unwarped volume of the toroidal sub-manifold 
 ${\mathbb{T}^2\over {\cal G}}$.}
   \label{instabeta3}
 \end{table}

The factor of ${1\over g_s^2}$ that appears in \eqref{notiyell3} for $n_2 = 2$ matches up with the similar dependence in \eqref{notiyell}, but now there is more: the generic dependence becomes 
${1\over g_s^{2(1 + n_2)/3}}$, implying that other powers of inverse $g_s$ should appear both in the sum as well as in the exponent. This differs from \eqref{notiyell} where the factor of ${1\over g_s^2}$ is universal over the whole range of ($\{l_i\}, n_i$). For large enough $n_2$, it appears that the exponential term is heavily suppressed killing the contributions to the energy-momentum tensor altogether as seen from the individual components of the energy-momentum tensor:

{\footnotesize
\bg\label{yellstop3}
&& \mathbb{T}^{({\rm np}; 3)}_{ab}(z) = \sum_k c_k \left[g_s^{4/3} + \sum_{n_2} {k \mathbb{V}_2 \over g_s^{2(1 + n_2)/3}} \left(g_s^{\theta + 4/3}\right)\right]
~{\rm exp} \left(-\sum_{n_2} {k \mathbb{V}_2 \over g_s^{2(1 + n_2)/3}} \cdot g_s^\theta\right) \\
&& \mathbb{T}^{({\rm np}; 3)}_{\mu\nu}(z) = \sum_k c_k \left[g_s^{-8/3} + \sum_{n_2} {k \mathbb{V}_2 \over g_s^{2(1 + n_2)/3}}\left(g_s^{\theta - 8/3}\right)\right]
~{\rm exp} \left(-\sum_{n_2} {k \mathbb{V}_2 \over g_s^{2(1 + n_2)/3}} \cdot g_s^\theta\right) \nonumber\\
&& \mathbb{T}^{({\rm np}; 3)}_{mn}(z) = \mathbb{T}^{({\rm np}; 3)}_{\alpha\beta}(z)  = \sum_k c_k \left[g_s^{-2/3} + \sum_{n_2}{k \mathbb{V}_2 \over g_s^{2(1 + n_2)/3}} \left(g_s^{\theta - 2/3}\right)\right]
~{\rm exp} \left(- \sum_{n_2} {k \mathbb{V}_2 \over g_s^{2(1 + n_2)/3}} \cdot g_s^\theta\right). \nonumber \nd}
However the scaling analysis shows that $\theta = {2\over 3}(2 + n_2)$, so the series could be summed as in \eqref{mcamahal} instead. Putting a bound like \eqref{mcabuddhi}, one could even get a convergent series (see footnote \ref{knauf}). Any remaining divergence can be controlled by $\mathbb{F}^{(1)}(x - z)$, 
$\mathbb{F}^{(1)}(z - z')$ and $\mathbb{F}^{(1)}(x - z')$  
as before, but now there is also the volume factors, although new subtleties appear because of the $n_2$ factor in \eqref{notiyell3} and \eqref{yellstop3}. Before we discuss how to control this, let us  express the generic form of the energy-momentum tensor from \eqref{yellstop3} in the following way:
\bg\label{yellpach}
\mathbb{T}^{({\rm np}; 3)}_{{\rm M}_a{\rm N}_a}(z) = \sum_k c_k \left[g_s^{l_a} 
+ \sum_{n_2} {k \mathbb{V}_2 \over g_s^{2(1 + n_2)/3}} \left(g_s^{\theta + l_a}\right)\right]
~{\rm exp} \left(-\sum_{n_2} {k \mathbb{V}_2 \over g_s^{2(1 + n_2)/3}} \cdot g_s^\theta\right), \nd
where the $g_s$ dependence in the exponential part for $\theta = {2\over 3}(2 + n_2)$ go as 
$g_s^{2/3}$, as before. The ${\rm M}_p$ dependence is now useful to quantify. If the derivative action do not act on \eqref{duitagra3}, most of the changes in {\red red} in \eqref{notiyell3} do not appear, and the energy-momentum tensor takes the following form:

{\footnotesize
\bg\label{notiyell4}
\mathbb{T}^{({\rm np}; 4)}_{\rm MN}(z) 
& = &  \sum_{\{l_i\}, n_i, k} c_k~{\bf g}_{\rm MN}(z)~ {\rm exp}\left[- k \int d^6 z' {\sqrt{{\bf {\red g}}_{\red 6}(z')}\over 
\red{g_s^{2/3}}} 
\left({ \mathbb{F}^{(1)}(z - z') 
\mathbb{Q}_{\rm T}^{(\{l_i\}, n_i)} (z') \over {\rm M}_p^{\sigma(\{l_i\}, n_i) - 6}}\right)
{\red \hat{\mathbb{V}}_{2}}\right]\nonumber\\
&+ &\sum_{\{l_i\}, n_i, k}\left({2k c_k ~{\red{ \mathbb{V}_{2}}} \over {\rm M}_p^{\sigma(\{l_i\}, n_i) - {11}}}
\int d^{11} x ~{\mathbb{F}^{(1)}(x - z)\over {\red g_s^{2/3}}}\cdot {\delta \left(\sqrt{{\bf {\red g}}_{\red 6}(z)}\mathbb{Q}_{\rm T}^{(\{l_i\}, n_i)}(z)\right)\over 
\delta {\bf g}^{\rm MN}(z)}\sqrt{{\bf g}_{11}(x) \over {\bf {g}}_{11}(z)}\right) \nonumber\\ 
&&~~~~~~~~~ \times
{\rm exp}\left[- k \int d^6 z' {\sqrt{{\bf {\red g}}_{\red 6}(z')}\over 
\red{g_s^{2/3}}} 
\left({ \mathbb{F}^{(1)}(x - z') 
\mathbb{Q}_{\rm T}^{(\{l_i\}, n_i)} (z') \over {\rm M}_p^{\sigma(\{l_i\}, n_i) - 6}}\right)
{\red \hat{\mathbb{V}}_{2}}\right], \nd}
whose behavior is very similar to what we had in \eqref{notiyell}. For example when $\theta = 4/3$, one may easily see that the exponential factor goes as ${\rm exp}\left(-k \hat{\mathbb{V}}_{2} g_s^{2/3}\right)$, with the integrand on the second term behaving similar to \eqref{notiyell}. This series can be controlled, as we discussed in much details earlier (see footnote \ref{knauf}). The interesting question then is the scenario where the derivatives act on \eqref{duitagra3}. Taking one derivative on \eqref{duitagra3}, brings down a factor of $-2y_\alpha {\rm M}_p^2$, but there is also a factor of inverse ${\rm M}_p$ from \eqref{chatchat}, so that the overall scaling is $-2y_\alpha {\rm M}_p$. This is good, but the worrisome feature is the {\it minus} sign, which will occur every time we take an {\it odd} number of derivatives. Fortunately this can be cured 
from the start when we sum the trans-series in \eqref{ginagers}. Recall that the aim of such a summation process is to convert any series with terms like ${\red \pm} {f(z) \over g_s^2}$ to 
${\rm exp}\left({\red -} {f(z) \over g_s^2}\right)$, so that when $g_s \to 0$, or $f(z) \to \infty$, the exponential factor becomes very small, irrespective of the sign of the individual terms in the series. To deal with this, let us divide the $n_2$ derivatives as $n_2 = (n_\Omega, n_Q)$ where $n_\Omega$ is the number of derivatives that act on \eqref{duitagra3}, or alternatively on ${\bf G}_{{\rm MN} ab}$ in \eqref{selahran}, and $n_Q$ is the number of derivatives that act on everything else in \eqref{selahran}. 
We can now insert a factor of $(-1)^{n_\Omega}$ in the exponential piece \eqref{lushreli}. This way, once we take $\theta = 4/3$, the energy-momentum tensor would scale with respect to ${\rm M}_p$ as 
${\red {\rm M}_p^{n_{{}_\Omega}} {\rm exp}\left(-{\rm M}_p^{n_{{}_\Omega}} g_s^{2/3}\right)}$ that becomes smaller as $n_2$ 
increases\footnote{A more acute question is what happens when the derivative action {\it removes} the ${\rm M}_p$ factor. This happens, for example when the first derivative action brings down a 
$-2y_\alpha {\rm M}_p^2$ factor, and the second derivative action removes $y_\alpha$. In fact the combined action also removes the ${\rm M}_p^2$ factor. However the exponential piece does retain the information of the 
derivative action that acts on the exponential factor as well as the other derivative action. This means, no matter how the derivatives act, there would always be a factor that goes to zero as:  
 \bg\label{dumplameye}
 \lim_{n_{{}_\Omega} \to \infty} {\rm M}_p^{n_{{}_\Omega}} {\rm exp}\left(-{\rm M}_p^{n_{{}_\Omega}} g_s^{2/3}\right) \to 0, \nonumber \nd 
 which for ${\rm M}_p \to \infty$ goes to zero even faster. The conclusion remains unchanged for $g_s < 1$. For $g_s \to 0$, we can always arrange ${\rm M}_p$ to go to infinity {\it faster} than some given power of $g_s$. For example imagine the exponential part is ${\rm exp}\left(-{\rm M}_p^{k_1} g_s^{k_2}\right)$, and let $g_s$ goes to zero as $g_s \to \epsilon$. Then as long as ${\rm M}_p$ goes to infinity as 
 ${\rm M}_p \to \epsilon^{-\kappa}$ with $\kappa > {k_2\over k_1}$, the exponential part vanishes. This way convergence can be attained. \label{kalingo}} 
Similar story could be developed for other values of $\theta > 0$, and show convergences there. 
We could also go back to \eqref{notiyell}, and insert two changes: one, use \eqref{duitagra3} and two, insert the warped volume of ${\mathbb{T}^2\over {\cal G}}$.  We will however continue to keep all functional dependences on ${\cal M}_4 \times {\cal M}_2$ for simplicity.  The result is:

{\footnotesize
\bg\label{notiyell5}
\mathbb{T}^{({\rm np}; 5)}_{\rm MN}(z) 
& = &  \sum_{\{l_i\}, n_i, k} c_k~{\bf g}_{\rm MN}(z)~ {\rm exp}\left[- k \int d^6 z' {\sqrt{{\bf {\red g}}_{\red 6}(z')}~
\red{g_s^{2(2 - n_2)/3}}} 
\left({ \mathbb{F}^{(1)}(z - z') 
\mathbb{Q}_{\rm T}^{(\{l_i\}, n_i)} (z') \over {\rm M}_p^{\sigma(\{l_i\}, n_i) - 6}}\right)
{\red {\mathbb{V}}_{{\bf T}^2}}\right]\nonumber\\
&+ &\sum_{\{l_i\}, n_i, k}\left({2kc_k ~{\red{{\rm M}_p^{2n_2}~ \mathbb{V}_{{\bf T}^2}}} \over {\rm M}_p^{\sigma(\{l_i\}, n_i) - {11}}}
\int d^{11} x ~{{\red (2z_\alpha)^{n_2}}~\mathbb{F}^{(1)}(x - z)\over {\red g_s^{2(1+ n_2)/3}}}\cdot {\delta \left(\sqrt{{{\red g}}_{\red 6}(z)}\mathbb{Q}_{\rm T}^{(\{l_i\}, n_i)}(z)\right)\over 
\delta {\bf g}^{\rm MN}(z)}\sqrt{{\bf g}_{11}(x) \over {\bf {g}}_{11}(z)}\right) \nonumber\\ 
&&~~~~~~~~~\times 
 {\rm exp}\left[- k \int d^6 z' {\sqrt{{\bf {\red g}}_{\red 6}(z')}~
\red{g_s^{2(2 - n_2)/3}}} 
\left({ \mathbb{F}^{(1)}(x - z') 
\mathbb{Q}_{\rm T}^{(\{l_i\}, n_i)} (z') \over {\rm M}_p^{\sigma(\{l_i\}, n_i) - 6}}\right)
{\red {\mathbb{V}}_{{\bf T}^2}}\right], \nd}
where ${\red{\bf g}_6}$ is the warped determinant of the metric for the base ${\cal M}_4 \times {\cal M}_2$, whereas ${\red g_6}$ is the un-warped determinant. $\mathbb{V}_{{\bf T}^2}$ is the un-warped volume of the orthogonal space ${\mathbb{T}^2 \over {\cal G}}$. Interestingly, the $g_s$ and the ${\rm M}_p$ behavior of this is similar to \eqref{notiyell3}, the KKLT instanton, and because of that, the series \eqref{notiyell5} should be convergent (in the sense discussed in footnote \ref{knauf}). Interestingly, once we remove the volume dependence, the result becomes:

{\footnotesize
\bg\label{notiyell6}
\mathbb{T}^{({\rm np}; 6)}_{\rm MN}(z) 
& = &  \sum_{\{l_i\}, n_i, k} c_k~{\bf g}_{\rm MN}(z)~ {\rm exp}\left[- k \int d^6 z' 
{\sqrt{{\bf {\red g}}_{\red 6}(z')} \over {\red g_s^{2n_2/3}}} 
\left({ \mathbb{F}^{(1)}(z - z') 
\mathbb{Q}_{\rm T}^{(\{l_i\}, n_i)} (z') \over {\rm M}_p^{\sigma(\{l_i\}, n_i) - 6}}\right)\right]\nonumber\\
&+&  \sum_{\{l_i\}, n_i, k}\left({2k c_k ~{\red{{\rm M}_p^{2n_2}}} \over {\rm M}_p^{\sigma(\{l_i\}, n_i) - {11}}}
\int d^{11} x ~{{\red (2z_\alpha)^{n_2}}~\mathbb{F}^{(1)}(x - z)\over {\red g_s^{2(3+ n_2)/3}}}\cdot {\delta \left(\sqrt{{{\red g}}_{\red 6}(z)}\mathbb{Q}_{\rm T}^{(\{l_i\}, n_i)}(z)\right)\over 
\delta {\bf g}^{\rm MN}(z)}\sqrt{{\bf g}_{11}(x) \over {\bf {g}}_{11}(z)}\right) \nonumber\\ 
&&~~~~~~~\times
{\rm exp}\left[- k \int d^6 z' {\sqrt{{\bf {\red g}}_{\red 6}(z')} \over {\red g_s^{2n_2/3}}} 
\left({ \mathbb{F}^{(1)}(x - z') 
\mathbb{Q}_{\rm T}^{(\{l_i\}, n_i)} (z') \over {\rm M}_p^{\sigma(\{l_i\}, n_i) - 6}}\right)\right].\nd}
Let us summarize what we have so far. Assuming no dependency on the coordinates of 
${\mathbb{T}^2 \over {\cal G}}$ there are two possible choices for $\Omega_{ab}$ given in \eqref{duitagra3}:
one with ${\rm B}_{11} = 0$ and the other with non-zero ${\rm B}_{11}$. For both cases, the $g_s$ scalings of the energy-momentum tensors for the BBS and the KKLT instanton gases behave differently depending on whether the orthogonal volumes of the internal sub-manifold are taken into account or not. For the BBS instanton gas, the energy-momentum tensor without switching on the volume of 
${\mathbb{T}^2 \over {\cal G}}$ is given in \eqref{notiyell}. Once we switch on the volume of 
${\mathbb{T}^2 \over {\cal G}}$, the result changes to \eqref{notiyell5}, with $n_2 = n_{{}_\Omega}$, where 
$n_{{}_\Omega}$ is the number of derivatives acting on \eqref{duitagra3}. On the other hand, for the KKLT instanton gases, once we switch on the volume of the sub-manifold ${\cal M}_2$, the energy-momentum tensor become \eqref{notiyell3}. If not, the result is:

{\footnotesize
\bg\label{notiyell7}
\mathbb{T}^{({\rm np}; 7)}_{\rm MN}(z) 
& = &  \sum_{\{l_i\}, n_i, k} c_k~{\bf g}_{\rm MN}(z)~ {\rm exp}\left[- k \int d^6 z' 
{\sqrt{{\bf {\red g}}_{\red 6}(z')} \over {\red g_s^{2n_2/3}}} 
\left({ \mathbb{F}^{(1)}(z - z') 
\mathbb{Q}_{\rm T}^{(\{l_i\}, n_i)} (z') \over {\rm M}_p^{\sigma(\{l_i\}, n_i) - 6}}\right)\right]\nonumber\\
&+ &\sum_{\{l_i\}, n_i, k}\left({2kc_k ~{\red{{\rm M}_p^{2n_2}}} \over {\rm M}_p^{\sigma(\{l_i\}, n_i) - {11}}}
\int d^{11} x ~{{\red (-2z_\alpha)^{n_2}}~\mathbb{F}^{(1)}(x - z)\over {\red g_s^{2n_2/3}}}\cdot {\delta \left(\sqrt{{\bf {\red g}}_{\red 6}(z)}\mathbb{Q}_{\rm T}^{(\{l_i\}, n_i)}(z)\right)\over 
\delta {\bf g}^{\rm MN}(z)}\sqrt{{\bf g}_{11}(x) \over {\bf {g}}_{11}(z)}\right) \nonumber\\ 
&&~~~~~~~\times 
{\rm exp}\left[- k \int d^6 z' {\sqrt{{\bf {\red g}}_{\red 6}(z')} \over {\red g_s^{2n_2/3}}} 
\left({ \mathbb{F}^{(1)}(x - z') 
\mathbb{Q}_{\rm T}^{(\{l_i\}, n_i)} (z') \over {\rm M}_p^{\sigma(\{l_i\}, n_i) - 6}}\right)\right], \nd}
which has a different $g_s$ scaling as expected. Once $n_2 \equiv n_{{}_\Omega} = 0$, we are basically dealing with ${\rm B}_{11} = 0$ in \eqref{duitagra3}. The results for all the cases are summarized in 
{\bf Tables \ref{instabeta3}} and {\bf \ref{instabeta4}}.

Before ending this section let us resolve couple other issues that we have been dragging along with our formulae so far. The first one is the appearance of an integral of the form:

{\footnotesize
\bg\label{Ebanks2}
\hskip-.1in\int d^{11}x \sqrt{-{\bf g}_{11}(x)} ~\mathbb{F}^{(1)}(x - z) ~{\rm exp}\left[- {\rm M}_p^{n_\Omega} g_s^{2/3}
\int d^6 y'~\mathbb{F}^{(1)}(x - y') ~\mathbb{A}_e(y')\right] \equiv g_s^{-14/3}~{\bf G}_e(z, g_s, n_2), \nd}
where $\mathbb{A}(y')$ contains the quantum series \eqref{selahran} etc., and $n_\Omega = n_2$. The integral over $d^{11} x$ is finite and well-defined, much like what we had in \eqref{Ebanks}. The extra exponential piece only makes the convergence better, both in terms of the integral and in terms of the choice $n_2 >> 1$, with the function ${\bf G}_e(z, g_s, n_2)$ vanishing for the latter case. The smoothness of the warp-factor $H(y)$ and the localized nature of the non-locality function $\mathbb{F}^{(1)}(x - y')$ 
are also the factors contributing to the convergence. On the other hand, the integral over $d^6y'$ is well-defined because of the compactness of the internal manifold ${\cal M}_4 \times {\cal M}_2$.

The second issue has to do with the appearance of the metric components in the non-perturbative energy-momentum tensors $\mathbb{T}_{\rm MN}^{({\rm np}; l)}$, {\it i.e.} the metric factor ${\bf g}_{\rm MN}(z)$ accompanying \eqref{notiyell}, \eqref{notiyell3}, \eqref{notiyell4}, \eqref{notiyell5}, \eqref{notiyell6} and 
\eqref{notiyell7}. They typically blow-up in the limit $g_s \to 0$, but do not contribute to the Einstein's EOMs. How do we take care of this?     
To proceed, we will follow the analysis in footnote \ref{luisis}. Let us define a function:
\bg\label{nicola} 
&&f(x) \equiv \sum_{n \ge 1} d_n x^n = d_1 x + d_2 x^2 + d_3 x^3 + d_4 x^4 + ........ \nonumber\\
&& d_n \equiv  \sum_{l \ge 1} {c_l (-l)^n \over n!} =  {c_1(-1)^n\over n!} + {c_2(-2)^n\over n!} + 
{c_3(-3)^n\over n!} + {c_4(-4)^n\over n!} + ......., \nd
with $(d_i, c_j)$ to be arbitrary constants. Once the set of $d_i$ are specified, the corresponding $c_j$ could be easily determined by inverting  a certain matrix whose details appear in footnote \ref{luisis}. We can now combine things together to express the following series:
\bg\label{walker}
x d_1 &=& {c_1(-1)^1\over 1!}~x + {c_2(-2)^1\over 1!}~x + {c_3(-3)^1\over 1!}~x + {c_4(-4)^1\over 1!}~x + ......
\nonumber\\ 
x^2 d_2 &=& {c_1(-1)^2\over 2!}~x^2 + {c_2(-2)^2\over 2!}~x^2 + {c_3(-3)^2\over 2!}~x^2 + {c_4(-4)^2\over 2!}~x^2 + ......
\nonumber\\  
x^3 d_3 &=& {c_1(-1)^3\over 3!}~x^3 + {c_2(-2)^3\over 3!}~x^3 + {c_3(-3)^3\over 3!}~x^3 + {c_4(-4)^3\over 3!}~x^3 + ......, \nd
and so on. The point of this obvious exercise was to justify one little thing, which becomes apparent when we add every term vertically down. Summing vertically down we easily get:

{\footnotesize
\bg\label{vajuan}
f(x) = c_1\left(e^{-x} - 1\right)  + c_2\left(e^{-2x} - 1\right)  + c_3\left(e^{-3x} - 1\right)  + c_4\left(e^{-4x} - 1\right) + ... = \sum_{k \ge 1} c_k\left(e^{-kx} - 1\right), \nd} 
which is all we need. The above result remains unchanged no matter what $c_i$ we choose, positive or negative. The issue of convergence of such a series (both in terms of $k$ and $(\{l_i\}, n_i)$)
has already been dealt with earlier, so we will not discuss it further here. If we now identify $f(x)$ with, say,  $\mathbb{U}^{(1)}_{(\{l_i\}, n_i)}(y)$ in
\eqref{ginagers} and $x$ by the integral structure therein, then the action ${\bf S}_2$ in \eqref{sheela2}, gets modified to the following indefinite integral structure:

{\footnotesize
\bg\label{sheelabr}
{\bf S}'_2 = \sum_{\{l_i\}, n_i, k} \int d^{11} x \sqrt{-{\bf g}_{11}} \sum_{r = 1}^\infty  c_k\left[
{\rm exp}\Big(- k {\rm M}_p^6 \int d^6 y \sqrt{{\bf g}_6(y)}
~\mathbb{F}^{(r)}(x - y) \mathbb{W}^{(r - 1)}\left(y; \{l_i\}, n_i\right)\Big) - 1\right], \nonumber\\ \nd}
in precisely the same way as in our simple exercise \eqref{vajuan}. In fact all the non-perturbative actions that we wrote as a trans-series should be modified in the aforementioned way. This addition of a counter-term with a relative {\it minus} sign in \eqref{sheelabr}, and subsequently in all other actions, removes the extra metric dependences from all the energy-momentum tensors, because:
\bg\label{maikam}
\lim_{g_s \to 0} \Big[{\bf g}_{\rm MN} ~{\rm exp}\left(-k g_s^{|\theta|}\right) - {\bf g}_{\rm MN}\Big] ~\to~0, \nd
for all $\theta > 0$, with similar cancellations for every $k$. To summarize then, the $g_s$ scalings of all the non-perturbative energy-momentum tensors are exactly as they appear in {\bf Tables \ref{instabeta3}} and {\bf \ref{instabeta4}} without any superfluous metric factors.

\begin{table}[tb]  
 \begin{center}
\renewcommand{\arraystretch}{1.5}
\begin{tabular}{|c|c|c|c|c|}\hline instanton type & $\mathbb{V}^a_2 \mathbb{V}^b_{{\bf T}^2}$  & ${g_s}$ scaling & $\mathbb{T}^{({\rm np})}_{\rm MN}$ & $\theta_{\rm min}$ \\ \hline\hline
BBS & $a = b = 0$ & ${g_s^{\theta + l_a}\over g_s^{2(3+n_2)/3}}$ & \eqref{notiyell6} & ${2\over 3}(n_2 + 4)$\\ \hline
Delocalized BBS & $a = 0, b = 1 $ & ${g_s^{\theta + l_a}\over g_s^{2(1+n_2)/3}}$ & \eqref{notiyell5} 
& ${2\over 3}(n_2 + 2)$
 \\ \hline
KKLT & $a = b = 0$ & ${g_s^{\theta + l_a}\over g_s^{2n_2/3}}$ & \eqref{notiyell7} & ${2\over 3}(n_2 + 1)$ \\ \hline
Delocalized KKLT & $a =1,  b = 0$ & ${g_s^{\theta + l_a}\over g_s^{2(1+n_2)/3}}$ & \eqref{notiyell3} 
& ${2\over 3}(n_2 + 2)$
\\ \hline
\end{tabular}
\renewcommand{\arraystretch}{1}
\end{center}
 \caption[]{The ${g_s}$ scalings of the four kind of wrapped M5-instantons that contribute to the non-perturbative energy-momentum tensor $\mathbb{T}^{({\rm np})}_{\rm MN}$, but now with the two-form 
 $\Omega$ as defined to be \eqref{duitagra3}.
Again,  the $g_s$ and $M_p$ scalings remain unchanged for the two cases with ${\rm B}_{11} = 0$ and ${\rm B}_{11} \ne 0$, if the derivatives do not act on \eqref{duitagra3}. The other parameter $\theta$, $\theta_{\rm min}$ and $l_a$ are defined as before.}
   \label{instabeta4}
 \end{table}


\subsection{Expectation values and the Schwinger-Dyson equations \label{sec3.3}}

With all the perturbative and non-perturbative quantum corrections at hand, including being equipped with the construction of the Glauber-Sudarshan states, we are ready to study the basics EOMs governing them. Some parts of the EOMs have already been discussed in details in \cite{desitter2}, so we will not take that path here. Instead we will analyze the EOMs using the Schwinger-Dyson equations \cite{dyson}. The full M-theory action, that includes all the corrections that we studied in section \ref{sec3.1} and \ref{sec3.2}, can be written as:
\bg\label{cixmel}
{\bf S}_{\rm tot} \equiv {\bf S}_1 + {\bf S}_2' + {\bf S}_3 + {\bf S}_4 + ... = {\bf S}_1 + {\bf S}_{\rm np} 
+ {\bf S}_{\rm b} + {\bf S}_{\rm top}, \nd
where ${\bf S}_1$ is the M-theory action in \eqref{sheela} that includes the infinite collection of perturbative local and non-local corrections, including the action for M2 and fractional M2 branes; ${\bf S}_{\rm np}$ is the action for the instanton gas studies in sections \ref{sec3.2.1} and \ref{sec3.2.3} that includes ${\bf S}_2'$ from 
\eqref{sheelabr} and other contributions that we elaborated in section \ref{sec3.2.3}; ${\bf S}_{\rm top}$ is the topological part of M-theory action that we studied in full details in \cite{desitter2}, so we don't discuss it here;
and ${\bf S}_{\rm b}$ is the action of the branes and surfaces, including their fermionic and higher order interactions, that do not appear in ${\bf S}_1$. As an example, we have uplifted-six-brane fermionic and higher order interactions in 
${\bf S}_3$ and ${\bf S}_4$ given as \eqref{sheela3} and \eqref{sheela4} respectively. The other wrapped brane actions may be thought of as coming from the non-local interactions and topological action that we discussed in ${\bf S}_1$ and ${\bf S}_{\rm top}$
from \eqref{sheela} and \cite{desitter2} respectively.  Note also that each of the pieces in \eqref{cixmel} has an {\it infinite} number of terms, so ${\bf S}_{\rm tot}$ is pretty much an exhaustive collection.

${\bf S}_{\rm tot}$ is also in some sense a Wilsonian action constructed by integrating out the UV modes  
in the solitonic background. As we elaborated earlier, such integrating out procedure is possible because our vacuum is supersymmetric and solitonic, so do not suffer from any pathologies attributed to vacua like Bunch-Davies and other similar avatars. Supersymmetry is broken spontaneously from switching on 
coherent states of non-self-dual G-fluxes as in \eqref{sosie}, so the positive cosmological constant $\Lambda$ appears   
from a conspiracy between these fluxes and quantum corrections in a way discussed in \cite{desitter2} with the zero point energy playing no part here. This is of course the advantage we get from our choice of vacuum, but here we want to inquire about the {\it stability} of the Glauber-Sudarshan state amidst the infinite set of quantum corrections emanating from ${\bf S}_{\rm tot}$. This is where the Schwinger-Dyson equations \cite{dyson} become immensely useful.  

The original formulation of the Schwinger-Dyson equations (SDEs) in \cite{dyson} provide relations between Green's functions in QFT as expectation values over states. In the gravitational and the flux sectors the SDEs imply:
\bg\label{angelwondr}
\int \left[{\cal D} g_{\rm MN}\right]~{\delta \over \delta g_{\rm PQ}} \equiv 0 \equiv 
\int \left[{\cal D} {\rm C}_{\rm MNP}\right]~{\delta \over \delta {\rm C}_{\rm PQR}}, \nd 
{\it i.e.} the integrals over total derivatives vanish, so it shouldn't be too hard to get them here. However before we get the required SDEs, 
recall that in \cite{desitter2, desitter3}
we carefully distinguished between warped ({\it i.e.} $g_s$ dependent) and un-warped ({\it i.e.} $g_s$ independent) parts of the metric. Question is, in the computation \eqref{mcapopo}, what metric was used? The answer is simple: for us there is only the solitonic background \eqref{betbab3} with the modes $\Psi_k$ in \eqref{hudson} for the gravitational sector and $\Upsilon_k$ (that we discuss below) for the 
flux sector\footnote{Not to be confused with the fermions ${\bf\Psi}$ and ${\bf\Upsilon}$ used in section \ref{sec3.2.2}!}. Everything else, in particular the background \eqref{vegamey3} and the corresponding time-dependent G-flux components, must appear as expectation values over the generalized
Glauber-Sudarshan states\footnote{Recall from section \ref{sec2.4} that we call the shifted interacting vacuum $\mathbb{D}(\sigma)\vert \Omega\rangle$ as the generalized Glauber-Sudarshan states to distinguish it from the original Glauber-Sudarshan states created out of the shifted harmonic vacuum 
$\mathbb{D}_0(\sigma) \vert 0\rangle$.}. 

Looking at the integral form of the result on the RHS of \eqref{mcapopo}, we see that  $\alpha_{\mu\nu}^{(\psi)}({\bf k}, t)$ appears as the expectation value. Since 
$\alpha_{\mu\nu}^{(\psi)}({\bf k}, t)$ captures the complete time-dependence of the corresponding metric components, our analysis in \eqref{mcapopo} has resulted in a fully {\it warped} metric components from the path integral. In retrospect, this is what should have been, so the apparent consistency is not much of a surprise, although one concern could be raised here: the ${\cal O}\left({g_s^a\over {\rm M}^b_p}\right)$
corrections accompanying the expected answer. Does that mean we deviate from the de Sitter background? 
The answer, as we shall soon see, is no: once we make the right choice of the displacement operator, these correction terms do not appear anymore. 
What is interesting however is that this choice also paves the way to achieve the fully warped metric components from the SDEs, as will be elaborated below.

There is yet another thing that needs elaboration before we proceed further, and has to do with the displacement operators $\mathbb{D}(\alpha)$ defined in \eqref{maristone}. The action of the displacement operator is \eqref{15movs}, but it hides the fact that there are both gravitational and four-form fields participating in the construction. This means the form of $\mathbb{D}(\alpha)$ cannot be as simple as 
\eqref{juanivan}, and the modified form should incorporate all the existing field components that form the generalized Glauber-Sudarshan states in our set-up. This is a tedious exercise, but we can make it simple by resorting to few definitions. Let $\{\alpha_{\rm MN}\}$ denotes the set of Fourier components in 
\eqref{cannon}; and $\{\beta_{\rm MNP}\}$ denotes the corresponding set for the three-form flux components 
${\bf C}_{\rm MNP}$, then $\mathbb{D}(\alpha)$ from \eqref{juanivan} may be modified to:

{\footnotesize
\bg\label{juanivan2}
\mathbb{D}(\alpha, \beta) &= & {\rm exp}\left[\int_{-\infty}^{+\infty}d^{11} k\Big( 
\alpha^{(\Psi)}_{\rm MN}({\bf k}, k_0) \widetilde{g}^{\ast{\rm MN}}({\bf k}, k_0) 
+ \beta^{(\Upsilon)}_{\rm MNP}({\bf k}, k_0) \widetilde{C}^{\ast{\rm MNP}}({\bf k}, k_0)\Big)\right]\\
&\times & {\rm exp}\left[-{1\over 2} \int_{-\infty}^{+\infty} 
d^{11}k \Big(\alpha^{(\Psi)}_{\rm MN}({\bf k}, k_0) \alpha^{\ast(\Psi){\rm MN}}({\bf k}, k_0)
+ \beta^{(\Upsilon)}_{\rm MNP}({\bf k}, k_0) \beta^{\ast(\Upsilon){\rm MNP}}({\bf k}, k_0)\Big)+ .... \right], 
\nonumber \nd}
where the tilde denote Fourier transforms, and the dotted terms are the higher order mixing between the various components coming from our generic definition of $a^\dagger_{\rm eff}$ in \eqref{adler}. This is arranged in a way that \eqref{juanivan2} is not unitary. The other components in \eqref{juanivan2} are 
$\Psi_k$ and $\Upsilon_k$ which are respectively the set of Schr\"odinger wave-functions in \eqref{hudson} and a similar set for the three-form flux components ${\bf C}_{\rm MNP}$. Note that ${\bf C}_{\rm MNP}$ are not gauge invariants, and neither are the metric components, so $\mathbb{D}(\alpha, \beta)$ would change the expectation values of the corresponding fields in the right way under gauge transformations. We also expect:
\bg\label{mcadumpla}
\langle \alpha, \beta \vert \alpha, \beta \rangle &= & \langle \Omega \vert \mathbb{D}^\dagger(\alpha, \beta)
\mathbb{D}(\alpha, \beta)\vert \Omega \rangle\nonumber\\
 &= & {\int \left[{\cal D}\{g_{\rm MN}\}\right] 
\left[{\cal D}\{{\rm C}_{\rm PQR}\}\right] ~e^{i{\bf S}_{\rm tot}}\mathbb{D}^\dagger(\alpha, \beta)
\mathbb{D}(\alpha, \beta) \over \int \left[{\cal D}\{g_{\rm MN}\}\right] 
\left[{\cal D}\{{\rm C}_{\rm PQR}\}\right] ~e^{i{\bf S}_{\rm tot}}}, \nd
where using some abuse of notation, we have taken $\mathbb{D}(\alpha, \beta)$ to denote both the operator and the field. Which is which should be clear from the context. The set $\{g_{\rm MN}\}$ denotes the set of metric components $({g}_{mn}, {g}_{\alpha\beta}, {g}_{ab}, {g}_{\mu\nu})$ and the set 
$\{{\rm C}_{\rm MNP}\}$ denotes the C-fields that appear from the G-flux components 
$({\rm G}_{mnpq}, {\rm G}_{mnp\alpha}, {\rm G}_{mn\alpha\beta}, {\rm G}_{mnpa}, {\rm G}_{mnab}, 
{\rm G}_{0ijm})$ and other permutations. These will be related to the components that we encountered in sections \ref{sec3.1} and \ref{sec3.2}, and also in \cite{desitter2, desitter3}. Finally, ${\bf S}_{\rm tot}$ is the fully interacting action written in \eqref{cixmel} and is responsible for creating the interacting vacuum $\vert\Omega\rangle$. 

There are still couple more issues that we need to clarify before we proceed further. First,
the way we have expressed $\vert \alpha, \beta\rangle$, it is clearly not normalized because 
$\mathbb{D}(\alpha, \beta)$ is not unitary. The {\it shifted} vacuum $\vert \alpha, \beta \rangle \equiv
\mathbb{D}(\alpha, \beta) \vert \Omega \rangle$, as we showed earlier using the simpler version \eqref{bwstefen}, does produce the expected answer in \eqref{mcapopo}, so we expect the same to hold for 
$\langle \{g_{\rm MN}\} \rangle_{(\alpha, \beta)}$. This is a straightforward exercise so we will not do it here, instead we want to point out that the expectation values of the G-flux components 
$\langle \{{\rm G}_{\rm MNPQ}\}\rangle_{(\alpha, \beta)}$ now reproduce the expected results from 
\cite{desitter2, desitter3}.

Secondly, as we cautioned earlier, because of the presence of metric and C-fields, which are not gauge invariant quantities, there should be Faddeev-Popov ghosts changing:
\bg\label{dlofvero}
{\bf S}_{\rm tot} \to {\bf S}_{\rm tot} - {\bf S}_{\rm ghost}, \nd
where the relative sign is chosen for later convenience.
If \eqref{dlofvero} is always true, then it will make 
the action even more complicated than \eqref{cixmel}. There are specific gauge choices that do not create propagating ghosts, when we take the metric and the C-field degrees of freedom separately, but it is not clear such a gauge exists when we take everything {\it together}. This would mean that we might have to venture beyond \eqref{cixmel}. Taking all these into considerations, the first set of Schwinger-Dyson equations resulting from \eqref{mcadumpla} then takes the following form:
\bg\label{SDEss}
&&\left\langle {\delta {\bf S}_{\rm tot}\over \delta \{g^{\rm MN}\}}\right\rangle_{(\alpha, \beta)} = 
\left\langle {\delta {\bf S}_{\rm ghost}\over \delta \{g^{\rm MN}\}}\right\rangle_{(\alpha, \beta)} - 
\left\langle {\delta \over \delta \{g^{\rm MN}\}}~{\rm log}\Big(\mathbb{D}^\dagger(\alpha,\beta)
\mathbb{D}(\alpha,\beta)\Big) \right\rangle_{(\alpha, \beta)}\\
&&\left\langle {\delta {\bf S}_{\rm tot}\over \delta \{{\rm C}^{\rm MNP}\}}\right\rangle_{(\alpha, \beta)} = 
\left\langle {\delta {\bf S}_{\rm ghost}\over \delta \{{\rm C}^{\rm MNP}\}}\right\rangle_{(\alpha, \beta)} - 
\left\langle {\delta \over \delta \{{\rm C}^{\rm MNP}\}}~{\rm log}\Big(\mathbb{D}^\dagger(\alpha,\beta)
\mathbb{D}(\alpha,\beta)\Big) \right\rangle_{(\alpha, \beta)}, \nonumber \nd
where {\it all} degrees of freedom appear on both sides of the two set of equations, making it a complicated set of coupled differential equations. Question is how to solve these equations to extract useful data for the generalized Glauber-Sudarshan states. 

First, even without solving anything we see that the SDEs' are expressed as expectation values over the generalized Glauber-Sudarshan states. This is already a good start as our earlier path integral approach in \eqref{mcapopo} showed us that the expectation value of the metric over the state $\vert\alpha\rangle$ does reproduce the 
de Sitter space-time. Secondly, the functional derivatives are taken with respect to the space-time metric and C-field components so it would be useful to bring \eqref{juanivan2} to the space-time integral format. This becomes:
\bg\label{juanivan3}
\mathbb{D}(\alpha, \beta) &= & {\rm exp}\left[\int_{-\infty}^{+\infty}d^{11} x \sqrt{-g_{11}}\Big( 
\alpha^{(\Psi)}_{\rm MN}(x){g}^{{\rm MN}}(x) 
+ \beta^{(\Upsilon)}_{\rm MNP}(x) {C}^{{\rm MNP}}(x)\Big)\right] \nonumber\\
&\times & {\rm exp}\left[-{1\over 2} \int_{-\infty}^{+\infty} 
d^{11}x \sqrt{-g_{11}}\left(\left\vert\alpha^{(\Psi)}_{\rm MN}(x)\right\vert^2 
+ \left\vert\beta^{(\Upsilon)}_{\rm MNP}(x)\right\vert^2\right) + ...... \right],  \nd
where we have used the normalization condition from footnote \ref{sarabig} to bring the second line in the right form. The metric determinant is of the solitonic background \eqref{betbab3} which means 
$\sqrt{-g_{11}} = h_2^{-1} h_1^{4/3}$. Plugging \eqref{juanivan3} in \eqref{SDEss} reproduces:
\bg\label{michelragi}
&& {\delta \over \delta \{{\rm C}^{\rm MNP}(z)\}}\left[{\rm log}\Big(\mathbb{D}^\dagger(\alpha,\beta)
\mathbb{D}(\alpha,\beta)\Big)\right] = 2 h_2^{-1}(z) h_1^{4/3}(z) ~\beta^{(\Upsilon)}_{\rm MNP}(z) + ....
\\
&& {\delta \over \delta \{{g}^{\rm MN}(z)\}}\left[{\rm log}\Big(\mathbb{D}^\dagger(\alpha,\beta)
\mathbb{D}(\alpha,\beta)\Big)\right] = \left(2 - g^{\rm M'N'}(z) g_{\rm M'N'}(z)\right) h_2^{-1}(z) h^{4/3}_1(z)~\alpha^{(\Psi)}_{\rm MN}(z) + .., \nonumber \nd
where the dotted terms are higher order mixing terms resulting from \eqref{adler}. Note that the results, at least to the order that we study here, do not depend directly on $\mathbb{D}(\alpha, \beta)$, but are proportional to the background metric \eqref{vegamey3} and the corresponding G-flux components. It is also interesting to ask how does the expectation value, computed in \eqref{meyepolce2}, change if two different generalized Glauber-Sudarshan states are used, for example $\vert \alpha_1, \beta_1\rangle$ and $\vert\alpha_2, \beta_2\rangle$. If we ignore $\beta_i$ for simplicity, then we ask how the expectation value 
$\langle \alpha_2 \vert {\bf g}_{\mu\nu} \vert \alpha_1\rangle$ differs from \eqref{meyepolce2}. In path integral form, the numerator takes the following form: 

{\footnotesize
\bg\label{meyepolcedul}
&&\int \left[{\cal D}g_{\mu\nu}\right] e^{iS} {\mathbb{D}}^\dagger(\alpha_2) \delta g_{\mu\nu}  
{\mathbb{D}}(\alpha_1)
 =  \left(\prod_{k} \int d\left({\bf Re}~\widetilde{g}_{\mu\nu}(k)\right) d\left({\bf Im}~\widetilde{g}_{\mu\nu}(k)\right)\right)
  {\rm exp}\left[{i\over V}\sum_{k} k^2 \vert \widetilde{g}_{\mu\nu}(k)\vert^2 + iS_{\rm sol} + ..\right] 
  \nonumber\\
&&~~~~~~~~~~~~~~~~~~~~\times
 {\rm exp} \left\{{1\over V} \sum_{k'} 
\left[{\bf Re}\left(\alpha^{(\psi)}_{(1)\mu\nu}(k') + \alpha^{(\psi)}_{(2)\mu\nu}(k')\right) 
- i{\bf Im}\left(\alpha^{(\psi)}_{(2)\mu\nu}(k') - \alpha^{(\psi)}_{(1)\mu\nu}(k')\right)\right]
{\bf Re}~\widetilde{g}^{\mu\nu}(k')\right\} \nonumber\\
&&~~~~~~~~~~~~~~~~~~~~\times
 {\rm exp} \left\{{1\over V} \sum_{k'} 
\left[{\bf Im}\left(\alpha^{(\psi)}_{(1)\mu\nu}(k') + \alpha^{(\psi)}_{(2)\mu\nu}(k')\right) 
+ i{\bf Re}\left(\alpha^{(\psi)}_{(2)\mu\nu}(k') - \alpha^{(\psi)}_{(1)\mu\nu}(k')\right)\right]
{\bf Im}~\widetilde{g}^{\mu\nu}(k')\right\} \nonumber\\
&&~~~~~~~~~~~~~~~~~~~~~ \times  {1\over V} \sum_{k''} ~\psi_{{\bf k}''}({\bf x}, y, z) e^{-ik''_0t}
\left({\bf Re}~\widetilde{g}_{\mu\nu}(k'') + i{\bf Im}~\widetilde{g}_{\mu\nu}(k'')\right)~{\rm exp}\left(-{1\over V}\sum_{k'} \vert\alpha^{(\psi)}_{\mu\nu}(k')\vert^2\right), \nonumber\\ \nd}
where $\delta g_{\mu\nu}$ implies we have ignored the solitonic part of the field $g_{\mu\nu}$.   
We see that the coefficients of ${\bf Re}~\widetilde{g}^{\mu\nu}(k')$ and ${\bf Im}~\widetilde{g}^{\mu\nu}(k')$ become {\it complex}. Interestingly the complex factor is 
$\alpha^{(\psi)}_{(2)\mu\nu}(k') - \alpha^{(\psi)}_{(1)\mu\nu}(k')$ and therefore vanishes when $\alpha_2 = \alpha_1$. The denominator will have a similar form as \eqref{meyepolcedul} except without the metric field. The integral can be easily performed, and here for illustrative purpose let us assume that  
$\widetilde{g}^{\mu\nu}(k')$ is real. Putting everything together we get\footnote{In this section both fields and operators of the solitonic background \eqref{betbab3} will be denoted by Roman letters {\it i.e.} $g_{\rm MN}$ and ${\rm C}_{\rm MNP}$, whereas the fields of the background \eqref{vegamey3} will be denoted by bold-faced letters {\it i.e.} ${\bf g}_{\rm MN}$ and 
${\bf C}_{\rm MNP}$. In this way connecting to variables from sections \ref{sec3.1} and \ref{sec3.2} will be easier.}:
\bg\label{mcaok}
&& \langle \alpha_2\vert {g}_{\mu\nu}\vert \alpha_1 \rangle = {\eta_{\mu\nu} \over h_2^{2/3}(y, {\bf x})} + 
{1\over 2} \int {d^{10} {\bf k}\over 2\omega^{(\psi)}_{\bf k}} ~
\hat{\alpha}^{(\psi)}_{\mu\nu}({\bf k}, t)\psi_{\bf k}({\bf x}, y, z)
+ {\cal O}\left({g_s^{c}\over {\rm M}_p^d}\right) \\
&& \hat{\alpha}^{(\psi)}_{\mu\nu}({\bf k}, t) \equiv
{\bf Re}\left(\alpha^{(\psi)}_{(1)\mu\nu}({\bf k}, t) + \alpha^{(\psi)}_{(2)\mu\nu}({\bf k}, t)\right) 
- i{\bf Im}\left(\alpha^{(\psi)}_{(2)\mu\nu}({\bf k}, t) - \alpha^{(\psi)}_{(1)\mu\nu}({\bf k}, t)\right), \nonumber \nd
which tells us that unless $\alpha^{(\psi)}_{(1)\mu\nu}({\bf k}, t) = \alpha^{(\psi)}_{(2)\mu\nu}({\bf k}, t) =
\alpha^{(\psi)}_{\mu\nu}({\bf k}, t)$ where $\alpha^{(\psi)}_{\mu\nu}({\bf k}, t)$ is the value from \eqref{cnelson}, the expectation value cannot produce a de Sitter space. Additionally, the inequality between
$\alpha^{(\psi)}_{(i)\mu\nu}({\bf k}, t)$ suggests that \eqref{mcaok} may not even be real. This means, an equality between $\alpha^{(\psi)}_{(i)\mu\nu}({\bf k}, t)$ only guaranties a de Sitter space when 
$\alpha^{(i)(\psi)}_{\mu\nu}({\bf k}, t)$ takes the value in \eqref{cnelson}, otherwise it will be another time-dependent space-time. 

Such a criterion is particularly useful when we evaluate the expectation values of the products of metric and G-flux components. A simple example would be the expectation value 
$\langle \alpha, \beta \vert g_{\mu\nu}(z) g^{\mu\nu}(z) \vert \alpha, \beta \rangle$, where 
$\vert\{\alpha\} \rangle \equiv \vert \alpha \rangle$ denotes the coherent states associated with the metric sector and $\vert\{\beta\}\rangle \equiv \vert \beta \rangle$ denotes the coherent states associated with the G-flux sector. In the {\it mixed} sector, as we discussed earlier, the coherent states may be denoted as  
$\vert\alpha, \beta\rangle$. For the simple case, once we concentrate on the gravitational sector, we 
expect the following decomposition:
\bg\label{aisha}
\langle g_{\mu\nu}(z) g^{\mu\nu}(z) \rangle_\alpha
 \equiv \langle \alpha \vert g_{\mu\nu}(z) g^{\mu\nu}(z) \vert \alpha \rangle = 
 \int {d^2\alpha'\over \pi} \langle \alpha \vert g_{\mu\nu}(z)\vert \alpha' \rangle  
 \langle \alpha' \vert g^{\mu\nu}(z)\vert \alpha\rangle, \nd
 where we have imposed the completeness property of the Glauber-Sudarshan states. Here $\vert\alpha\rangle$ is related to $\alpha^{(\psi)}_{\mu\nu}({\bf k}, t)$ from \eqref{cnelson}, but $\vert \alpha'\rangle$ could in principle be arbitrary. This means $\langle \alpha' \vert g_{\mu\nu}(z)\vert \alpha\rangle$ from 
 \eqref{mcaok} isn't necessarily a de Sitter space, unless $\alpha' = \alpha$. Thus the decomposition 
 \eqref{aisha} implies:
 \bg\label{millerS}
\langle \vert g_{\mu\nu}(z)\vert^2 \rangle_\alpha \equiv
\langle g_{\mu\nu}(z) g^{\mu\nu}(z) \rangle_\alpha = \left\vert \langle g_{\mu\nu}(z) \rangle_\alpha\right\vert^2 + c\sum_{\alpha' \ne \alpha} \vert\langle \alpha \vert g_{\mu\nu}(z)\vert \alpha' \rangle\vert^2 \nd
where the sum is over backgrounds of the form \eqref{mcaok}, with $\alpha_1 = \alpha$ and $\alpha_2 = \alpha'$, that deviate from the de Sitter space; and $c$ is a constant that is required to convert the integral in \eqref{aisha} to a sum.
In a similar vein, any powers of metric or G-flux components, or even mixed powers of metric and flux components would have at least a decomposition of the form \eqref{millerS}. The sum in \eqref{millerS} involve terms with $\alpha' > \alpha$ as well as with $\alpha' < \alpha$. They come with opposite signs in \eqref{mcaok}, so it will be worthwhile to evaluate this directly from the path integral. The integral that we are looking for now is:

{\footnotesize
\bg\label{meyepolceteen}
&&\int \left[{\cal D}g_{\mu\nu}\right] ~e^{iS} {\mathbb{D}}^\dagger(\alpha)\vert g_{\mu\nu}\vert^2 {\mathbb{D}}(\alpha)
 =  \left(\prod_{k} \int d\left({\bf Re}~\widetilde{g}_{\mu\nu}(k)\right) d\left({\bf Im}~\widetilde{g}_{\mu\nu}(k)\right)\right)
  {\rm exp}\left[{i\over V}\sum_{k} k^2 \vert \widetilde{g}_{\mu\nu}(k)\vert^2 + iS_{\rm sol} + ...\right] \nonumber\\
&&\times ~{1\over V^2}~
 {\rm exp} \left[{2\over V} \sum_{k'} ~
\left({\bf Re}~\alpha^{(\psi)}_{\mu\nu}(k')~{\bf Re}~\widetilde{g}^{\mu\nu}(k') + {\bf Im}~\alpha^{(\psi)}_{\mu\nu}(k')~{\bf Im}~\widetilde{g}^{\mu\nu}(k')\right) + ...\right]
~{\rm exp}\left(-{1\over V}\sum_{k'} \vert\alpha^{(\psi)}_{\mu\nu}(k')\vert^2\right)
\nonumber\\
&& \times   \sum_{k'', k'''} \psi_{{\bf k}''}({\bf x}, y, z) \psi_{{\bf k}'''}({\bf x}', y', z') 
e^{-i(k''_0t + k_0''' t')}
\Big({\bf Re}~\widetilde{g}_{\mu\nu}(k'') + i{\bf Im}~\widetilde{g}_{\mu\nu}(k'')\Big)
\Big({\bf Re}~\widetilde{g}^{\mu\nu}(k''') + i{\bf Im}~\widetilde{g}^{\mu\nu}(k''')\Big), \nonumber\\ \nd}
which is basically the numerator of the expectation value \eqref{aisha}, except that we have ignored the solitonic part of the metric. This can be easily rectified. Note that we have separated the two metric components over space and time so that short distance singularities may be avoided. we will also avoid {\it summing} over repeated indices to avoid overcomplicating the integral. This means the tensorial property of the metric is not much of a concern here, and assuming this to be the case, the integral 
\eqref{meyepolceteen} may be evaluated with the aid of a few notations. Let $(k, k', k'', k''') \equiv (k_p, k_n, k_m, k_l)$ and $\widetilde{g}_{\mu\nu}(k) \equiv \Phi(k_p) \equiv \Phi_p$. We will assume 
${\bf Re}~\alpha^{(\psi)}_{\mu\nu}(k') ={\bf Im}~\alpha^{(\psi)}_{\mu\nu}(k') \equiv \alpha_n$ for simplicity that can be easily relaxed. As will be clear, none of these assumptions are necessary, and more importantly do not effect the final conclusion, so over-complicating the analysis will lead to the same conclusion as with the simpler version that we choose here. As an exercise, the reader could verify this in details.  We will also go to the Euclidean formalism so that $ik^2 \to -k^2 \equiv -k^2_p$. With these changes, the integral \eqref{meyepolceteen}, now becomes:

{\footnotesize
\bg\label{meyline}
{\rm Num}\left[\langle\vert g_{\mu\nu}\vert^2\rangle_\alpha\right] &=& 
\left(\prod_{p} \int d\left({\bf Re}~\Phi_p\right) d\left({\bf Im}~\Phi_p\right)\right)
  {\rm exp}\left[-{1\over V}\sum_{p} k_p^2 \left(\left({\bf Re}~\Phi_p\right)^2 + \left({\bf Im}~\Phi_p\right)^2
  \right) + ... \right] \nonumber\\
&\times &
 {\rm exp} \left[{2\over V} \sum_{n} ~\alpha_n
\left({\bf Re}~\Phi_n + {\bf Im}~\Phi_n\right) + ...\right]
~{\rm exp}\left(-{1\over V}\sum_{n} \vert \alpha_n \vert^2\right) \nonumber\\
& \times & {1\over V^2}  \sum_{m, l} \psi_{m}({x}, y, z) \psi_{l}({x}', y', z') 
\Big({\bf Re}~\Phi_m + i{\bf Im}~\Phi_m\Big)
\Big({\bf Re}~\Phi_l + i{\bf Im}~\Phi_l\Big), \nd} 
where $\psi_m(x, y, z) \equiv \psi_{{\bf k}''}({\bf x}, y, z) e^{-ik_0'' t}$. The form of the integral is somewhat similar to the integral one would encounter when computing the two-point function. However there is a crucial difference: the presence of ${\rm exp}\left[{2\over V} \sum_{n} ~\alpha_n\left({\bf Re}~\Phi_n + {\bf Im}~\Phi_n\right)\right]$ factor that was responsible in \eqref{meyepolce}, \eqref{meyepolce2} and \eqref{meyepolce3} to give  non-zero results for one-point functions by {\it shifting} the vacuum. Here, and because of this, the integral 
in \eqref{meyline} cannot just be the result that we know for the two-point function. There will be more contributions that typically {\it vanish} in the usual computation of the two-point function. To quantify this, let us express the contributions from the last line of \eqref{meyline} as four sectors:

{\footnotesize
\bg\label{chuchitk}
&&(+~+):~~ {\bf Re}~\Phi_m~{\bf Re}~\Phi_l - {\bf Im}~\Phi_m~{\bf Im}~\Phi_l 
+ i\left({\bf Im}~\Phi_m~{\bf Re}~\Phi_l + {\bf Re}~\Phi_m~{\bf Im}~\Phi_l\right)\nonumber\\
&&(+~-):~~ {\bf Re}~\Phi_m~{\bf Re}~\Phi_l + {\bf Im}~\Phi_m~{\bf Im}~\Phi_l 
+ i\left({\bf Im}~\Phi_m~{\bf Re}~\Phi_l - {\bf Re}~\Phi_m~{\bf Im}~\Phi_l\right)\nonumber\\
&&(-~+):~~ {\bf Re}~\Phi_m~{\bf Re}~\Phi_l + {\bf Im}~\Phi_m~{\bf Im}~\Phi_l 
- i\left({\bf Im}~\Phi_m~{\bf Re}~\Phi_l - {\bf Re}~\Phi_m~{\bf Im}~\Phi_l\right)\nonumber\\
&&(-~-):~~ {\bf Re}~\Phi_m~{\bf Re}~\Phi_l - {\bf Im}~\Phi_m~{\bf Im}~\Phi_l 
- i\left({\bf Im}~\Phi_m~{\bf Re}~\Phi_l + {\bf Re}~\Phi_m~{\bf Im}~\Phi_l\right), \nd}
where we have used $\Phi(-k_m) \equiv \Phi_{-m}  = \Phi^\ast_m$. In the usual computation in QFT, all the imaginary pieces ({\it i.e.} the coefficients of $i$ and {\it not} ${\bf Im}~\Phi_m$) in \eqref{chuchitk} cancel out because they form linear terms in a gaussian integral. Clearly this cannot be the case now. Similarly, all the real pieces in the $(+~+)$ sector also cancel out because of the relative {\it minus} sign. The behavior of the 
$(-~+)$ and $(-~-)$ sector would be similar to the  $(+~+)$ and $(+~-)$ sector so we could concentrate only on the first two. However for each of the two sectors we could either have $m \ne l$ or $m = l$. The result of the integrals for each of the sector then yields the following:
\bg\label{halimadi}
&&(+~+), (m = l): ~~ {i\over 2} \left(\prod_p~{1\over k_p^2}\right) \sum_{m} ~{\alpha^2_m \over 
k_m^4}~ \psi_m(x, y, z)~ \psi_m(x', y', z')\\
&&(+~+), (m \ne l): ~~ {i\over 2} \left(\prod_p~{1\over k_p^2}\right) \sum_{m, l} ~{\alpha_m \alpha_l\over 
k_m^2 k_l^2}~\psi_m(x, y, z) ~\psi_l(x', y', z')\nonumber\\
&&(+~-), (m \ne l): ~~ {1 \over 2} \left(\prod_p~{1\over k_p^2}\right) \sum_{m, l} ~{\alpha_m \alpha_l\over 
k_m^2 k_l^2}~\psi_m(x, y, z) ~\psi^\ast_l(x', y', z')\nonumber\\
&&(+~-), (m = l): ~~ {1 \over 2} \left(\prod_p~{1\over k_p^2}\right) \sum_{m} \left({\alpha^2_m\over 
k_m^4} - {2\over k_m^2}\right)~\psi_m(x, y, z) \psi^\ast_m(x', y', z'), \nonumber \nd
where one may easily verify that, when $\alpha_p = 0$, all the contributions vanish, except for one term from the $(+~-)$ sector with $m = l$. In fact this term is exactly the {\it propagator} for the gravitons as may be seen from the following computation:
\bg\label{propag}
 \sum_m ~ {\psi_m(x, y, z) \psi^\ast_m(x', y', z') \over k_m^2} = \int d^{11}k ~
 {\psi_k({\bf x}, y, z) \psi^\ast_k({\bf x}', y', z') e^{-ik_0(t - t')} \over k^2 + i\epsilon}, \nd
 where the infinite product coefficient in \eqref{halimadi} is cancelled by the denominator of the path integral.
The spatial wave-function $\psi_k$ appears from \eqref{hudson} and contributes to \eqref{cannon}, with the $i\epsilon$ factor taking care of the residue at the poles in the usual way.
 
The results in \eqref{halimadi}, confirms the generic expectation \eqref{millerS}, and the first term which is the square of the expectation value $\langle g_{\mu\nu}\rangle_\alpha$ is basically the first term in the 
sector $(+~-)$ for $m = l$. In the language of $\langle g_{\mu\nu}\rangle_\alpha$, the squaring would involve two integrals $d^{11} k d^{11} k'$, but $(k, k')$ are related by $\delta^{11}(k + k')$ (recall we are in the $(+~-)$ sector hence $k = -k'$ and not $k = k'$), so we do get the first term from $(+~-)$ with $m = l$ correctly. Note that $\alpha_m^2 = {1\over 2} \alpha_m^2 \vert 1 + i\vert^2$, so the results fit well, confirming in turn the decomposition \eqref{millerS}. Generalizing this, we expect, for example:
\bg\label{einstobeta}
\left\langle {\rm R}_{\rm MN} - {1\over 2} g^{\rm MN} {\rm R} \right\rangle_{(\alpha, \beta)} &= & 
\langle {\rm R}_{\rm MN}\rangle_{(\alpha, \beta)} - {1\over 2} \langle g^{\rm MN} \rangle_{(\alpha, \beta)} 
\langle {\rm R} \rangle_{(\alpha, \beta)} + ..... \\
& = & \langle {\rm R}_{\rm MN}\rangle_{(\alpha, \beta)} - {1\over 2} \langle g^{\rm MN} \rangle_{(\alpha, \beta)}\langle g^{\rm PQ}\rangle_{(\alpha,\beta)} \langle {\rm R}_{\rm PQ} \rangle_{(\alpha, \beta)} + .....,
\nonumber \nd
where the dotted terms would be the {\it sum} of the terms where we allow
 intermediate states like $(\alpha', \beta'), 
(\alpha'', \beta'')$ etc. to appear with the condition that $\alpha \ne (\alpha', \alpha'')$ and $\beta \ne (\beta', \beta'')$. Again, since we are in the gravitational sector, we won't need the information for $\beta$ in 
\eqref{einstobeta}, so we will suppress it. This means we can express the expectation value of the Ricci curvature in the following suggestive way:
\bg\label{scixor}
\langle {\rm R}_{\rm MN} \rangle_\alpha & = & -{1\over 2} \left[\partial_{\rm P} \partial_{\rm Q} \langle g_{\rm MN} \rangle_\alpha + \partial_{\rm M} \partial_{\rm N} \langle g_{\rm PQ} \rangle_\alpha -
\partial_{({\rm M}} \partial_{|{\rm P}|} \langle g_{{\rm N}){\rm Q}} \rangle_\alpha\right]\langle g^{\rm PQ}\rangle_\alpha \\
& & + {1\over 2}\left[{1\over 2} \partial_{\rm M} \langle g_{\rm PQ}\rangle_\alpha 
\partial_{\rm N} \langle g_{\rm RS}\rangle_\alpha  + 
\partial_{\rm P} \langle g_{\rm MQ}\rangle_\alpha 
\partial_{[{\rm R}} \langle g_{|{\rm N}|{\rm S}]}\rangle_\alpha\right]  \langle g^{\rm PR}\rangle_\alpha
\langle g^{\rm QS}\rangle_\alpha \nonumber\\
&& -{1\over 4}\left[\partial_{({\rm M}}\langle g_{{\rm N}){\rm Q}}\rangle_\alpha -\partial_{\rm Q}
\langle g_{\rm MN}\rangle_\alpha\right] \left[2\partial_{\rm Q}\langle g_{\rm RS}\rangle_\alpha
- \partial_{\rm S}\langle g_{\rm PR}\rangle_\alpha \right] \langle g^{\rm PR}\rangle_\alpha
\langle g^{\rm QS}\rangle_\alpha + .... \nonumber \nd 
where the symbol $|{\rm P}|$ stands for the index neutral to symmetrization or anti-symmetrization, and the dotted  terms are the ones that have $(\alpha', \alpha'',..)$ intermediate states with none of them equal to $\alpha$. 
The results \eqref{scixor} and \eqref{einstobeta} convey something very important, once we note that 
$\langle g_{\rm MN}\rangle_\alpha = {\bf g}_{\rm MN}$ from \eqref{meyepolce2}, where ${\bf g}_{\rm MN}$ is precisely the {\it warped}, {\it i.e.} $g_s$ dependent, metric from \cite{desitter2, desitter3}. The two equations, 
\eqref{scixor} and \eqref{einstobeta}, and especially \eqref{einstobeta}, tell us that the expectation value of the Einstein tensor over the generalized Glauber-Sudarshan states has a part that is exactly the Einstein tensor {\it computed using the metric \eqref{vegamey3}}! 

The above conclusion is important so let us summarize what we have so far. Given a solitonic background
\eqref{betbab3} and the corresponding G-flux components to support it, we can construct generalized Glauber-Sudarshan states over it. Expectation values of the metric and the G-flux components on these states give us the {\it time-dependent} background \eqref{vegamey3}, and the corresponding {\it time-dependent} G-flux components. Not only that, it now appears that the expectation values of the metric and the flux EOMs have parts that are precisely the EOMs for the metric \eqref{vegamey3} and the corresponding G-flux components. In other words, we can quantify the above statements by first noting:
\bg\label{cassie}
&& \left\langle {\delta {\bf S}_{\rm tot}\over \delta \{g^{\rm MN}\}}\right\rangle_{\sigma} =
{\delta {\bf S}^{(\sigma)}_{\rm tot}\over \delta \langle g^{\rm MN}\rangle_\sigma} + 
\sum_{\sigma' \ne \sigma} \left\langle {\delta {\bf S}_{\rm tot}\over \delta \{g^{\rm MN}\}}\right\rangle_{(\sigma \vert \sigma')} \nonumber\\
&&\left\langle {\delta {\bf S}_{\rm tot}\over \delta \{{\rm C}^{\rm MNP}\}}\right\rangle_{\sigma}=
{\delta {\bf S}^{(\sigma)}_{\rm tot}\over \delta \langle{\rm C}^{\rm MNP}\rangle_\sigma} + 
\sum_{\sigma'\ne \sigma}
\left\langle {\delta {\bf S}_{\rm tot}\over \delta \{{\rm C}^{\rm MNP}\}}\right\rangle_{(\sigma\vert \sigma')}, \nd
where ${\bf S}^{(\sigma)}_{\rm tot} \equiv {\bf S}_{\rm tot}\left(\langle g_{\rm PQ} \rangle_\sigma, 
\langle {\rm C}_{\rm PQR} \rangle_\sigma\right)$, which is basically the RHS of \eqref{sheela} defined in terms of the warped metric and flux components, ${\bf g}_{\rm MN}, {\bf C}_{\rm MNP}$ and ${\bf G}_{\rm MNPQ}$. 
The perturbative quantum terms will then be \eqref{selahran}, and the non-perturbative terms will be as elaborated in section \ref{sec3.2}.  The other quantities are defined as follows: $\sigma \equiv (\alpha, \beta)$ and $(\sigma\vert \sigma')$ denote the intermediate generalized Glauber-Sudarshan states $\{\sigma'\}$. As such the sum in \eqref{cassie} is over those states with the condition that they do not equal $\{\sigma\}$, at least not all of them. Combining \eqref{cassie} with \eqref{SDEss}, then leads to the following set of equations:
\bg\label{casguv1}
&& {\delta {\bf S}^{(\sigma)}_{\rm tot}\over \delta \langle g^{\rm MN}\rangle_\sigma} = 
{\delta {\bf S}^{(\sigma)}_{\rm tot}\over \delta \langle{\rm C}^{\rm MNP}\rangle_\sigma} = 0, \nd
which are exactly the EOMs that we encountered, and solved, in \cite{desitter2} and \cite{desitter3}! Appearance of these EOMs, while a bit surprising, should have been anticipated because there is always going to be a sector that produces the metric \eqref{vegamey3} $-$ and the corresponding G-flux components to support it $-$ as a solution to {\it some} EOMs. The reason is simple: the background \eqref{vegamey3} appears from the most probable value in the generalized Glauber-Sudarshan state. Such a system should be supported by minimizing an action if it has to survive in eleven (or ten in IIB)-dimensional space-time. The only action that we have here is \eqref{sheela}, so it is not much of a surprise that we get \eqref{casguv1}. However what is surprising that the Schwinger-Dyson's equations lead to two other set of equations:    
\bg\label{casguv2}
&&  \sum_{\sigma' \ne \sigma} \left\langle {\delta {\bf S}_{\rm tot}\over \delta \{g^{\rm MN}\}}\right\rangle_{(\sigma \vert \sigma')} = \left\langle {\delta {\bf S}_{\rm ghost}\over \delta \{g^{\rm MN}\}}\right\rangle_{\sigma} - \left\langle {\delta \over \delta \{g^{\rm MN}\}}~{\rm log}\Big(\mathbb{D}^\dagger(\sigma)
\mathbb{D}(\sigma)\Big) \right\rangle_{\sigma}\\
&& \sum_{\sigma'\ne \sigma}
\left\langle {\delta {\bf S}_{\rm tot}\over \delta \{{\rm C}^{\rm MNP}\}}\right\rangle_{(\sigma\vert \sigma')}= 
\left\langle {\delta {\bf S}_{\rm ghost}\over \delta \{{\rm C}^{\rm MNP}\}}\right\rangle_{\sigma} - 
\left\langle {\delta \over \delta \{{\rm C}^{\rm MNP}\}}~{\rm log}\Big(\mathbb{D}^\dagger(\sigma)
\mathbb{D}(\sigma)\Big) \right\rangle_{\sigma},  \nonumber \nd
that cleanly isolates the ghost action from \eqref{casguv1}  so that it appears only in \eqref{casguv2}. The LHS of \eqref{casguv2} requires us to take expectation values over states that generically do not  lead to de Sitter spaces (as we saw in some simple computations above). More so, the LHS could even be complex. The RHS has ghost action that are expressed using fermionic variable, whose variations with respect to the metric and the flux components could lead to complex quantities. The remaining terms should be balanced by the variations of the  ${\rm log}\Big(\mathbb{D}^\dagger(\sigma)
\mathbb{D}(\sigma)\Big)$ part (note that $\mathbb{D}(\sigma)$ is not unitary). Despite encouraging signs of consistency, the two set of equations in \eqref{casguv2} are in fact very hard to verify because of our ignorance of the complete behavior of either the ghost action  or the displacement operator 
$\mathbb{D}(\sigma)$ from \eqref{maristone}. Luckily however, the consistency of our analysis do not rely much on the solutions of \eqref{casguv2}, as long as \eqref{casguv1} has solutions. We will therefore leave 
the analysis of \eqref{casguv2} for future work and concentrate on the solutions of \eqref{casguv1}.

Finding the solutions to \eqref{casguv1} is made easier because of our earlier works \cite{desitter2} and 
\cite{desitter3}, where we studied the EOMs in great details. Since the readers could get most of the analysis from these two papers, we don't want to repeat them here. Instead we would like to emphasize the role played by the non-perturbative terms that shape the solutions of \eqref{casguv1} using the computations of section \ref{sec3.2}. 
To see this we will start with one of the space-time EOM:

{\footnotesize
\bg\label{khaner2}
{\delta {\bf S}^{(\sigma)}_{\rm tot}\over \delta \langle g^{\rm 00}\rangle_\sigma} = 
\langle {\rm R}_{00}\rangle_{(\sigma|\sigma)} - {1\over 2} \langle g_{00} \rangle_\sigma \langle {\rm R}\rangle_{(\sigma|\sigma)} - \langle {\rm T}^{({\rm f})}_{00}\rangle_{(\sigma|\sigma)} 
- \langle {\rm T}^{({\rm b})}_{00}\rangle_{(\sigma|\sigma)} - \langle {\rm T}^{({\rm p})}_{00}\rangle_{(\sigma|\sigma)} - \langle {\rm T}^{({\rm np})}_{00}\rangle_{(\sigma|\sigma)} = 0, \nonumber\\ \nd}
where $(\sigma|\sigma)$ denote the expectation values are taken over the generalized Glauber-Sudarshan states 
$\sigma \equiv (\alpha, \beta)$ with the condition that the {\it intermediate} states are also $\sigma$. 
$\langle {\rm T}^{({\rm q})}_{00}\rangle_{(\sigma|\sigma)}$ is the expectation value of the energy-momentum tensors for ${\rm q} = ({\rm f, b, p, np})$, {\it i.e.} fluxes, branes, perturbative and non-perturbative quantum corrections respectively. The first two terms of \eqref{khaner2} then becomes:

{\footnotesize
\bg\label{khaner22}  
\langle {\rm R}_{00}\rangle_{(\sigma|\sigma)} - {1\over 2} \langle g_{00} \rangle_\sigma \langle {\rm R}\rangle_{(\sigma|\sigma)} = {\bf R}_{00} - {1\over 2} {\bf g}_{00} {\bf R} = {1\over g_s^2}\left(d_0(y) + 
\sum_{n \ge 0} f_n(y) ~g_s^{2n/3}\right)\eta_{00}, \nd}
where the middle bold-faced terms are the Ricci curvature, metric and the Ricci scalar computed for the 
background \eqref{vegamey3} and are therefore $g_s$ dependent\footnote{This is a bit subtle. From \eqref{mcapopo} we know that $\langle g_{00} \rangle_\sigma$ is not just ${\bf g}_{00}$ from \eqref{vegamey3}, but has ${\cal O}\left({g_s^a \over {\rm M}_p^b}\right)$ corrections (see footnote \ref{joelle} and \ref{ladkill}). However since the space-time metric, that goes as $g_s^{-8/3}$, is 
dominant over any perturbative corrections, we can safely ignore it here. We will come back to this subtle point soon.}. 
On the extreme right, we show the $g_s$ dependence of these terms with $(d_0(y), f_n(y))$ being some functions of the coordinates of the base manifold ${\cal M}_4 \times {\cal M}_2$ of \eqref{anonymous} and may be extracted from section 4.1.4 (case 1, equations (4.69) and (4.70)) of \cite{desitter2}. For us, we will only worry about $(d_0, f_0)$, and they take the following values:
\bg\label{khanerdui}
d_0 \equiv 3 \Lambda + {{\rm R}\over 2 H^4}, ~~~~~ f_0 \equiv -{4\over H^6} \left[\left(\partial H\right)^2 - 
{\square H^4\over 8 H^2}\right], \nd  
where $H(y) = h^{1/4}(y)$ is the warp-factor in \eqref{vegamey3}; $\Lambda$ is the cosmological constant; 
$(\partial H)^2 = \partial^{\rm M} H \partial_{\rm M} H$ with ${\rm M} \in {\cal M}_4 \times {\cal M}_2$;
$\square$ is the Laplacian over the internal manifold;
and ${\rm R}$ is the Ricci scalar computed using the internal metric components $(g_{mn}(y), g_{\alpha\beta}(y), g_{ab}(y))$ {\it without} any $g_s$ or $H(y)$ dependences. Both the terms in \eqref{khanerdui} have inverse $g_s^2$ dependences.  
Interestingly, as already pointed out in \cite{desitter2}, the {\it inverse} 
$g_s$ dependence is very important. It is easy to show that both the energy-momentum tensors for the 
G-flux components and for the M2-branes, the inverse $g_s$ terms appear naturally (see for example equations  (4.71) and (4.74) of \cite{desitter2}), so the question is to find them for the perturbative and the non-perturbative quantum terms. This is where the hard work of sections \ref{sec3.1} and \ref{sec3.2} pays off. 

The perturbative corrections to the energy-momentum is easier to handle so we will address it first. In some sense we already have the answer in \eqref{novharr}. Here we want to see the lowest order, {\it i.e.} $n = 0$ case in \eqref{khaner22}. This would be the choice $q = 2$ in \eqref{novharr}, which means:
\bg\label{subtagra}
\langle {\rm T}^{\rm p}_{00}\rangle_{(\sigma|\sigma)} \equiv \mathbb{C}^{(2, 2p)}_{00}, \nd
where $p$ determines the {\it moding} of the G-flux components as given in \eqref{prcard} (see \cite{desitter2} for more details on this). It turns out, $p \ge {3\over 2}$, so the choice of $q = 2$ in \eqref{novharr} 
only provides a countable number of terms because of the following constraint:
\bg\label{subkotha}
\mathbb{N}_1 + \mathbb{N}_2 + (p + 2)\mathbb{N}_3 + (2p + 1) \mathbb{N}_4 + (p - 1) \mathbb{N}_5 = 2,
\nd
where $\mathbb{N}_i \in \mathbb{Z}$ are defined as in \eqref{cibelmoja}. In our case it appears that the contributions from \eqref{subtagra} can only {\it renormalize} the contributions from the G-fluxes. This was already noticed in \cite{desitter2}, so it is not a new observation. It is also easy to see that for $p = 0$, {\it i.e.} for time-independent G-flux  components, there are an {\it infinite} number of solutions to \eqref{subkotha} with 
no $g_s$ hierarchy. Unfortunately such terms also lack ${\rm M}_p$ hierarchy as was shown in 
\cite{desitter3} with the choice of $\Omega_{ab}$ as in \eqref{duitagra3}, implying that there may not be an effective field theory description with time-{\it independent} G-flux components for the background 
\eqref{vegamey3}. 

The non-perturbative corrections are more interesting, for they involve non-trivial contributions from the quantum series \eqref{selahran}. As we studied in sections \ref{sec3.2.1} and \ref{sec3.2.3}, the non-perturbative corrections typically appear from the BBS and KKLT type instantons, that have their roots in the non-local counter-terms that we discussed in \cite{desitter2}. There are also non-perturbative corrections from fermionic condensates discussed in section \ref{sec3.2.2}. All of these may be combined together to form:
\bg\label{karinak}
\langle {\rm T}^{({\rm np})}_{00}\rangle_{(\sigma|\sigma)} \equiv \mathbb{T}^{({\rm BBS})}_{00} + 
\mathbb{T}^{({\rm dBBS})}_{00} + \mathbb{T}^{({\rm KKLT})}_{00} + \mathbb{T}^{({\rm dKKLT})}_{00}
+ \mathbb{T}^{({\rm ferm})}_{00}, \nd
where dBBS and dKKLT stand for delocalized BBS-type and delocalized KKLT-type instanton contributions to the energy-momentum tensors, and the last term in \eqref{karinak} is the fermionic contribution. All of their contributions, except the fermionic one, have been summarized in {\bf Table \ref{instabeta4}}, so we can easily extract what we need for our case. In the following we will denote the $g_s$ scalings of each of these contributions:

{\footnotesize
\bg\label{mindik}
&& \mathbb{T}^{({\rm BBS})}_{00}= \mathbb{T}^{({\rm np}; 6)}_{00}(z, \theta)~\delta\left(\theta - {8\over 3} - {2n_2\over 3}\right), ~~~~~~
\mathbb{T}^{({\rm dBBS})}_{00}= \mathbb{T}^{({\rm np}; 5)}_{00}(z, \theta)~\delta\left(\theta - {4\over 3} - {2n_2\over 3}\right) \nonumber\\
&& \mathbb{T}^{({\rm KKLT})}_{00}= \mathbb{T}^{({\rm np}; 7)}_{00}(z, \theta)~\delta\left(\theta - {2\over 3} - {2n_2\over 3}\right), ~~~~
\mathbb{T}^{({\rm dKKLT})}_{00}= \mathbb{T}^{({\rm np}; 3)}_{00}(z, \theta)~\delta\left(\theta - {4\over 3} - {2n_2\over 3}\right), \nd}
where the explicit expressions for $\mathbb{T}^{({\rm np}; 6)}_{00}(z, \theta), \mathbb{T}^{({\rm np}; 5)}_{00}(z, \theta), \mathbb{T}^{({\rm np}; 7)}_{00}(z, \theta)$ and $\mathbb{T}^{({\rm np}; 3)}_{00}(z, \theta)$ appear in \eqref{notiyell6}, \eqref{notiyell5}, \eqref{notiyell7}, and \eqref{notiyell3} respectively; and $\theta$ is one-third the expression in \eqref{subkotha}. All of these contribute to order $g_s^{-2}$ to the non-perturbative energy-momentum tensor \eqref{karinak}, with $n_2$ being the number of derivatives along direction ${\cal M}_2$ in \eqref{anonymous}. The number of quantum terms contributing may again be extracted from the higher powers of the G-flux and the curvature tensors
in \eqref{selahran}, with the constraint:
\bg\label{subkotha2}
\mathbb{N}_1 + \mathbb{N}_2 + (p + 2)\mathbb{N}_3 + (2p + 1) \mathbb{N}_4 + (p - 1) \mathbb{N}_5 = 2n_2 + 2l, \nd
where $l = 4$ for the BBS-type instantons, $l = 2$ for the delocalized BBS-type and the delocalized KKLT-type instantons, and $l = 1$ for the KKLT-type instantons. All of these depend on $n_2$, the number of derivatives along ${\cal M}_2$, and this could be arbitrary. However because of the exponential suppressions for higher values of $n_2$, as discussed in footnote \ref{kalingo}, the series in $n_2$ is convergent. The series in $k$, as they appear in the individual expressions of the non-perturbative energy-momentum tensors, is also convergent as discussed in footnote \ref{knauf}. Thus in the end, the non-perturbative instantons contribute finite quantum corrections to the Schwinger-Dyson equation \eqref{khaner2}. 

The fermionic contributions mostly appear from \eqref{mutachat2}. However it will be interesting to bring it in the form of an expectation value over the coherent states $\sigma = (\alpha, \beta, ..)$ where the dotted terms now include the generalized Glauber-Sudarshan states for fermions. These fermionic coherent states are easy to construct in the same vein as  the bosonic states, but we will not do so here. It will suffice to know that they exist and contribute in the same way as before. In fact existence of such a generalized Glauber-Sudarshan states will imply:
\bg\label{minkithai}
\langle \bar{\Psi} \Psi \rangle_{\sigma} = \sum_{p, p'} \bar{\bf\Psi}^{(p)} {\bf\Psi}^{(p')} 
\left({g_s\over H}\right)^{(p + p' - 4)/3}, \nd
which may be derived from \eqref{chand}, and $\bar{\Psi}^{(p)}$ has no $g_s$ dependence. The readers should also note the notational difference: $\bar{\Psi}\Psi$ on the LHS of \eqref{minkithai} is the fermionic condensate over the solitonic background 
\eqref{betbab3}, whereas $\bar{\bf\Psi} {\bf\Psi}$  and $\bar{\bf\Psi}\Omega_{{\rm MN}ab}{\bf\Psi}$ are the fermionic bilinears over the background \eqref{vegamey3}.    
In fact \eqref{minkithai} is all we need to interpret the fermionic contributions to the energy-momentum tensor in \eqref{mutachat2}. All the fermionic bilinears appear from the corresponding {\it condensates} 
over the generalized Glauber-Sudarshan states and contribute as: 
\bg\label{dorarma}
\mathbb{T}^{({\rm ferm})}_{00} = \mathbb{T}^{({\rm np}; 2b)}_{00}(z, \theta)~\delta\left(\theta_k - {4\over 3}\right), \nd
where $\theta_k$ as in \eqref{desmey} and the functional form for $\mathbb{T}^{({\rm np}; 2b)}_{00}(z, \theta)$ is given in \eqref{mutachat2}. As noted earlier, \eqref{mutachat2} or \eqref{dorarma} can now allow contributions like $\left(\bar{\bf\Psi}{\bf\Psi}\right)^q$, for $q \ge 0$ to the energy-momentum tensor. Such contributions make
the constraint a bit different from \eqref{subkotha} and 
\eqref{subkotha2}, in the following way:
\bg\label{subkotha3}
n_1 + n_2 + \mathbb{N}_1 + (p - 1) \mathbb{N}_5 + 2q(p - 2) = 4, \nd
with $\mathbb{N}_i$ as in \eqref{cibelmoja}, and $(n_1, n_2)$ are the derivatives along ${\cal M}_4$ and 
${\cal M}_2$ respectively. The {\it difference} alluded to above is not just the difference in form of \eqref{subkotha3}, but also the choice for $p$ which now takes $p \ge {5\over 2}$ compared to $p \ge {3\over 2}$ earlier. Interestingly, as noted in section \ref{sec3.2.2}, when $q = 0$, $p$ is bounded below by $p \ge {3\over 2}$. 

On the other hand, contributions directly from F-theory seven-branes appear from \eqref{mutachat} only for the right embeddings, otherwise they only contribute perturbatively. Non-trivial embeddings will require the seven-branes to have some orientations along the $(a, b)$ directions. If this is the case, one has to make sure that such embeddings are {\it stable}, and do not revert back to the standard embeddings. For our case we will avoid complicating the analysis, and only take standard embeddings of the seven-branes. As such \eqref{mutachat} do not contribute non-perturbatively. This would imply what we have so far should be enough to analyze the Schwinger-Dyson equation \eqref{khaner2}, except that we will need one more definition before we write down the space-time EOM. This is the G-flux expectation value:
\bg\label{toulaH}
\langle {\rm G}_{\rm MNPQ}\rangle_\sigma &=& {\int\left[{\cal D}{\rm G}_{\rm MNPQ}\right] ~e^{i{\bf S}_{\rm tot}}
~\mathbb{D}^\dagger(\sigma)~{\rm G}_{\rm MNPQ}(x, y, z) ~\mathbb{D}(\sigma)\over 
\int\left[{\cal D}{\rm G}_{\rm MNPQ}\right] ~e^{i{\bf S}_{\rm tot}}
~\mathbb{D}^\dagger(\sigma)~\mathbb{D}(\sigma)} \\
& = & {\bf G}_{\rm MNPQ} \equiv \sum_{p, q} ~{\cal G}^{(p, q)}_{\rm MNPQ}(y) \left({g_s\over H}\right)^{2p/3} {\rm exp}\left(-{qH^{1/3} \over g_s^{1/3}}\right) + {\cal O}\left({g_s^c\over {\rm M}_p^b}\right), \nonumber \nd
where $p \ge {3\over 2}$ and $q \ge 0$ with $({\rm M, N}) \in {\cal M}_4 \times {\cal M}_2 \times {\mathbb{T}^2\over {\cal G}}$ in \eqref{anonymous}. The result is straight-forward but some care is needed to interpret all the sides of \eqref{toulaH}. The LHS is the expectation value of the {\it operator} ${\rm G}_{\rm MNPQ}$ over the generalized Glauber-Sudarshan states 
$\sigma \equiv (\alpha, \beta)$. The path integral formula on the RHS has only fields, so ${\rm G}_{\rm MNPQ}$ is a field integrated in the standard way with one key difference: the {\it modes} are to be selected from $\Upsilon_{\bf k}$ now, or more appropriately the $\gamma_{\bf k}(x, y, z)$ modes from \eqref{prcard}, instead of the ones from \eqref{hudson}. The part of $\mathbb{D}(\sigma)$ that is relevant for the computation in 
\eqref{toulaH} is $\mathbb{D}(\beta)$, and ${\bf S}_{\rm tot}$ is the total action from \eqref{cixmel} but one has to use the variables for the solitonic background \eqref{betbab3}. The second line is very close to the expected G-flux components\footnote{Following the notations of \cite{desitter2}, and the fact that for $q > 0$
in \eqref{toulaH} the exponential term vanishes when $g_s \to 0$, we will only consider the components 
${\cal G}^{(p, 0)}_{\rm MNPQ} \equiv {\cal G}^{(p)}_{\rm MNPQ}$ in all our analysis, unless mentioned otherwise.}
that we want for supporting a background like \eqref{vegamey3}, with the additional $g_s$ corrections that are sub-dominant for $g_s < 1$. 
Note that the expectation value leads to time-dependent G-flux components, reinforcing our earlier conclusion that the time-dependences of the expectation values and the existence of the Glauber-Sudarshan states  go hand in hand.

The energy-momentum tensors from the fluxes then follow similar path laid out earlier once \eqref{toulaH} is established. For us the concern is $\langle\mathbb{T}^{({\rm f})}_{00}\rangle_{(\sigma|\sigma)}$. Using the input \eqref{toulaH}, the expectation value of the energy-momentum tensor takes the following form:
\bg\label{cabrera}
\langle\mathbb{T}^{({\rm f})}_{00}\rangle_{(\sigma|\sigma)} = \sum_n \left(h_n^{(1)}(y) 
+ {h_n^{(2)}(y)\over g_s^2}  + {h_n^{(3)}(y)\over g_s^4}\right) \eta_{00} ~g_s^{2n/3}, \nd
where $n \ge 0$ and the functional form for $h_n^{(q)}$ may be derived from \cite{desitter2}.   Since we are looking for $g_s^{-2}$ scalings, we only require the functional forms for $h_0^{(2)}(y)$ and $h_3^{(3)}(y)$, or more appropriately, the functional form for the sum $h_0^{(2)}(y) + h_3^{(3)}(y)$. This is easy to work out, and the result is:

{\footnotesize
\bg\label{relcha}
h_0^{(2)}(y) + h_3^{(3)}(y) = -{1 \over 16 H^8} \left({\cal G}^{(3/2)}_{mnab}{\cal G}^{(3/2)mnab} + 
2{\cal G}^{(3/2)}_{m\alpha ab}{\cal G}^{(3/2)m\alpha ab} + {\cal G}^{(3/2)}_{\alpha\beta ab}
{\cal G}^{(3/2)\alpha\beta ab}\right), \nd}
where $(m, n) \in {\cal M}_4;  (\alpha, \beta) \in {\cal M}_2$ and $(a, b) \in {\mathbb{T}^2\over {\cal G}}$.  The other G-flux component do not participate to this order in $g_s$. This is all consistent with what we had in 
\cite{desitter2}, and more so, we seem to be getting everything from expectation value over the generalized Glauber-Sudarshan states. Finally, we can also add up all the perturbative and non-perturbative quantum terms from 
\eqref{subtagra}, \eqref{mindik} and \eqref{dorarma} to get the final expression for their contributions to the energy-momentum tensors:

{\footnotesize
\bg\label{meychel}
\langle\mathbb{T}^{({\rm Q})}_{00}\rangle_\sigma \equiv  \langle\mathbb{T}^{({\rm p})}_{00}\rangle_{(\sigma|\sigma)} + \langle\mathbb{T}^{({\rm np})}_{00}\rangle_{(\sigma|\sigma)} 
 &= &   \mathbb{T}^{({\rm np}; 6)}_{00}~\delta\left(\theta - {8\over 3} - {2n_2\over 3}\right)
 +  \mathbb{T}^{({\rm np}; 7)}_{00}~\delta\left(\theta - {2\over 3} - {2n_2\over 3}\right) \\
 &+& \left(\mathbb{T}^{({\rm np}; 3)}_{00} + \mathbb{T}^{({\rm np}; 5)}_{00}\right)~ \delta\left(\theta - {4\over 3} - {2n_2\over 3}\right) + \mathbb{T}^{({\rm np}; 2b)}_{00}~ \delta\left(\theta_k - {4\over 3}\right) + \mathbb{C}^{(2, 2p)}_{00}, \nonumber \nd}
where $\theta$ is one-third the expression in \eqref{subkotha}; and $n_2$ is the number of derivatives along ${\cal M}_2$ of \eqref{anonymous}. The most prominent contribution comes from the first term in 
\eqref{meychel}, which is from the BBS type instanton gas; and as we go to larger values of $n_2$, the contributions become increasingly smaller. At each level of $n_2$, there are finite (and hence countable) number of terms, so the system is very well defined.  Therefore plugging \eqref{meychel}, \eqref{relcha}, 
\eqref{khanerdui} and \eqref{khaner22}, in the Schwinger-Dyson equation \eqref{khaner2}, we get:

{\footnotesize
\bg\label{ragipstar}
{\delta {\bf S}^{(\sigma)}_{\rm tot}\over \delta \langle g^{\rm 00}\rangle_\sigma} = 6\Lambda + {{\rm R}\over H^4} -{\square H^4\over H^8} + \eta^{00} \langle\mathbb{T}^{({\rm Q})}_{00}\rangle_\sigma + 
2\left(h_0^{(2)} + h_3^{(3)}\right) -{2{\rm T}_2(n_4 + \bar{n}_4) \over H^8 \sqrt{g_6}}~\delta^8(y - y_o)= 0, 
\nd}
where ${\rm R}$ and $g_6$ are respectively the Ricci scalar and the metric-determinant of the internal six-manifold ${\cal M}_4 \times {\cal M}_2$ from 
\eqref{anonymous} without the warp-factor $H$ or the $g_s$ factors, and $\Lambda$ is the four-dimensional cosmological constant in the IIB side. The last term appears from M2 and 
$\overline{\rm M2}$-branes with the coefficients defined as in \eqref{sheela}. The equation \eqref{ragipstar} is an {\it exact} equation in the sense that almost all possible contributions have been taken into account.

\noindent In the following we will try to quantify briefly the above statement as most of this was already demonstrated rigorously in \cite{desitter2} (see for example the discussions in section 4.3.1 of \cite{desitter2}). However one puzzle appears now that has to do with the form of the expectation values in say \eqref{mcapopo}, \eqref{toulaH} and in the following:

{\footnotesize
\bg\label{clarafoy}
\langle {g}_{\alpha\beta}\rangle_\sigma &=& F_1(t)~ g_{\alpha\beta}(y)~H^2(y)~ g_s^{-2/3} + .. = 
\sum_{k\ge 0} D_k \left({g_s\over H}\right)^{2(k - 1)/3} H^{4/3}(y)~ g_{\alpha\beta}(y)
+ {\cal O}\left({g_s^c\over {\rm M}_p^b}\right)\nonumber\\
\langle {g}_{mn}\rangle_\sigma &= & F_2(t)~ g_{mn}(y)~H^{2}(y) ~g_s^{-2/3}+ .. = 
\sum_{k\ge 0} C_k \left({g_s\over H}\right)^{2(k - 1)/3} H^{4/3}(y) ~g_{mn}(y)+ {\cal O}\left({g_s^c\over {\rm M}_p^b}\right), \nd}
where the corrections are sub-leading in the limit $g_s < 1$. The analysis follows exactly the same procedure we applied for $\langle {\bf g}_{\mu\nu} \rangle_\sigma$ in \eqref{camber} and 
\eqref{mcapopo} (the difference $\sigma$ from $\alpha$ is irrelevant as we are in the gravitational sector). 
The aforementioned puzzle here is that the expectation values themselves have 
${\cal O}\left({g_s^a \over {\rm M}_p^b}\right)$ corrections taking us away, albeit in the sub-leading sense, from an exact de Sitter background.  How are these corrections accommodated in the Schwinger-Dyson equations? Additionally, how is the four-dimensional Newton's constant in the IIB side kept time-independent? 

This is subtle so we need to tread carefully.  The ${\cal O}\left({g_s^a \over {\rm M}_p^b}\right)$ corrections that appear to the expectation values, for example to \eqref{mcapopo}, \eqref{toulaH} and \eqref{clarafoy}, come from the wave-function of the interacting vacuum in \eqref{siortegas}. For the present purpose, we can generalize it to 
$\Psi_\Omega^{(\sigma)}\left(g_{\rm MN}, {\rm C}_{\rm MNP}, t\right)$. The wave-function has three parts: (a) the Glauber-Sudarshan wave-function for the harmonic-vacuum, (b) the correction coming from $\delta\mathbb{D}$ and (c) the {\it integrated} effect from the full interacting Hamiltonian ${\bf H}_{\rm int}$. The integration is from $-T = -\infty$ (in a slightly imaginary direction) till the present epoch $t$ (or $\sqrt{\Lambda} t$). On the other hand, the Schwinger-Dyson equation, in say \eqref{ragipstar}, makes sense only for $-{1\over \sqrt{\Lambda}} < t < 0$. What happens in these two regimes?

One answer could be that the integrated effect on the wave-function \eqref{siortegas} {\it cancels} out completely so that there are no ${\cal O}\left({g_s^a \over {\rm M}_p^b}\right)$ corrections to either 
\eqref{vegamey3} or to \eqref{toulaH}. Such a conclusion would be consistent with the result we get from the Schwinger-Dyson equation, and in turn will confirm the outcome (1) given towards the end of section \ref{sec2.4}.

Unfortunately such a conclusion is very hard to prove because we have no control on the dynamics for 
$t < -{1\over \sqrt{\Lambda}}$. All we can say here is that the integrated effect of ${\bf H}_{\rm int}$ on the wave-function \eqref{siortegas} appears to cancel out in the regime $-{1\over \sqrt{\Lambda}} < t < 0$. This will lead to the outcome (2) in section \ref{sec2.4}, although there is a possibility is that maybe \eqref{casguv1} is not completely correct. Could it be possible that the Schwinger-Dyson equations \eqref{SDEss}
actually decompose to the following:

{\footnotesize
\bg\label{casguvo}
&&   {\delta {\bf S}^{(\sigma)}_{\rm tot}\over \delta \langle g^{\rm MN}\rangle_\sigma} = 
 \left\langle {\delta {\bf S}_{\rm ghost}\over \delta \{g^{\rm MN}\}}\right\rangle_{(\sigma)} - 
\left\langle {\delta \over \delta \{g^{\rm MN}\}}~{\rm log}\Big(\mathbb{D}^\dagger(\sigma)
\mathbb{D}(\sigma)\Big) \right\rangle_{\sigma}-  \sum_{\sigma' \ne \sigma} \left\langle {\delta {\bf S}_{\rm tot}\over \delta \{g^{\rm MN}\}}\right\rangle_{(\sigma \vert \sigma')}\\
&&{\delta {\bf S}^{(\sigma)}_{\rm tot}\over \delta \langle{\rm C}^{\rm MNP}\rangle_\sigma}= 
\left\langle {\delta {\bf S}_{\rm ghost}\over \delta \{{\rm C}^{\rm MNP}\}}\right\rangle_{\sigma} - 
\left\langle {\delta \over \delta \{{\rm C}^{\rm MNP}\}}~{\rm log}\Big(\mathbb{D}^\dagger(\sigma)
\mathbb{D}(\sigma)\Big) \right\rangle_{\sigma} -  \sum_{\sigma'\ne \sigma}
\left\langle {\delta {\bf S}_{\rm tot}\over \delta \{{\rm C}^{\rm MNP}\}}\right\rangle_{(\sigma\vert \sigma')},  \nonumber\nd}
instead of \eqref{casguv1}? There is something interesting about the set of equation in \eqref{casguvo}: the extra 
${\cal O}\left({g_s^a\over {\rm M}_p^b}\right)$ corrections to the expectation values \eqref{mcapopo}, 
\eqref{toulaH} and \eqref{clarafoy} can now be compensated by the RHS of \eqref{casguvo} such that 
\eqref{ragipstar} remains unchanged {\it without} any extra ${\cal O}\left({g_s^a\over {\rm M}_p^b}\right)$ factors. All EOMs that we studied in \cite{desitter2} remain as they were without any extra factors. This is good, but there is an issue with \eqref{casguvo} when we compare with 
\eqref{michelragi}. The terms of \eqref{michelragi} starts with warped metric and G-flux components. If we look at the $g_{00}$ part of the solution, then the LHS of \eqref{casguvo} starts with terms with $g_s$ dependence as $g_s^{-2}$, whereas the corresponding terms in \eqref{michelragi} start with $g_s^{-8/3}$. Thus they cannot be matched. While this doesn't disprove the existence of an equation like \eqref{casguvo}, the {\it non-existence} of \eqref{casguv1} would be much more puzzling: even if the metric and the G-flux components receive
${\cal O}\left({g_s^a\over {\rm M}_p^b}\right)$ corrections, the {\it corrected} metric and the flux components should satisfy equations similar to \eqref{casguv1}, otherwise there is a possibility that such configurations cannot be supported in eleven-dimensional space-time. All of these then appears to indicate that \eqref{casguvo} {\it cannot} be the right EOMs, and our earlier EOMs, \eqref{casguv1} and \eqref{casguv2}, should still be the correct ones here. This is despite the fact that we never realize a configuration like \eqref{vegamey3}, and the corresponding G-flux components, by directly solving supergravity EOMs here. 

We should then look for a solution to the conundrum that not only allows SDEs like \eqref{casguv1}, but also shows that there may not be any extra ${\cal O}\left({g_s^a\over {\rm M}_p^b}\right)$ corrections to the expectation values themselves. To find this, let us go back to 
the functional form for $a_{\rm eff}$ and 
$a^\dagger_{\rm eff}$ from \eqref{adler}. The effective annihilation operator $a_{\rm eff}$ 
was defined via the action $a_{\rm eff} \vert\Omega \rangle = 0$, {\it i.e.} $a_{\rm eff}$ annihilates the interacting vacuum $\vert\Omega\rangle$. However as discussed in footnote \ref{bribeac}, this is not enough to fix the form for $a_{\rm eff}$ unambiguously implying, in turn, that the form for the displacement operator 
$\mathbb{D}(\sigma)$, where we use $\sigma = (\alpha, \beta, ..)$ instead of $\alpha$, described via 
\eqref{15movs} or \eqref{maristone}, cannot be fixed unambiguously either. More importantly, it is 
$\delta\mathbb{D}(\sigma)$, defined in \eqref{shelort} that suffers ambiguity because $\mathbb{D}_0$, which is the displacement operator for the harmonic vacuum, or $\hat{\mathbb{D}}_0(\sigma)$, which is the non-unitary version of $\mathbb{D}_0(\sigma)$, are unambiguously fixed. The functional form for 
$a_{\rm eff}$ is defined by $c_{lmn}$ parameters, and if each $(l, m, n)$ go from 1 to $N$, where $N$ is arbitrarily large, then there are at least $N^3$ number of parameters here. We can make this more precise by assigning the following functional form for $a_{\rm eff}$:

{\footnotesize
\bg\label{adlerr}
&&a_{\rm eff}({\bf k}) = a_{\bf k} + \sum_{\{i_p\}} \int c_{i_1....i_n}({\bf k}; {\bf k}_1, .., {\bf k}_n; t) \prod_{i = 1}^n a_{{\bf k}_i}^{(i_i)} ~
\delta^{10}\left(\sum_{l = 1}^n {\bf k}_l - {\bf k}\right) d^{10}{\bf k}_i + {\cal O}\left[{\rm exp}\left(c_{i_1....i_n}\right)\right]\\
&&~~~ + \sum_{\{j_p\}} \int d_{j_1....j_m}({\bf k}; {\bf k}_1, .., {\bf k}_m; t) \prod_{j = 1}^m a_{{\bf k}_i}^{\dagger(j_i)} ~
\delta^{10}\left(\sum_{l = 1}^n {\bf k}_l - {\bf k}\right) d^{10}{\bf k}_j + {\cal O}\left[{\rm exp}\left(d_{j_1....j_n}\right)\right] + {\rm permutations}, \nonumber \nd}
where the {\it permutations} involve various symbolic permutations of the $a_{{\bf k}_i}^{(j_i)}$ of 
$a_{{\bf k}_l}^{\dagger(j_l)}$ both in the polynomial and the exponential forms, and the $j_i$ superscript denote the creation or the annihilation operator for a 
given component of $g_{\rm MN}$ or ${\rm C}_{\rm MNP}$. The coefficients $c_{i_1....i_n}$ and 
$d_{i_1....i_n}$ are functions of ${\bf k}_i$ and $t$ and also of the string coupling in the solitonic background, and thus should be related to $c_{lmn}$ in \eqref{adler}. The above form \eqref{adlerr} is the most generic annihilation operator one could write for an interacting theory, although one can easily see that imposing $a_{\rm eff}({\bf k}) \vert\Omega\rangle = 0$ cannot fix the forms of $c_{i_1....i_n}$ and 
$d_{i_1....i_n}$ unambiguously. Interestingly however, the {\it number} of variables appearing in \eqref{adlerr} seems to be 
similar to the number of variables that would appear in the interacting Hamiltonian ${\bf H}_{\rm int}$. This means, $\delta \mathbb{D}(\sigma)$ defined as \eqref{shelort} will also have exactly the same number of variables as in ${\bf H}_{\rm int}$, and we can, in turn, use this information to fix the form of 
$\delta\mathbb{D}(\sigma)$ as:

{\footnotesize 
\bg\label{sisortegass}
&&\int [{\cal D}g'_{\rm MN}][{\cal D}{\rm C}'_{\rm PQR}]\langle g_{\rm MN}, {\rm C}_{\rm PQR} \big\vert \delta\mathbb{D}(\sigma(t))\big\vert g'_{\rm MN}, {\rm C}'_{\rm PQR}\rangle 
\Psi_0(g', {\rm C}')
=  -  \sum_n {(-i)^n\over n!}\int \left[{\cal D}g'{\cal D}\hat{g}\right]_{\rm MN} \left[{\cal D}{\rm C}'
{\cal D}\hat{\rm C}\right]_{\rm PQR}\nonumber\\
&& ~~~\times \langle\mathbb{D}_0(\sigma(t))\rangle \int_{-T}^t dt_1.....dt_n 
\langle \hat{g}_{\rm MN}, \hat{\rm C}_{\rm PQR}\big\vert \mathbb{T} \left\{\prod_{i = 1}^n
\int d^{10} x_i {\bf H}_{\rm int}\left(t_i, {\bf x}_i, y_i, z_i\right)\right\}\big\vert g'_{\rm MN}, {\rm C}'_{\rm PQR}\rangle
\Psi_0\left(g', {\rm C}'\right), \nonumber\\ \nd}
with $T \to \infty$ in a slightly imaginary direction and we have defined $\langle\mathbb{D}_0(\sigma(t))\rangle$ as $\langle\mathbb{D}_0(\sigma(t))\rangle = 
\langle g_{\rm MN}, {\rm C}_{\rm PQR}\vert \mathbb{D}_0(\sigma(t))\vert \hat{g}_{\rm MN}, 
\hat{\rm C}_{\rm PQR}\rangle$ and $\Psi_0(g', {\rm C}') \equiv \Psi_0({\rm g}'_{\rm MN}, {\rm C}'_{\rm PQR})$ is the vacuum state wave-function. 
In the second line we have used $\mathbb{D}_0(\sigma)$ instead of $\mathbb{D}(\sigma)$
because both $\delta\mathbb{D}(\sigma)$ and ${\bf H}_{\rm int}$ are already proportional to powers of the string coupling, so $\delta\mathbb{D}(\sigma){\bf H}_{\rm int}$ would be highly sub-leading. The equation 
\eqref{sisortegass} can be exactly solved despite the complicated nature of it, and we get\footnote{There is of course a constant of proportionality accompanying \eqref{thaipa} which is the overlap integral 
$\langle \Omega \vert 0 \rangle$ that we ignore here. There has to be a {\it non-zero} overlap, but other than that this is essentially a constant.}:
\bg\label{thaipa}
\boxed{\mathbb{D}(\sigma, t) = \mathbb{D}_0(\sigma, t) ~{\rm exp} \left(i \int_{-T}^t d^{11} x~{\bf H}_{\rm int}\right)} \nd
where ${\bf H}_{\rm int}$ is the full interacting Hamiltonian that we discussed in sections \ref{sec3.1} and 
\ref{sec3.2}, and $T \to \infty(1- i\epsilon)$. The above identification easily justifies  
$\delta \mathbb{D}(\sigma)$ to be proportional to powers of ${\bf H}_{\rm int}$ with no zeroth order terms. The degrees of freedom also match, and \eqref{thaipa} satisfies \eqref{sisortegass} to all orders in string coupling. The wave-function of the shifted interacting vacuum now satisfies:
\bg\label{duibonort}
\Psi^{(\overline{\alpha})}_{\Omega}\left(g_{\mu\nu}, t\right)  =  {\rm exp}\left[\int_{-\infty}^{+\infty} d^{10}{\bf k} ~{\rm log}
\left(\Psi^{(\overline{\alpha})}_{\bf k}\left(\widetilde{g}_{\mu\nu}({\bf k}), t\right)\right)\right], \nd
which is exactly the Glauber-Sudarshan wave-function with one key difference: the 
$\sigma \equiv (\overline{\alpha}, \overline{\beta})$ appearing in \eqref{thaipa} is not exactly $(\alpha, \beta)$ from \eqref{cannon2}, \eqref{cnelson}, \eqref{prcard} or \eqref{kwinter} {\it yet}. This is also reflected in 
\eqref{duibonort} where the vacuum wave-function in the configuration space is shifted by 
$\overline{\alpha}^{(\psi)}_{\mu\nu}$ instead of $\alpha^{(\psi)}_{\mu\nu}$ from \eqref{cannon2} and \eqref{cnelson}, for example.
For computational purpose, especially in the path integrals, we can replace $\mathbb{D}_0(\sigma)$ in \eqref{thaipa} by the non-unitary part 
$\hat{\mathbb{D}}_0(\sigma, t)$, as we have done so earlier. In fact including \eqref{thaipa} in the path integral computation, say in \eqref{meyepolce}, \eqref{meyepolce2} and \eqref{meyepolce3}, one can easily show that there are {\it no} 
${\cal O}\left({g_s^a\over {\rm M}_p^b}\right)$ corrections to the results anymore. This is because, by taking for example \eqref{mcapopo}, we can rewrite it as:

{\footnotesize
\bg\label{ranimukh}
\langle {\bf g}_{\mu\nu}(x, y, z)\rangle_{\alpha} = {\eta_{\mu\nu} \over h_2^{2/3}(y, {\bf x})} + 
{\bf Re}\left[\int {d^{10} {\bf k}\over 2\omega^{(\psi)}_{\bf k}} ~
\left(\overline{\alpha}^{(\psi)}_{\mu\nu}({\bf k}, t) + {\cal O}\left({g_s^{c}\over {\rm M}_p^d}\right)\right)\psi_{\bf k}({\bf x}, y, z)\right] =  
{\eta_{\mu\nu}\over \left(\Lambda\vert t\vert^2 \sqrt{h}\right)^{4/3}}, \nonumber\\ \nd}
where we expect $\overline{\alpha}^{(\psi)}_{\mu\nu}({\bf k}, t) + {\cal O}\left({g_s^{c}\over {\rm M}_p^d}\right) 
\equiv {\alpha}^{(\psi)}_{\mu\nu}({\bf k}, t)$ {\it i.e.} \eqref{cannon2} and \eqref{cnelson}. Note that the  
$({\bf x}, y, z)$ dependence only comes from the spatial wave-function $\psi_{\bf k}({\bf x}, y, z)$, but the temporal dependences come from two sources, $\overline{\alpha}^{(\psi)}_{\mu\nu}({\bf k}, t)$ and the 
${\cal O}\left({g_s^a\over {\rm M}_p^b}\right)$ corrections. Similar identification
would extend to $\overline{\beta}$ as well as \eqref{kimbas} and 
\eqref{clarafoy} too. Therefore \eqref{vegamey3} will continue to be the exact answer that we get from the Glauber-Sudarshan states in the interval $-{1\over \sqrt{\Lambda}} < t < 0$, leading to option (2) given towards the end of section \ref{sec2.4}. The time evolution of such a state appearing from \eqref{thaipa} then may be expressed as:
\bg\label{mom}
\vert\sigma, t\rangle \equiv
\vert \overline{\alpha}, \overline{\beta}; t\rangle = {\rm exp}\left[-i\int_{-1/\sqrt{\Lambda}}^t d^{11} x\Big({\bf H}_0 + 
{\bf H}_{\rm int}\Big)\right] \mathbb{D}_0(\overline{\alpha}, \overline{\beta}, 0)\vert 0 \rangle, \nd
where ${\bf H}_0$ is the free part of the M-theory Hamiltonian. Such a temporal evolution tells us that the 
initial state $\vert \sigma, 0\rangle$ undergoes some {\it squeezing}, implying that the number of gravitons (and also the flux quanta, although we will ignore them here) changes with respect to time in the following way:
\bg\label{gendafool}
N^{(\psi)}(t) = \int_{-{\rm M}_p}^{+{\rm M}_p} d^{10}{\bf k} \left\vert \alpha^{(\psi)}_{\mu\nu}({\bf k}, \omega_{\bf k})
{\rm exp}\left(-i\omega^{(\psi)}_{\bf k}t\right) + {\cal O}\left({g_s^{c}\over {\rm M}_p^d}\right)\right\vert^2, \nd
where we used \eqref{cnelson2} to express the Fourier modes from \eqref{cannon2}. 
The modes are also integrated till 
${\rm M}_p$ as the interactions responsible for \eqref{ranimukh} and \eqref{gendafool} appear from integrating out the modes from 
$\Lambda_{\rm UV}$ till ${\rm M}_p$. One may easily see that \eqref{gendafool} is time-dependent even to zeroth order in $g_s$, although it does reduce to 
\eqref{triplets} in the limit $g_s \to 0$ ({\it i.e.} for the free case)\footnote{A puzzle could arise at this stage related to the form \eqref{gendafool}. Doesn't the integrand in \eqref{gendafool} appear from 
$\left\vert\overline{\alpha}^{(\psi)}_{\mu\nu}({\bf k}, t)\right\vert^2$, and therefore should be time-{\it independent} as in \eqref{triplets} earlier? The answer is yes, for the choice of the mode, but {\it no} for the temporal independence because of the following reason. While the temporal dependence of 
${\alpha}^{(\psi)}_{\mu\nu}({\bf k}, t)$ is indeed of the form ${\rm exp}\left(-i\omega^{(\psi)}_{\bf k}t\right)$ and is therefore eliminated in \eqref{triplets}, the temporal dependence of $\overline{\alpha}^{(\psi)}_{\mu\nu}({\bf k}, t)$ is highly nontrivial as evident from \eqref{mom}. This justifies our identification 
$\overline{\alpha}^{(\psi)}_{\mu\nu}({\bf k}, t) = {\alpha}^{(\psi)}_{\mu\nu}({\bf k}, t) + {\cal O}\left({g_s^{c}\over {\rm M}_p^d}\right)$ in \eqref{gendafool}, thus resolving the apparent puzzle. \label{mardaani}}.
This change in the number of graviton appears exclusively from the interacting part of the M-theory Hamiltonian, and is an essential feature to reproduce the exact background \eqref{vegamey3}\footnote{Interestingly, the choice of the vacuum \eqref{duibonort} resolves another issue related to the {\it fluctuations} over the coherent de Sitter space viewed as an Agarwal-Tara state \eqref{ferelisa}. The 
${\cal O}(g_s)$ factors in \eqref{kimbas} and \eqref{sabinde} no longer appear as they are absorbed in 
$\overline{\alpha}^{(\psi)}_{\mu\nu}({\bf k}, t)$ to convert it to ${\alpha}^{(\psi)}_{\mu\nu}({\bf k}, t)$. The 
$C_{m{\bf k}}^{(\psi)}$ coefficients of \eqref{ckijwala}, that appears in $f({\bf k}, k_0)$ of \eqref{sabinde}
as $C_{m{\bf k}}^{(\psi)} \equiv C^{(\psi)}_{m}(k) \equiv \delta_{m2}C^{(\psi)}(k)$, can now be easily mapped to either the Bunch-Davies or the $\alpha$-vacua. More details on this will appear elsewhere.}.

The {\it exactness} alluded to above suggests that the lowest order EOMs appearing from the SDEs should fix the form of both the de Sitter metric as well as the internal manifold. How is this possible in the light of higher order $g_s$ corrections? The answer is not too hard to see. Consider for example \eqref{khaner22}. 
The RHS of the equation is determined from $d_0(y)$ and $f_n(y)$, and the lowest order EOMs fix the forms of $d_0(y)$ and $f_0(y)$ in \eqref{khanerdui}. Imagine this fixes the Ricci scalar ${\rm R}$ and the cosmological constant $\Lambda$ (of course other lowest order EOMs should participate to achieve the goal). Once we go to the next order in $g_s$, the function $f_n(y)$ in \eqref{khaner22} develops higher order $g_s$ corrections from the $F_i(t)$ factors in \eqref{betta3} or the M-theory uplift \eqref{vegamey3}.
We therefore conclude the following:
\newline
\newline
\noindent\fbox{%
    \parbox{\textwidth}{%
The fact that the {\it full} non-K\"ahler internal metric over the space 
${\cal M}_4 \times {\cal M}_2$ appears from taking the expectation values over the generalized Glauber-Sudarshan states is a consequence of two underlying conspiracies: one, the choice of the modes 
$\Big(\eta_{\bf k}({\bf x}, y, z, t), \xi_{\bf k}({\bf x}, y, z, t)\Big)$ along directions ${\cal M}_2$ and 
${\cal M}_4$ respectively; and two, the choice of the Glauber-Sudarshan states $\vert\sigma\rangle 
\equiv \mathbb{D}(\sigma) \vert \Omega \rangle$ with $\vert\Omega\rangle$ being the full interacting vacuum in M-theory, and $\mathbb{D}(\sigma)$ satisfying \eqref{thaipa}.}}

\vskip.15in

\noindent Putting \eqref{mcapopo} and \eqref{clarafoy} together, leads to the emergence of the full metric \eqref{vegamey3} from expectation values over these states.  Finally, the coefficients 
$C_k$ and $D_k$ in \eqref{clarafoy}, may be easily derived from the following equation:
\bg\label{clarmoj}
\sum_{\{k_i\}} ~D_{k_1} C_{k_2} C_{k_3} \left({g_s\over H}\right)^{2(k_1 + k_2 + k_3)/3} = 1, \nd
by going order by order in powers of $g_s/H$ with $C_0 = D_0 = 1$ and $k \in {\mathbb{Z}\over 2}$. For example going to next order $g_s^{1/3}$, we get $D_{1/2} = -2 C_{1/2}, D_1 = 3C^2_{1/2} - 2C_1$, etc. All of these keep the four-dimensional Newton's constant time-independent in the IIB side, once we impose 
\eqref{thaipa}, although one question arises: What about {\it renormalization} or {\it running} of the four-dimensional Newton's constant? Could this happen here? 

To answer this and other related questions, we have to get
back to the discussion that we left-off before \eqref{clarafoy} and ask what happens once we go to higher orders in $g_s$. First, $f_n(y)$ contains all informations of the $F_i(t)$ factors, so higher order in $g_s$ will switch on higher order terms in $C_k$ and $D_k$ from \eqref{clarafoy}. In fact we are looking at terms that scale as $\left({g_s\over H}\right)^{(n-6)/3}$ with $n > 0$. The quantum contributions to the energy-momentum tensor to any $n$ can be written from \eqref{meychel} as:

{\footnotesize
\bg\label{meychel2}
\langle\mathbb{T}^{({\rm Q}| n)}_{00}\rangle_\sigma &\equiv&  \langle\mathbb{T}^{({\rm p}| n)}_{00}\rangle_{(\sigma|\sigma)} + \langle\mathbb{T}^{({\rm np}| n)}_{00}\rangle_{(\sigma|\sigma)} 
 =    \mathbb{T}^{({\rm np}; 6)}_{00}~\delta\left(\theta - {8+n \over 3} - {2n_2\over 3}\right)
 +  \mathbb{T}^{({\rm np}; 7)}_{00}~\delta\left(\theta - {2+n\over 3} - {2n_2\over 3}\right) \nonumber\\
 &+& \left(\mathbb{T}^{({\rm np}; 3)}_{00} + \mathbb{T}^{({\rm np}; 5)}_{00}\right)~ \delta\left(\theta - {4+n\over 3} - {2n_2\over 3}\right) + \mathbb{T}^{({\rm np}; 2b)}_{00}~ \delta\left(\theta_k - {4+n\over 3}\right)  + \mathbb{C}^{(n+2, 2p)}_{00}, \nd}
where we see that as we go to higher $n$, the quantum terms become increasingly more involved, although now there are two suppression factors: higher $n$ are suppressed by powers of $g_s$, and higher $n_2$ are suppressed by exponentially decaying factors. The story should now be clear. As we go to higher order in $g_s$, (a) higher orders in G-fluxes, {\it i.e.} $p > {3\over 2}$ in \eqref{toulaH}, (b) higher orders in $F_i(t)$ factors, {\it i.e.} $(C_k, D_k)$ for $k > 0$ in \eqref{clarafoy}, and (c) higher orders in quantum corrections, {\it i.e.} 
$n > 0$ in \eqref{meychel2}, are {\it simultaneously} switched on. The equations, up to the next two orders in $g_s$,  governing these modes are now:
\bg\label{keishala}
2\eta^{00} \mathbb{T}_{00}^{(Q| 1)} = \eta^{ii} \mathbb{T}_{ii}^{(Q| 1)}, ~~~
C_{1/2}^2 = 3\left(2\eta^{00} \mathbb{T}_{00}^{(Q| 2)} - \eta^{ii} \mathbb{T}_{ii}^{(Q| 2)}\right), \nd
where the repeated indices are summed over. The above two trace equations imply that the higher order 
quantum terms are balanced against the higher order terms from $F_i(t)$ factors, {\it keeping the lowest order background \eqref{vegamey3} intact}. The above two equations also imply delicate balancing as 
$C_{1/2}$ is a constant but the quantum terms are classified by \eqref{meychel2}. The fluxes, to this order, cancel out, so they do not contribute to the trace equations. Similar story unfolds along the $(m, n)$ directions because the quantum terms therein are of the form:

{\footnotesize
\bg\label{meychel3}
\langle\mathbb{T}^{({\rm Q}| s)}_{mn}\rangle_\sigma 
 &= &   \mathbb{T}^{({\rm np}; 6)}_{mn}~\delta\left(\theta - {8+ s \over 3} - {2n_2\over 3}\right)
 +  \mathbb{T}^{({\rm np}; 7)}_{mn}~\delta\left(\theta - {2+s\over 3} - {2n_2\over 3}\right)\\
 &+& \left(\mathbb{T}^{({\rm np}; 3)}_{mn} + \mathbb{T}^{({\rm np}; 5)}_{mn}\right)~ \delta\left(\theta - {4+s\over 3} - {2n_2\over 3}\right) + \mathbb{T}^{({\rm np}; 2b)}_{mn}~ \delta\left(\theta_k - {4+s\over 3}\right) + \mathbb{C}^{(s+2, 2p)}_{mn}, \nonumber \nd}
with $s \ge 0$; and 
where $\mathbb{T}^{({\rm np}; r)}_{mn}$ for $r = 6, 5, 7, 3$ and $2b$ are defined in \eqref{notiyell6}, \eqref{notiyell5}, \eqref{notiyell7}, \eqref{notiyell3}, and \eqref{mutachat2} respectively. The other two variables 
$(\theta_k, \theta)$ are in \eqref{desmey} and one-third the function in \eqref{subkotha} respectively.  Note, despite similar classification with respect to $(\theta, \theta_k)$, the quantum terms are in general different from \eqref{meychel2}. Similarly the contributions from the G-fluxes are also different from \eqref{cabrera} and \eqref{relcha}; and may be written as:
\bg\label{knivesout}
\langle\mathbb{T}_{mn}^{({\rm f})}\rangle_{(\sigma|\sigma)} = \sum_{s\ge 0} \left({\rm T}_{mn}^{(1|s)}(y) + 
g_s^2 ~{\rm T}_{mn}^{(2|s)}(y) + {{\rm T}_{mn}^{(3|s)}(y)\over g_s^2}\right) g_s^{2s/3}, \nd
where the functional form for ${\rm T}_{mn}^{(r|s)}(y)$ may be extracted from eq. (4.12) of \cite{desitter2}. Interestingly now, because of the fact that $p \ge {3\over 2}$ in \eqref{toulaH}, the lowest order $s = 0$ contributions only appear from ${\rm T}_{mn}^{(3|0)}(y)$ and not from ${\rm T}_{mn}^{(1|0)}(y)$. Therefore combining \eqref{meychel3} and \eqref{knivesout} together, we get the SDE 
${\delta {\bf S}^{(\sigma)}_{\rm tot} \over \delta\langle g^{mn}\rangle_\sigma} = 0$
satisfied by the unwarped ({\it i.e.} $g_s$ and $H(y)$ independent) internal metric component $g_{mn}$ along ${\cal M}_4$ in 
\eqref{anonymous} as:
\bg\label{boneout}
{\rm R}_{mn} - {1\over 2} g_{mn} {\rm R} - 6\Lambda H^4 g_{mn} = 
\langle\mathbb{T}_{mn}^{(Q|0)}\rangle_\sigma + 
{\rm T}_{mn}^{(3|3)}, \nd
where the last two terms appear from \eqref{meychel3} and \eqref{knivesout} respectively, $H(y)$ is the warp-factor; and $\Lambda$ is the cosmological constant. The metric $g_{mn}(y)$ is clearly non-K\"ahler
with the Ricci scalar satisfying the relation ${\rm R} = -\left(\mathbb{T}^{(Q|0)} + {\rm T}^{(3|3)}\right) - 
24\Lambda H^4$ where the term in the bracket is the trace of the RHS of \eqref{boneout}. This means some part of the internal curvature does get contribution from the four-dimensional cosmological constant $\Lambda$ in the IIB side. 

What happens when we go to higher orders in $g_s$? Here it would mean going to orders $g_s^{1/3}$, 
$g_s^{2/3}$ and beyond. Something interesting happens now. To the two higher orders in $g_s$, the Schwinger-Dyson equations reveal the following two equations:
\bg\label{polortega}
g_{mn} = {1\over \mathbb{A}_1(y)}\left(\langle\mathbb{T}_{mn}^{(Q|1)}\rangle_\sigma + 
{\rm T}_{mn}^{(3|{7\over 2})}\right), ~~~~
g_{mn} = {1\over \mathbb{A}_2(y)}\left(\langle\mathbb{T}_{mn}^{(Q|2)}\rangle_\sigma + 
{\rm T}_{mn}^{(3|{4})}\right), \nd
where $\mathbb{A}_1(y)$ and $\mathbb{A}_2(y)$ are two functions that may be read from eq (4.19) and 
eq (4.24) respectively of \cite{desitter2}. On the RHS of the two equations, both the quantum and the flux terms are at {\it higher} orders. The quantum terms are $\langle\mathbb{T}_{mn}^{(Q|1)}\rangle_\sigma$ 
and $\langle\mathbb{T}_{mn}^{(Q|2)}\rangle_\sigma$ from \eqref{meychel3}; and the flux terms are 
${\rm T}_{mn}^{(3|{7\over 2})}$ and ${\rm T}_{mn}^{(3|{4})}$ from \eqref{knivesout}. Whereas on the LHS are the {\it zeroth} order unwarped metric components. If we go to even higher orders in $g_s$, we get similar equations. This implies that as we go to higher order in $g_s$, higher order terms in quantum, G-flux and
$F_i(t)$ are switched on in such a way that the unwarped metric $g_{mn}$ remains intact. This is our stability criterion and it occurs in the following way.
\newline
\newline
\noindent\fbox{%
    \parbox{\textwidth}{%
The higher order G-flux components with $p > {3\over 2}$ in \eqref{toulaH} and higher order $F_i(t)$ components with 
$k > 0$ in \eqref{clarafoy}, {\it balance against} the higher order quantum terms, for example with $n \ge 1$ in \eqref{meychel2} and $s \ge 1$ in \eqref{meychel3}, to keep the lowest order Schwinger-Dyson equation, for example \eqref{ragipstar} and \eqref{boneout}, unchanged. This balancing act happens to {\it all} orders in $g_s$ and ${\rm M}_p$ such that the background \eqref{vegamey3} along-with the supporting G-flux components remain uncorrected to arbitrary orders in  ${g_s^a\over {\rm M}_p^b}$ provided the choice 
\eqref{thaipa} is considered.}}

\vskip.1in

\noindent Such a balancing criterion is interesting but question is what it implies for the running of the four-dimensional Newton's constant? The four-dimensional Newton's constant is of course time-independent for the solitonic vacuum, but for the background \eqref{betta3}, or it's M-theory uplift \eqref{vegamey3}, it depends crucially on the un-warped metric components $g_{mn}$ and $g_{\alpha\beta}$ (recall 
$F_1(t)F^2_2(t) = 1$ so it introduces no time dependence). Our discussion above shows that both the internal components of the metric {\it do not} receive ${\cal O}\left({g_s^a\over {\rm M}_p^b}\right)$ corrections. Does that mean the four-dimensional Newton's constant do not get renormalized? The answer turns out to be the opposite: there does appear to be finite renormalization of the Newton's constant. To see this let us go back to the metric equation \eqref{boneout}. On the RHS there are flux contributions from 
${\rm T}_{mn}^{(3|3)}$ and quantum contributions from $\langle\mathbb{T}_{mn}^{(Q|0)}\rangle_\sigma$.
The quantum contributions to the energy-momentum tensor are determined by $(\theta, \theta_k)$ that can be easily read up from \eqref{meychel3}. Once we know the values for $\theta_k$ and $\theta$, they will fix the {\it number} of quantum terms from \eqref{selahran} that contributes. What this does no tell us is the exact {\it coefficients} of the quantum terms contributing to the energy-momentum tensor. Since these coefficients appear from integrating out the high energy modes over the solitonic vacuum, their precise values might depend on what energy scale we are in. This means the RHS of  \eqref{boneout} could have some dependence on the energy scale, implying that the metric factor, and therefore the volume of the internal six manifold (i.e the base in \eqref{anonymous}), {\it might} have some scale dependence. From here it appears that the four-dimensional Newton's constant could in principle get renormalized accordingly (although it will be time-independent). Of course there is a possibility that the contributions of the quantum terms are such that the volume of the  six-manifold do not change, implying no renormalization of the four-dimensional Newton's constant. This would then be an interesting and surprising conclusion, although to verify either of these conclusions would require us to work out the precise coefficients of the quantum terms contributing to the energy-momentum tensor. Such a computation is clearly beyond the scope of this work, and will hopefully be dealt in near future.

\subsection{Dynamical moduli stabilization and supersymmetry breaking \label{sec3.4}}

Let us briefly discuss  how moduli stabilization could work in a set-up like ours. On the solitonic background, the metric configuration is given by \eqref{betbab3}. Let ${\rm G}^{(0)}_{\rm MNPQ}(y)$ and 
${\rm G}^{(0)}_{0ij{\rm M}}(y)$ $-$ where $({\rm M, N})$ and $(i, j)$ denote the coordinates of eight-manifold 
\eqref{anonymous} and two spatial directions respectively $-$ be the G-flux components to support the metric configuration \eqref{betbab3}. The system is governed by an interacting Hamiltonian ${\bf H}_{\rm int}$ which, as we saw earlier, has an infinite number of local and non-local, including their perturbative, non-perturbative and topological, interactions. 
Switching on such interactions would fix {\it all} the K\"ahler and the complex structure moduli of the eight manifold (similar stabilization will occur on the dual IIB side also). Once the moduli are fixed at the solitonic vacuum, we can study the fluctuations and from there construct the Glauber-Sudarshan state. The metric and the G-flux components of the de Sitter space are then:
\bg\label{WW}
&&\langle g_{\mu\nu}\rangle_\sigma = g_s^{-8/3} \eta_{\mu\nu}, ~~     
 \langle g_{\alpha\beta}\rangle_\sigma = g_s^{-2/3} H^2(y) F_1(t)g_{\alpha\beta}, ~~ 
  \langle g_{mn}\rangle_\sigma = g_s^{-2/3} H^2(y) F_2(t) g_{mn}  \nonumber\\
   &&  \langle g_{ab}\rangle_\sigma = g_s^{4/3}\delta_{ab}, ~~ \langle {\rm G}_{\rm MNPQ}\rangle_\sigma = \sum_{p \ge 3/2} {\cal G}^{(p)}_{\rm MNPQ} 
   \left({g_s\over H}\right)^{2p\over 3}, ~~  \langle {\rm G}_{0ij{\rm M}}\rangle_\sigma = 
   \sum_{p \in {\mathbb{Z}\over 2}} {\cal G}^{(p)}_{0ij{\rm M}}\left({g_s\over H}\right)^{{2\over 3}(p - 6)}, \nonumber\\ \nd     
where $({\rm M, N})$ are the coordinates of the eight-manifold, $(m, n) \in {\cal M}_4$, 
$(\alpha, \beta) \in {\cal M}_2$ and $(a, b) \in {\mathbb{T}^2\over {\cal G}}$. It is also known that 
${\cal G}^{(0)}_{0ij{\rm M}} = -  \partial_{\rm M}\left({\epsilon_{0ij}\over H^4}\right)$.
The above set of relations tell us that the Glauber-Sudarshan state would allow the internal moduli to vary accordingly with $g_s$ and $F_i(t)$ in a controlled way described above, and there would be no Dine-Seiberg \cite{dines} runaway. This is what we referred to as the {\it dynamical moduli stabilization} earlier. 

The next question is how to quantify the supersymmetry breaking in our set-up. To do this we will have to work out the SDEs for the flux sector given in \eqref{casguv1}. We will not work out all the flux equations here, as most are presented in \cite{desitter2}, but will suffice ourselves with one set of equations given by the following SDE:
\bg\label{jhuliMM}
{\delta {\bf S}^{(\sigma)}_{\rm tot} \over \delta \langle {\rm C}^{012}\rangle_\sigma} = 0. \nd
The flux EOM from \eqref{jhuliMM} is a bit more non-trivial to work out because we need to consider the contributions of the quantum terms from the topological sector also. Nevertheless, after some careful manipulations, the result may be presented in the following way:

{\footnotesize
\bg\label{evaBegone}
&& -b_1 \square H^4
+ {1\over \sqrt{g_6}}\sum_{\{k_i\}} \partial_{N}\Big(\sqrt{g_6} ~H^8 {\cal G}^{(k_3)}_{012M} g^{MN}\Big) 
C_{k_1} {D}_{k_2} ~\delta(k_1 + k_2 + k_3 - 3)\\
&& ~~= b_2 \sum_{\{k_i\}} {\cal G}^{(3/2)}_{N_1... N_4} \left(\ast_8{\cal G}^{(3/2)}\right)^{N_1...N_4}
+ {1 \over \sqrt{g_6}} [{\mathbb Y}_8]
+ {b_3\over \sqrt{g_6}}\sum_{\{k\}} \partial_{N}\Big(\sqrt{g_6}\left(\mathbb{Y}^{(k)}_4\right)^{012N}\Big)
\delta\left(\theta - {8\over 3}\right), \nonumber
\nd}
where $(b_2, b_3)$ are numerical constants and $b_1 \equiv 
\sum_{\{k_i\}} C_{k_1} {D}_{k_2}~\delta(k_1 + k_2 - 3)$ with $(C_k, D_k)$ are defined as in 
\eqref{clarafoy}. $\mathbb{Y}_8$ is the eight-form defined in \eqref{casguv1} and we take the simplified version with $a_3 = a_4 = a_5 = ...= 0$, with the index-free notation $[{\mathbb Y}_8]$ being defined as the contraction of $\mathbb{Y}_8$ with 
the un-warped ({\it i.e.} $g_s$ independent) epsilon tensor. $\mathbb{Y}^{(k)}_4$ are the quantum terms from the topological sectors 
and are classified by $\theta = {8\over 3}$, where $\theta$ is
one-third the function in \eqref{subkotha}. These topological terms may be formally presented in the same way as we did in section \ref{sec3.1}, but we won't do it here. The readers may look up our earlier work 
\cite{desitter2} for details on this. Finally, the flux-component ${\cal G}^{(k_3)}_{012M}$ may be easily read up from \eqref{WW}.   

Let us now compare \eqref{evaBegone} with SDE from the gravitational sector, namely \eqref{ragipstar}. Both these equations are written in terms of $\square H^4$ and square of the G-flux components, which appears in \eqref{ragipstar} via \eqref{relcha}. There are however few differences, which are crucial: the quadratic part of the G-flux components in \eqref{evaBegone} appear with a Hodge star, plus the quantum contributions are a bit different\footnote{In the construction of  $\mathbb{Y}^{(k)}_4$ we have only taken the infinite set of {\it perturbative} terms from \eqref{selahran}. There is of course the whole non-perturbative sector, similar to what we had in section \ref{sec3.2} and thus  
affecting the topological interactions, that we do not consider in \eqref{evaBegone}. Thus to compare the quantum terms of \eqref{evaBegone} with the ones in \eqref{ragipstar} we will have to introduce the non-perturbative corrections. This is technically challenging, but we do know that their contributions will be {\it finite}, just like what we had in \eqref{ragipstar}. More details on this will appear elsewhere.}
from \eqref{ragipstar}.  Multiplying  \eqref{ragipstar} by $b_1$ and 
subtracting  it from \eqref{evaBegone}, will remove the $\square H^4$, and we can easily see that:
\bg\label{Mgaru}
\left\vert {\cal G}^{(3/2)}_{{\rm MN}ab} - \left(\ast_8{\cal G}\right)^{(3/2)}_{{\rm MN}ab} \right\vert > 0, \nd
signalling the breaking of supersymmetry; with $({\rm M, N})$ restricted to, and the Hodge star defined using the un-warped metric of, 
${\cal M}_4 \times {\cal M}_2$ (i.e the metric components $g_{mn}$ and $g_{\alpha\beta}$). One could also express \eqref{Mgaru} as in \eqref{sosie} (or as in footnote \ref{ADM}), but both of these would eventually become \eqref{Mgaru}. 

We can quantify the supersymmetry breaking even further  by analyzing the fermionic terms on the seven-branes as studied in section \ref{sec3.2}. Our aim here would be to show how \eqref{Mgaru} actually breaks supersymmetry. For this, let us consider the fermionic action from \eqref{sheela4}. The relevant terms may be arranged together from \eqref{sheela4} to take the following suggestive form:

{\footnotesize
\bg\label{adastra}
{\bf S}'_4 = {\bf T}_7 \int  d^7 \sigma \sqrt{-{\bf g}_7}\Bigg[\left({\rm tr}_{\rm adj}~\bar{\bf \Psi} {\bf \Psi}\right)^q
{\bf G}^{{\rm MN}ab} + {\rm tr}_{\rm adj}~
\bar{\bf\Psi}\Big(e_{11} \hat{\Omega}^{{\rm MN}ab} + e_{12} \hat{\Omega}^{'{\rm MN}ab}\Big){\bf\Psi}\Bigg]
\Bigg({\bf G}_{{\rm MN}ab} - \left(\bf{\ast_8}{\bf G}\right)_{{\rm MN}ab}\Bigg),\nonumber\\ \nd}
where the {\bf bold} faced fields are extracted from the expectation values as in \eqref{WW}, and therefore $\bf{\ast_8}$ is now defined with respect to the $g_s$ dependent metric components. The other quantities appearing in 
\eqref{adastra} are defined from \eqref{mink} and \eqref{mcatrason}, and in fact the second term in the action appears from $\vert {\bf G}^{\rm tot}_{{\rm MN}ab}\vert^2$ with ${\bf G}^{\rm tot}_{{\rm MN}ab}$ as in 
\eqref{lebantagra}. The choice of the relative sign is motivated from \cite{fermions} although the analysis here is very different (we also have branes and not anti-branes here\footnote{It is interesting to note here that the mass term coming from $\vert {\bf G}_{{\rm MN}ab} - \left(\bf{\ast_8}{\bf G}\right)_{{\rm MN}ab}\vert$ could in principle be related to switching on $(0, 4)$ fluxes over the eight-manifold \eqref{anonymous}, much along the lines of \cite{fermions, beno}. The eight-manifold doesn't have to be a complex manifold as long as it has an almost complex structure. More details on this will be presented elsewhere.}). Taking $q = 1$ in \eqref{adastra} and 
dimensionally reducing the above integrand to four space-time dimensions will provide a {\it mass term} to the fermion coming from ${\bf \Psi}(y^m, g_s)$ with the mass term being proportional to \eqref{Mgaru} once we take the lowest order $g_s$ components from \eqref{WW}. On the other hand, over the solitonic background \eqref{betbab3}, the fluxes remain self-dual and we see that no mass term is generated. This mass term in the time-dependent case of course breaks supersymmetry, but here we see that it also a 
function of the coordinates of the internal six-manifold ${\cal M}_4 \times {\cal M}_2$. Thus instead of using 
the mass term for the fermions to contribute to the vacuum energy, we can interpret \eqref{adastra} as another interaction in the theory. This way the contributions to the cosmological constant would only appear from the fluxes and the quantum terms, exactly as we have advocated earlier (see for example eq. (4.192) in \cite{desitter2}).

In determining the supersymmetry breaking condition in \eqref{Mgaru} and \eqref{adastra}, we have  kept a subtlety under the rug related to the connection to the cosmological constant $\Lambda$. As discussed above, the cosmological constant 
is an {\it emergent} quantity in our model, meaning that its value is determined by the fluxes, branes and the quantum corrections, and is not a quantity that we add to the EOMs by hand. On the other hand, the supersymmetry breaking condition is also determined in terms of 
fluxes and quantum corrections as may be seen by subtracting \eqref{ragipstar} from \eqref{evaBegone}, or directly from \eqref{adastra}. Does this mean that the supersymmetry breaking scale is determined by the cosmological constant (or the Hubble parameter)? The answer is {\it no}, because the cosmological constant appears from an {\it integrated} condition as shown in \cite{nogo, desitter2} and is therefore suppressed by the unwarped volume of ${\cal M}_4 \times {\cal M}_2$ in the following way:

{\footnotesize
\bg\label{ethanresto}
\Lambda = {1\over 12 \mathbb{V}_6} [\mathbb{T}^{(Q)}]^i_i - {1\over 48 \mathbb{V}_6 H^4}
\left(2 [\mathbb{T}^{(Q)}]^a_a + [\mathbb{T}^{(Q)}]^{\rm M}_{\rm M}\right) - {5\over 384 \mathbb{V}_6 H^8}
\langle {\cal G}^{(3/2)}_{{\rm MN}ab} {\cal G}^{(3/2){\rm MN} ab}\rangle_{\rm av} -{n_4 {\rm T}_2 \over 
6\mathbb{V}_6 H^8}, \nd}
where the repeated indices are summed over with $({\rm M, N}) \in {\cal M}_4 \times {\cal M}_2$ and 
$\mathbb{V}_6$ is the un-warped volume of the six-manifold. We have taken the warp-factor $H(y) =$ 
constant for simplicity and $(n_4, {\rm T}_2)$ are the data for the integer and fractional M2-branes from 
\eqref{sheela}. We have also defined:

{\footnotesize
\bg\label{dchimika}
[\mathbb{T}^{(Q)}]^{\rm M}_{\rm M} \equiv \int d^6 y \sqrt{g_6} ~g^{\rm MN} 
\langle \mathbb{T}^{(Q)}_{\rm MN}\rangle_\sigma, ~~~
\langle {\cal G}^{(3/2)}_{{\rm MN}ab} {\cal G}^{(3/2){\rm MN} ab}\rangle_{\rm av} \equiv 
\int d^6 y \sqrt{g_6} ~{\cal G}^{(3/2)}_{{\rm MN}ab} {\cal G}^{(3/2){\rm MN} ab}, \nd}
where $g_{\rm MN}$ is the un-warped metric, and $\langle \mathbb{T}^{(Q)}_{\rm MN}\rangle_\sigma$
is the expectation value of the energy-momentum tensor over the Glauber-Sudarshan state $\vert \sigma\rangle$ similar to what we showed in \eqref{meychel}. This means they contain all the perturbative and the non-perturbative contributions, and are thus classified accordingly. Additionally, the relative signs in 
\eqref{ethanresto} are important, and as long as the quantum terms along the two spatial directions 
{\it dominate} over all the other negative terms, the cosmological constant will be positive. The question is how big it can be?

In a moduli stabilized scenario, the un-warped volume of the six-manifold, $\mathbb{V}_6$,
is fixed to a large value so that supergravity description may be valid. As such this implies that the cosmological constant should be very small. However at this stage one might question the fact that both the G-flux components and the quantum terms also appear as integrated over the eight-manifold in 
\eqref{ethanresto}. Shouldn't the volume factors {\it cancel} out? The answer is again {\it no}, because the G-flux components that appear above are of the form ${\cal G}^{(3/2)}_{{\rm MN} ab}$ which are highly {\it localized} fluxes and therefore the global behavior do not effect them. The quantum series are also constructed out of these flux and metric components (the latter could also be taken to be localized functions) so there indeed appears a genuine volume suppression in the expression of the four-dimensional cosmological constant $\Lambda$ in \eqref{ethanresto}. 
On the other hand, there is no such suppression factor in the supersymmetry breaking condition, therefore it appears that the supersymmetry breaking scale should be much larger than the cosmological constant (or the Hubble scale). In addition to that, there are other differences, namely in the exact arrangement of the flux and the quantum terms in \eqref{Mgaru} and \eqref{ethanresto}, confirming that the two quantities cannot be similar.  

Finally, let us ask what happens when we go to the strong coupling limit of type IIB. Recall that our analysis is done in the type IIB side at the constant coupling limit of F-theory \cite{shalsen} where we allow constant dilaton and vanishing axion fields. We can now S-dualize the IIB background which will simply change the type IIB metric by a constant factor (if $\varphi_b$ denotes the constant dilaton in the IIB side, then under a S-duality the metric changes by a constant multiplicative factor proportional to $e^{-\varphi_b}$, with the dilaton changing by $\varphi_b \to -\varphi_b$). This means we are dealing with exactly similar background 
as before! Lifting this to M-theory will then reproduce the metric and the flux components from the expectation values over a similar Glauber-Sudarshan state just like we had earlier, implying that the type IIB strong coupling configuration  mirrors the weak coupling scenario to a great extent. All the conclusions in the presence of time-dependent degrees of freedom $-$ and therefore the pathologies in the absence of time dependences $-$ will carry over to the strong coupling side as before.  The quantum break time, 
{\it i.e.} where the type IIA strong coupling sets in, will change to:
\bg\label{qotombo}
-{1\over e^{\varphi_b} \sqrt{\Lambda}} ~ < ~ t ~ < 0, \nd
where $t$ is now an appropriate scaled temporal coordinate that keeps the metric configuration unchanged (other spatial coordinates need to be scaled in a similar way). Thus at both strong and weak coupling, time-dependent degrees of freedom appear to be essential to allow for a four-dimensional EFT description to be valid, although it seems like the situation at {\it unit} type IIB string coupling should remain out of the reach of our analysis. This is not quite so, because the type IIA coupling $g_s^2 \equiv g_b^2 {\rm H}^2 \Lambda \vert t\vert^2 << 1$ even if the type IIB coupling $g_b \ge 1$ by choosing the appropriate temporal domain as in \eqref{qotombo}. This necessitates the temporal dependence of the degrees of freedom to allow for a four-dimensional EFT description at any values of the IIB coupling, although   
there could also be potential issues once the axio-dilaton goes away from the constant coupling scenario. Such a picture might lead to a time-dependent axio-dilaton already in the type IIB side which in turn would imply a new kind of analysis in the M-theory side. Indulging in an analysis like this might reveal some interesting physics but appears to be an un-necessary complication at this stage, and therefore we will avoid discussing it further here\footnote{In addition to that, imagine that the IIB coupling takes the maximum value at $y = y_1$, with $y \in {\cal M}_4 \times {\cal M}_2$, and the warp-factor ${\rm H}(y)$ peaks at $y = y_2$, then we can still be at the weak-coupling limit in the type IIA side as long as 
$g^2_s \equiv g^2_b(y_1) {\rm H}^2(y_2) \Lambda\vert t\vert^2 << 1$. Thus, despite deviating away from the constant-coupling scenario, the analysis may still be easily tractable.}.

All the above set of computations would hopefully convince the readers that the Glauber-Sudarshan state indeed captures all the essential properties of a de Sitter space. In the following section we will discuss other properties of the Glauber-Sudarshan state that will further reinforce the fact that representing de Sitter space as a Glauber-Sudarshan is not only solid but appears to be {\it essential} to allow for a stable, non-supersymmetric state to exist in string theory.

\section{Properties of the Glauber-Sudarshan state \label{sec4}}
As mentioned in the Introduction, there has been a recent slew of papers arguing against the validity of long-lived de Sitter spacetimes from string theoretic derivations. Although the so-called de-Sitter conjecture was motivated by the difficulty of finding meta-stable de Sitter vacua in string theory \cite{String_No_dS}, soon evidence for it came when starting from the distance conjecture \cite{SDC}, and invoking the Bousso covariant entropy bound \cite{Bousso_Bound}, for a causal patch in de Sitter spacetime \cite{vafa22}. The distance conjecture, having been tested more extensively in string theory constructions \cite{SDC_test}, put the de Sitter conjecture on a much firmer footing. It has since also been shown that one can arrive at the de Sitter conjecture starting from the distance conjecture by assuming the species bound \cite{Hebecker_Wrase}. However, what was still lacking is a deeper quantum gravity argument, revealing the underlying reason why such a conjecture, claiming the absence of meta-stable de Sitter spacetimes in string theory, should be taken seriously. One such argument came in the form of the `no eternal inflation' principle \cite{No_EI} and  another from the so-called `trans-Planckian censorship conjecture' (TCC) \cite{tcc}. A related argument also came in the form of the \textit{quantum breaktime of de Sitter spacetimes} after which the interactions break down the semiclassical description of de Sitter space with a causal horizon \cite{dvaliG}. Although these arguments differ amongst themselves regarding the time of validity of a consistent meta-stable de Sitter spacetime (which we shall discuss later on), together they establish more fundamental evidence for the de Sitter conjecture, albeit at the cost of refining the original conjecture by allowing for short-lived meta-stable de Sitter spacetimes.  

In the following, we shall establish how our solution manages to escape the swampland by focussing on the TCC since it is the most concrete principle from which the de Sitter conjecture can be derived\footnote{Indeed, the de Sitter conjecture is more vague and invokes some $\mathcal{O}(1)$ numbers which can be explicitly fixed only when referring to the TCC \cite{tcc} (or some similar principle).}. More generally, we shall also point out that the time scales on which our solution can be trusted is completely compatible with the allowed lifetime of a (short-lived) metastable de Sitter spacetime, as per the swampland. We shall then focus on how radiative corrections, which typically forces one to fall into the \textit{quantum swampland} \cite{Quantum_swampland}, is also  naturally avoided by our solution. Finally, we shall turn to old arguments where it was established that the symmetries of a de Sitter spacetime should necessarily break after some time, and \textit{not} be eternal, for it to have a finite entropy and show how our solution automatically complies with this restriction.

\subsection{Trans-Planckian censorship \& navigating out of the swampland \label{sec4.1}}
Typically, the above-mentioned arguments lead to an upper bound on the lifetime of any de Sitter vacua which is far shorter than those associated with stringy constructions. One of these arguments -- the TCC \cite{tcc} -- is an elevation of the old `trans-Planckian problem' of inflationary cosmology \cite{martin} to the level of a hypothesis. To understand this problem in detail, and how our solution eventually manages to avoid it, let us recall that in free quantum field theory on Minkowski spacetime, one starts by canonically quantizing the fields described on a Fock space. Even for an expanding background, cosmological perturbations can be similarly quantized since at linear order (\textit{i.e.} only considering the quadratic Hamiltonian), each of the Fourier modes evolve independently. Thus, ignoring non-Gaussianities, one needs to quantize a set of harmonic oscillators described on a Fock space, as in the case of flat spacetime. However, the novelty lies in the fact that the mass of these oscillators are time-dependent due to the time-dependence of the dynamical background. In other words, the Fourier modes are quantized in terms plane wave modes which have a constant wavelength in comoving coordinates. However, this implies that the physical wavenumber of these modes are redshift with time, given by $p=k/a(t)$, where $a(t)$ is the scale factor of the universe and $p$ and $k$ stand for the physical and comoving wavenumber, respectively. 

The most commonly understood manifestation of the trans-Planckian problem occurs when one takes a macroscopic classical perturbation today and `evolves' it backwards in time. The physical wavelength associated with this mode gets \textit{blueshifted} due to the expansion of spacetime. If one allows for the quasi-de Sitter phase of expansion to last for a significant long amount of time, one would find that a classical perturbation mode, visible in the sky today, actually originated from a physical wavelength smaller than the (four-dimensional) Planck length $\ell_{\rm Pl}$. Of course, for this to be true, one would have to assume that the field variables on quasi-de Sitter spacetime can exist as an effective field theory on scales smaller than $\ell_{\rm Pl}$, which is manifestly problematic from our understanding of quantum gravity. Note that the pinnacle of success of inflation lies in explaining macroscopic perturbations, which source the structure formation of the universe, as originating from quantum vacuum fluctuations. From this point of view, the TCC simply turns around this crowning glory of inflation to \textit{posit} that any accelerated phase of expansion can only be valid for a finite amount of time and not be semi-infinite in the past. The upper limit on the duration of this accelerating background is set by requiring that any mode, with a physical wavelength equal to or smaller than $\ell_{\rm Pl}$, should never cross the Hubble horizon ($\H^{-1}$) so that it does not decohere and become part of the classical perturbations.

On the contrary, an immediate obstruction to such a way of thinking comes from the following -- as one traces back a classical perturbation mode, linear perturbation theory breaks down much before the physical energy, corresponding to the given mode, becomes of the order of ${\rm M}_{\rm Pl}$, the four-dimensional Planck mass\footnote{We symbolize this differently from ${\rm M}_p$ which was used to denote the eleven-dimensional Planck mass.}. In other words, even before the physical wavelength of a given perturbation mode can get to $\mathcal{O}(\ell_{\rm Pl})$, the linear perturbation would become comparable to the magnitude of the background variable, thereby breaking down perturbation theory. This conclusion makes sense if one considers the fact that the universe is extremely inhomogeneous and anisotropic on Planck scales and, therefore, one cannot use the approximation of linear perturbation theory and quantize fluctuations in terms of its Fourier modes. Indeed, as long as one considers the expanding background within an effective field theory description of gravity, it must break down on wavelengths smaller than $\ell_{\rm Pl}$ since such energies would collapse parts of space-time into black holes (or a collection of them)  \cite{Kaloper_no_hair}. Therefore, from this point of view, it might indeed seem conceivable that the argument presented above for the trans-Planckian problem should never arise since it stretches the effective field theory of linear perturbations beyond its realm of validity. 

However, one must keep in mind that the above heuristic idea is not the main theoretical argument behind the TCC and should rather be viewed as an intuitive understanding of the problem. The principle conceptual difficulty is that of non-unitarity of the Hilbert space of the perturbations \cite{martin,non_unitary}, as can be understood as follows. Recall that one needs to impose a UV cut-off even for quantum field theory on flat spacetime, for the purposes of renormalization, to get physically meaningful answers. In the case of gravity, the UV cut-off is not just a computation tool but is rather a physical one, given by ${\rm M}_{\rm Pl}$. In analogy with Minkowski spacetime, one would then expect that imposing such a cutoff would get rid of the trans-Planckian problem and give us a well-defined, decoupled effective field theory of inflation below scales of $\mathcal{O}\({\rm M}_{\rm Pl}\)$. However, this is precisely where the main difficulty associated with expanding backgrounds show up. The problem is that for such spacetimes, the UV cut-off must be fixed in physical coordinates while the Fourier modes have wavelengths which are expanding in those coordinates. Therefore, a mode whose physical wavelength is above ${\rm M}_{\rm Pl}$ to begin with, during inflation, might have its wavelength red-shifted to energies below ${\rm M}_{\rm Pl}$ and thus would be part of the low-energy effective field theory. In this way, more and more modes would redshift from the UV into the Hilbert space of system modes describing perturbations on top of an expanding background and make it time-dependent. A time-dependent Hilbert space, having to accommodate more degrees of freedom to explain physical phenomena with time-evolution, is a classic sign of non-unitarity creeping into the theory. Of course, this is a problem associated with any expanding background. What is special for (quasi-)de Sitter setups is that some of these UV modes can eventually cross the Hubble radius and thus become observable at late times. From this point of view, one can view the TCC as a requirement of keeping this non-unitarity hidden behind the Hubble horizon so that even if these trans-Planckian modes somehow get generated, they never become part of our low-energy system.

Having set up the trans-Planckian problem elaborately, let us now explain how our solution manages to evade it. First, let us give some estimates for the relevant time-scales involved in the problem. The TCC postulates that any trans-Planckian mode should never cross the Hubble radius so that it cannot decohere and become classical. This can be mathematically formulated as an upper bound on inflation, given by \cite{tcc}
\begin{eqnarray}\label{TCC1}
	N < \ln\(\dfrac{{\rm M}_{\rm Pl}}{\H_{\rm f}}\)\,, 
\end{eqnarray}
where $N$ is the number of $e$-foldings of inflation and $\H_{\rm f}$ is the value of the Hubble parameter at the end of inflation. For a meta-stable de Sitter spacetime, this translates into an upper bound for the lifetime of such a solution, given by \cite{tcc}\footnote{$\H$ denotes the Hubble parameter in this section, and should not be mixed with the warp-factor, $H(y)$, from the earlier sections.}:
\begin{eqnarray}\label{TCC2}
	T < \frac{1}{\H} \ln\(\dfrac{{\rm M}_{\rm Pl}}{\H}\)\,.
\end{eqnarray}
Thus, according to the TCC, the lifetime of any metastable de Sitter spacetime should be bounded as above for it to avoid the trans-Planckian problem. On the other hand, several arguments for assigning a finite entropy to de Sitter spacetimes leads to a bound of the form \cite{ArkaniHamed:2007ky}:
\begin{eqnarray}\label{Quantum_Break_time}
	T < \frac{1}{\H}\; S_{\rm dS} = \frac{1}{\H} \;\(\dfrac{{\rm M}_{\rm Pl}}{\H}\)^2\,.
\end{eqnarray}
A similar bound was also derived by treating de Sitter as a coherent state on top of a Minkowski vacuum in a toy model, the upper limit coming from the `quantum break-time' of the system, after which the interaction terms lead to the breaking of the semi-classical description of the system \cite{dvaliG}. There has been a fierce debate recently as to which of these two time-scales should be treated as the maximum allowable lifetime of a consistent metastable de Sitter vacuum \cite{Dvali_decoupling}. For completeness, let us point out that the crucial argument that quantum modes become classical after they cross the Hubble horizon has indeed been challenged in \cite{TCC_Horizon_crossing}, pointing out mechanisms (such as that of parametric resonance) which can `classicalize' it even within the Hubble radius. Moreover, several arguments from string theory, such as the distance conjecture or the weak gravity conjecture, also gives rise to a refined version of the TCC  \cite{Refined_TCC} with an $\mathcal{O}(1)$ number appearing on the RHS of \eqref{TCC1} and \eqref{TCC2}. In light of this, what can be unambiguously stated is that there is a time-scale beyond which any consistent description of a de Sitter spacetime should break down, albeit the upper limit on the lifetime is still under contention.

However, note that in our case the amount of time we can trust our de Sitter solution as described by a Sudarshan-Glauber state, before the system becomes strongly-coupled, is given by $|T|< 1/\H$. As shown in \eqref{spamca}, after this time, the string coupling becomes $g_s \sim 1$ and the system is strongly-coupled, bringing the validity of our solution into question. In this sense, we can be agnostic about the debate regarding the lifetime of a de Sitter vacua since the time for which we can trust our coherent state to give a de Sitter description is smaller than both the time-scales mentioned above. Note that this is quite a remarkable observation, by itself, since de Sitter \textit{vacua} which appear in string theory typically have a much longer lifetime \cite{tcc,Lifetime_dS}, such as in the KKLT \cite{kklt} and LVS scenarios \cite{LVS}\footnote{The reason for this is that we do not depend on some gravitational decay channel such as through the Coleman-de Luccia tunneling. Rather, it is the system becoming strongly-coupled that determines the time for which we can trust our solution.}. 

Although the above heuristic arguments are interesting, let us now come to the real reason why our de Sitter solution remains unscathed by the above-mentioned trans-Plackian problem and, therefore, the TCC. The crucial underlying reason is precisely the fact that it is not a \textit{vacuum} solution but rather a coherent state on top of a (warped-) Minkowski vacuum. To understand this better, let us revert to the effective field theory (EFT) approach to the trans-Planckian problem of inflation. Typically, for an EFT on a flat spacetime, one would expect that effects of trans-Planckian physics would be suppressed by factors of $\mathcal{O}\(\H^2/{\rm M}_{\rm Pl}^2\)$ \cite{Dvali_decoupling}. However, the non-unitarity associated with expanding backgrounds manifests itself in a way such that there are parts of parameter space in an EFT of inflation, in which there are trans-Planckian effects larger than this, namely violating expectations of de-coupling of inflation from Planck scale physics \cite{EFT_Inflation}. So how does decoupling work for perturbations in inflation? Indeed, it is known that if one makes the following assumptions \cite{martin}:
\begin{enumerate}
	\item The microscopic structure of space-time, on Planck scales, is Lorentz invariant, and 
	\item The perturbations (or the expansion of the field modes on the de Sitter spacetime) are in their local vacuum,
\end{enumerate}
only then does the effects of decoupling kick in and the trans-Planckian problem mentioned above goes away. In this case, one can show that the probability of producing a trans-Planckian mode in the theory is exponentially suppressed, given by $e^{-{\rm M}_{\rm Pl}^2/\H^2}$, due to a Boltzmann factor. Although the above-mentioned assumptions seem very strong for a classical de-Sitter spacetime, say, with a Bunch-Davies vacuum, it is \textit{exactly} what we have in our construction of de Sitter as a Glauber-Sudharshan state. Our solitonic vacuum is indeed supersymmetric, warped-Minkowski and satisfies both of the above criteria.

At this point, the acute reader might ask how does our de Sitter solution solve the unitarity problem mentioned above? As was clear from the discussion above, the trans-Planckian difficulties arise as an effect of having time-dependent frequencies associated with the perturbation modes. However, as was manifestly shown earlier  in \eqref{sabinde}, the time-dependencies of the frequencies of perturbations in our case are actually \textit{artifacts} of Fourier transforms over a de Sitter state, viewed as a Glauber-Sudarshan state. Another way of seeing that the trans-Planckian problem cannot arise in our paradigm is due to the fact that the perturbations on top of our Glauber-Sudarshan state can be expressed as an Aggarwal-Tara state \textit{over the same Minkowski vacuum}. In other words, the dS perturbations with time-dependent frequencies can be rewritten as (infinite) linear combination of perturbations, with time-independent ones, as shown in \eqref{jlewis} and \eqref{sabinde}. This is the crucial reason why we are able to write down a well-defined Wilsonian effective action and never have to resort to the TCC for solving the puzzle of the trans-Planckian modes as they are necessarily decoupled, as they should be for the low-energy effective action. Additionally since the UV modes in the Minkowski space are exponentially suppressed by the Boltzmann factor 
${\rm exp}\left(-{k^2\over \Lambda^2_{\rm UV}}\right)$, and since we create quantum modes in de Sitter from fluctuations over the Minkowski space-time, we naturally expect that trans-Planckian modes to also be suppressed in the resulting de Sitter case as mentioned earlier.
All of this fits in nicely with the intuitive expectation that the trans-Planckian problem finds its resolution in a UV-complete theory of inflation, which is the case for our system. Our warped-Minkowski vacuum, on which the de Sitter coherent state is constructed, arises in string theory, thereby resolving these trans-Planckian problems without having to hide them behind the Hubble horizon as was proposed by the TCC. 

\subsection{The choice of vacuum and the quantum swampland \label{sec4.2}}
There is a different point of view proposed as to why de Sitter solutions might indeed be in a swampland, heuristically relating the instabilities of de Sitter spacetime, coming from field theoretic arguments, to the de Sitter conjecture \cite{Quantum_swampland}. The basic idea is that even if one is able to find an effective potential which gives rise to a de Sitter vacua within a stringy construction, at the classical level, then radiative loop corrections would necessarily destroy it leading one into a quantum version of the swampland. It is so because even for classical solutions which obey the equation of state $p =-\rho$, the one-loop effective potential essentially breaks this, yielding $w\neq 1$, unless one assumes the Bunch-Davies vacuum for the fields on top of the de Sitter spacetime. However, it has been argued that the Bunch-Davies is a rather unnatural choice for the vacuum \cite{dS_instability} and thus should be discarded. On the other hand, any other sensible choice of the vacuum necessarily leads to the \textit{leaking} of the cosmological constant, and one gets into the quantum swampland.

At first sight, one might wonder if the choice of the quantum vacuum should play any serious role in the search of de Sitter vacua in string theory. After all, in the absence of full-fledged string loop calculations, \textit{i.e.} quantum correction in spacetime, as can only arise in string field theory, the main focus has been to derive effective potentials which can support a de Sitter solution using stringy effects. In fact, one might even wonder if a classical (or non-perturbative) de Sitter vacua which is ruled out by the swampland can even be resurrected by employing these radiative loop corrections. However, as has been shown in  \cite{Quantum_swampland}, these quantum loop corrections cannot improve the stability of the solution (although they can affect the value of the cosmological constant). More interestingly, only examining quantum field theoretic calculations on de Sitter spacetimes, one can establish a relation between them and the swampland. 

The crucial realization behind the argument for the quantum swampland is the non-uniqueness of the vacua for de Sitter space. Typically, one chooses the Bunch-Davies vacuum by expanding a field in its momentum modes, picking one of these modes and blue-shifting it backwards until the effects of de Sitter spacetime can be ignored. At this point, one can safely pick the unique Minkowski vacuum. This procedure can be repeated for all the momentum modes to arrive at the Bunch-Davies, or the Euclidean, vacuum. However, as already mentioned in our discussion on the TCC, this procedure cannot work if there is a fundamental UV cutoff since one cannot trace a given mode beyond this energy scale. Moreover, the Bunch-Davies is not the only de Sitter-invariant vacuum; rather, there is a whole family of vacua which respects the symmetries of de Sitter spacetime called the $\alpha$-vacua \cite{alpha_vacua}. There are also complementary ways of showing the rather fine-tuned and contrived nature of the Bunch-Davies vacuum, especially when considering a causal patch of de Sitter instead of global de Sitter, as is appropriate for our solution \cite{dS_instability}. Having argued that the naive choice of the Bunch-Davies vacuum is a probably a wrong one, \cite{Quantum_swampland} shows that as soon as one  chooses a different vacuum (such as an instantaneous Minkowski vacuum \cite{Danielsson:2002kx}), the cosmological constant leaks and leads to a quantum decay of the de Sitter spacetime.

Let us now understand how our de Sitter solution can never run into these problems associated with radiative corrections. As shown explicitly, the Bunch-Davies has a very different interpretation in our construction as a (generalized) Agarwal-Tara state. Our de Sitter is itself a built as a coherent state on top of a solitonic Minkowski solution, and perturbations on top it is viewed as a GACS on top of a Minkowski state. Firstly, choosing different coefficients in the definition of our GACS would lead to a different vacuum state for the fluctuation modes. Just as it was shown how one can reproduce the Bunch-Davies state starting from our generalized Agarwal-Tara state \eqref{kimbas}, we can also reproduce other de Sitter-invariant vacua, such as the $\alpha$-vacua, for different choices of the $C_{n {\bf k}}^{(\psi)} \(t\)$ in \eqref{ckijwala}. However, remember that for us it is always the solitonic background, with the corresponding interacting vacuum $\left|\Omega\(t\)\right\rangle$, on which we build both our Glauber-Sudarshan state and the Agarwal-Tara state for fluctuations on top it. Therefore, loop corrections do not spoil our solution since these radiative effects are all calculated with respect to a Minkowski background and \textit{not} a de Sitter one. More to the point, our de Sitter solution is constructed once we build the Glauber-Sudarshan state having taken all types of quantum corrections -- perturbative and non-perturbative, local and nonlocal -- into account. Essentially, we do not build an effective vacuum spacetime with an equation of state $w=-1$ by using some stringy quantum effects, as is the case, say, for the famous KKLT solution. Rather, our de Sitter solution is created as a coherent state in the presence of all sorts of quantum corrections embodied by our interaction Hamiltonian, ${\bf H}_{\text{int}}$, the description which is valid for a specific amount of time. The radiative corrections having been calculated for our interacting vacuum on flat spacetime do not lead to the same pathologies as they do for de Sitter space. In fact, the main argument for the quantum swampland was based on the ambiguity of choosing the vacuum in de Sitter space. However, in our case, there is only one clear vacuum in our theory  -- the \textit{interacting vacuum} $\left|\Omega\right\rangle$ due to the action of ${\bf H}_{\text{int}}$ on our solitonic vacuum -- and we build both our Glauber-Sudarshan and Agarwal-Tara states on top of this. Consequently, the loop corrections do not affect the stability of our solution as long as $g_s \ll 1$ as evidenced from our effective action. Let us assert that we do not have a rolling `quintessence' type solution which, incidentally, might also suffer from these loop corrections but rather our de Sitter solution is free from the quantum swampland, by construction, since it is built as a Glauber-Sudarshan state on top of a warped Minkowski background.

\subsection{Finite entropy of the de-Sitter solution \label{sec4.3}}
In spite of the Gibbons-Hawking entropy being a natural extension of the Bekenstein-Hawking entropy associated with a black hole (locally, the horizon of de Sitter is identical to that of a Schwarzschild black hole), the finiteness of the entropy of de Sitter space has been a long-standing puzzle. Since black holes are local objects, occupying a finite region of space, it is natural to finite entropy to  a black hole which, in turn, implies that the a finite number of states can describe the black hole system. On the other hand, the spatially flat slices of de Sitter space has infinite volume. And yet, as demonstrated by Gibbons and Hawking through Euclidean partition functions \cite{Gibbons_Hawking}, an inertial observer in de Sitter space detects thermal radiation at the temperature:
\begin{eqnarray}
	T_{\rm dS} = \dfrac{1}{2 \pi \ell}\,,
\end{eqnarray}
where, $\ell^2 \sim 1/\Lambda^2$ is the length scale related to the cosmological constant. One can then use the first law of thermodynamics to deduce that the entropy corresponding to de Sitter horizon, from the Gibbons-Hawking temperature\footnote{The first law can be expressed as $\(\partial S/\partial M\) = T^{-1}$; however, a priori,  there is no definition of the mass corresponding to the de Sitter horizon. The way out of this is to realize that we need only a mass differential and this was solved by introducing a negative mass. We gloss over these subtleties as they are quite well-known and have been discussed exhaustively in the literature (see \cite{dS_review} for an overview).}, to be: 
\begin{eqnarray}
	S_{\rm dS} = \dfrac{A}{4 G} \sim \dfrac{\pi \ell^2}{G}\,, 
\end{eqnarray}
with the horizon area $A \sim \pi \ell^2$, with $G$ is the Newton's constant. The finiteness of the de Sitter entropy stands out as a crucial test for any quantum gravity theory. Before describing how our description of de Sitter as a Glauber-Sudarshan state in string theory manages to explain the finite entropy of the resulting de Sitter spacetime, let us quickly review some of the known features of $S_{\rm dS}$.

Firstly, it is clear from the above discussion that the finiteness of entropy must somehow be related to the fact that any single observer has access to only a finite volume of de Sitter space. Note that this reference to an observer is crucial for the discussion of de Sitter space unlike in the case of black holes. We emphasize that the entropy is only finite since the cosmological horizon ensures that a given observer only can ever send signals to a finite portion of the universe. From this simple observation, one can draw the minimal conclusion that any effective field theory breaks down when one has $\e^{S_{\rm dS}}$ states behind the horizon. Indeed, if one turns off gravity by taking $G \rightarrow 0$, while still maintaining the same curved space geometry, the entropy \textit{does} go to infinity and one has a perfectly valid EFT description \cite{ArkaniHamed:2007ky}. 

Next, we need to understand the serious consequences one has for the underlying quantum gravity theory if indeed $S_{\rm dS}$ is to be finite. It has been noted that eternal de Sitter space has several conflicts with having a finite entropy. In \cite{Trouble_dS}, a thermofield double picture was developed to explain the finite thermal entropy of a causal patch of de Sitter, using arguments from complementarity. However, the main conclusion of this work was to show that the symmetries of de Sitter spacetime were incompatible with the finiteness of $S_{\rm dS}$. More explicitly, it was found that the Hamiltonian (as a generator of the de Sitter symmetry group) can only have a countable spectrum, which is required to have discrete energy eigenvalues and a finite entropy, if the symmetries are violated on time-scales in which this discreteness become significant. The relevant time-scale in this case would be the Poincar\'e recurrence time, $t_p \sim \e^{S_{\rm dS}}$, but the important point for us is that this provides a concrete argument against the existence of eternal de Sitter spacetimes coming from the finiteness of $S_{\rm dS}$. This is quite a remarkable finding since, classically, one only requires that de Sitter is the solution of Einstein's equations with a positive cosmological constant and can be eternal in the future\footnote{The Friedmann equation gives a constant Hubble parameter $3{\rm M}_{\rm Pl}^2 {\rm H}^2 = \Lambda$ for an eternal de Sitter space, sourced by a constant positive cosmological constant.} and yet, this is ruled out once the finiteness of $S_{\rm dS}$, itself a semiclassical result, is taken into consideration.

A stronger requirement for quantum gravity, due to the finiteness of $S_{\rm dS}$, would be the restriction that the Hilbert space is finite dimensional \cite{Finite_Hilbert_dS} and given by $\mathcal{N}\sim \e^{S_{\rm dS}}$, where $\mathcal{N}$ is the number of states on the Hilbert space, once the covariant entropy bound is taken into account. In other words, ruling out entropies larger than $S_{\rm dS}$ imposes a fundamental cutoff on the Hilbert space of the quantum gravity theory. This is the so-called $\Lambda-N$ correspondence  \cite{Lambda_N_correspondence}, with the number of degrees of freedom, $N$, being given by $N=\log~{\mathcal{N}}$, $\mathcal{N}$ being the number of states on the Hilbert space. A small but finite $\Lambda$, necessary for a finite $S_{\rm dS}$, ensures that the fundamental theory would have a very large, and yet finite, number of degrees of freedom. However, it should be emphasized that this interesting, and strong restriction, relating the size of the Hilbert space with the cosmological constant  requires additional conditions (such as a future asymptotic de Sitter region \cite{Bousso_review_dS_Entropy})\footnote{Note that having a positive $\Lambda$ is not \textit{sufficient} to guarantee the above conclusion due to the failure of having a covariant entropy bound in some cases.}. A final point to emphasize, which has already been mentioned earlier and is extremely relevant for our discussion, is that there exists other arguments which show that the finiteness of entropy results in an upper limit on the lifetime of the de Sitter space \cite{ArkaniHamed:2007ky}.

From our review of topics above, it should be clear to the reader that we want to first focus on the fact that our description of de Sitter, as a Glauber-Sudharshan state, can only be trusted for a finite amount of time given by $T < 1/{\rm H}$ \eqref{spamca}. Therefore, our solution automatically satisfies, at least, the necessary criterion of \cite{Trouble_dS} for having a finite entropy, as this time-scale is much smaller than the Poincar\'e recurrence time $t_p$. As an aside, let us also mention that our solution is completely free from problems such as that of Boltzmann brains (see \cite{BB} for details). This is so because the time-scale, characteristic of Boltzmann brains, is given by: 
\begin{eqnarray}
	T_{\rm BB} = \frac{1}{\rm H}\, \e^{S_{\rm E}}\,,
\end{eqnarray}
where $S_E$ is the instantonic action corresponding to a Boltzmann brain. However, given the time-limit after which our system becomes strongly-coupled, this would mean $S_E \leq 1$ can only be created by quantum fluctuations which, in turn, rules out any feasible Boltzmann brain mechanism.

Although we have argued that since our de Sitter solution has an upper time-limit, it is likely to have a finite $S_{\rm dS}$, we are yet to specify the microscopic mechanism through which such an entropy is generated. Note that it is a major challenge of any fundamental theory to calculate statistical entropy of de Sitter space (See, for instance, \cite{Maldacena:1998ih} for pioneering work in this direction for lower dimensional de Sitter space). In our case, we wish to interpret the entropy of the resulting de Sitter spacetime as an \textit{entanglement entropy} due to the interaction between the modes of the metric (and G-flux) fluctuations which give rise to the Glauber-Sudarshan state on top of the solitonic vacuum. Although we shall not give the detailed calculation, which we defer to future work, we can nevertheless point out why we expect such an entanglement entropy to exist and, moreover, why it should be finite. 

Let us begin with the latter point first. The number of gravitons\footnote{Of course, we should also mention G-flux particles but, in order to keep the discussion less complicated, we shall only focus on the part of the metric fluctuations of the Glauber-Sudarshan state ($\left|\alpha\right\rangle$ in our notation) instead of focussing on the full coherent state $\left|\alpha, \beta\right\rangle$. Note that this is done only for convenience and the discussion easily generalizes to the full case even if explicit computations become more tortuous in that case.} in a given coherent state is, of course, infinite once we allow for modes with all possible momenta. This is explicitly shown in \eqref{triplets} for our Glauber-Sudarshan state. However, as also mentioned earlier, one must have a short-distance cut-off in order to have a well-defined Wilsonian effective action. Indeed, one can also impose a physically-relevant infrared cut-off in the case of the coherent state giving rise to de Sitter. However, let us come to this point later. The first question is what does a finite number of gravitons in the Glauber-Sudarshan state have to do with a finite entropy?

Before we answer this question, note that we are not the first to show a relation between a coherent state description of the causal patch of de Sitter and it having a finite entropy. In \cite{dvaliG}, a coherent state constituting of soft gravitons was proposed as the ``quantum-corpuscular'' description of de Sitter space. Although not formulated from any fundamental theory (such as string theory), this still leads to a time-limit \eqref{Quantum_Break_time} after which such a semiclassical description of de Sitter stops being valid, as mentioned earlier. The number of gravitons in such a state can be found by using the number operator: $N = \langle N| \hat{N} |N\rangle$, where the coherent state $\left|N\right\rangle = \Pi_{k} |N(k)\rangle$ should be calculated over all frequencies: 
\begin{eqnarray}
	|N(k)\rangle = {\rm exp}\left(-{N(k)\over 2}\right) 
	\sum_{n_k = 0}^{\infty} \dfrac{[N(k)]^{n_k/2}}{\sqrt{n_k!}} \;|n_k\rangle\,.
\end{eqnarray} 
This would also, naturally, be an infinite number but the authors of \cite{dvaliG} make the crucial assumption that constituent gravitons of the coherent state satisfy the condition that the dominant wavelength is the one set by the Hubble radius $H^{-1}$! This leads to a gravitation occupation number given by, $N \sim {\rm M}_{\rm Pl}^4/\Lambda$, which coincides numerically with the de Sitter entropy $S_{\rm dS}$. First, notice that the finite occupation number of this coherent state picture comes from the \textit{requirement} that de Sitter is a good semiclassical, mean-field description only as long as the occupation number is related to the cosmological constant as given above. On the other hand, our Glauber-Sudarshan state, describing de Sitter in full string theory, has a finite occupation number since we require to have a short-distance cutoff. This is an essential difference and we do not assume that only gravitons of a specific frequency contribute to the coherent state. 

Given a coherent state with a finite number of gravitons, $N$, one can calculate how many states can correspond to having such a description. Of course, if these gravitons were purely non-interacting, then the number of states would be given by $N^\gamma$, $\gamma$ being the number of states (or polarization) of individual gravitons. However, it was heuristically argued in \cite{dvaliG}, that if the gravitons do interact, then the number of states would go as $\xi^N$, where $\xi$ denotes the number of states of individual distinguishable ``flavors'' of gravitons, the latter being an effect of interaction between gravitons. This would give a leading order entropy given as $S \sim N$, with logarithmic corrections. We refer the interested reader to \cite{dvaliG} for details of this estimate. 

\begin{figure}[h]
\centering
\begin{tabular}{c}
\includegraphics[width=\textwidth]{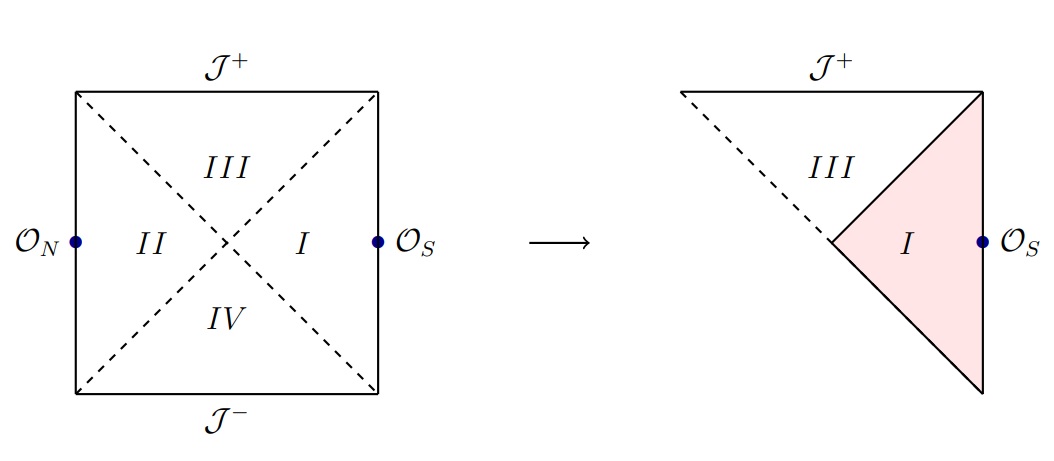}
\end{tabular}
\caption{The figure on the left shows the Penrose diagram for global four-dimensional de Sitter space. The poles $\mathcal{O_N}$ and $\mathcal{O_S}$	are time-like lines. The dashed lines denote the past and future horizons for an inertial observer at $\mathcal{O_S}$. Time runs from the past $\mathbf{\mathcal J^-}$ to the future $\mathbf{\mathcal J^+}$ conformal boundaries. \newline
			On the right, we have the Penrose diagram for the flat slicing of de Sitter, as is applicable for our solution. The causal patch of an observer, sitting at $\mathcal{O_S}$, is denoted by the shaded region (static patch). We trace over the modes in Region III to get the entropy corresponding to region I. The Hubble horizon separates regions III from I.}
\label{Fig:1}
\end{figure}

Although we agree with the general argument of having interacting gravitons, let us show how it leads to a finite de Sitter entropy in a more rigorous way. Recall that the total number of gravitons in our Glauber-Sudarshan state is given by \eqref{triplets}:
\begin{eqnarray}
	N^{(\psi)} = \int^{\Lambda_{\rm UV}} \d^{10}\,{\bf k}\; \left|\alpha^{(\psi)}_{\bf k}(0)\right|^2\,,
\end{eqnarray}
where we have explicitly introduced a short-distance UV cut-off $\Lambda_{\rm UV}$ to replace the $\infty$ appearing in \eqref{triplets}. Firstly, as mentioned earlier, this number is finite in our construction by virtue of the fact that the effective action must come out of integrating the high-energy UV modes and \textit{not} by requiring that the coherent state is packed with gravitons of a specific wavelength. However, in addition to the UV cut-off, an interesting question now arises whether there is any IR cut-off for us. Recall that our fluctuations, underlying the state $|\alpha\rangle$ which give rise to de Sitter space, are over a warped-Minkowski vacuum and we do not have any inherent preference for the IR cut-off. Nevertheless, we are interested in calculating the entropy corresponding to a causal patch of de Sitter whereas our solution represents the full de Sitter spacetime in the so-called flat slicing (see {\bf Figure \ref{Fig:1}}). Therefore, it is natural to identify the IR cut-off with the Hubble scale, \textit{i.e.} $\Lambda_{\rm IR} = \H$. However, there is no way to ``integrate'' out the IR degrees of freedom to get an effective action for modes in-between $\Lambda_{\rm UV} < k < \Lambda_{\rm IR}$. It is well-known that in this case, the standard treatment requires the description of the modes of interest in terms of the density matrix obtained after \textit{tracing out} the IR modes. As mentioned, we shall sketch the outline of this calculation below, following the notation of \cite{EE}, while deferring the details to later work.

For our quantum system, we break up the effective Hamiltonian, corresponding to the Wilsonian effective action, into the following form: 
\begin{eqnarray}
	{\bf H} = {\bf H}_{\rm sys} \otimes  \mathbbm{1} +\mathbbm{1}\otimes  {\bf H}_{\rm IR} + {\bf H}_{\rm int}\,,
\end{eqnarray}
where we have denoted the ${\bf H}_{\rm sys}$ as the Hamiltonian corresponding to the perturbations modes $\Lambda_{\rm UV} < k < \Lambda_{\rm IR}$ while  ${\bf H}_{\rm sys}$ refers to modes with $k < \Lambda_{\rm IR}$. The interaction Hamiltonian is the same as the one introduced in \eqref{novocaine}. Effectively, one breaks up the Hilbert space of all the modes which make up the de Sitter coherent state into $\mathcal{H} = \mathcal{H}_{\rm sys} \otimes \mathcal{H}_{\rm IR}$. The crucial point for us is that these modes, although complicated due to the underlying solitonic vacuum, still has a time-dependence of the form $e^{i\omega_{\bf k}t}$. Of course, this was important for us to begin with in the construction of the Glauber-Sudarshan state but here, its significance lies in the fact that we shall be able to use time-independent perturbation theory because of this. We shall see this shortly. 

Let us first denote the free vacuum state of the theory, before considering the ${\bf H}_{\rm int}$ as: 
\begin{eqnarray}
	\left|0,0\right\rangle =  \left|0\right\rangle_{\rm sys} \otimes  \left|0\right\rangle_{\rm IR}\,,
\end{eqnarray}
\textit{i.e.} as a (tensor) product of the individual vacuum states. However, once we turn on interactions, the only vacuum available to us is the interacting vacuum given in \eqref{novocaine}, as we have emphasized many times. Starting with this interacting vacuum, which can be written as:
\begin{eqnarray}\label{Int_Vac_Entropy}
	\left|\Omega\right\rangle = \left|0,0\right\rangle + \sum_{n\neq 0} A_n \left|n,0\right\rangle + \sum_{N\neq 0} B_N \left|0,N\right\rangle +\sum_{n,N\neq 0} C_{n,N} \left|n,N\right\rangle\,,
\end{eqnarray}
we want to trace out the IR modes. In the above, we have denoted energy eigenstates, corresponding to system and IR modes, by $\left|n\right\rangle$ and $\left|N\right\rangle$, respectively. This is where we shall use perturbation-theory to calculate the co-efficients $A, B, C$. In fact, the density matrix corresponding to the system modes, can be written in terms of the matric elements of $C$ alone (up to leading order) \cite{EE}:
\begin{eqnarray}\label{red_density}
	\rho_{\rm sys} = {\rm tr}_{\rm IR}\; \left|\Omega\right\rangle \left\langle\Omega\right| = 
	\begin{pmatrix}
		1 - |C|^2 & 0 \\
		0 & CC^\dagger\,.
	\end{pmatrix}	
\end{eqnarray}
$C$ can be formally expressed as
\begin{eqnarray}
	C_{n,N} = \dfrac{\left|\left\langle n,N\right| {\bf H}_{\rm int} \left|0,0\right\rangle \right|^2}{E_0 +\tilde{E}_0 -E_n -\tilde{E}_N} +\, \cdots
\end{eqnarray}
where the $\cdots$ above refer to higher order corrections in $\mathcal{O}(g_s)$. Given this reduced density matrix $\rho_{\rm sys}$, we can calculate the quantum von Neumann entropy corresponding to it, given by
\begin{eqnarray}
	S_{\rm ent} = -{\rm tr}\, \(\rho_{\rm sys} \log \rho_{\rm sys}\)\,,
\end{eqnarray} 
in terms of the matrix elements given above. The above result holds true for arbitrary dimension. As mentioned, we do not wish to do this explicit calculation here which would not only involve the ${\bf H}_{\rm int}$, including all the quantum terms described in \eqref{cixmel}, but also require including the full Glauber-Sudarshan state corresponding to both metric and G-flux fluctuations. Nevertheless, our main result can be understood as follows.

Firstly, we note that we give a precise microscopic origin to the de Sitter entropy in our formalism by relating it to the quantum entanglement entropy of the mode functions. We stress that this is not the usual entanglement entropy one sometimes calculate for fields on de Sitter space  \cite{dS_Entanglement} but rather that of the modes which give rise to the de Sitter spacetime itself. The main reason why we are able to do such an identification is simply because our de Sitter space comes into existence on considering fluctuations of the metric (and G-flux components) over a solitonic vacuum. The entanglement entropy is the entropy corresponding to the coupling between these modes themselves, which build up the Glauber-Sudarshan state itself. In other words, the interactions between the gravitons and flux-particles, which constitute our de Sitter coherent state $\left|\alpha, \beta\right \rangle$, is responsible for the origin of this entanglement entropy and it is thus natural to associate it with $S_{\rm dS}$. However, note that entanglement is purely a quantum property and thus we give a statistical explanation of $S_{\rm dS}$ in our formalism for a de Sitter spacetime in string theory.

Secondly, the inquisitive reader might question our sketch of the standard derivation above, given that we have used $\left|\Omega\right\rangle$ to compute the entanglement entropy, and not the Glauber-Sudarshan state corresponding to it $\mathbb{D}\(\alpha(t)\) \left|\Omega\right\rangle$, in \eqref{Int_Vac_Entropy}, before tracing out the IR modes. This is a very pertinent point; however, the simple calculation involving the vacuum suffices in this case as it has been shown that the entanglement entropy corresponding to \textit{any} coherent state is exactly the same as that for the vacuum state \cite{EE_coherent}, a conclusion that lends itself to any dimensional spacetime. 

Next, we arrive at the question of the finiteness of the entanglement entropy. This is the key property which would allow us to identify it with $S_{\rm dS}$. Note that in the absence of any interactions, \textit{i.e.} setting ${\bf H}_{\rm int}=0$ would result in the entanglement entropy going to infinity, as is expected for a free field theory. On the other hand, given a finite interaction term, and the fact that we have a UV cut-off $\Lambda_{\rm UV}$, ensures us that this quantity remains finite. In our case, the full interacting action of M-theory was meticulously spelt out in the previous section and, in fact, it was emphasized at the very outset of our construction of the Glauber-Sudarshan state that interactions were absolutely crucial for such a description to emerge. There is simply no free underlying theory available to us when we consider G-fluxes and obtain our solitonic background. If we turn off interactions, no construction of a de Sitter spacetime is possible and thus the entanglement entropy would simply have no meaning of being associated with $S_{\rm dS}$. What we stress is that we need not invoke any physical intuition, such as that of the coherent state being built out of gravitons of any specific wavelength, in order to obtain a finite entropy of de Sitter space. Furthermore, the intuition that interactions between the graviton constituents lead to the entropy corresponding to $S_{\rm dS}$ was rigorously shown to be associated with the entanglement entropy arising due to the graviton mode-couplings between those of the causal patch and those which are traced out in the far-infrared. 

Finally, we need to justify our choice of tracing over the IR modes in order to get our entanglement entropy. This brings us back to something which we mentioned at the very beginning of this section: The finiteness of entropy is related to the fact that an observer in the static patch only has access to a finite part of de Sitter space. In other words, for any inertial observer, the entropy corresponds to the region of de Sitter hidden behind the cosmological horizon. Therefore, it makes perfect sense for us to trace out over modes, which correspond to the region that is \textit{not} in causal contact with the observer. What determines this region, and therefore the IR cut-off, is the cosmological constant which, in our case, is emergent as a balance between fluxes and quantum corrections \cite{desitter2}. We emphasize that the tracing out carried out over here is \textit{not} the usual one used for thermofield-double systems -- starting with some (Hartle-Hawking type) Euclidean state which is entangled between left and right static patches and then tracing over the hidden region to obtain a thermal density matrix \cite{Trouble_dS}. Rather, our state is always the Glauber-Sudarshan state created over the interacting vacuum, for which we trace over the modes which have wavelengths larger than our IR cut-off. The physical reason behind choosing the cutoff is that, in the resulting de Sitter space, an observer in the static patch can only causally interact with the region that is not hidden behind the horizon.

Of course, we did not do the explicit calculation for the entanglement entropy taking the full interacting Lagrangian of M-theory into consideration, as well as carrying out the actual trace over the IR degrees of freedom. However, one should be convinced by now that such a procedure is, in principle, possible due to the time-dependence of our mode functions ${\bf \Psi}_{\bf k}({\bf x},y, z,t)$ even if their algebraic forms are complicated due to the solitonic background. Furthermore, the result of such a calculation would give us a finite answer. What would then be left is to equate this result to the de Sitter entropy $S_{\rm dS}$ (as one-quarter the horizon area) to find what it predicts for the allowed wavelength distribution of the gravitons comprising our Glauber-Sudarshan state. There is one final hiccup in our argument which shows up in the form of a well-known problem against interpreting $S_{\rm dS}$ as an entanglement entropy, since the latter $S_{\rm ent}$ would not only depend on the cutoff $\Lambda_{\rm UV}$ but also on the number of species. This latter dependence is what is known as the species puzzle \cite{Species_Puzzle}. However, if one introduces the species bound \cite{Species_bound}, then the cut-off which shall appear in the calculation of $S_{\rm ent}$ shall exactly be this effective cutoff, as demanded by the species bound. As goes the usual argument, if one has a length scale cutoff $\ell_{\rm eff} > \sqrt{N} \ell_{\rm Pl}$ in the resulting theory, it is natural that this would appear as the cutoff for the wavelength of the gravitons populating our Glauber-Sudarshan state. However, a pertinent question would be if there are a lot of light species present in our formalism such that this effective cutoff is reduced by an unacceptably huge amount? And this is precisely where our dynamical moduli stabilization comes to the rescue. As explained earlier, this means that the moduli are fixed at every instant of time in such a way that, at every instant of time, the Dine-Seiberg runaway is stopped. This ensures that we do not have exponentially light states appearing in our setup at any time and thus manage to avoid the species puzzle.

Finally, note that the entropy corresponding to the reduced density matrix in \eqref{red_density} would be the leading order result, with higher order corrections in $g_s$ to follow. Thus, after setting this leading order result to $S_{\rm dS}$, we shall also be able to systematically calculate the higher order corrections to the semiclassical result. This is as it should be for any truly microscopic understanding of $S_{\rm dS}$ and is only possible since we have a UV complete theory -- string theory -- describing our background.

\section{Discussions and conclusions}

In this work we investigated the realization of de Sitter space from string theory. There were many attempts to perform such constructions with various degrees of success, that included ingredients like fluxes, non-perturbative effects from instantons, orientifolds  and anti-branes. Our approach differs from all the previous attempts as we consider the appearance of the four-dimensional de Sitter metric from a
Glauber-Sudarshan (coherent) state in string theory. The foundation of our construction is four dimensional Minkowski vacuum obtained from string theory or M-theory compactifications as in \eqref{betbab3}. The latter realization, {\it i.e.} the uplift to M-theory, is only for convenience as it aids in making some of the computations easier to perform. The vacuum, either in IIB or in M-theory, is supersymmetric and stable and is well understood even in the presence of quantum corrections. These corrections, while necessary to realize the supersymmetric background itself, convert the vacuum from a free to an interacting one. This {\it interacting} vacuum forms the basis of  our subsequent constructions in the paper. For example,  
instead of switching to another vacuum which has the potential to be a non-supersymmetric de Sitter, we realize our de Sitter space by using a {\it displacement} operator on the interacting vacuum itself. One of our main result of the paper is the precise identification of the displacement operator. 

Displacing the interacting vacuum appropriately creates the necessary coherent or the Glauber-Sudarshan state from which one could extract the precise four-dimensional de Sitter metric $-$ along-with the metric information of the internal six-manifold $-$ by taking the expectation values of the various components of the metric operator over the Glauber-Sudarshan state. Additionally, it provides the information of the fluxes that are required to support the background metric configuration, either in the IIB or in the M-theory side. While the fluxes are supersymmetric over the Minkowski vacuum, they  break supersymmetry when we take the expectation values. A precise demonstration of these facts forms the basis of section \ref{sec2} of the paper.

 Another pleasant surprise of our approach is that not only the de Sitter space arises form the Glauber-Sudarshan state but also do the {\it fluctuations} over the de Sitter space. As an extension of electromagnetism inspired concepts, the fluctuations over a de Sitter space appear as a generalized Agarwal-Tara state where the photon-added coherent state is promoted to a generalized graviton (and flux)-added coherent state. This state is also identified for various components of the metric, details of which are shown in 
 sections \ref{sec2.3}, \ref{sec2.4} and \ref{sec2.5}.
 
 Many of the computations, demonstrating the aforementioned details, are performed in two ways throughout the paper: one, using state and operators and their expectation values; and two, using the path integral approach. We show that the latter approach is much more powerful and ties up with many well-known concepts in string theory, like the action of the vertex operators, flux induced non-K\"ahlerity etc. Additionally, some of the limitations of the state-operator formalism, for example being restricted to mostly on-shell computations or the appearance of un-necessary divergences,  are effectively eliminated in the path integral approach.   

The path integral approach also allowed us to choose between the three possible outcomes of our analysis, namely: (a)  retainment of the exact classical behavior, as expectation values, over an indefinite period of time during the temporal evolution of the system, (b) persistence of the exact classical behavior
for a certain interval of time beyond which the dominance of the full quantum behavior becomes prominent, or (c) appearance of the  perturbative corrections to the expectation values at least in some well defined temporal domain debarring the system to exhibit the full classical behavior anywhere in the domain.

This temporal domain, which is sometimes referred to as the {\it quantum break time}, is an important limitation of the system. In our case it is governed by the interval beyond which strong coupling effect sets in (at least from M-theory point of view which we use to analyze the dynamics) and we lose quantitative control of the dynamics, thus effectively eliminating option (a) above.  It is therefore option (b) that eventually appears consistent from our path integral approach, as any perturbative corrections appearing from option (c) would have implied time-dependent Newton's constant, non-exactness of the four-dimensional de Sitter solution, 
and other possible pathologies. A precise demonstration of these appear in sections \ref{sec2.4} and
\ref{sec2.5}. An important result of these sections is that 
the Agarwal-Tara state accurately reproduces the fluctuations over the 
Glauber-Sudarshan de Sitter space. The same state also allows an interpretation of the trans-Planckian issue, as resulting from the time dependent frequencies that we observe in mode expansions of the fluctuations over a de Sitter vacuum, to be simply an artifact of the Fourier transforms over the de Sitter space viewed as a Glauber-Sudarshan state. This shift in the viewpoint of de Sitter from being a {\it vacuum} to a {\it state} is crucial in resolving the trans-Planckian problem and thereby allowing Wilsonian effective action to be defined at {\it all} energy scales.

Stability of the Glauber-Sudarshan is also an important criterion on which all of our construction relies on. 
The analysis of the stability requires two different sets of computations, one, analyzing all possible perturbative and non-perturbative corrections affecting the system, and two, analyzing the equations that govern the dynamical evolutions of the expectation values of the metric and the flux components over the Glauber-Sudarshan state. The perturbative corrections have been discussed in details in \cite{desitter2, desitter3} and in section \ref{sec3.1} and \ref{sec3.2} we elaborate on all possible non-perturbative corrections, including the ones from the instantons, and the world-volume fermions on the seven-branes. 

The equations governing the dynamical evolutions of the expectation values are the Schwinger-Dyson's equations (SDEs). Interestingly, the SDEs reproduce all the M-theory EOMs in the presence of the aforementioned perturbative and the non-perturbative corrections.  These equations may be classified order by order in type IIA string coupling $g_s$ (which becomes a time-dependent quantity) with the lowest order equations determining the de Sitter background and the fluxes supporting it. The stability then works in the following way. The higher order flux components and the higher order metric components, {balance against} the higher order quantum terms to keep the lowest order SDEs unchanged. This balancing act happens to {\it all} orders in $g_s$ and ${\rm M}_p$ such that the de Sitter background 
along-with the supporting flux components remain uncorrected to arbitrary orders in  
${g_s^a\over {\rm M}_p^b}$. Such a stability criterion also guarantees option (b) mentioned above, namely the dominance of the exact classical behavior in the temporal domain whose boundary is dictated by the onset of type IIA strong coupling. The moduli are stabilized already at the vacuum level, and therefore the  dynamical evolution of the metric components also govern the  dynamical evolution of the moduli themselves disallowing, in turn, the Dine-Seiberg runaway at every stage of the evolution as long as we restrict the dynamics within the allocated temporal domain.

Having explained how our solution is able to go past the technical difficulties which have been pointed out for realizing four-dimensional de Sitter space in string theory, we went on to explore some of the properties of the constructed Glauber-Sudarshan state in section \ref{sec4}. The first thing we did was to show how this construction is able to bypass the so-called swampland conjectures and, in particular, the trans-Planckian censorship conjecture. This turns out to be yet another implication of being able to interpret the fluctuations on top of de Sitter as a state built out of the underlying Minkowski vacuum, namely the Agarwal-Tara state. More interestingly, this gives us a simple way to argue against the age-old instabilities of de Sitter spacetime against radiative corrections due to the choice of the vacuum associated with the mode functions for de Sitter space. During the temporal regime for which our solution is under quantitative control, these instabilities never show up as the artifacts of the time-dependent frequencies get explained as mentioned above. Finally, we find the remarkable result that the interpretation of de Sitter as a Glauber-Sudarshan state also helps in the microscopic understanding of the entropy associated with the de Sitter horizon. This is so because the modes which are responsible for the creation of the coherent state are themselves part of a highly interacting theory and are, therefore, necessarily entangled between themselves. Any entanglement between quantum modes must result in a nontrivial von Neumann entropy, which we reinterpret as the entropy associated with the resulting de Sitter state. The crucial role of the interactions can be easily understood on taking the limit in which they go to zero: Such a limit not only results in the entanglement entropy going to infinity but also ensures there is no longer a Glauber-Sudrashan state any longer, thus reducing the resulting space-time to the warped-Minkowski one. A detailed calculation for this entanglement entropy and its relation to the famous semiclassical result of the one-quarter horizon area is left for future work and we only show how it is going to be finite in our case. Of course, the calculation of the entanglement entropy needs to be done using perturbation theory, to evaluate the matrix elements, and is thus, in principle, able to extend beyond the semiclassical result, as it should be for the calculation on entropy of de Sitter from a microscopic theory. 

In retrospect our identification of de Sitter space to the Glauber-Sudarshan state in full string theory should not come as a big surprise, as a familiar lore in string theory identifies {\it every} curved background as some condensates of gravitons by exponentiating the vertex operator, much like the way discussed in 
footnote \ref{vertex}. The special case here is, because of the temporal dependence of the metric components, the Glauber-Sudarshan state could be defined on-shell, at least to a large extent. For generic curved background, with or without time-dependences, this may not always be possible and the Glauber-Suarshan state should be defined {\it off-shell}\footnote{Even the simpler case of a Higgs vacuum cannot be defined as an on-shell Glauber-Sudarshan state. Additionally, as we saw in the decomposition \eqref{cnelson}, there are small off-shell pieces from the time-independent parts of \eqref{betbab3}. Once the system becomes completely off-shell a decomposition like \eqref{cnelson} could still be used, but now it's only the second part of \eqref{cnelson} that would be relevant. See also \cite{dvalisol}.}. This off-shell formalism of the coherent states is in concordance with the expectation from string field theory where similar constructions show up, although the analysis gets technically challenging.  Thus instead of going far off-shell, it will be interesting to ask if such on-shell description can still be given for the inflationary models in string theory \cite{inflationB} as they are close to the de Sitter space that we discussed here. We believe this is possible, and more details will be presented in near future.

\section*{Acknowledgements:} We would like to thank Robert Brandenberger, Maxim Emelin, 
Shahin Sheikh-Jabbari and Sav Sethi for useful discussions. The work of KD is supported in part by the  Natural Science and Engineering Research Council of Canada (NSERC).
SB is supported in part by funds from NSERC, from the Canada Research Chair program, by a McGill Space Institute fellowship and by a generous gift from John Greig.


\end{document}